\documentclass[twocolumn]{aastex631}
\usepackage{amsmath}
\usepackage{savesym}
\savesymbol{tablenum}
\usepackage{siunitx}
\usepackage[utf8]{inputenc}
\restoresymbol{SIX}{tablenum}

\usepackage{graphicx}
\usepackage{placeins}
\usepackage{float}
\usepackage{txfonts}
\usepackage{hyperref}
\usepackage[load-configurations=abbreviations]{siunitx}
\usepackage{CJK}
\newcommand{\lp}{\left(}
\newcommand{\rp}{\right)}
\usepackage{lipsum}

\usepackage{enumitem}
\newcommand{\lb}{\left[}
\newcommand{\rb}{\right]}

\clearpage

\newcommand{\flux}{$\rm erg~cm^{-2}~s^{-1}$}

\begin{document}
\begin{CJK*}{UTF8}{gbsn}
\title{EP250827b/SN 2025wkm: An X-ray Flash-Supernova Powered by a Central Engine and Circumstellar Interaction}

\correspondingauthor{Gokul P. Srinivasaragavan}\email{gsriniv2@umd.edu}
\author[0000-0002-6428-2700]{Gokul P. Srinivasaragavan}
\affiliation{Department of Astronomy, University of Maryland, College Park, MD 20742, USA}
\affiliation{Joint Space-Science Institute, University of Maryland, College Park, MD 20742, USA}
 \affiliation{Astrophysics Science Division, NASA Goddard Space Flight Center, 8800 Greenbelt Rd, Greenbelt, MD 20771, USA}
 \affiliation{Division of Physics, Mathematics and Astronomy, California Institute of Technology, Pasadena, CA 91125, USA}

\author[0000-0002-4562-7179]{Dongyue Li}
\affiliation{National Astronomical Observatories, Chinese Academy of Sciences, Datun Road A20, Beijing 100012, China}

\author[0000-0002-9364-5419]{Xander J. Hall}
\affiliation{McWilliams Center for Cosmology and Astrophysics, Department of Physics, Carnegie Mellon University, 5000 Forbes Avenue, Pittsburgh, PA 15213}

\author[0000-0003-3115-2456]{Ore Gottlieb}
\affil{Department of Physics and Kavli Institute for Astrophysics and Space Research, Massachusetts Institute of Technology, Cambridge, MA 02139, USA}
\affiliation{Center for Computational Astrophysics, Flatiron Institute, 162 5th Avenue, New York, NY 10010, USA}
\affil{Department of Physics and Columbia Astrophysics Laboratory, Columbia University, Pupin Hall, New York, NY 10027, USA}
\author[0000-0001-9915-8147]{Genevieve~Schroeder}
\affiliation{Department of Astronomy, Cornell University, Ithaca, NY 14853, USA}

\author[0000-0002-2412-5751]{Heyang Liu}
\affiliation{National Astronomical Observatories, Chinese Academy of Sciences, Datun Road A20, Beijing 100012, China}

\author[0000-0002-9700-0036]{Brendan O'Connor}
   \affiliation{McWilliams Center for Cosmology and Astrophysics, Department of Physics, Carnegie Mellon University, 5000 Forbes Avenue, Pittsburgh, PA 15213}

\author{Chichuan Jin}
\affiliation{National Astronomical Observatories, Chinese Academy of Sciences, Datun Road A20, Beijing 100012, China}
\affiliation{School of Astronomy and Space Science, University of Chinese Academy of Sciences, Beijing 100049, China}
\affiliation{Institute for Frontier in Astronomy and Astrophysics, Beijing Normal University, Beijing 102206, China}

\author[0000-0002-5619-4938]{Mansi Kasliwal}
\affiliation{Division of Physics, Mathematics and Astronomy, California Institute of Technology, Pasadena, CA 91125, USA}

\author[0000-0002-2184-6430]{Tom\'as Ahumada}
\affiliation{Cerro Tololo Inter-American Observatory/NSF NOIRLab, Casilla 603, La Serena, Chile}

% ...



\begin{abstract}
We present the discovery of EP250827b/SN 2025wkm, an X-ray Flash (XRF) discovered by the Einstein Probe (EP), accompanied by a broad-line Type Ic supernova (SN Ic-BL) at $z = 0.1194$. EP250827b possesses a prompt X-ray luminosity of $\sim 10^{45} \, \rm{erg \, s^{-1}}$, lasts over 1000 seconds, and has a peak energy $E_{\rm{p}} < 1.5$ keV at 90\% confidence. SN 2025wkm possesses a double-peaked optical light curve (LC), though its bolometric luminosity plateaus after its initial peak for $\sim 20$ days, consistent with a central engine injecting additional energy into the explosion. Its spectrum transitions from a blue to red continuum with clear blueshifted broad absorption features consistent with a SN Ic-BL classification. We do not detect any transient radio emission and rule out the existence of an on-axis, energetic jet $\gtrsim 10^{50}~$erg assuming a typical LGRB circumburst constant density ($n \approx 10^{-3}$--$10^{-1}~{\rm cm}^{-3}$) and microphysical parameters ($\epsilon_{\rm e} = 0.1$ and $\epsilon_{\rm B} = 0.01$). In the model we invoke, the collapse gives rise to a long-lived magnetar, potentially surrounded by an accretion disk. Magnetically--driven winds from the magnetar and the disk mix together and break out with a velocity $\sim 0.35c$ and interact with an extended circumstellar medium with radius $\sim 10^{13}$ cm, generating X-ray breakout emission through non-thermal free-free processes. The disk outflows and magnetar winds power blackbody photospheric emission as they cool adiabatically and thermalize, producing the first SN peak. The spin-down luminosity of the magnetar and radioactive decay of $^{56}$Ni powers the late-time emission. We end by discussing the landscape of XRF-SNe within the context of EP's recent discoveries.
\end{abstract}

\section{Introduction}
\label{sec:intro}
Unlike gamma-ray bursts (GRBs), whose populations are relatively well characterized \citep{Woosley1993, Piran2004, Berger2014}, the origins of extragalactic fast X-ray transients (EFXTs) are still a mystery. EFXTs are short flashes of soft X-ray emission ranging from minutes to hours, discovered in the energy range 0.3 -- 10 keV. The majority of historical satellites (e.g., \textit{Swift}, \textit{Fermi}) were primarily sensitive to the gamma-ray sky, resulting in the under-exploration of the EFXT discovery space \citep{Polzin2023}. The few historical satellites sensitive to soft X-rays (HETE-2, $\sim 0.5 -- 10$ keV;  Beppo-Sax, $\sim$ 2 -- 25 keV; MAXI, $\sim$ 0.5 -- 30 keV) found a small number of EFXTs \citep{Sakamoto2005, Heise01, Negoro2016}, but studies determined their intrinsic rates are likely comparable to those of classical GRBs \citep{Sakamoto2005, Heise01}. Additional EFXTs were discovered through searches of historical archives \citep{Vazquez2022, Vasquez2025}, but the lack of real-time follow-up made their characterizations difficult.

Many EFXTs have accompanying high-energy gamma-ray emission; however, some have intrinsic peak energies (the energy in units of eV where the $\nu F_\nu$ spectrum reaches its maximum flux and turns over) $\lesssim 30 $ keV, which is lower than those of classical GRBs (hundreds of keV) and higher than those of supernovae (SNe; tens of eV). These EFXTs with low peak energies are known as X-ray Flashes (XRFs). Past soft X-ray missions such as BeppoSAX and HETE-2 detected XRFs at rates at least comparable to if not greater than those of classical GRBs \citep{Amati2002, Sakamoto2005}, suggesting a possible different progenitor origin. Some possibilities include high-$z$ \citep{Heise01} or off-axis \citep{Rhoads1997, Meszaros1998} GRBs, baryon-loaded GRBs with low Lorentz factors known as ``dirty fireballs'' \citep{Dermer1999}, supernova (SN) shock breakout or cooling \citep{Colgate1974,Balberg2011, Nakar2012, WaxmanSBO}, tidal disruption events (TDEs; \citealt{Vazquez2022}), off-axis GRBs \citep{Sarin2021}, magnetars after binary neutron star mergers \citep{Zhang2013, Sun2017, Xue2019, Lin2022, Vazquez2024}, or new, never-before-seen classes of transient phenomena.

We note here that since the physical origins for XRFs are still an open question, the cutoff of $\sim$ 30 keV is based primarily on historical observational studies, rather than a more physically-driven definition regarding their progenitor systems. \citet{Amati2002} defines XRFs as events that triggered the 2 -- 25 keV detector on Beppo-Sax, and not the higher energy gamma-ray detector. \citet{Sakamoto2005} defines XRFs through the fluence ratios between $S_X$(2 -- 30 keV)/$S_\gamma$(30 -- 400 keV), and finds that there is a correlation between the peak energies of bursts and their fluence ratios. The border they find between XRFs and ``X-ray Rich" GRBs is at 30 keV, and this is the cutoff we use for this work.

The Einstein Probe (EP; \citealt{Yuan2022,Yuan2025}), or Tianguan mission is changing the landscape of EFXT and XRF science, with its wide-field, soft X-ray capabilities. EP's all sky monitor (ASM) Wide-field X-ray Telescope (WXT) has an instantaneous field of view of 3850 deg$^2$ and operates from 0.5 to 4 keV. With over 10 times the field of view and more than one order of magnitude greater sensitivity than MAXI \citep{MAXI}, the only other currently operational X-ray ASM, EP has revolutionized the study of the soft X-ray time-domain sky. EP also possesses two conventional X-ray focusing telescopes (Follow-up X-ray Telescope, FXT) operating from 0.3 -- 10 keV that can provide arcsecond localizations.

In its first 1.5 years of operations, EP has already discovered numerous EFXTs, as well as a smaller sample of XRF candidates. EP240801a was a confirmed XRF with $E_{\rm{peak}} \sim 15 $ keV and was interpreted as either an off-axis or intrinsically weak jet \citep{Jiang2025}. EP241021a had no accompanying $\gamma$-ray emission \citep{Shu2025}, along with a luminous optical and radio counterpart \citep{Busmann2025, Wu2025, Gianfagna2025, Shu2025, Yadav2025, Quirola2025}. There are many proposed interpretations for EP241021a, including afterglow emission from wobbling jets \citep{Gottlieb2022,Gottlieb2025}, refreshed GRB shocks accompanying a low-luminosity GRB \citep{Busmann2025}, a compact star merger producing a compact object or binary compact object system \citep{Wu2025, Sun2025}, an intermediate mass black hole tidal disruption event \citep{Shu2025}, a collapsar origin with extra energy input from a central engine \citep{Quirola2025}, or an off-axis jet \citep{Gianfagna2025}. 

An even smaller subset of XRFs have detected associated SNe (XRF-SNe). These events are very important probes of the massive stellar deaths accompanying XRFs, giving a stronger indication of their progenitor system characteristics. Prior to EP, there were five such events -- XRF020903 \citep{Bersier2006, Soderberg2005}, XRF030723 \citep{Tominaga2004, Fynbo2004}, XRF060218/SN 2006aj (e.g., \citealt{Modjaz2006, Pian2006, 2006Natur.442.1014S, Ferrero2006, Mirabal2006, Kenji2007, Irwin2016, Sollerman2006, Campana2006, Mazzali2006}), XRF080109/SN 2008D (e.g., \citealt{Mazzali2008, Chevalier2008, Maund2009, Malesani2009, Modjaz2009, Soderberg2008}), and XRF100316D/SN 2010bh (e.g., \citealt{Cano2011, Bufano2012, margutti2013, Starling2011, Chornock2010}). 

Although GRB-SNe have relatively uniform properties \citep{cano2017, Srinivasaragavan2024}, studies of these XRF-SNe indicated they were diverse, giving more evidence that they may originate from different progenitors than classical GRBs. In addition, though all historical spectroscopically confirmed GRB-SNe and XRF-SNe were broad-lined Type Ic SNe (SNe Ic-BL; \citealt{Galama1998,Hjorth2003, cano2017}), the one exception was XRF080109/SN 2008D, which showed strong He features leading to a Type Ib SN classification \citep{Chevalier2008, Mazzali2008, Malesani2009, Modjaz2009}. This diversity within the XRF-SN population may be due to a variety of factors - differing surrounding CSM environments implying heterogenous mass-loss histories (e.g., \citealt{Soderberg2008, Srinivasaragavan2025b}, the existence of binary or tertiary systems \citep{Rastinejad2025}, along with varying central engine powering mechanisms (e.g., \citealt{Sun2025, Quirola2025}) have all been theorized to be the source of this diversity. However, no statistically robust conclusions have been made to date due to the small sample size of events. 

EP has already found three speectroscopically confirmed XRF-SNe, dramatically increasing the rate of XRF-SN discoveries. EP250304a possessed no associated $\gamma$-ray emission \citep{GCN39600}, and showed spectroscopic evidence that its optical counterpart was a SN Ic-BL \citep{GCN39851}. Two other EP XRF-SNe have been studied in detail in the literature. EP240414a/SN 2024gsa \citep{Srivastav2025, VanDalen2025, Sun2025, Zheng2025, Hamidani2025}  has $E_{\rm{peak}}$ $< 1.3$ keV \citep{Sun2025}, making it an XRF. Its optical counterpart was a SN Ic-BL that possessed a mysterious early red peak. The origin of this peak had several theories in the literature, some of which included interaction of a GRB jet with a dense circumstellar medium \citep{VanDalen2025}, shock cooling emission following interaction with a dense circumstellar medium \citep{Sun2025}, an afterglow from a mildly relativistic cocoon \citep{Hamidani2025} or off-axis jet \citep{Zheng2025}, afterglow emission from wobbling jets \citep{Gottlieb2022,Gottlieb2025}, or refreshed shocks from a GRB \citep{Srivastav2025}. In addition,  \citet{VanDalen2025} suggested that EP240414a had resemblances to luminous fast blue optical transients  (LFBOTS; \citealt{Drout2014, Pursiainen2018, Prentice2018, Margutti2019, Perley2019, Ho2023a}) due to its rise time, though its red colors indicate otherwise. 

EP250108a/SN 2025kg was also an XRF, with $E_{\rm{peak}} < 1.8$ keV \citep{Li2025}, and its associated SN 2025kg was a SN Ic-BL \citep{Srinivasaragavan2025b, Rastinejad2025} that possessed a blue peak prior to the SN peaking. Similary to EP240414a/SN 2024gsa, there were varying interpretations for the progenitor system of this event, including cooling emission of black hole disk outflows \citep{Gottlieb2025}, a black hole-driven jet that is stifled by its progenitor star's envelope, leading to a shocked cocoon \citep{Eyles-Ferris2025}, a similar shocked cocoon interacting with an extended circumstellar medium (CSM; \citealt{Srinivasaragavan2025b}), and a magnetar-powered explosion \citep{Li2025, Aguilar2025, Zhu2025}. 

Recently, \citet{Quirola2025} argues that the late-time emission of EP241021a may be consistent with SNe Ic-BL, though no SN features were found in the spectrum at late-times, and the emission peaks at an order of magnitude higher than what is expected for SNe Ic-BL. We direct the readers to \citet{Quirola2025} for a further discussion of EP241021a as a possible XRF-SN, but do not consider it as an XRF-SN for the rest of this work.

In this Letter, we present the discovery EP250827b/SN 2025wkm at $z = 0.1194$, the fourth XRF-SN discovered by EP, and the third studied in detail in the literature. We note that EP250827b/SN 2025wkm was discovered through a different method than the previous three XRF-SNe, made possible by the Zwicky Transient Facility's (ZTF; \citealt{Bellm2019, Graham2019, Dekany2020,Masci2019}) shadowing of EP's public observing schedule, which resulted in an optical counterpart association to a subthreshold EFXT (see \S \ref{ZTFdiscovery} for more details). We discovered both the EFXT EP250827b as well as its optical counterpart SN 2025wkm, but note that \citet{GCN41670} reported the first classification as a SN Ic-BL.

The Letter is structured as follows: in \S \ref{Observations} we present the observations of EP250827b/SN2025 wkm; in \S \ref{Analysis} we present analysis of the X-ray prompt emission, associated SN, and radio observations; in \S \ref{xrayanalysis} we present analytical arguments to describe the X-ray prompt emission, in \S \ref{LCmodel} we present light curve (LC) modeling used to characterize the SN LC; in \S \ref{Discussion} we discuss the implications our results; and in \S \ref{Conclusion} we summarize our findings. Throughout this paper we utilize a flat $\Lambda$CDM cosmology with $\Omega_{\rm m}=0.315$ and $H_{0} = 67.4$~km~s$^{-1}$~Mpc$^{-1}$ \citep{Planck18} to convert the redshift to a luminosity distance and correct for the Milky Way extinction of $E(B-V)_{\rm{MW}} = 0.05$ mag \citep{Schlafly2011}, using the \citet{ccm1989} extinction law with $R_v = 3.1$.

\section{Observations} 

In this section, we present the discovery and observations of EP250827b/SN 2025wkm. 

\label{Observations}

\subsection{X-ray Prompt Emission}
\label{xrayobs}
\begin{figure*}
\begin{center}
\begin{tabular}{cc}
    \includegraphics[trim=0.1in -0.3in 0.1in 0.9in, clip=1, scale=0.25]{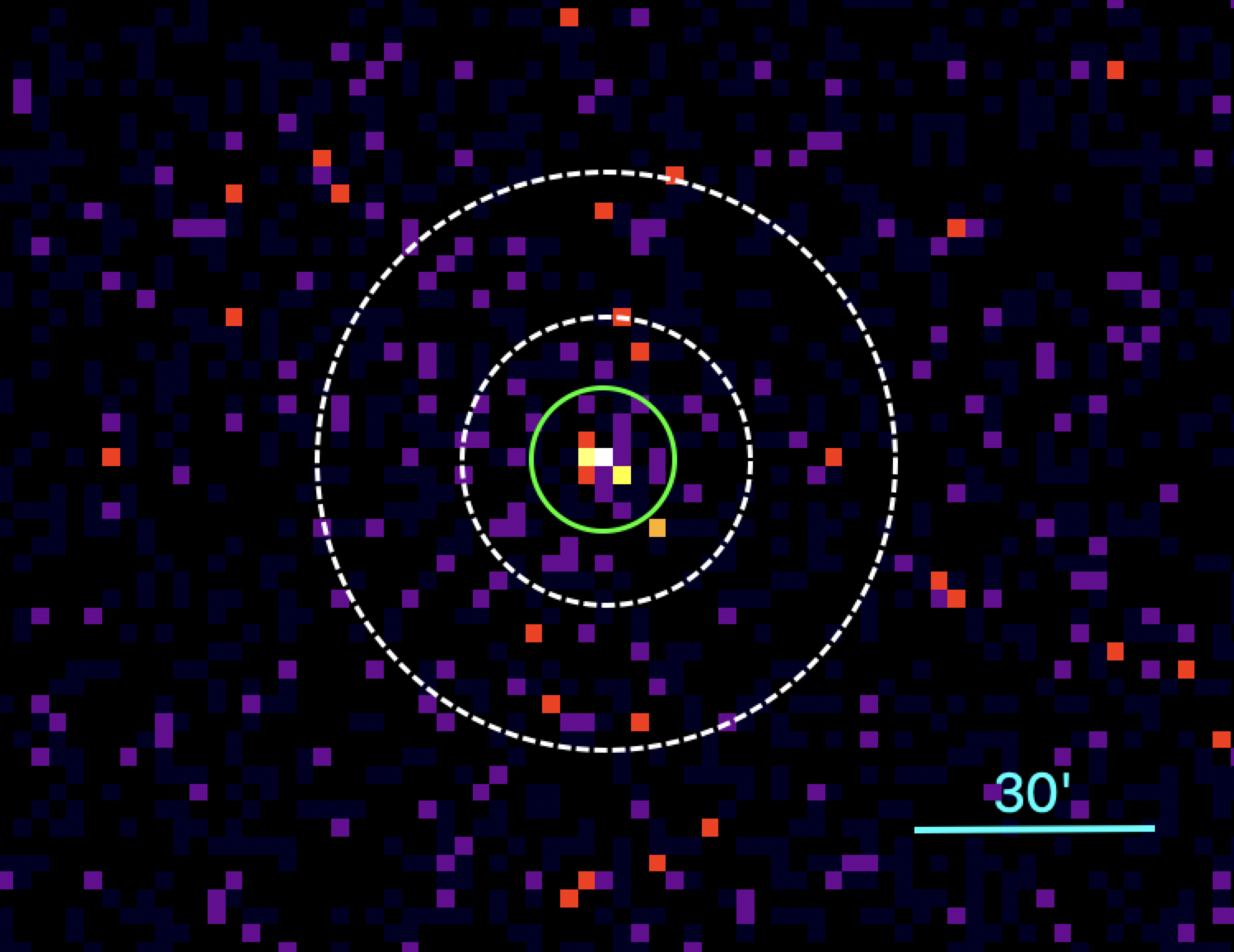}&
    \includegraphics[trim=0.1in 0.6in 0.1in 0.6in, clip=1, scale=0.22]{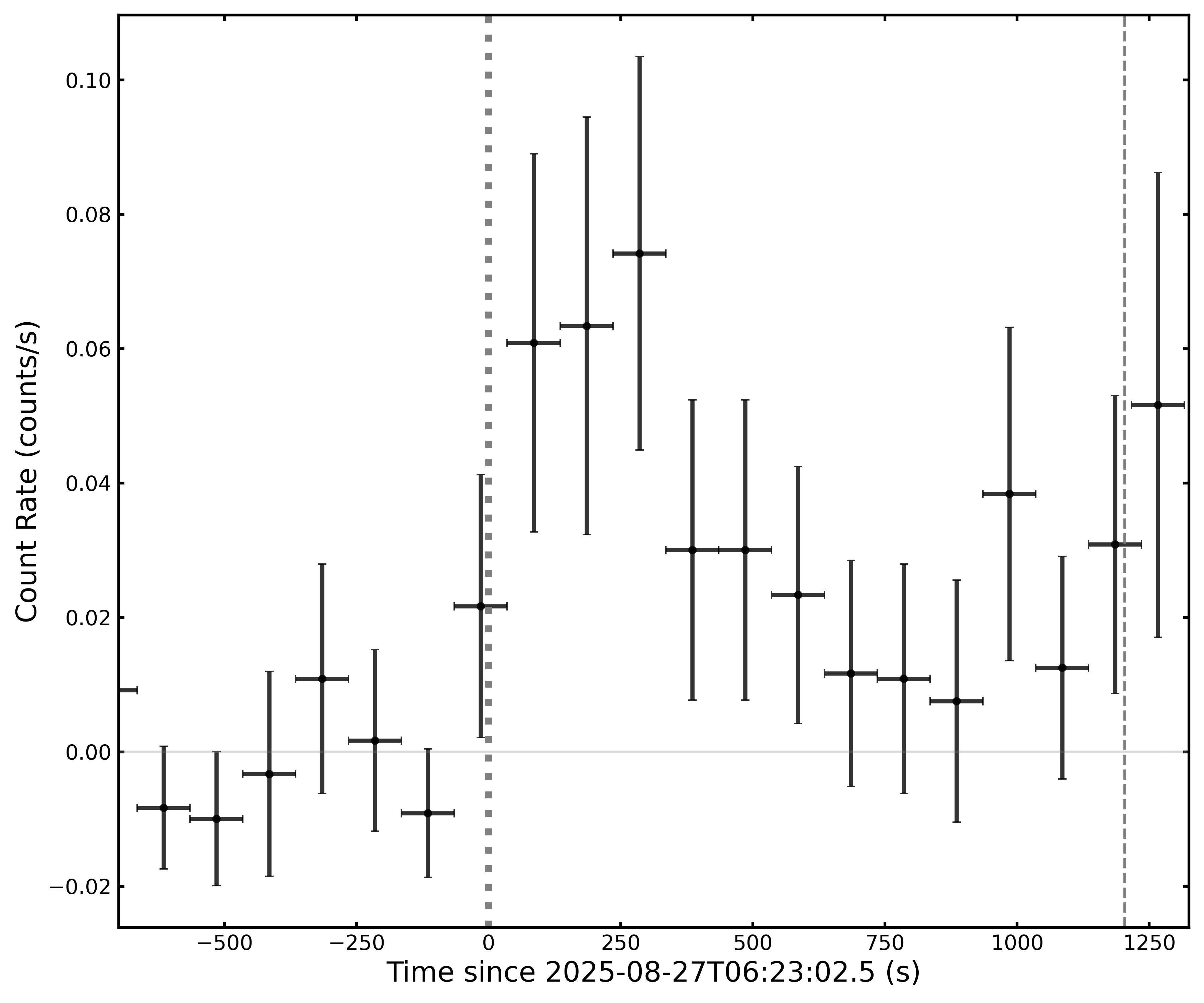}
\end{tabular}
\end{center}
    \caption{{\it Left}: The EP-WXT 0.5-4 keV image of EP250827b. The green circle represents the 9 arcmin aperture used for extracting the spectrum and the light curve of EP250827b. The dashed annulus with inner and outer radius of 18 arcmin and 36 arcmin indicates the background region. {\it Right:} The EP-WXT 0.5-4 keV background-subrtracted light curve of EP250827b. The vertical dotted and dashed line indicate the start time, and the $T_{90}$ of the flare respectively, derived with the Bayesian block method. The horizontal gray line shows the zero-count-rate level.}
    \label{fig:wxt_detection}
\end{figure*}

EP250827b/SN~2025kg was detected by the WXT at RA = 36.581$^\circ$, Dec = 37.499$^\circ$ (J2000; \citealt{GCN.41635}), with a positional uncertainty of 2.4 arcminutes in radius at 90\% confidence level, incorporating both statistical and systematic errors, and reported by the offline archive search pipeline; see the WXT 0.5-4 keV image in left panel of Figure~\ref{fig:wxt_detection}. The detection has a significance of 6.8$\sigma$, calculated using the Li-Ma formula \citep{Li1983}. The source was not reported to the public NASA General Coordinates Network (GCN), as was the case for every other XRF-SN discovery. This is because the X-ray transient was not an onboard trigger, and was flagged as a subthreshold event in the EP data stream. Its confirmation as a real EFXT was only done after ZTF discovered its optical counterpart (see \S \ref{ZTFdiscovery}).

The WXT data were processed using the WXT Analysis Software (WXTDAS; Liu et al., in prep) with the latest calibration database (WXTCALDB V2.35). The CALDB was initially constructed from ground calibration experiments \citep{Cheng2025} and subsequently refined through a series of in-flight calibration campaigns\footnote{In particular, the in-orbit effective area if found to be broadly consistent with the ground calibration result, showing systematic uncertainty arounf 10\% (90\% C.L). The energy response of the CMOS detector (i.e. the gain coefficient and energy resolution) is also in line with ground measurement, with slight variation of (1-2)\%. No significant degradation in these parameters has been detected after nearly two years of on-orbit operations.} conducted after launch (Cheng et al., in prep). 
Photon positions were reprojected onto the celestial coordinate system, and the pulse-invariant value—representing energy in channel units—was computed for each event based on the bias and gain parameters stored in the CALDB. After flagging bad and flaring pixels and assigning event grades, we selected events with grades 0--12 and no anomalous flags to generate a cleaned event list and a corresponding image in the 0.5--4\,keV band for subsequent source detection. The light curve and spectrum of the source and background within a specified time interval were extracted using a circular source region with a radius of 9~arcmin and an annular background region with inner and outer radii of 18~arcmin and 36~arcmin, respectively. 

The temporal properties of the source, including the start time ($t_0$), total duration ($T_{100}$), and the $T_{90}$ duration, were derived following a standard analysis procedure adapted for EP fast X-ray transients (Wu et al. in prep.). The start time of the transient was objectively determined by applying the Bayesian block algorithm to the net light curve, defined as the beginning of the first time bin with a signal-to-noise ratio (SNR) greater than 3. The redefined start time of the event is 2025-08-27T06:23:02.5 (UTC), which we refer to hereafter as $t_0$. The $T_{90}$  duration was calculated as the time interval containing 90\% of the total net counts (from 5\% to 95\% of the cumulative counts). The uncertainty on $T_{90}$  was estimated using a Monte Carlo approach, wherein the net counts in each light-curve bin were independently resampled based on their Poisson statistics to generate multiple realizations of the net light curve. For each realization, $T_{90}$  was recomputed from the accumulated counts, and the final uncertainty was derived from the distribution of the resulting $T_{90}$  values. The measured $T_{90}$ for EP250827b is $T_{90} = 1203.6^{+48.0}_{-55.5}$\,s. Furthermore, inspection of the light curve, as shown in the right panel of Figure~\ref{fig:wxt_detection}, indicates that the flare was still in progress at the end of the observation. Consequently, the derived duration is likely a lower limit, underestimating the true event duration.

\subsection{X-ray Follow-up Observations}

\subsubsection{\textit{Swift}-XRT}

The \textit{Neil Gehrels Swift Observatory} X-ray Telescope \citep{Burrows2005} observed SN 2025wkm on 2025-08-31 (PI: M. Coughlin), 2025-09-05, and 2025-09-09  (PI: X. Hall; Table~\ref{tab:ep_log}), but did not detect any source \citep{hall_EP250827b_2025}. We used the Living Swift XRT Point Source Catalogue \citep[LSXPS;][]{LSXPS} upper limit server\footnote{\url{https://www.swift.ac.uk/LSXPS/ulserv.php}} to derive $3\sigma$ upper limits, shown in Table \ref{tab:ep_log}, in the energy range of 0.3 -- 10 keV, assuming a photon index $\Gamma = 2$.

\subsubsection{Einstein Probe FXT}

Follow-up observations were conducted with the EP-FXT instrument over a period spanning from 4.8 to 26.8 days after the trigger. The FXT, one of the primary payloads aboard EP. It comprises two co-aligned modules (FXT-A and FXT-B), each equipped with 54 nested Wolter-I paraboloid--hyperboloid mirror shells. The source and background regions in the FXT data were processed using dedicated FXT data analysis software (FXTDAS V1.20\footnote{\url{https://epfxt.ihep.ac.cn/analysis}}). A summary of the EP observations is provided in Table~\ref{tab:ep_log}. No X-ray sources were detected in any of the four EP-FXT observations. We estimated the 90\% confidence upper flux limits in the energy range of 0.5--10~keV assuming the same spectral model as used in the WXT analysis, shown in table \ref{tab:ep_log}.

\begin{deluxetable*}{lllc}[htb!]
\label{tab:ep_log}
\tablecaption{Log of X-ray observations. Fluxes are unabsorbed, assuming an absorbed power-law (\texttt{tbabs * ztbabs * powerlaw}, \citealt{Willingale2013}). }
\tablewidth{0pt} 
\tablehead{\colhead{ObsID} & \colhead{Start Time}  & \colhead{Exposure} & \colhead{Flux (~\flux})\\
\colhead{} & \colhead{(UTC)}  & \colhead{(s)}
}
\tabletypesize{\normalsize} 
\startdata 
EP-WXT  & & &  (0.5 -- 4 keV)\\
\hline
11916650962 & 2025-08-27 06:10:21  & 2063 & $3.4^{+6.0}_{-1.2}\times10^{-11}$  \\
\hline
EP-FXT & & & (0.5 -- 10 keV)\\
\hline
06800000865 & 2025-09-01 02:35:49 & 2995 & $ < 1.1 \times 10^{-13}$ \\
06800000881 & 2025-09-06 02:29:39 & 5997 & $< 1.1 \times 10^{-14}$\\
06800000898 & 2025-09-11 18:21:49 & 6027 & $< 3.9 \times 10^{-14}$\\
06800000912 & 2025-09-23 02:09:59 & 4141 & $< 2.6 \times 10^{-14}$\\
\hline
Swift & & & (0.3 -- 10 keV) \\
\hline
03000057001 & 2025-08-31 21:09:31 & 4030 & $< 1.1 \times 10^{-13}$ \\
03000057002 & 2025-09-05 02:34:00 & 1160 & $< 3.3 \times 10^{-13}$  \\
03000057003 & 2025-09-09 18:20:00 & 1525 & $< 2.3 \times 10^{-13}$\\
\enddata 
\end{deluxetable*}

\subsection{Gamma-Ray Constraints}
There was no Fermi-GBM \citep{Meegan+2009} onboard trigger around $t_0$, and during the duration of the event. While the location of the transient was visible above Earth to Fermi/GBM for this full duration, Fermi entered the South Atlantic Anomaly about 100 s after $t_0$ and thus has no data after this time. The GBM targeted search \citep{Goldstein2019}, developed to search for GRB-like signals between 64 ms and 32.768 s in duration, was run in the time interval [$t_0$-50; $t_0$+100] s, finding no signal consistent with the EP transients, neither temporally nor spatially. Using the ``soft'' spectral template (Band function with $E_{\rm{p}}$ = 70 keV, $\alpha$ = -1.9, $\beta$ = -3.7), we derived a flux upper limit of $3.52\times 10^{-9}$ erg cm$^{-2}$ s$^{-1}$, in the energy band 10-1000 keV.

\subsection{Radio Observations}
\label{VLA}
We observed the field of EP\,250827b/SN 2025wkm with the NSF's Karl G. Jansky Very Large Array (VLA) on 2025-09-13 at a mid time of 13:03:23 UT ($\sim 17.3~$days post burst) at a mid frequency of 10~GHz (4~GHz bandwidth) under program 25B-363 (PI Perley). We used J0251+4315 for complex phase calibration and 3C48 for flux and gain calibration. We reduced and imaged the data using the Common Astronomy Software Applications \citep[\texttt{CASA};][]{2007ASPC..376..127M} VLA Calibration Pipeline\footnote{\url{https://science.nrao.edu/facilities/vla/data-processing/pipeline}} and VLA Imaging Pipeline\footnote{\url{https://science.nrao.edu/facilities/vla/data-processing/pipeline/vipl}}.
At the location of SN 2025wkm, we detect a $\sim 6 \sigma$ source. We measure the flux density of the source using the \texttt{imtool} tool from \texttt{pwkit} \citep[][]{2017ascl.soft04001W}, and find $F_\nu = 28 \pm 5~\mu$Jy (beam size of $0.7\arcsec \times 0.6 \arcsec$). 

We initiated an additional epoch at 10~GHz on 2025 September 25 at a mid-time of 04:24:50 UT  ($\sim 28.9~$days post burst), and measure the flux density of the source to be $F_\nu = 29 \pm 5~\mu$Jy (beam size of $1.1\arcsec \times 0.6 \arcsec$). Given the lack of evolution of the radio source, we do not associate this source with SN\,2025wkm, and rather assume the radio emission arises from the host galaxy (see \S \ref{sec:RadioAnalysis} and Figure \ref{fig:RadioLuminosity} for estimates on how radio emission from the transient would result in temporal evolution over the two epochs). If the radio emission can be attributed solely to star formation, this would indicate a radio star formation rate of ${\rm SFR}_{\rm radio} \approx 3~M_\odot \, {\rm yr^{-1}}$, similar to the optically derived ${\rm SFR}$ of $2.83 \pm 0.04 \, M_\odot \, {\rm yr^{-1}}$ (\S~\ref{spectra}). Given that the beam sizes of the images are similar to the apparent size of the host galaxy, we cannot disentangle whether any excess radio emission is present at the location of EP250827b/SN 2025wkm. Thus, we use our host galaxy detections as upper limits on the radio emission associated with EP250827b/SN 2025wkm for subsequent analysis (\S~\ref{sec:RadioAnalysis}).

\subsection{Photometric Observations}
Here we present the photometric observations of EP250827b/SN 2025wkm. A full log of photometric and spectroscopic observations are presented in the Appendix, in Tables \ref{appendix:phot_log} and \ref{spectratable}. Photometric uncertainties reported include both statistical errors and systematic errors from template subtraction

\subsubsection{ZTF Detection of Optical Counterpart}
\label{ZTFdiscovery}
Starting in March 2025, ZTF began using its wide field of view (47 deg$^2$) to shadow the regions observed by the EP's WXT \citep{Ahumada2025GCN39791}. The EP observing schedule is publicly available, and after requesting the corresponding fields, the ZTF scheduler adds to its plan those areas that have two or more EP observations.

The goal of this project is to co-discover sources detected by EP by crossmatching them with ZTF alerts, which includes sources reported publicly to GCNs, along with subthreshold sources only available to the ZTF+EP collaboration team. Using Kowalski \citep{kowalski} and Fritz, the SkyPortal \citep{VanderWalt2019, Coughlin2023skyportal} instance of ZTF, the alerts are crossmatched within the EP detection error circles and sent to human scanners, who evaluate the temporal and spatial differences to assess whether the crossmatches are plausible optical counterparts. 

On August 27th, 2025 (MJD 60914.4), during routine scanning of the ZTF-EP crossmatches, ZTF25abmpngy/SN 2025wkm was discovered. The spatial and temporal offsets from EP250827b were 0.27 arcmin and +0.23 days, respectively. It was originally detected in the $r$ band at r = 20.33 $\pm$ 0.1 mag \citep{GCN.41635}, and subsequent detections showed a persistent optical source. The last upper limit from ZTF at the position of the transient was obtained on MJD 60913.42, 0.98 days before the first detection, with g $>$ 21.37 mag. The EP team reported that the associated EFXT was a subthreshold detection, and they confirmed that it was a real transient after the detection of the possible optical counterpart. This is why there was a 5 day latency between the reported GCN \citep{GCN.41635} and the detection of the EFXT. The relationship between latency of EFXT detections and the discovery of optical counterparts and spectroscopic redshift measurements is explored in \citet{Oconnor2025b}.

We compute the probability of chance coincidence with the following equation
\begin{equation}
    P_1 \approx R_\text{SN} V_\text{eff} \Delta t,
\end{equation}
where $P_1$ is the probability of one SN given the SN rate $(R_\text{SN})$, the effective volume $(V_\text{eff})$, and the time between the EP event and the ZTF observation ($\Delta t$). Given the positional error of EP-WXT ($2.4^{\prime}$), the time delay ($0.23\text{d}$) the intrinsic brightness of SN 2025wkm (${-}19.5$ AB mags), the average ZTF depth \citep[20.4 AB mags;][]{Masci2019}, and the volumetric rate of Ic-BL SN \citep[$3.5\times10^{-7} \text{yr}^{-1}\text{Mpc}^{-3}h^3_{70}$;][]{pessi_supernova_2025}, we find the chance of coincidence to be $\log_{10}(P_1) \sim -7.1$.

\subsubsection{The Spectral Energy Distribution Machine (SEDM)}
We performed observations with the Spectral Energy Distribution Machine \citep[SEDM;][]{sedm, Rigualt2019}, mounted on the 60-inch telescope at Palomar Obsevatory. We took images in the $g$, $r$, and $i$-band throughout the duration of our campaign. Standard reduction techniques were applied to the data \citep{Kim2022}. We used the Pan-STARRS1 catalog \citep[PS1;][]{Chambers2016,2020ApJS..251....7F} for photometric calibration. We performed image subtraction with FPipe \citep{FrSo2016} which uses templates from PS1 imaging  \citep{Chambers2016,2020ApJS..251....7F}.

\subsubsection{Fraunhofer Telescope at Wendelstein Observatory (FTW)}
We performed observations with the Three Channel Imager (3KK; \citealt{lang2016wendelstein}) instrument mounted on the FTW \citep{2014SPIE.9145E..2DH} in the $g'$, $r'$, $i'$, $z'$, and $J$ bands. The optical CCD and NIR CMOS data were reduced using a custom pipeline developed at Wendelstein Observatory \citep{2002A&A...381.1095G, Busmann2025}. For the astrometric calibration of the images, we used the Gaia EDR3 catalog \citep{Gaia2021, 2021A&A...649A...2L, gaiaEDR3}. We used the Pan-STARRS1 catalog \citep[PS1;][]{Chambers2016,2020ApJS..251....7F} for the optical photometric calibration and the 2MASS catalog \citep{Skrutskie2006} for the $J$ band. Tools from the AstrOmatic software suite \citep{SourceExtractor, 2006ASPC..351..112B, 2002ASPC..281..228B} were used for the coaddition of each epoch's individual exposures. We used the Saccadic Fast Fourier Transform (SFFT; \citealt{hu_image_2022}) algorithm for image subtraction. For subtraction templates we use PS1 imaging for the $g$, $r$, $i$, and $z$ bands \citep{Chambers2016,2020ApJS..251....7F}. For the $J$ band, we use a synthetic host subtraction using \texttt{GALSYNTHSPEC}\footnote{\url{https://github.com/robertdstein/galsynthspec}} with host modeling performed using Prospector \citep{prospector}.

\subsubsection{TRT}
We performed observations with the 70cm telescope of Thai-Robotic Telescope located at Sierra Remote Observatories, California, United States (TRT-SRO). Images observed were processed through standard procedures and combined using the Image Reduction and Analysis Facility \citep[\texttt{IRAF;}][]{1986SPIE..627..733T}. The result flux was calibrated with nearby Pan-STARRS1 field stars \citep[][]{chambers2019panstarrs1surveys}, with the correction from ref \footnote{\url{https://classic.sdss.org/dr4/algorithms/sdssUBVRITransform.php}}. The logs of photometric observations and results are listed in Table \ref{appendix:phot_log} in the Appendix.

\subsubsection{ALT}
We performed observations with the 100cm A, B and C telescopes of the JinShan project, located at Altay, Xinjiang, China (ALT-100A, ALT-100B and ALT-100C). The data were processed using standard procedures with the Image Reduction and Analysis Facility \citep[\texttt{IRAF;}][]{1986SPIE..627..733T}. Aperture photometries were then conducted on the stacked images subtracting Pan-STARRS1 field images \citep[][]{chambers2019panstarrs1surveys}, with zero points measured using the nearby Pan-STARRS1 catalogs. 

\subsubsection{NOT}
We performed observations using the Alhambra Faint Object Spectrograph and Camera (ALFOSC\footnote{\href{http://www.not.iac.es/instruments/alfosc}{{http://www.not.iac.es/instruments/alfosc}}}) mounted on the 2.56~m Nordic Optical Telescope (NOT) located at the Roque de los Muchachos Observatory on La Palma (Spain). The data were processed using standard procedures with the Image Reduction and Analysis Facility \citep[\texttt{IRAF;}][]{1986SPIE..627..733T}. Aperture photometries were then conducted on the stacked images subtracting Pan-STARRS1 field images \citep[][]{chambers2019panstarrs1surveys}, with zero points measured using the nearby Pan-STARRS1 catalogs.


\subsubsection{LCO}
Through the Global Supernova Project \citep{2017AAS...23031803H}, we obtained the \textit{BVgri}-band images with the network of 1.0\,m telescopes of the Las Cumbres Observatory (LCO) \citep{Brown2013}. \textit{BV} and \textit{gri} point-spread function (PSF) photometry was calibrated to Vega and AB magnitudes, respectively. PSF photometry was performed using AutoPhot \citep{Brennan2022d}. 

\subsubsection{LT}
We performed observations with the Liverpool Telescope (LT; \citealt{Steele2004}) IO:O camera in SDSS \textit{ugriz} filters. Raw data was processed by the automatic LT IO:O reduction pipeline. For images in \textit{griz} filters, image subtraction and PSF photometry was conducted using a custom pipeline, using Pan-STARRS1 as reference and for image calibration. \textit{u}-band images were calibrated using the SDSS Extended Northern+Equatorial u'g'r'i'z' Standards \footnote{\url{https://www-star.fnal.gov}} taken on each night of the observations. The \textit{u}-band magnitudes were obtained via aperture photometry, and corrected to AB magnitude system \footnote{\url{https://www.sdss4.org/dr17/algorithms/fluxcal}}; host subtraction was not performed.

\subsubsection{TNOT}
We performed observations with the Tsinghua–Nanshan Optical Telescope (TNOT), an equatorial-mount telescope located at the Nanshan Station of the Xinjiang Astronomical Observatory (XAO), Chinese Academy of Sciences (CAS). The system comprises an 80\,cm ASA800 Ritchey--Chr\'etien optical tube mounted on an ASA DDM500 premium equatorial mount. The field of view is $26.5'\times26.5'$. Multi-band follow-up observations of EP250827b/SN~2025wkm were obtained with TNOT and reduced with IRAF (bias subtraction and flat-fielding). Instrumental magnitudes were measured with AutoPhot \citep{Brennan2022d} and calibrated against the Gaia Synthetic Photometry catalog \citep{GaiaCollaboration2023A&A...674A..33G}.

\subsubsection{OSN}
Additional observations were gathered at the 1.5m OSN telescope at Observatorio de Sierra Nevada (Granada, Spain), using BVRI filters. 
Magnitudes were derived using DAOPHOT and calibrated Pan-STARSS  secondary standards (in the Sloan system) and using the transformation  equations from \citet{Kostov2018}.

\subsubsection{Swift-UVOT}
We observed with the Ultra-Violet Optical Telescope \citep{swift_uvot}, on board the \textit{Neil Gehrels Swift Observatory}. Observations were conducted in three epochs, across all UV filters (U, UVW1, UVM2, UVW2). The data were reduced through \texttt{uvotredux}\footnote{\url{https://github.com/robertdstein/uvotredux}}, using the standard \texttt{HEASOft} tools \citep{HEASOFT}. We performed a synthetic host subtraction using \texttt{GALSYNTHSPEC}\footnote{\url{https://github.com/robertdstein/galsynthspec}}, with host modeling performed using Prospector \citep{prospector}. The transient lies on top of a small (red) host galaxy, with no significant offset to the center of this host (0.2"). As such, we follow the common procedure for estimating host galaxy contribution of nuclear transients such as Tidal Disruption Events (see e.g. \citealt{Hammerstein2023} for details).

\subsection{Spectroscopic Observations}
Here we present the spectroscopic observations of EP250827b/SN 2025wkm.

\subsubsection{Gemini}

We performed observations with the Gemini Multi-Object Spectrographs (GMOS) at Gemini North under programs GN-2025B-Q-130 (PI: A. Ho), GN-2025B-FT-104 (PI: B. O'Connor), and GN-2025B-Q-125 (PI: G. Srinivasaragavan). Longslit spectroscopy of the optical transient was acquired on 2025-08-31, 2025-09-02, 2025-09-04, 2025-09-07, 2025-09-12, and 2025-10-03. We made use of both the B480 and R400 gratings. The data were reduced and analyzed using the \texttt{DRAGONS} software \citep{Labrie2019}.

\subsubsection{NGPS}
We performed observations with the Next Generation Palomar Spectrograph \citep[NGPS;][]{Jiang:2018SPIE10702E..2LJ, Kasliwal:2024TNSAN.340....1K} located on the Palomar Observatory 200-inch telescope. Spectra were obtained on 2025-09-02, 2025-09-17, 2025-09-24, and 2025-10-08 (PI: M. Kasliwal and K. Das). On UT 2025-09-02, 2025-09-17, and 2025-10-08 spectra were obtained with a 1.5" wide slit and 2x3 spatial $\times$ spectral binning. On UT 2025-09-24 the spectrum was obtained with a 1.0" wide slit and 2x2 spatial $\times$ spectral binning. While NGPS will eventually have four arms and obtain data from 310 nm-1040 nm at the time of these observations only the R and I channels (555nm-1040nm) were available. The data were reduced using standard methods with a custom pipeline developed for NGPS. 

\subsubsection{NOT}

We performed observations using the Alhambra Faint Object Spectrograph and Camera (ALFOSC\footnote{\href{http://www.not.iac.es/instruments/alfosc}{{http://www.not.iac.es/instruments/alfosc}}}) mounted on the 2.56~m Nordic Optical Telescope (NOT) located at the Roque de los Muchachos Observatory on La Palma (Spain), one under the program P71-812 (PI: J.P.U.Fynbo) and two under the program NOIRLab-20255B-353868 (PI: T. Ahumada). The first spectrum under program P71-812 was obtained with a $1.3^{\prime\prime}$ wide slit, and the later two spectra under program NOIRLab-20255B-353868 were obtained with a 1.0`` wide slit, all using grism \#4. The observations were taken on 2025-09-03, 2025-09-17 and 2025-09-27, respectively. The first spectrum was reduced using a standard procedure with the Image Reduction and Analysis Facility \citep[\texttt{IRAF;}][]{1986SPIE..627..733T}, and the later two spectra were reduced in a standard manner using a custom fork of \texttt{PypeIt} \citep{pypeit:zenodo,pypeit:joss_arXiv,pypeit:joss_pub}.

\subsubsection{The Low Resolution Imaging Spectrometer (LRIS)}
We performed observations using LRIS \citep{lris}, mounted in the Keck I telescope, to acquire spectroscopy  on 2025-10-22 (PI: Y.J. Qin). We utilized the 400/8500 red grating and the long 1.0`` slit, with blue grism 400/3400. We then processed these data using \texttt{lpipe} \citep{lpipe}.

\subsubsection{GTC}
We obtained long-slit spectroscopy using the enhanced Optical System for Imaging and low-Intermediate-Resolution Integrated Spectroscopy (OSIRIS+) spectrograph on the 10.4m Gran Telescopio Canarias (GTC). Observations were conducted at two epochs to monitor the spectral evolution of the source (PI: Castro-Tirado). The first epoch was obtained on 2025-09-05 (i.e., ~9.0 days post-burst) using the R1000B grism with a 1.23″ slit and 1000s exposure time. The second epoch was secured on 2025-09-14 ($\Delta$t = 18.0 days; 9.0 days after our first observation), again with R1000B (1000s), supplemented by a higher-resolution R2500I spectrum to cover the redder part up to 10,000 $\AA$.

Data reduction followed standard procedures using a custom OSIRIS+ pipeline based on IRAF tasks, including bias subtraction, flat-fielding, wavelength calibration, optimal extraction, and flux calibration. The last step was performed using observations of the spectrophotometric standard star G191-B2B obtained with a 2.52″ slit on the same night.

\begin{figure}
    \centering
    \includegraphics[width=\linewidth]{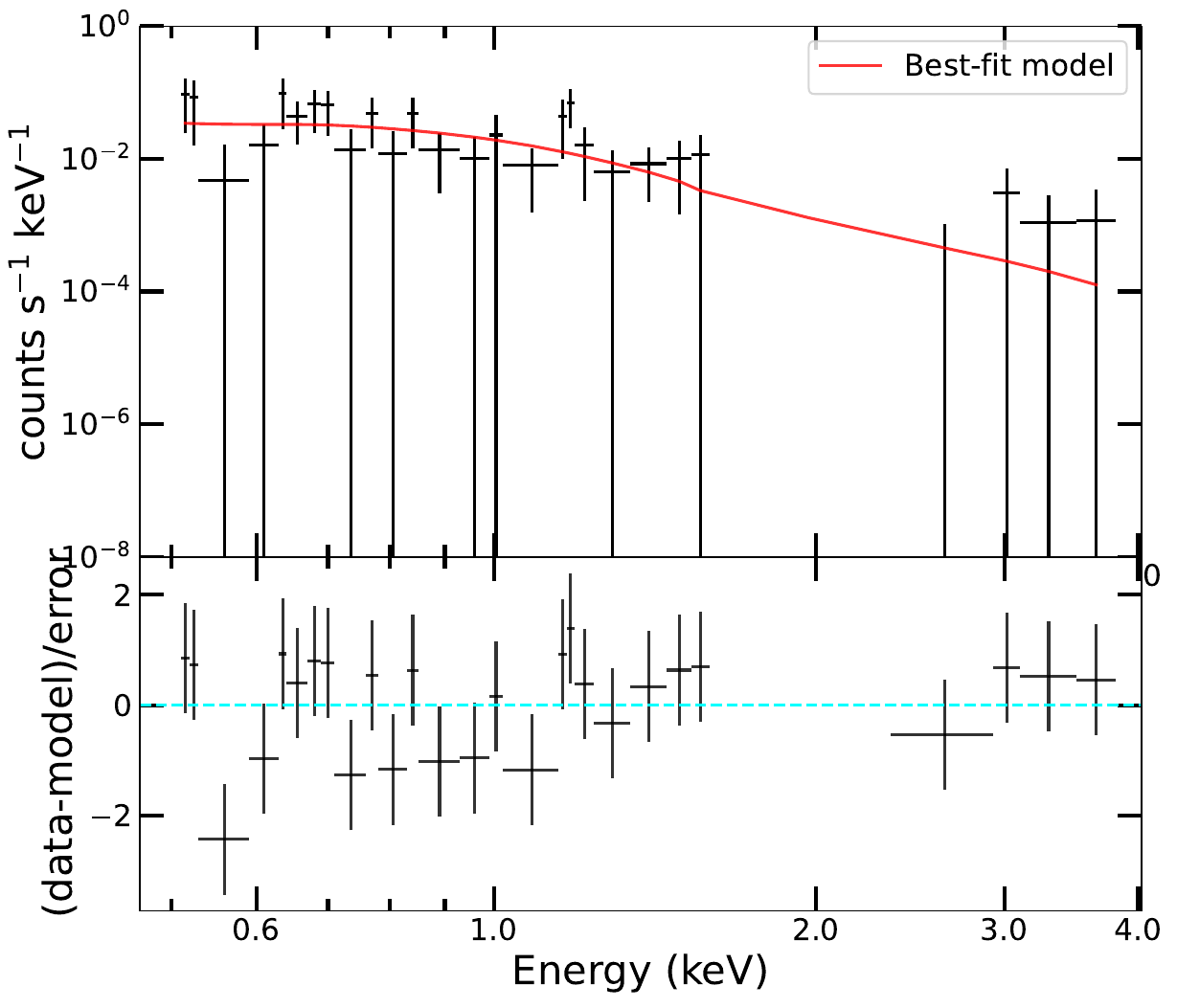}
    \caption{The EP-WXT observed spectrum and the predicted best-fit absorbed power-law model. Data are presented as the count rate spectrum with 1$\sigma$ uncertainties based on Poisson statistics for visualization purposes}.
    \label{ep_xray_fitting}
\end{figure}

\begin{deluxetable*}{cccccccc}[htb!]
\label{ep_xray_fitting_results}
\tablecaption{Spectral fitting results for the prompt X-ray emission of EP250827b/SN 2025wkm, as observed by EP-WXT during the interval 2025-08-27T06:10:21 to 2025-08-27T06:44:44 UTC.}
\tablehead{\colhead{Model} & \colhead{$\alpha_{\rm{X}}$} & \colhead{$\beta_{\rm{X}}$} & \colhead{$E_{\rm peak}$} & \colhead{$T_{BB}$} & \colhead{$N_{\rm H,int}$} & \colhead{$F_{\rm unabs}$ 0.5 -- 4 keV} & \colhead{CSTAT/(d.o.f)} \\
\colhead{} & \colhead{} & \colhead{} & \colhead{(keV)} & \colhead{(keV)} & \colhead{(10$^{21}$~cm$^{-2}$)} & \colhead{(10$^{-11}$~erg~cm$^{-2}$~s$^{-1}$)} & \colhead{}
}
\tabletypesize{\normalsize} 
\startdata 
Power-law & $-$3.3$^{+1.0}_{-1.7}$ & - & - & - & $<$7.1 & 3.4$^{+6.0}_{-1.2}$ & 27.8/24 \\
Blackbody & - & - & - & 0.22$^{+0.03}_{-0.03}$ & $<$2.1 & 1.9$^{+0.3}_{-0.3}$ & 28.3/24 \\
Broken Power-law & $-$1 (Fixed) & $-$3.3$^{+1.1}_{-1.7}$ & $<$1.5 & - & $<$7.0  & 3.4$^{+5.8}_{-1.2}$ & 27.8/23 \\
\enddata 
\tablecomments{Errors represent the 1$\sigma$ uncertainties. The upper limits are at the 90\% confidence level.}
\end{deluxetable*}
\section{Analysis}
\label{Analysis}
In this section we present analysis of EP250827b's X-ray emission, SN 2025wkm's LC, and the radio observations of the source. 

\subsection{X-ray Analysis} 
The WXT spectrum was analyzed with \texttt{XSPEC} (v12.14.0; \citealp{Arnaud1996}) using an absorbed power-law (\texttt{tbabs * ztbabs * powerlaw}), and applying W-statistics\citep{Wachter1979}. In this model, the \texttt{tbabs} component accounts for the fixed Galactic absorption (5.4 $\rm \times10^{20}~cm^{-2}$) estimated using the method from \citet{Willingale2013} and \texttt{ztbabs} represents intrinsic absorption ($N_{\rm H,ins}$) at the redshift of $z = 0.1194$.  The \texttt{powerlaw} represents a power-law spectrum of the form $N(E) = K \times (E/1~{\rm keV})^{\alpha}$, where $K$ is the normalization and $\alpha$ is the photon index. This fit yielded a photon index of $\alpha = -3.3^{+1.0}_{-1.7}$, and the intrinsic absorption cannot be well constrained ($N_{\rm H,ins} < 7.1 \times 10^{21} {\rm cm}^{-2}$; 90\% confidence) with an acceptable fit statistic of CSTAT/d.o.f. = 27.8/24 \citep{Cash1979}. The corresponding average unabsorbed flux in the 0.5-4.0 keV band is $3.4^{+6.0}_{-1.2}\times10^{-11}$~\flux, while the peak flux\footnote{To properly determine the peak flux while accounting for statistical fluctuations, the light curve was adaptively binned. Consecutive temporal bins were grouped until a minimum signal-to-noise ratio (S/N) of 5 was achieved per bin and the signal-to-noise ratio is calculated using the Li-Ma formula. The peak flux was then extracted from this dynamically binned light curve.} reaches $2.5^{+4.7}_{-2.0}\times 10^{-10}$~\flux. The best-fit results are shown in Figure \ref{ep_xray_fitting}. 

For comparison, an absorbed blackbody model (\texttt{tbabs * ztbabs * bbody}) was also tested. The Akaike Information Criterion (AIC) for this model is 32.3 (AIC = 2k + C-stat, where k is the number of free parameters), which yields a $\Delta$AIC $=0.5$ relative to the power-law.  This indicates that the two models are statistically indistinguishable. It similarly gave poorly constrained intrinsic absorption, with a 90\% upper limit of $N_{\rm H,ins} < 2.1 \times 10^{21}~\rm cm^{-2}$ and a temperature of $0.22 \pm 0.03$ keV. 

The soft spectrum, as characterized by the single power-law model, suggests a spectral peak energy ($E_{\rm peak}$) near or below the WXT's lower energy bound of 0.5 keV. To quantify this, we used an absorbed broken power-law model (\texttt{tbabs * ztbabs * bknpower}), fixing the first power-law index ($\alpha_{\rm{X}}$) to the typical GRB value of $-1$, motivated by the possible connection between EP X-ray transients and GRBs \citep{Aryan2025}. The second index ($\beta_{\rm{X}}$) was fitted to be $-3.3^{+1.1}_{-1.7}$, consistent with the single power-law model. While the $E_{\rm peak}$ itself could not be well constrained, we obtained a 90\% confidence upper limit of $E_{\rm{peak}} < 1.5$ keV \footnote{We attempted to fit the spectrum using the absorbed broken power-law model, fixing the $\alpha_{\rm X}$ to values between $-$0.8 and $-$1.2. Under these assumptions, the derived 90\% confidence upper limits on $E_{\rm peak}$ fall within the range of $-$1.4~keV to $-$1.7~keV.}. This model provided a comparable fit statistic to the single power-law (CSTAT/(d.o.f.) = 27.7/23), despite the additional free parameter. The AIC values for all three models differ by less than 2 ($\Delta$AIC $\leq$ 2 ), indicating that they are statistically indistinguishable. All fitting results are summarized in Table~\ref{ep_xray_fitting_results}, and the uncertainties for each parameter are calculated using the \texttt{XSPEC}  \texttt{error} command.

\begin{figure}
    \centering
    \includegraphics[width=0.9\linewidth]{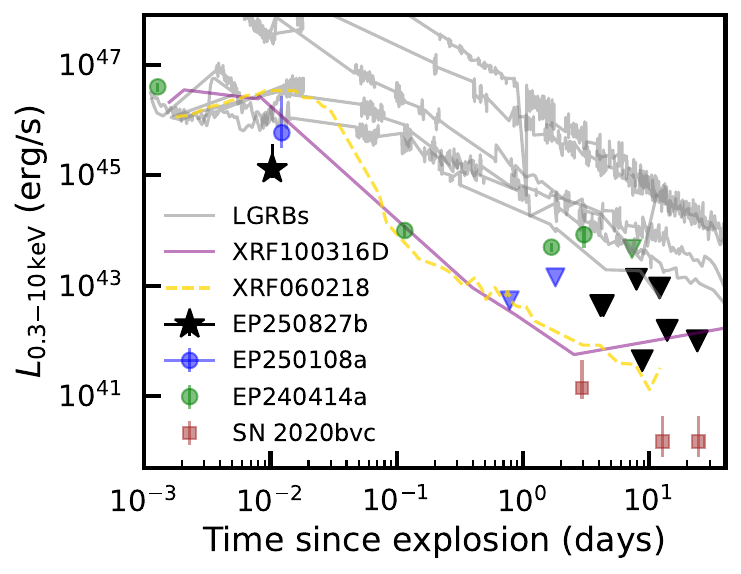}
    \caption{Comparison of 0.3 -- 10 keV (observer frame) X-ray upper limits of EP250827b/SN 2025kg (black points) to the X-ray LCs of nearby LGRBs detected by \textit{Swift}-XRT \citep{Evans2009},  XRF060218a/SN 2006aj \citep{Campana2006}, XRF100316D/SN 2020bh \citep{Starling2011}, EP250108a/SN 2025kg \citep{Srinivasaragavan2025b}, EP240414a/SN 2024gsa \citep{Sun2025}, and double-peaked SN Ic-BL SN 2020bvc \citep{Ho2020b}, with times in the rest frame. The WXT detections for the three EP events  are also shown, though we show the luminosity in the 0.5 -- 4 keV range, where the time of detection is estimated as the end-time of the X-ray detection.  }
    \label{Xrayfig}
\end{figure}

In Figure \ref{Xrayfig}, we show the X-ray light curve of EP250827b compared to other XRF-SNe, along with SN 2020bvc \citep{Ho2020b, Izzo2020}, which was a similar double-peaked SN Ic-BL that was discovered optically, without the use of a high-energy trigger. We also show the X-ray afterglow emission from a few nearby ($z \lesssim 0.4$) LGRBs taken from the \textit{Swift}-XRT archive \citep{Evans2009} as reference. We see that EP250827b's prompt emission is fainter than those of XRF060218 and XRF100316D, and on the lower end of EP250108a's error bars. The upper limits following the prompt emission are consistent with XRF060218's X-ray afterglow LC, as well as detections corresponding to SN 2020bvc.

\subsection{Ultraviolet, Optical, and Near-Infrared Light Curve Analysis}
\label{LCanalysis}
\begin{figure*}
    \centering
    \includegraphics[width=\linewidth]{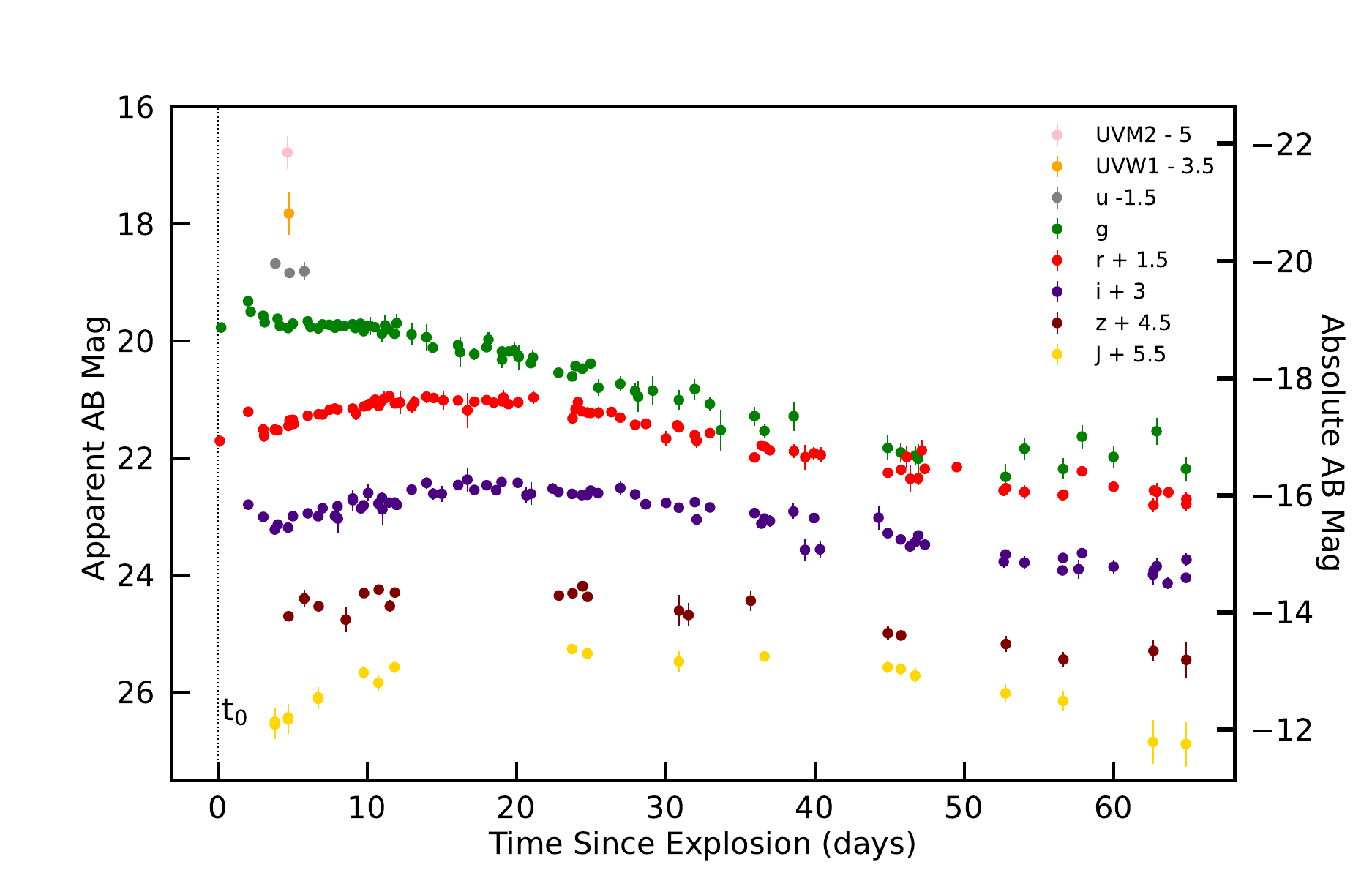}
    \caption{Light Curve of EP250827b/SN 2025wkm in UVM2, UVW1, and $grizJ$ bands, corrected for Milky Way extinction, in the observer frame. We note that we also have an additional epoch of BVRI data, but do not show it in the LC for visual purposes. We show the time of explosion $t_0$ with a vertical dotted line. The photometry is compiled using multiple different telescopes, and the process used to compile the LC is described in \S \ref{LCanalysis}. A full table of the photometric observations, along with the different telescopes used for every epoch is presented in Table \ref{appendix:phot_log} in the Appendix.}
    \label{LCfigure}
\end{figure*}
We show the light curve (LC) of EP250827b/SN 2025wkm in Figure \ref{LCfigure}, in the UVM2, UVW1, and $grizJ$ bands. We constructed the LC by combining all available photometry from all facilities, retaining each instrument's native pass-band. We removed outliers by discarding measurements more than $2\sigma$ from the data distribution by comparing the observed data point to the nearest two points. When a single facility obtained multiple measurements within the same night, we consolidated them into one nightly point per band using an inverse variance weighted average. This occurred most frequently for ZTF, where chip-gap coverage often produces repeated visits within a night. In the following analysis, we account for the Milky Way extinction of $A_{\rm{V}} = 0.155$ mag. We describe constraints on the host galaxy extinction in \S \ref{spectra}, but do not account for it in our analysis, as we only derive an upper limit. 

The LC possesses a clear initial peak in the $gri$ bands at $\sim$ 2 days after $t_0$, assumed to be the time of explosion for the rest of this work. The LC then declines in every band, and peaks again at progressively later times in the $grizJ$ bands, though sparser coverage in the $z$ and $J$ bands do not make it clear where the exact time of the peak is. The first optical detections in $r$ and $g$ band are 3.2 hours after $t_0$ in the observer frame, allowing for constraints on the rise of the initial first peak, unlike the LC for EP250108a/SN 2025kg \citep{Srinivasaragavan2023, Eyles-Ferris2025, Li2025}, whose first optical photometry point was during the decline phase.

After the second photometry point in $r$ and $g$ band at $t_0 + 2$ days, the LC is already declining, and the decline continues until $\sim t_0 + 3$ days. Therefore, the LC most likely peaks prior to $t_0+2$ days. This corresponds to a lower limit on the absolute magnitude of the first peak  of  $M_g = -19.31 \pm 0.06$ and $M_r = -18.93 \pm 0.03$. The first peak has clear blue colors, where $g-r \sim - 0.4$ mag. The luminosity and colors of the first peak are very similar to those of EP250108a/SN2025kg \citep{Srinivasaragavan2025b, Eyles-Ferris2025}, though the first peak declines on a much quicker timescale than for EP250108a/SN 2025kg ($\sim$ 5 days in $r$ band and $\sim$ 8 days in g band). 

\begin{figure}
    \centering
    \includegraphics[width=0.99\linewidth]{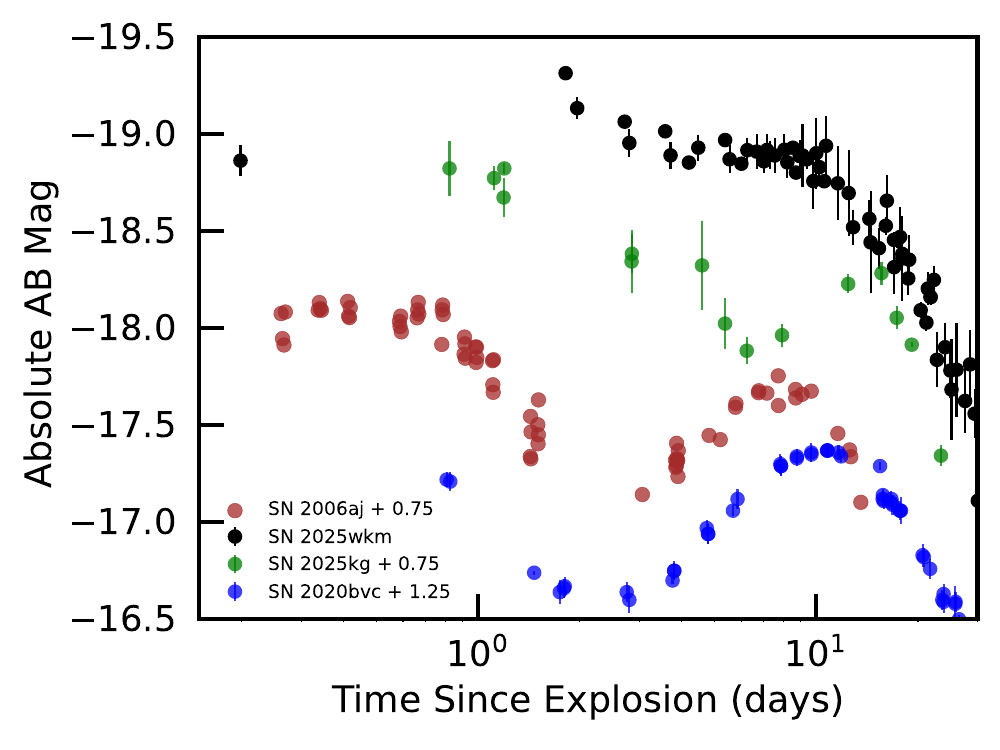}
    \caption{Rest-frame $g$-band LC of SN 2025wkm, in comparison to other double-peaked SN Ic-BL, including the $g$-band LCs of EP250108a/SN 2025kg \citep{Srinivasaragavan2025b}, SN 2020bvc \citep{Ho2020b}, and the B-band LC of XRF060218/SN 2006aj \citep{Modjaz2006}.}
    \label{gcompare}
\end{figure}

We spectroscopically confirm a SN Ic-BL classification during the second phase of the optical LC (more details in \S \ref{spectra}). SN 2025wkm rises to maximum light during its second peak on a timescale of $\sim 14$ days at an absolute magnitude of $M_r \sim -19.2 $ mag, which again is similar to the $r$-band peak properties of SN 2025kg \citep{Srinivasaragavan2025b, Rastinejad2025, Li2025}. However, the second peak in $g$ band is not as pronounced, and the $g$ band LC plateaus until around $\sim 12$ days at $M_g \sim -19$ mag before it starts declining. This is unlike SN 2025kg, whose $g$ band LC had a clear peak 18 days after explosion, with $M_g = -18.95 \pm 0.06$. In Figure \ref{gcompare}, we show the $g$ and B-band LCs of double-peaked SNe Ic-BL in the literature. SN 2025wkm clearly has a broader second peak than every other event that has been observed, with a pronounced plateau rather than a decline into a second rise. The peak magnitude in $r$ band corresponding to the second peak is within the median range reported for SNe Ic-BL in sample papers ($M_r = -18.5 \pm 0.9$; \citealt{Taddia2018, Srinivasaragavan2024b}), and is consistent with the average seen in the GRB-SN population \citep{cano2017}.

An empirical correlation between the first peak and second peak in double-peaked stripped-envelope SNe exists -- $M_2 = 0.8 \times M_1 - 4.7$, where $M_1$ and $M_2$ are the absolute magnitudes of the first and second peak, respectively, in $r$ band \citep{Kaustav2024}. Substituting $M_1$ into the expression, the second peak should have a brightness $M_2 \sim  -19.3$ mag. This is consistent with the observed second peak brightness. In the sample of 54 stripped-envelope SNe this relation was tested on, only two were SNe Ic-BL \citep{Kaustav2024}. The mechanism used to describe the first peak in these SNe was shock cooling emission from SN ejecta interacting with an extended CSM (\citealt{Piro2021}). EP250108a/SN 2025kg did not follow this correlation, though it still followed the overall trend, where the first peak is more luminous than the second peak \citep{Srinivasaragavan2025b}.

\subsection{Bolometric Luminosity Light Curve}
\label{bolsection}
\begin{figure}
    \centering
    \includegraphics[width=1\linewidth]{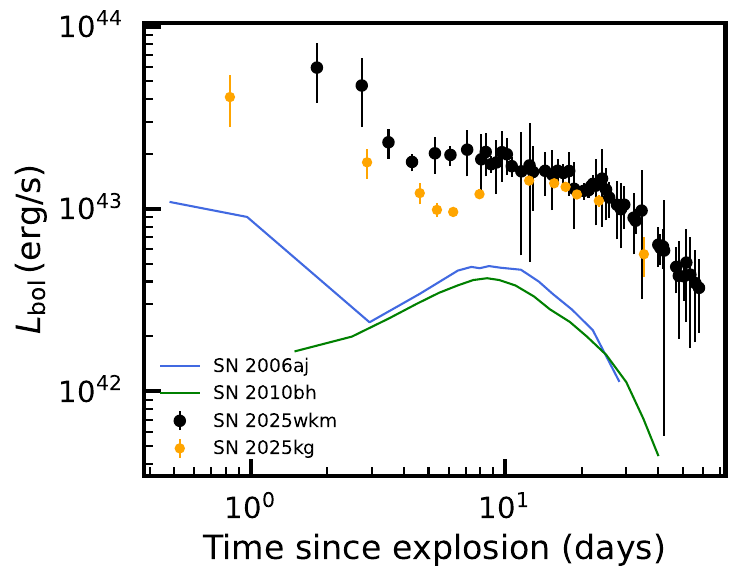}
    \caption{Bolometric luminosity LC of SN 2025wkm in the rest frame, computed through fitting blackbodies to SEDs binned every 0.75 days. We also show the bolometric luminosity LCs of SN 2025kg \citep{Srinivasaragavan2025b} SN 2006aj \citep{Modjaz2006, Pian2006}, and SN 2010bh \citep{Olivares2012}.}
    \label{bolLC}
\end{figure}

\begin{figure}
    \centering
    \includegraphics[width=1\linewidth]{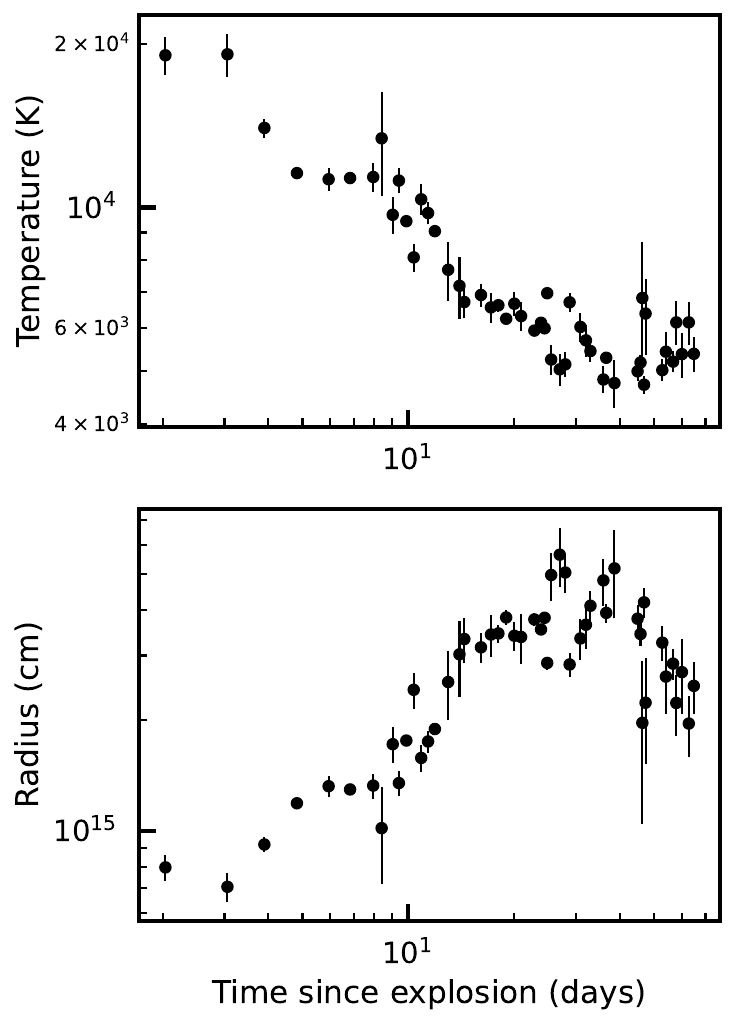}
    \caption{Temperature and Radius evolution as a function of time in the observer frame, from fitting blackbodies to SEDs binned every 0.5 days.}
    \label{temprad}
\end{figure}

Due to exquisite $gri$-band photometric coverage across the LC, along with the supplementation of one UV epoch, and multiple epochs in $z$ and $J$ band, we are able to construct a bolometric luminosity LC through fitting a blackbody model across different epochs. We bin the data every 0.5 days, and fit a blackbody to bins with at least 3 photometric band measurements. We do not compute bolometric luminosities for epochs where there are less than 3 photometric band measurements. We utilize the blackbody fitting tool in \texttt{Redback} \citep{Sarin2024} to perform this analysis, and show the bolometric luminosity LC in Figure \ref{bolLC}, along with the derived temperatures and radii in Figure \ref{temprad}. 

We also show comparisons to the bolometric LCs of EP250108a/SN 2025kg, XRF060218/SN 2006aj \citep{Modjaz2006, Bianco2014, Brown2014}, and XRF100316D/SN 3020bh \citep{Olivares2012}. We note that the bolometric LC of EP250108a/SN 2025kg was computed using a different method, as there was only sparse $i$ band coverage, and no coverage in any other bands. \citet{Srinivasaragavan2025b} utilized bolometric correction coefficients from \citet{Lyman2014} to compute the bolometric luminosity LC, which were measured by fitting the SEDs of a large sample of stripped-envelope SNe with broadband coverage. The bolometric luminosity LC of XRF060218/SN 2006aj snf XRF100315D/SN 2010bh was computed using a similar method that we used, utilizing blackbody fits to photometry. 

We find that the shape of SN 2025wkm's bolometric luminosity LC differs from those of the others. SN 2010bh does not display a double-peaked bolometric luminosity LC. This is likely due to observations not being at early enough to detect the initial shock cooling peak, as its shock cooling phase evolved more rapidly than the other events (see \S \ref{CSMtable} for more).  Both SN 2006aj and SN 2025kg display a decline at the beginning of the LC, which transitions into a clear rise that peaks at around 10 days after explosion and ends with a second decline. SN 2025wkm follows the trend of a decline at the beginning of the LC; however, there is not a clear second rise and decline like the other two events. Instead, the bolometric luminosity LC plateaus until $\sim$ 20 days, and then declines at a similar rate to the other two events. This difference in shape, especially at earlier epochs prior to 10 days, may be due to brightness differences between the first two peaks in the LC when compared to other events. However, the sustained plateau may originate from an additional energy source such as a central engine, which we investigate later in \S \ref{magnetaranalysis}. We note that we only have a $r$ and $g$ band point 2 hours after $t_0$, which does not allow us to compute a bolometric luminosity point at the beginning of the LC to capture the rise. Therefore, the peak bolometric lumonisity we derive during the first peak is a lower limit, which is $L_{\rm{bol, peak}} = 5.4 \pm 1.7 \times 10^{43} \, \rm{erg} \, \rm{s^{-1}}$. This is comparable to the peak luminosity of SN 2025kg, but higher than SN 2006aj by a factor of $\sim 5$. 

In Figure \ref{temprad}, we find that the blackbody temperature plateaus over the first $\sim 3$ days and peaks at $\sim 20,000$ K, declines until $\sim 20$ days, and plateaus around 5000 K for the remaining epochs. The blackbody radius decreases over the first $\sim 3$ days, increases around $\sim$ 20 days, and then plateaus around $4 \times 10^{15}$ cm. The decline in temperature until $\sim 6$ days after peak light is very similar to the behavior exhibited by samples of SNe Ic-BL presented in \citet{Taddia2018} and \citet{Srinivasaragavan2024b}. However, the radii of the SNe in these samples exhibited a decline after $\sim 10$ days after peak light, while the radius plateaus in SN 2025wkm. We note that small-scale variability in the temperature and radius is spurious and is an artifact of the fitting procedure.

In addition, the early-time behavior within the three days is unlike what is seen for most SNe Ic-BL. The plateau in temperature and decline in radius may be an artifact of the fitting procedure due to degeneracies between the blackbody temperature and radius. However, it is physically possible if a central engine is heating up inner parts of the ejecta, while outer parts are already optically thin. In this scenario, we expect the temperature to increase while the radius recedes. This scenario is consistent with progenitor scenarios we test in \S \ref{LCmodel}. The blackbody radius we derive at $\sim 2$ days is around $7.9 \times 10^{14}$ cm, which corresponds to an average velocity over the first 2 days of $\sim 0.15 c$. This is comparable to the average velocity of EP250108a/SN 2025kg of $\sim 0.1c$ over the first 3.9 days.
\subsection{Spectral Analysis}
\label{spectra}
\begin{figure}
    \centering
    \includegraphics[width=0.9 \linewidth]{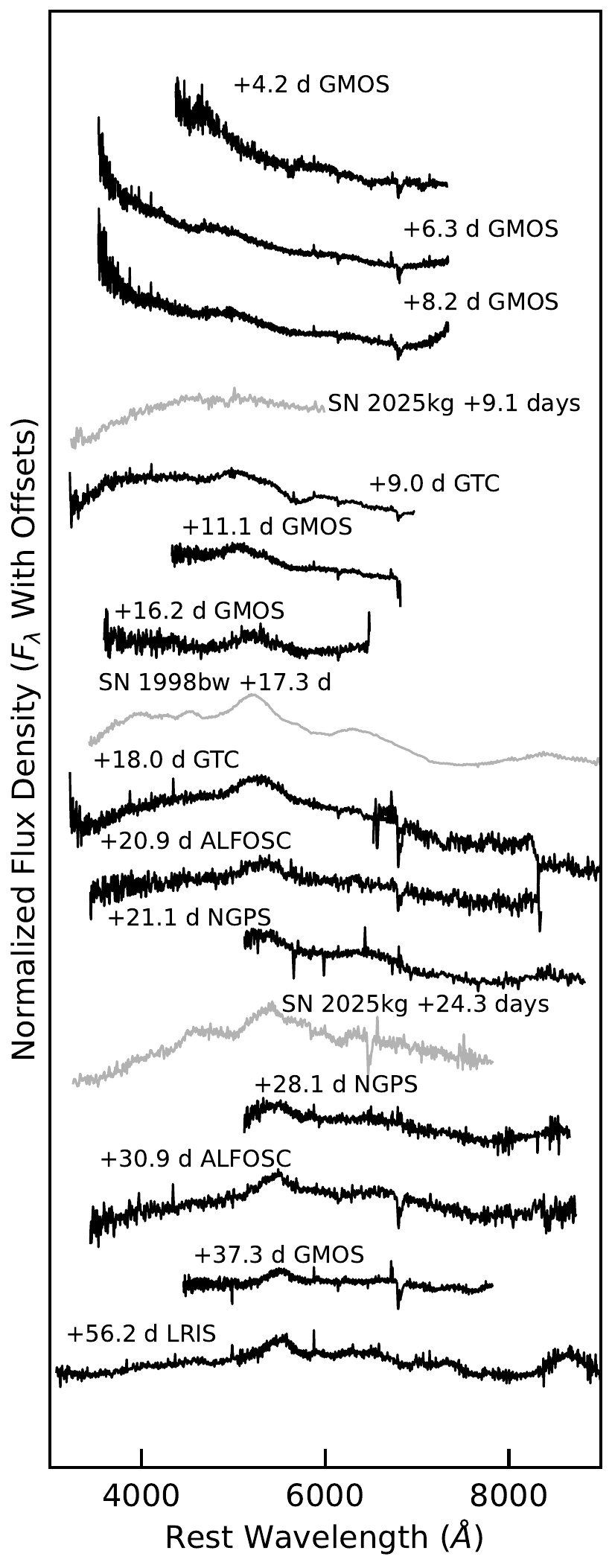}
    \caption{Spectra series of EP250827b/SN 2025wkm, with host galaxy emission lines clipped, along with the phase of the spectrum and the instrument it was taken on. The spectra start as a blue, mostly featureless continuum, and then evolve to redder SN spectra, with clear, broad absorption features. The lack of H and He and broad features lead to a classification of SN Ic-BL. We also show selected spectra of EP250108a/SN 2025kg at certain phases from \citet{Srinivasaragavan2025b}, as well as the best-matching SN Ic-BL SN 1998bw \citep{patat2001}}. Some spectra have been smoothed for viewing purposes. The GMOS and ALFOSC spectra are not corrected for telluric features.
    \label{spectrafigure}
\end{figure}

\begin{figure*}
    \centering
    \includegraphics[width=0.9\linewidth]{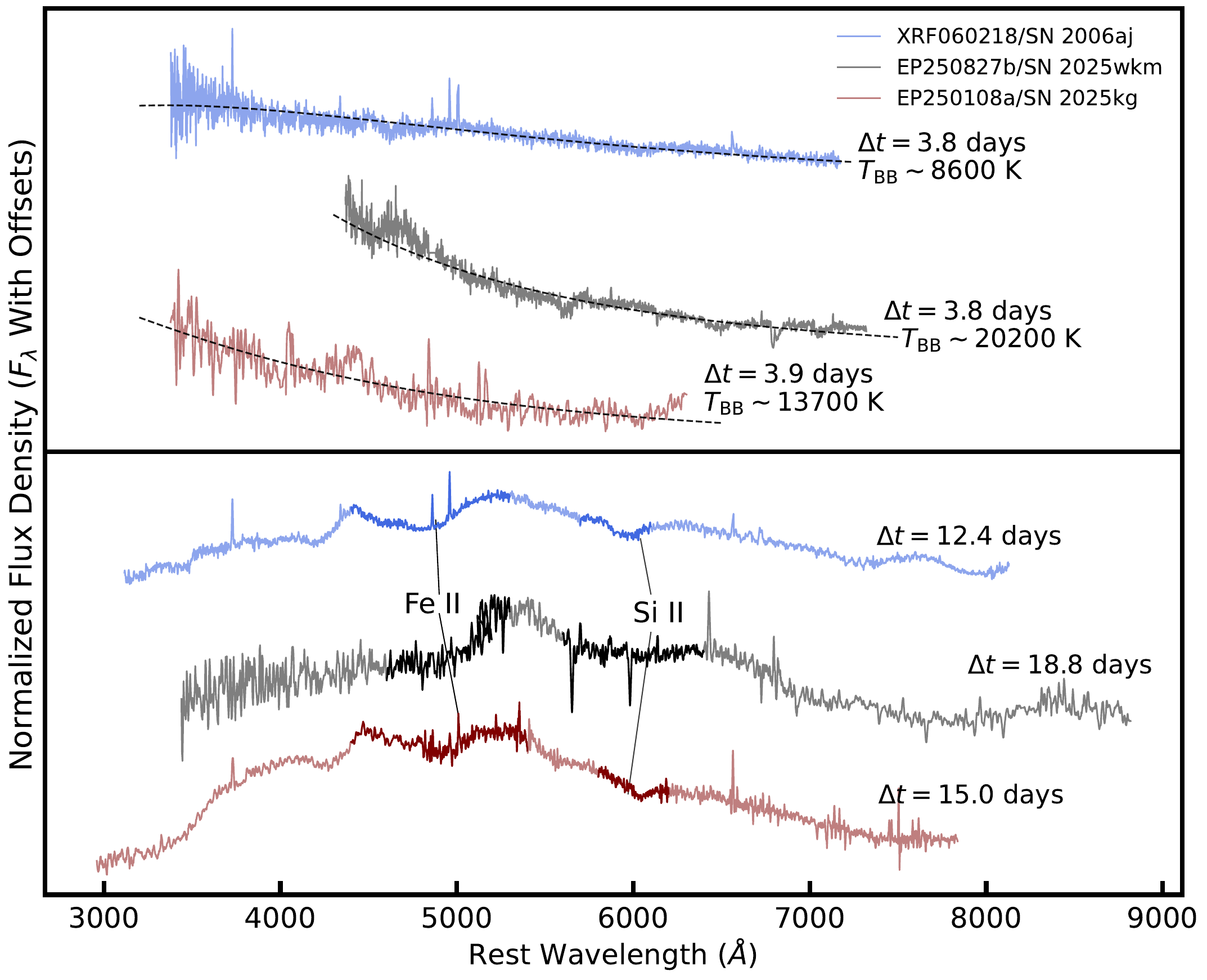}
    \caption{Close-up spectra of EP250827b/SN 2025wkm at 3.8 days and 18.8 days $+ t_0$ (rest-frame). Comparison spectra of XRF060218/SN 2006aj \citep{Modjaz2006} and EP250108a/SN 2025kg \citep{Srinivasaragavan2025b} are also shown, at similar phases. We show these two events as comparison because their X-ray prompt emission characteristics are similar to that of EP250827b/SN 2025wkm, along with the double-peaked nature of their SNe Ic-BL LCs. Some host galaxy emission lines are clipped for viewing purposes. In the top panel, we also fit a blackbody spectrum to each of the spectra, and show the derived rest-frame blackbody temperature ($T_{\rm{BB}}$). EP250827b/SN 2025wkm displays an extremely blue continuum in its first spectrum, along with a broad absorption feature $\sim 4500 \AA$ corresponding to blueshifted Fe II at $\sim 40,000 \, \rm{km \, s^{-1}}$. The blackbody temperature derived is $T_{\rm{BB}} \sim 20,200$ K, which is significantly hotter than $T_{\rm{BB}}$ for the compared events. In the bottom spectra, we see clear broad absorption features corresponding to blueshifted Fe II and Si II, and the lack of clear H and He absorption lead to a SN Ic-BL classification. We also show the Fe II and Si II absorption features bolded in the comparison spectra as reference. }
    \label{closeupspectra}
\end{figure*}
We obtained ten spectra of EP250827b/SN 2025wkm up to $t_0 + 37.3$ days and present them in Figure \ref{spectrafigure}, with host emission lines clipped. The early-time spectra up to $t_0 + 8.2$ days display a clear hot, blue continuum. A close-up of the first spectrum is shown in Figure \ref{closeupspectra}, along with a comparison to spectra of XRF060218/SN 2006aj \citep{Modjaz2006} and EP250108a/SN 2025kg \citep{Srinivasaragavan2025b} at a similar rest-frame phase. We fit a blackbody to each of the spectra, and find that SN 2025wkm possesses a much higher temperature than the other two events, with $T_{\rm{BB}} = (2.2 \pm 0.6) \times 10^3$ K, which is consistent with the temperature found from fitting the SED at a similar phase in \S \ref{bolsection}. Therefore, we find that SN 2025wkm stays blue for a longer time period than SN 2025kg. Figure \ref{spectrafigure} shows a spectrum of SN 2025kg 9.1 days after explosion if it exploded at $z = 0.1194$, and the spectrum shows a much redder continuum than SN 2025wkm does at $t_0 + 8.2$ days. 

By the fourth spectrum at $t_0+11.1$ days, we find that the continuum reddens significantly, and broad absorption features in the spectra clearly develop, corresponding to blueshifted Fe II and Si II. We show a closer look at the SN spectra in Figure \ref{closeupspectra}, and show the broad Fe II and Si II absorption features in bold. These features, along with a lack of H and He features, indicate that SN 2025wkm is a SN Ic-BL. We run the spectrum obtained at $t_0 + 11.1$ days through the SuperNova Indentification Code \citep{SNID}, and find that the best-match spectrum is SN Ic-BL 1998bw, allowing us to confirm its classification as a SN Ic-BL.

We then measure the velocity of the 5169 $\AA$ Fe II line for every spectrum using the open source code SESNspectraLib\footnote{https://github.com/metal-sn/SESNspectraLib} \citep{Modjaz2016, Liu2016}. The Fe II line is a good proxy for the photospheric expansion velocity $v_{\rm{ph}}$ \citep{Modjaz2016}. We smooth each of the spectra using SESNspectraPCA\footnote{https://github.com/metal-sn/SESNspectraPCA}. SESNspectraLib then calculates the blueshift of the Fe II line at 5169 $\AA$ relative to a standardized SN Ic spectroscopic template at a similar phase. The uncertainty is estimated through adding the uncertainty on the velocity of the mean SN Ic template in quadrature with the uncertainty on the relative blue-shift. The uncertainty is dominated by the strength of the blueshifted Fe II absorption feature relative to the continuum. 

\begin{figure}
    \centering
    \includegraphics[width=0.9\linewidth]{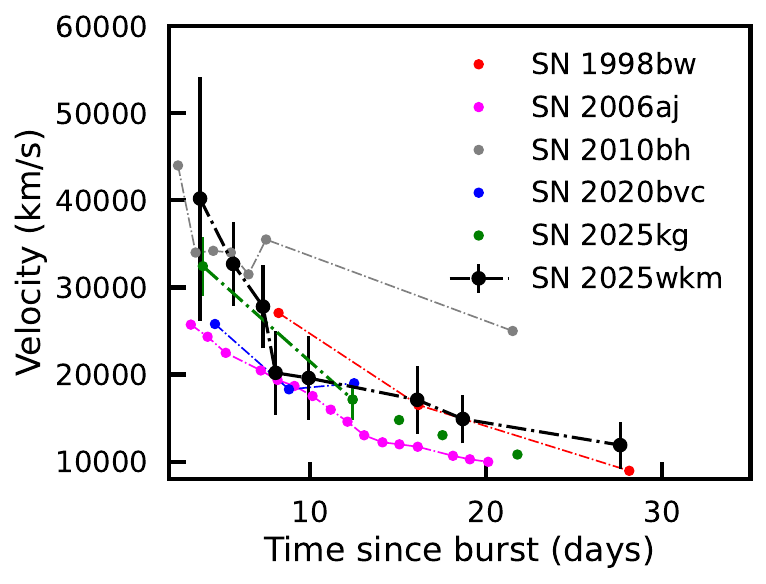}
    \caption{Photospheric velocity evolution for SN 2025wkm, compared to other prominent XRF and GRB--SNe (SN 1998bw, \citealt{Iwamoto98}; SN 2006aj, \citealt{Mazzali2006};  SN 2010bh, \citealt{Chornock2010}; SN 2025kg, \citealt{Srinivasaragavan2025b}) and double-peaked SN Ic-BL 2020bvc, \citealt{Ho2020b}), with times in the rest frame.  }
    \label{velocityfigure}
\end{figure}

We derive velocity measurements for 6 out of the 10 spectra, and they are presented in Table \ref{velocitytable}. In addition, we show the velocity evolution over time in Figure \ref{velocityfigure}, compared to SN 2006aj \citep{Modjaz2006}, SN 2010bh \citep{Chornock2010}, SN 2025kg \citep{Srinivasaragavan2025b}, as well as SN 2020bvc \citep{Ho2020b, Izzo2020}. We see that SN 2025wkm possesses higher velocities at early times than SN 1998bw, SN 2006aj, and SN 2020bvc, and comparabe velocities to SN 2025kg and SN 2010bh. At later times, the velocity evolution flattens out and reaches broadly consistent values with the rest of the events with the exception of SN 2010bh, which continues to display a velocity $\sim 25,000 \, \rm{km \, s^{-1}}$ even at later phases.

\citet{Srinivasaragavan2025b} recognized a possible blueshifted Fe II feature in their spectrum of EP250108a/SN 2025kg taken at $3.9$ rest-frame days after explosion (shown in Figure \ref{closeupspectra}), corresponding to a velocity of $\sim 30,000$ km s$^{-1}$. However, they stressed that this line identification was not robust, in part due to low signal-to-noise in their spectra. In addition, they discussed that at the blackbody temperature they derived $\sim 13,700$ K, Fe has mostly transitioned from Fe II to Fe III. In addition, the optical LC during the time of the spectral epoch had significant contributions from a component not dominated by radiaoctive decay \citep{Srinivasaragavan2025b, Eyles-Ferris2025, Li2025}, which would make it surprising that a photospheric Fe II line was present in the spectrum. 

However, our identification of a blueshifted Fe II feature in the spectrum of EP250827b/SN 2025wkm at $t_0+4.2$ days (a close-up shown in Figure \ref{closeupspectra}) is more robust. Though the temperature that we derive is still extremely high $\sim 20,200$ K, the signal-to-noise is much higher than that of the spectrum of EP250108a/SN 2025kg taken at a similar epoch, and the broad absorption feature at the beginning of the spectrum is seen much more clearly. A similar broad absorption feature at slightly redder wavelengths is also seen in XRF060218/SN 2006aj's early-time spectrum shown in the same Figure. Figure \ref{spectrafigure} shows that this absorption feature clearly shifts to redder wavelengths over time. In addition, though SN 2025wkm also displays an inital first peak in its LC not due to radioactive decay, the first peak declines much more rapidly than in SN 2025kg (see Figure \ref{bolLC}), and it is therefore plausible that a photospheric line may be present at this stage in the LC's evolution.

We then try to estimate the host-galaxy extinction by measuring the equivalent width (EW) of the Na I absorption doublet (5890, 5896 $\AA$), using the NGPS spectrum at $t_0 + 28.1$ days. This feature is a proxy for the amount of host-galaxy extinction, with various relations presented in the literature \citep{Stritzinger2018, Osmar2023}. The relation from \citet{Stritzinger2018} is $A_{\rm{V}}^{\rm{host}}\rm{[mag]} = 0.78(\pm0.15) \times EW [\AA]_{\rm{Na \, I}}$, while the relation from \citet{Osmar2023} is  $A_{\rm{V}}^{\rm{host}}\rm{[mag]} = 0.02 + 0.73 \times EW [\AA]_{\rm{Na \, I}} \pm 0.29$. However, \citet{Poznanski2011} showed that a large scatter exists in these correlations when using low-resolution spectra, and quantitative relations must be viewed conservatively. Therefore, rather than overestimating the extinction using the empirical relations, we use the value derived through this method as a conservative upper limit for the host-galaxy extinction \citep{Srinivasaragavan2024b, Srinivasaragavan2025b}. 

The larger value derived between the \citet{Stritzinger2018} and \citet{Rodriguez2024} relations is $A_{\rm{V}}^{\rm{host}} < 0.73$ mag. This value is consistent with the extinction inferred from the Balmer decrement assuming Case B recombination \citep{Osterbrock1989} of $A_{\rm{V}}^{\rm{host}} \sim 0.3$ mag. If the host extinction is close to this upper limit and is significant, then the analysis performed in \ref{LCanalysis} would be impacted as the overall brightness of the LC would increase. For example, if $A_{\rm{V,host}} \sim 0.5$ mag, $M_g$ would decrease by $\sim 0.6$ mag and $M_r$ would decrease by $\sim 0.4$ mag, making EP250827b/SN 2025wkm brighter than EP250108a/SN 2025kg \citep{Srinivasaragavan2025b, Rastinejad2025}. However, given the lack of additional constraints, we do not try to further quantify the impact that host galaxy extinction may have on our analysis.

\begin{figure}
    \centering
    \includegraphics[width=0.99\linewidth]{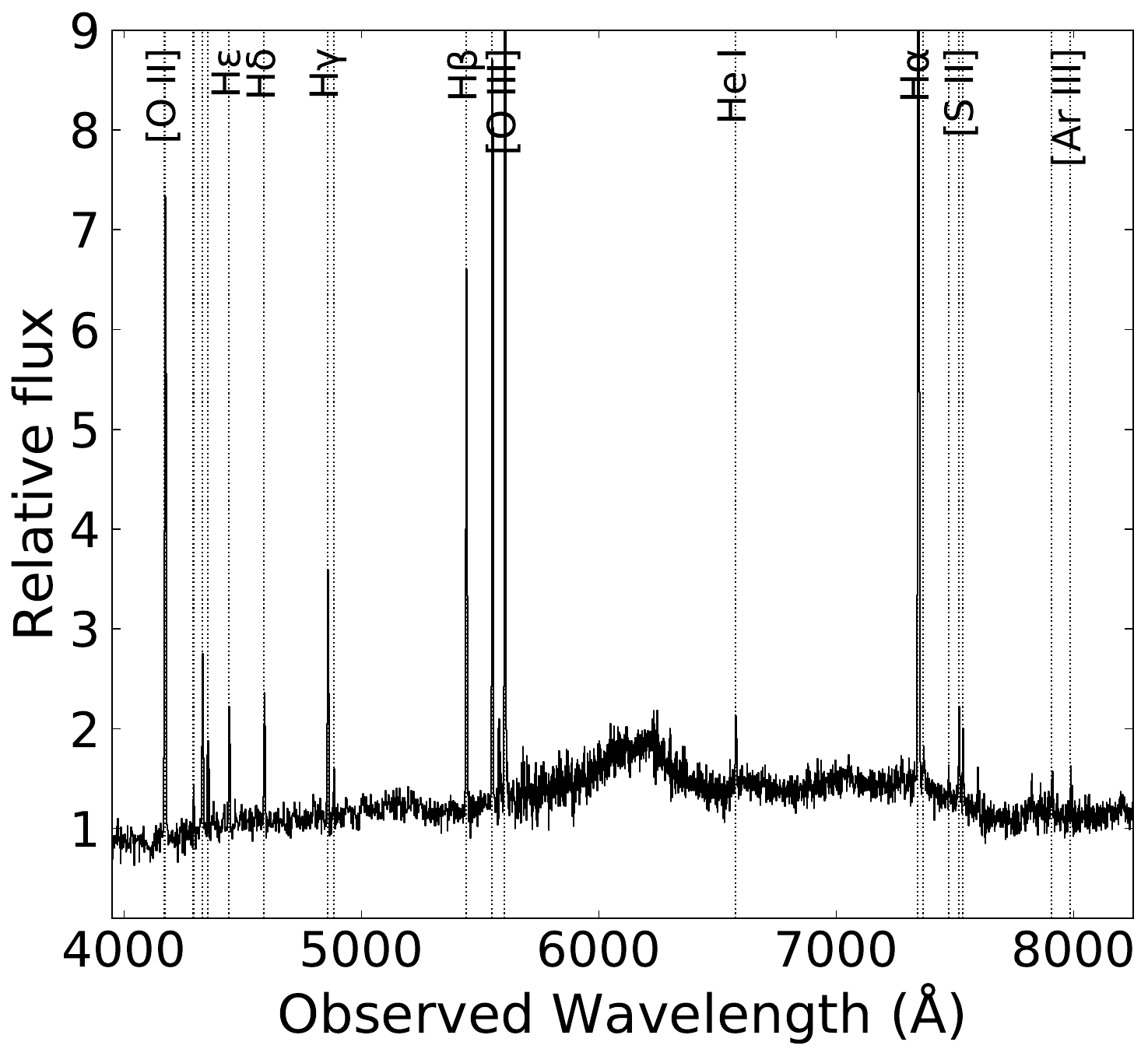}
    \caption{Late-time Keck LRIS spectrum on 2025-10-22 (corresponding to 56.2 days). We identify numerous narrow emission lines produced by star formation within the host galaxy. These lines yield a redshift of $z=0.1194$. We note that the narrow galaxy lines are clipped in Figure \ref{spectrafigure}. }
    \label{hostfigure}
\end{figure}

We also show the Keck LRIS spectrum obtained at $t_0+56.2$ days, without host-galaxy emission lines clipped, in Figure \ref{hostfigure}. The host galaxy has an abundance of 
nebular and recombination features characteristic of a star-forming galaxy: the [O\,\textsc{ii}] doublet at $\lambda\lambda3726,3729$; Balmer lines from H$\eta$ through H$\alpha$ (H$\eta$~3835, H$\zeta$~3889, H$\epsilon$~3970, H$\delta$~4102, H$\gamma$~4340, H$\beta$~4861, H$\alpha$~6563); He\,\textsc{i} at $\lambda\lambda3889,5876,6678,7065$; high-excitation [Ne\,\textsc{iii}] at $\lambda\lambda3869,3967$; [O\,\textsc{iii}] including the auroral line near $\lambda4363$ and the strong nebular lines at $\lambda\lambda4959,5007$; [N\,\textsc{ii}]~$\lambda6583$ adjacent to H$\alpha$; the [S\,\textsc{ii}] doublet at $\lambda\lambda6716,6731$; and [Ar\,\textsc{iii}]~$\lambda7136$. These emission lines allowed us to make a redshift measurement of $z = 0.1194 \pm 0.0002$. Additional narrow lines are shown in Figure \ref{hostfigure}. We estimate the star formation rate using the equivalent width of the observed H$\alpha$ line \citep{Kennicutt1998}, and find a $\rm{SFR_{H\alpha}} = 2.83 \pm 0.04 \, \rm{M_\odot \, yr^{-1}}$. This is consistent with the SFR derived from radio observations in \S \ref{VLA}.

\begin{deluxetable}{lr}[htb!]
\tablecaption{Photospheric velocity measurements of SN\,2025wkm. }
\label{velocitytable}
\tablewidth{0pt} 
\tablehead{\colhead{Time (days)} & \colhead{$v_{\rm{ph}}\, \rm{(km \, s^{-1})}$}}
\tabletypesize{\normalsize} 
\startdata 
4.2 & $40200 \pm 14000$ \\
6.3 & $32700 \pm 4800$ \\
8.2 & $27800 \pm 4800$ \\
9.0 & $20200 \pm 4800$ \\
11.1 & $19600 \pm 4800$ \\
18.0 & $17100 \pm 3900$ \\
20.9 & $15900 \pm 2800$ \\
30.9 & $11900 \pm 2700$ \\
57.2 & $8300 \pm 2000$ \\
\enddata 
\end{deluxetable}

\subsection{Radio Analysis}
\label{sec:RadioAnalysis}
\begin{figure}
    \centering
    \includegraphics[width=0.99\linewidth]{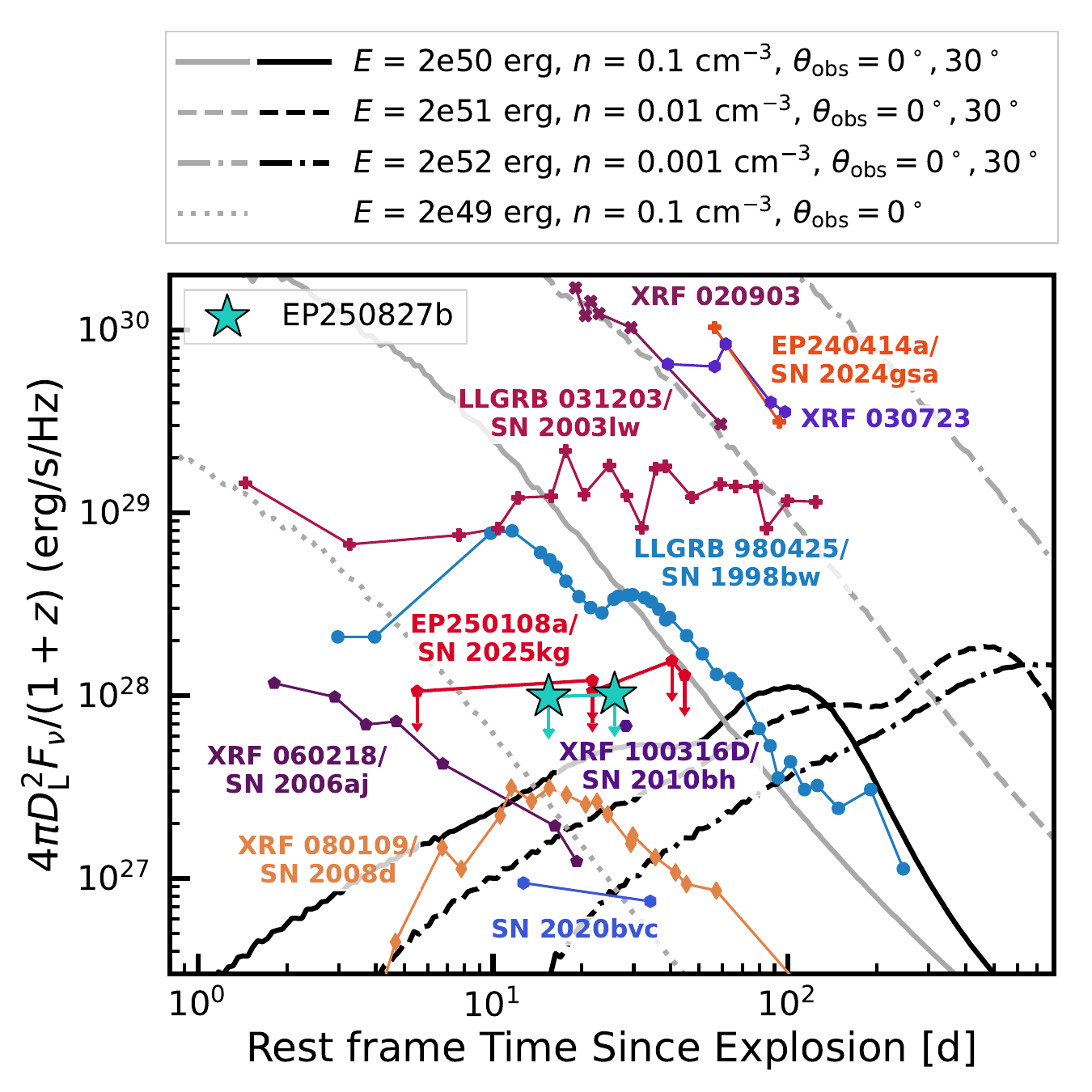}
    \caption{{The 10 GHz radio light curve of EP250827b (blue stars), where we use the detection of the host galaxy as upper limits on the radio emission. Also shown are the radio light curves of  LLGRB 980425/SN 1998bw \citep{1998Natur.395..663K, 1998ApJ...497..288W}, XRF 020903 \citep{2004ApJ...606..994S}, XRF 030723 \citep{Chandra2012}, LLGRB 031203/SN 2003lw \citep{2004Natur.430..648S}, XRF 060218/SN 2006aj \citep{2006Natur.442.1014S}, XRF 080109/SN 2008d \citep{2008Natur.453..469S}, XRF 100316D/SN 2010bh \citep{2013ApJ...778...18M}}, the SNe Ic-BL 2020bvc \citep{2020ApJ...902...86H}, and EP events EP240414A/SN 2024gsa \citep{Bright2025} and EP250108a/SN 2025kg \citep{Srinivasaragavan2025b}. Lines represent predicted radio emission for an on-axis ($\theta_{\rm obs} = 0^\circ$, grey) or off-axis ($\theta_{\rm obs} = 30^\circ$, black) afterglow generated with the \texttt{FIREFLY} code \citep{2024ApJ...976..252D}, for various pairs of jet energy $E$ and circumburst density $n$. }
    \label{fig:RadioLuminosity}
\end{figure}

The detection of EP250827b in conjunction with the discovery of the SN 2025wkm is reminiscent of nearby GRBs with associated SNe Ic-BL \citep{Soderberg+2006,Campana2006,Woosley2006, Modjaz2016, cano2017}, and could imply that EP250827b was a GRB, either on-axis or moderately off-axis. With this context as a backdrop, we explore what constraints our radio observations can place on the progenitor of EP250827b. 

The lack of detected radio emission associated with EP250827b/SN 2025wkm results in luminosity limits of $\sim 10^{28}~{\rm erg~s}^{-1}{\rm Hz}^{-1}$ (Figure~\ref{fig:RadioLuminosity}), $\sim 3$ orders of magnitude lower than typical on-axis GRBs \citep{Chandra2012}, indicating that EP250827b was not produced by a typical on-axis GRB. However, many of the nearby ($z <0.3$) GRB-SNe have belonged to the class of low-luminosity GRBs (LLGRBs) and XRFs. LLGRBs have isotropic equivalent luminosities $\lesssim 10^{49} \, \rm{erg \, s^{-1}}$ \citep{Liang2007}, and may have similar origins to XRFs, though some may have higher peak energies. LLGRBs such as the prototypical GRB-SN LLGRB 980425/SN 1998bw \citep{Galama1998} and XRFs often have much lower radio luminosities than typical GRBs. Our luminosity limits are $\sim 3$--$5 \times$ lower than the radio emission associated with LLGRB 980425/SN 1998bw at a similar frequency and epoch \citep{1998Natur.395..663K, 1998ApJ...497..288W}. We are also able to rule out emission similar to XRF 020903, LLGRB 031203/SN 2003lw, and XRF 030723 \citep{2004ApJ...606..994S, 2004Natur.430..648S, Chandra2012} as well as EP240414A/SN 2024gsa \citep{Bright2025}, as our limits are $\sim 2$ orders of magnitude lower. However, we are unable to rule out low-luminosity ($\sim 10^{27}~{\rm erg~s}^{-1}{\rm Hz}^{-1}$), fast-fading radio emission, such as that associated with XRF060218/SN 2006aj, XRF 080109/SN 2008d, XRF 100316D/SN 2010bh \citep{2006Natur.442.1014S, 2008Natur.453..469S, 2013ApJ...778...18M} and the double-peaked SN Ic-BL 2020bvc \citep{Ho2020b}, at a similar frequency and epoch. Similar luminosity limits, and as a result constraints, were achieved for the recent EP250108a/SN 2025kg \citep{Srinivasaragavan2025b}. 

If EP250827b was the result of an off-axis GRB, we may expect to see late-rising radio emission as the GRB jet decelerates and spreads. Several studies searching for late-rising radio emission following SNe Ic-BL have been performed, with some promising candidates  \citep{2003ApJ...599..408B, Soderberg+2006, Corsi2016, Corsi2024, 2025ApJ...995...61S}.
We compare our limits to GRB afterglow models generated by the \texttt{FIREFLY} code \citep{2024ApJ...976..252D}, where we assume a jet opening angle of $\theta_{\rm j} \approx 7^\circ$, electron powerlaw index $p = 2.133$, and fractions of energy imparted on the electrons and magnetic field of $\epsilon_{\rm e} = 0.1$ and $\epsilon_{\rm B} = 0.01$, respectively, similar to values derived from afterglow modeling of LGRBs \citep{2002ApJ...571..779P, 2003ApJ...597..459Y, 2010ApJ...711..641C, Cenko2011, 2015ApJ...799....3R, 2021ApJ...911...14K, 2022ApJ...940...53S}. We vary the values of the total jet energy $E$ and circumburst density $n$ to determine what combinations are compatible with our radio limits. 

Based on these generated models, we find that our limits are consistent with a GRB observed  $\theta_{\rm obs}\approx 30^\circ$ off-axis with typical GRB energies of $E \approx 10^{50}$--$10^{52}~$erg  \citep[e.g.][]{2001ApJ...562L..55F, 2003ApJ...594..674B, 2014PASA...31....8G}, assuming the circumburst density is $n \approx 10^{-3}$--$10^{-1}~{\rm cm}^{-3}$, with higher energies requiring lower densities to be consistent with our limits. Additionally, a low energy on-axis ($\theta_{\rm obs} = 0^\circ$) GRB ($E\approx 10^{49}~$erg) is still consistent with our limits, even within a moderately high density environment ($n \approx 0.1~{\rm cm}^{-3}$), as such emission is expected to fade beyond detection prior to our first observation. We note that these estimates are based on only one set of parameters for $p$, $\theta_{\rm j}$, $\epsilon_{\rm e}$, and $\epsilon_{\rm B}$, and assume a uniform circumburst medium. Higher values of $p$ would not only produce brighter emission, but also would result in faster fading emission post peak. Higher values of $\epsilon_{\rm B}$ would similarly produce brighter emission, whereas lower values of $\epsilon_{\rm B}$ would produce dimmer emission \citep[e.g.][]{2025ApJ...995...61S}. As a result, our limits on $E$ and $n$ would be even more constraining for an off-axis GRB with $p > 2.133$ and/or $\epsilon_{\rm B} > 0.1$. 

The jet structure also affects the radio emission behavior, where wider $\theta_{\rm j}$ (as typically seen in in LLGRBs; \citealt{Liang2007}), would result in a later peak \citep{Sari1999}, and a shallower rise, resulting in brighter emission at earlier times. Therefore, our limits would be even more constraining for a jet with $\theta_{\rm j} > 7^\circ$. Additionally, a wind-like circumburst density would produce brighter emission at earlier times \citep{2022Univ....8..588Z}, and as a result our limits on $E$ would be more constraining for a wind-like environment. If future radio observations resulted in a detection, it would be possible to constrain at least some of these various assumed parameters \citep[e.g.][]{2025ApJ...995...61S}. 
Given the radio brightness of the host galaxy, in order to confidently detect a late rising radio afterglow, the emission must rise above $\sim 45~\mu$Jy ($\sim 2\times10^{28}~{\rm erg~s}^{-1}{\rm Hz}^{-1}$), which we predict will occur $\gtrsim 1~$year post burst for energies $E \gtrsim 10^{51}~$erg. 

Overall, EP250827b is not consistent with a a typical on-axis GRB with a radio afterglow luminosity of $\sim 10^{31}~{\rm erg~s}^{-1}{\rm Hz}^{-1}$ and energy of $10^{50}$--$10^{52}~$erg. However, a lower energy on-axis jet ($\sim 10^{49}~$erg) or a higher energy ($\sim 10^{50}$--$10^{52}~$erg) off-axis ($\theta_{\rm obs}\sim 30^{\circ}$) GRB is still consistent with our non-detections. Deep, late-time ($\gtrsim 1~$year) radio follow-up of EP250827b may reveal a rising radio component, which would support the off-axis GRB scenario. 

In addition, we note here that a magnetar central engine is favored by the optical light curve analysis (see \S \ref{magnetaranalysis}). Radio emission from an embedded magnetar wind nebula is expected to arise in these systems (e.g., \citealt{Metzger2014, Omand2018, Murase2016}). However, the emission is expected to be strongly absorbed by the SN ejecta at $\sim$weeks \citep{Metzger2014, Eftekhari2021}, and thus our 10 GHz limits primarily constrain external-shock scenarios from jets or cocoons, while meaningful wind magnetar nebula constraints require higher frequency or later epoch observations.

\section{Estimating the X-ray Prompt Emission}
\label{xrayanalysis}
In this section, we test two different models to see if they can reproduce the X-ray prompt emission described in \S \ref{xrayobs}. In particular, we check if the parameters needed to reproduce a luminosity of $\sim 10^{45} \rm{\, erg \, s^{-1}}$, timescale of $ \gtrsim  1000$ seconds, and peak energy $< 1.5$ keV are reasonable for the system. 


\subsection{Cocoon Shock Breakout}

We consider whether a mildly-relativistic ($\beta > 0.4$) cocoon shock breakout (from a stellar surface or from an extended circumstellar medium, CSM) can explain an X-ray transient with characteristic photon energy $\sim 1~\mathrm{keV}$, luminosity $L_X \sim 10^{45}\,\mathrm{erg\,s^{-1}}$, and duration $\Delta t \gtrsim 10^3\,\mathrm{s}$. We show that neither the cocoon of a successful jet nor a choked jet-cocoon can satisfy all observables simultaneously.

In relativistic, radiation-mediated shocks, the photon spectrum is pair-regulated: copious $ e^\pm $ pairs set the comoving temperature at $ T \gtrsim 50\,\mathrm{keV} $, almost independently of the breakout conditions \citep{Budnik2010,Nakar2012}, as long as the shock is not in thermal equilibrium, meaning that the electrons and ions are decoupled. \citet{Weaver1976} and \citet{Katz2010} showed that for mildly relativistic shocks ($\beta > 0.6$), there is not enough time for the plasma to generate blackbody photons in the downstream portion of the shock, meaning that cocoon must be out of thermal equilibrium. 

The observed temperature is then Doppler-boosted by a factor of order the shock Lorentz factor, giving $ T_{\rm obs} \gtrsim 50\,\Gamma\,\mathrm{keV} $. A mildly relativistic shock, as expected for a cocoon \citep{Gottlieb2018b}, is therefore pair-regulated and would produce shock breakout emission with $ T_{\rm obs} \gtrsim 100\,\mathrm{keV} $, about $\sim 2$ orders of magnitude higher than the observed emission. Off-axis viewing or de-boosting cannot rescue this picture, as the Doppler factor that would lower the apparent temperature also suppresses the luminosity by orders of magnitude, yielding values far below the measured $ L_X \sim 10^{45}\,\mathrm{erg\,s^{-1}} $. For a shock velocity of $ \beta \sim 0.35 $ in the relevant density range, the downstream temperature may reach the observed value of $ \sim 1\,\mathrm{keV} $ \citep[e.g.,][]{Ito2020,Levinson2020,Irwin2025}. We calculate the shock breakout emission from the subrelativistic ejecta in \S\ref{breakoutCSM}.

We emphasize that a choked-jet cocoon, although it may be subrelativistic by the time of breakout, is unlikely to be the source of the observed subrelativistic X-ray emission. Observations of SNe Ic-BL indicate that the subrelativistic ejecta component cannot originate from the jet-cocoon system itself and instead requires an additional component that dominates the energy budget, regardless of whether the jet successfully escapes or becomes choked \citep{Eisenberg2022}. \citet{Gottlieb2025} demonstrated that this component is naturally produced by the remnant accretion disk that accompanies jet launching and provides the bulk of the energy at subrelativistic velocities, although a subrelativistic magnetar wind could yield a similar outcome. In systems where the jet is choked, this conclusion becomes even stronger: the weaker jet generates a weaker cocoon, while the accretion disk or magnetar continues to power a massive, wide-angle outflow. The disk or magnetar-driven wind, therefore, inevitably dominates both the subrelativistic ejecta and the associated breakout-like X-ray emission. 

\subsection{SN Ejecta Shock Breakout With an Extended CSM}
\label{breakoutCSM}
Here we use the same treatment presented in \citet{Haynie2021} to calculate subrelativistic shock breakout emission of ejecta in an extended CSM environment. We assume that diffusion is the dominant rate-limiting process. The rise time of the breakout emission is 

\begin{equation}
    t_r = \frac{R_{\rm{e}}^2}{R_d v_t}
\end{equation}
where $R_{\rm{e}}$ is the radius of the extended CSM and $v_t$ is the characteristic velocity of the shock. $R_d$ is
\begin{equation}
    R_d = \frac{\kappa D v_t}{c}
\end{equation}
where $\kappa$ is the opacity, and $D$ is the mass loading factor, 
\begin{equation}
   D = \frac{\dot M}{4\pi v_t}.
\end{equation}
After integrating this expression with respect to the mass, using $M_{\rm{e}} \sim 4\pi R_{\rm{e}} D$ we substitute this into the expression for $R_d$ and get 
\begin{equation}
    R_d = \frac{\kappa v_t}{c} \frac{M_{\rm{e}}}{4\pi R_{\rm{e}}}.
\end{equation}
Then substituting this expression into $t_r$, we find 
\begin{equation}
\label{timescale}
    t_r = \frac{4 \pi R_{\rm{e}}^3 c}{\kappa M_{\rm{e}} v_t^2}.
\end{equation}

In the diffusion dominated regime, the luminosity of the shock breakout is 
\begin{equation}
    L_{\rm{SBO}} = \frac{E_{\rm{SBO}}}{t_r} \approx 4\pi D v_t^3,
\end{equation}
where $E_{\rm{SBO}}$ is the energy of the shock breakout. 

The rise time of the shock breakout should be similar to the rest-frame timescale of the observed prompt emission, if the prompt emission is generated through this mechanism. Therefore, we solve Equations 5 and 6 in parallel, for $R_{\rm{e}}$ and $M_{\rm{e}}$, taking a characteristic value of $\kappa = 0.2 \, \rm{cm^2 \, g^{-1}}$, and $v_t \sim 0.35c$, below the threshold where pair production begins to affect the emission. This velocity is consistent with the high ejecta velocity inferred from the optical spectra at early times ($\sim 0.15c$ at $t_0+3.8$ days). It further supports, on one hand, the absence of a relativistic jet (consistent with the lack of a bright X-ray afterglow or radio detection), and, on the other hand, the presence of a central engine, such as a magnetar or an accretion disk. Solving Equations 6 and 9, we derive $R_{\rm{e}} \sim 1 \times 10^{13}$ cm and $M_{\rm{e}} \sim 6 \times 10^{-6} \, \rm{M_\odot}$.

This is a very low-density CSM at $R_{\rm{e}} \sim 10^{13} $ cm, of $\sim 3\times10^{-12}$ g cm$^{-3}$. Qualitatively, this is consistent with the upper limit on the peak energy of the X-ray prompt emission, as it would take a very low-density CSM to push the prompt emission peak energy less than 1.5 keV, given such a high initial shock velocity of $0.35c$. \citet{Shiode2014} determined that in compact Wolf-Rayet stars (the progenitors of SNe Ic-BL), wave excitation and damping during the Si burning phase in the months -- decades before collpase can inflate $10^{-3} - 1 \, \rm{M_\odot}$ of material to $\sim 100$'s of $\rm{R_\odot}$. The extended radius we derive is consistent with their predictions, though EP250827b/SN 2025wkm's mass-loss rate is significantly lower than predicted by the models presented in \citet{Shiode2014}.

We now compute the expected peak energy from this breakout emission to check if it is consistent with the observed peak energy upper limit. If the radiation is thermalized, we expect the shock breakout spectrum to resemble a blackbody, where 
\begin{equation}
    T_{\rm{BB}} \approx \left(\frac{L_{\rm{SBO}}}{4\pi R_{\rm{e}}^2 \sigma_{\rm{SB}}}\right)^{1/4}
\end{equation}
where $\sigma_{\rm{SBO}}$ is the Stefan-Boltzmann constant. Substituting our known values into this expression, we find $T_{\rm{BB}} \sim 3.5 \times 10^5$ K, or a peak energy of 0.03 keV. However, this temperature only holds if the radiation is thermalized and in true thermal equilibrium. To see if this is the case, we compute the thermal coupling coefficient in an expanding gas from \citet{Nakar2010}, 

\begin{eqnarray}\label{EQ eta Def}
\nonumber
  \eta 
   &\approx&  \frac{7 \cdot 10^{5} {\rm ~s~}}{\min\{t,t_d\}}
    \left(\frac{\rho}{10^{-10}{\rm g/cm^{3}}}\right)^{-2} \left(\frac{kT_{BB}}{100 eV}\right) ^{7/2}  ,
\end{eqnarray}
where $k$ is the Boltzmann constant, and ${\min\{t,t_d\}}$ is the minimum time between the time we are computing the coefficient and the diffusion time. We compute this coefficient at the end of the prompt emission timescale of $t \sim 1000$ s. If $\eta < 1$, then the system can be approximated as being in thermal equilibrium, and the observed temperature $T = T_{\rm{BB}}$. But, if $\eta > 1$, then free-free emission will dominate the emission, and the spectrum will transition to an optically thin regime that is also modified by Comptonization of photons by neighboring electrons \citep{Nakar2010}. Substituting our known values, taking into account that the density is increased by a factor of seven due to compression of the shock \citep{WaxmanSBO}, we derive $\eta = 234$. Therefore, free-free emission dominates, and a Wien spectrum can be used to represent this system at high energies, where the true temperature becomes determined by
\begin{eqnarray}
    T{\xi(T)^2} = T_{\rm{eq}}{\eta^2}
\label{eq21}
\end{eqnarray}
where $\xi(T)$ is the Comptonization correction factor, given by 
\begin{equation}\label{EQ xi}
  \xi(T) \approx \max\left\{1,\frac{1}{2}\ln[y_{max}]\left(1.6+\ln[y_{max}]\right)\right\},
\end{equation}
where $y_{\rm{max}}$ is the Compton parameter, 
\begin{equation}\label{EQ ymax}
    y_{max} \equiv \frac{kT}{h\nu_{min}}=3  \left(\frac{\rho}{10^{-9}~{\rm g/cm^{-3}}}\right)^{-1/2} \left(\frac{T}{\rm 100
    eV}\right)^{9/4}. 
\end{equation}
Therefore, we solve Equation \ref{eq21} for $T$, substituting our known values find $T \sim 0.7$  keV. This is consistent with the peak energy upper limit derived in \S \ref{xrayanalysis}, making the SN ejecta shock breakout in an extended CSM scenario self-consistent over all observables. We note that if we had obtained spectroscopy of the transient at $< 4$ days after $t_0$, we could have searched for signatures of CSM interaction in the spectra (e.g., broad H$\alpha$, flash ionization features) to confirm the scenario we hypothesize here. This highlights the need to obtain very early-time spectroscopy of XRF-SNe in order to characterize their progenitor systems with accuracy.

\section{Modeling the Optical Emission}
\label{LCmodel}
Here we utilize numerical techniques to test different models that can reproduce the optical LCs of EP250827b/SN2025wkm. As described in \S \ref{LCanalysis}, the LC displays a prominent first peak whose origin is not known, and a second peak/plateau that is due to the the late-time SN emission.

\subsection{Modeling the Late-time SN Emission}
In order to model the late-time SN emission, we fit two different models to the LC after $\sim 7$ days, when different emission mechanisms that contribute to the first peak are mostly negligible when compared to the SN emission. This is the approach that has been historically used when modeling double-peaked stripped envelope SNe (e.g., \citealt{Kaustav2024}), and also the approach used when modeling the similar double-peaked SN 2025kg in association with EP250108a \citep{Rastinejad2025}. However, we note that it is often hard to quantify exactly when the first peak ends and the second peak begins to dominate. This is especially true in the case of EP250827b's g-band LC, as we note a plateau after the initial decay in \S \ref{LCanalysis}. Therefore, the statistical uncertainties that we report associated with parameters derived from fitting the second peak are likely an underestimate of the true uncertainties, which arise from systematic biases from fitting the second peak separately.

The two models are a radioactive decay model from \citet{Arnett1982}, commonly used to model SNe Ic-BL \citep{Taddia2018, Srinivasaragavan2024b}, and a magnetar model, where the spindown of a magnetar powers the SN \citep{Omand2024}. We perform the fits for these models through their implementations in \textsc{Redback} \citep{Sarin2024}, which is an open-source electromagnetic transient Bayesian inference software. We utilize \texttt{bilby} \citep{Ashton2019} and \texttt{Dynesty} \citep{Dynesty} to perform the sampling and derive posteriors. We also fit the models assuming a Gaussian likelihood with a systematic error added in quadrature to the statistical errors on the photometry, $\sigma_{\rm{sys}}$, to capture systematic uncertainties from combining photometry from multiple telescopes. We also allow the host-galaxy extinction $A_{V\rm{,host}}$ to vary as a free parameter, with a prior between 0 and 0.72 mag (the upper limit found in \S \ref{spectra}). We assume a blackbody spectral template for each of the models, as we find this to be a reasonable approximation of the SED even to late times through our fitting procedure in \S \ref{bolsection}.

\subsubsection{Radioactive Decay Model}
We begin by fitting the second peak of the LC to the \citet{Arnett1982} radioactive decay model. This model assumes that the peak bolometric luminosity of the SN is equal to the instantaneous heating rate from the decay of $^{56}$Ni and $^{56}$Co, assuming spherical symmetry and further radioactive inputs \citep{Valenti2008}. We account for gamma-ray leakage in the late-time LC \citep{Clocchiatti1997}, and therefore do not assume full gamma-ray trapping. The main parameters in this model that describe the bolometric LC are the nickel mass ($M_{\rm{Ni}}$) and characteristic photon diffusion time scale ($\tau_m$). $M_{\rm{Ni}}$ sets the peak luminosity, while $\tau_m$ relates to the rise ande decline time of the SN. The kinetic energy ($E_{\rm{KE}}$) and ejecta mass ($M_{\rm{ej}}$) of the SN are related through 

\begin{equation}
 M_\mathrm{ej} = \frac{\tau_\mathrm{m}^2\beta c v_\mathrm{sc}}{ {2}\kappa_\mathrm{opt}}\textrm{\! ,}
 \label{eq4}
\end{equation}
and 
\begin{equation}
  E_\mathrm{KE}  = \frac{3 v_\mathrm{sc}^2 M_\mathrm{ej}}{10}\textrm{,}
 \label{eq:vsc}
\end{equation}
\newline
where $\beta = 13.8$ \citep{Valenti2008}, $c$ is the speed of light, $\kappa_{\mathrm{opt}}$ is a constant, average optical opacity, and $v_{\rm{sc}}$ is the photospheric velocity $v_{\rm{ph}}$ at peak light.

The free parameters in our fit are $M_{\rm{ej}}$, the fraction of Nickel in the ejecta $f_{\rm{Ni}}$, the ejecta velocity $v_{\rm{ej}}$, the optical opacity $\kappa_{\rm{opt}}$, the gamma-ray opacity $\kappa_\gamma$, and the temperature where the photosphere begins to recede $T_{\rm{floor}}$. These parameters all have broad uniformed priors in the fitting procedure. We report the main explosion properties in Table \ref{radioactivetable}. The full corner plots are presented in the Appendix. We derive an extremely high $f_{\rm{Ni}} \sim 0.6$ from our modeling, which is not physically well motivated - it is expected that events powered purely by radioactive decay possess $f_{\rm{Ni}} < 0.5$ \citep{kasen2010}. Though we noted earlier that the true uncertainties are likely higher than the reported statistical errors, our result of a large $f_{\rm{Ni}}$ is consistent with the prolonged plateau in the bolometric luminosity LC reported in \S \ref{bolsection}, as events powered purely by radioactive decay are not expected to possess this plateau \citep{Arnett1982}.

\begin{figure}
    \centering
    \includegraphics[width=0.99\linewidth]{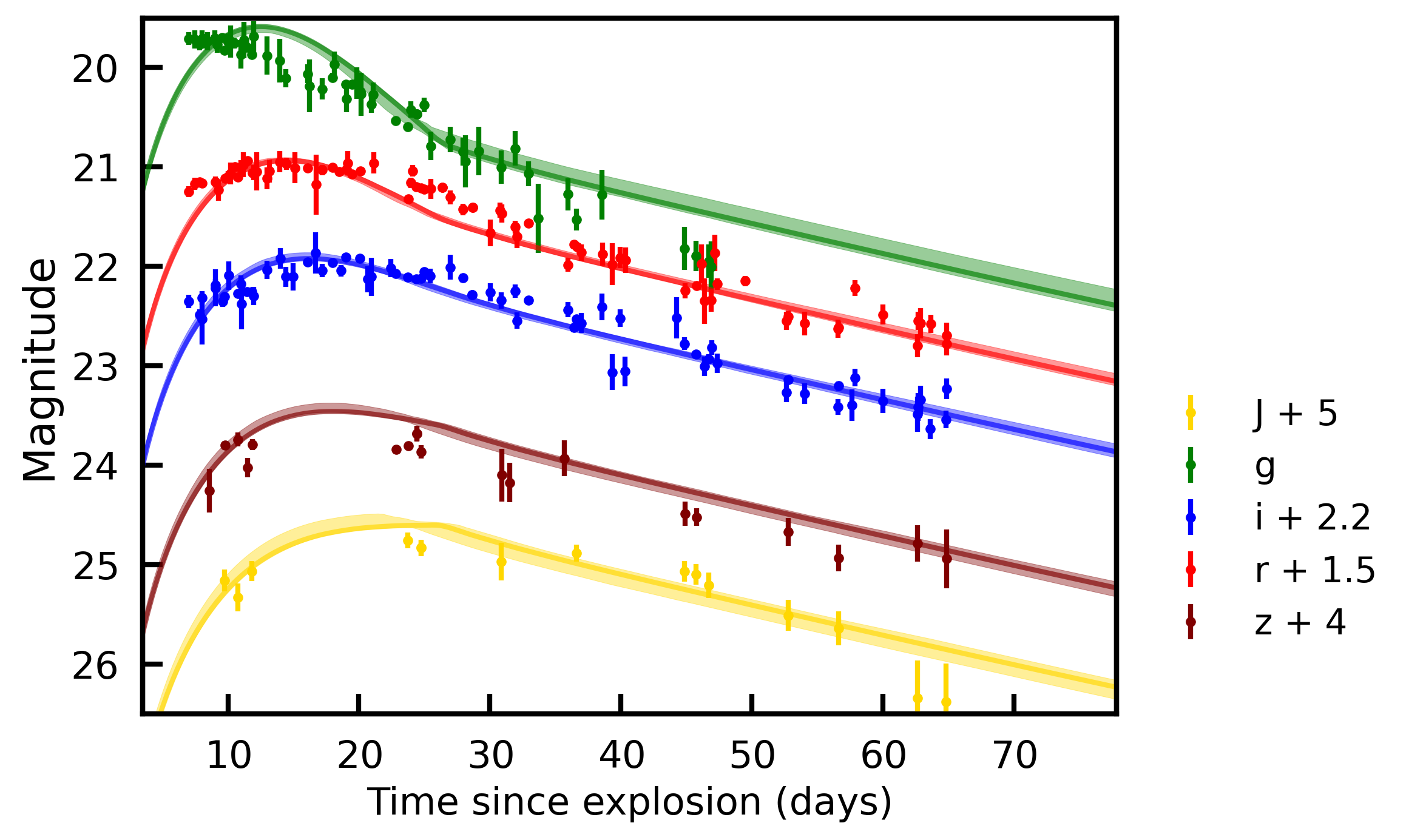}
    \caption{Fitting of radioactive decay model of \citet{Arnett1982} to the photometry of SN 2025wkm after 7 days, with the Nickel fraction set to a physically reasonable value $f_{\rm{Ni}} = 0.3$. We show the maximum likelihood fit and the 90\% credible interval.}
    \label{radioactivefigure}
\end{figure}

Events that have a higher $f_{\rm{Ni}}$ must be powered by an additional mechanism.  In Figure \ref{radioactivefigure}, we show the maximum likelihood fit and 90\% confidence interval along with the observed photometry when setting the Nickel fraction to a reasonable value $f_{\rm{Ni}}=0.3$.


\begin{deluxetable}{lr}[htb!]
\label{radioactivetable}
\tablecaption{Explosion properties of SN\,2025wkm from the \citet{Arnett1982} model} 
\tablewidth{0pt} 
\tablehead{\colhead{Parameter} & \colhead{Median}}
\tabletypesize{\normalsize} 
\startdata 
$M_{\rm{Ni}} \, (M_\odot)$ & $0.65^{+0.17}_{-0.14}$  \\
$E_{\rm{KE}}$ ($10^{51}$\,erg) & $2.12^{+0.47}_{-0.39}$ \\
$M_{\rm{ej}}$ ($M_\odot$) & $1.04^{+0.18}_{-0.08}$ \\
\enddata 
\end{deluxetable}
Some caution should be used in using the inferred explosion properties from simplified models using the Arnett model (designed for thermonuclear supernovae) on type Ib/c supernovae~\citep{Arnett17}. \citet{Khatami2019} showed that the time-dependent diffusion limits are not accurately taken into account for in Arnett-like models. Therefore, systems with longer diffusion times or more centralized heating cannot be as accurately modeled through the  \citet{Arnett1982} model. \cite{Niblett25} ran a series of radiation-transport models studying Type Ic supernovae.  The models assumed a velocity distribution for the ejecta based on a single model from a 1-dimensional hydrodynamic explosion~\citep{Fryer18}.  The light-curve calculations mapped this 1-dimensional hydrodynamics explosion into calculations assuming ballistic (i.e. homologous) outflow conditions.  A prescription to incorporate shock heating from the forward shock that converts kinetic energy to thermal energy was added to this basic homologous transport calculations.

The series of models in \cite{Niblett25} did not form a broad grid varying all the explosion conditions.  Models studied the amount of $^{56}$Ni and its level of mixing.  A smaller set of models studied the role of ejecta mass and total energy.  A small number of shock heating models were also included.  But this suite of models did not include many other initial condition properties.  For example, these models use a single explosion profile describing the velocity distribution with ejecta mass.  As has been seen in kilonova explosions, this velocity profile can have a dramatic effect on the observed light-curves~\citep{Fryer24}.   In addition, the sparsity of the suite of models from \cite{Niblett25} was not sufficient to identify direct fits to the our broadband light-curve.  But we did find reasonable fits with total ejecta mass of $2-4.5\,M_\odot$ and a total $^{56}$Ni yield of $0.2\,M_\odot$:  e.g. models ni0.2f1m0.5, ni0.2f0.25m0.5, ni0.2f1m1, ni0.2f0.25m1, ni0.2f1m1sig0.93 - see \cite{Niblett25} for the detailed description of these models.  For larger $^{56}$Ni masses, these radiation transport models predict a too-bright late-time light curve.   With a modest amount of shock heating, the required $^{56}$Ni mass could be below $<0.2\,M_\odot$.

\subsubsection{Magnetar-powered Model}
\label{magnetaranalysis}
Some SNe Ic-BL possess explosion energies $\sim 10$ times higher than normal SNe, or a Nickel fraction with respect to the $M_{\rm{ej}}$ of $f_{\rm{Ni}} > 0.5$ \citep{Taddia2018, Srinivasaragavan2024b}. These SNe Ic-BL cannot be modeled purely by the radioactive decay of $^{56}\rm{Ni}$. A popular model that has been invoked is the magnetar model, where the spindown of a magnetar remnant left behind after core-collapse powers part of the SN LC (e.g., \citealt{kasen2010, Woosley2010, Inserra2013, Dessart2019, Chen2016, Hsu2021, Yu2017}). We direct the readers to \citet{kasen2010} and \citet{Woosley2010} for more details on the model, but describe the basic picture here, following the description presented in \citet{Omand2024}. The spin-down energy gets emitted as a particle wind that is highly magnetized. This wind expands relativistically and creates a shock when it collides with the inner ejecta, and the reverse shock bounces back and shocks the expanding wind itself. This reverse shock accelerates the particles in the wind to extremely high, ultrarelativistic velocities. These particles then emit non-thermal radiation, mostly synchrotron and inverse Compton scattering \citep{Gaensler2006}, and is known as a pulsar wind nebula. The pulsar wind nebula then accelerates the SN ejecta, and the winds get mixed with the SN ejecta. This leads to the winds thermalizing, increasing the temperature and therefore the bolometric luminosity \citep{kasen2010}. 

\begin{figure}
    \centering
    \includegraphics[width=0.99\linewidth]{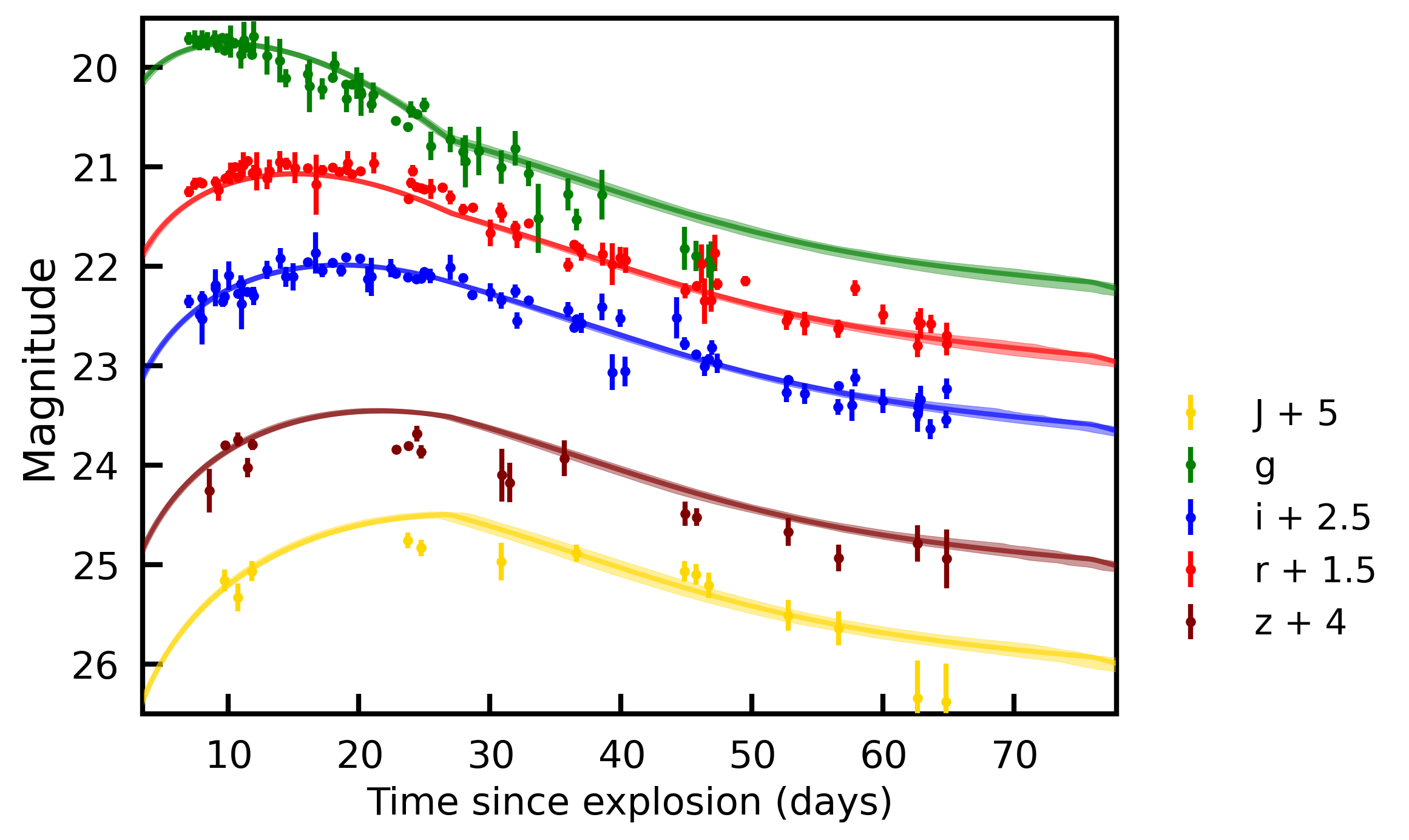}
    \caption{Fitting of the magnetar model from \citet{Omand2024} to the photometry of SN 2025wkm after 7 days. We show the maximum likelihood fit and the 90\% credible interval.}
    \label{magnetarSNfigure}
\end{figure}

This magnetar model has historically been invoked to describe superluminous SNe (SLSNe; \citealt{Nicholl2017}). \citet{Omand2024} relax the assumptions made in previous parameter estimation codes using this model and present a semi-analytic magnetar model based on previous magnetar-driven kilonova models (e.g., \citealt{Yu2013, Metzger2019, Sarin2022}), with an aim of uniting SNe Ic-BL and SLSNe under one theoretical framework. We use the implementation of this model in \texttt{Redback} to fit the main SN LC after 7 days. We direct the reader to \citet{Omand2024} for the full set of equations describing this model. The main free parameters are $L_0$ , the initial spin-down luminosity of the magnetar, and $t_{\rm{SD}}$, the spin down time. Using these parameters, we can derive the initial magnetic field of the magnetar $B$, along with the spin period $P_0$ through the relations 

\begin{align}
    L_0 = & 2.0 \times 10^{47} P_{0, -3}^{-4} B_{14}^2  \, \rm{erg \, s^{-1}}\label{eqn:l0scale}, \\
    t_{\rm SD}= & 1.3 \times 10^5 P_{0, -3}^2 B_{14}^{-2} \left( \frac{M_{\rm NS}}{1.4 M_\odot} \right) \, \rm{s}, \label{eqn:tsdscale}
\end{align}
where $P_{0, -3}= P_0/10^{-3}$ in seconds and $B_{14} = B/10^{14}$ in Gauss \citep{Omand2024}. $M_{\rm{NS}}$ is the mass of the neutron star, which we assume is 1.4 $\rm{M_\odot}$. We also assume a moment of inertia of $1.3 \times 10^{45} \rm{\, g \, cm^2}$ (see \citealt{Omand2024} for a more detailed description of these assumptions). We show the maximum likelihood fit and 90\% confidence interval along with the observed photometry in Figure \ref{magnetarSNfigure}, and report the main explosion properties in Table \ref{magnetaretable}. The full corner plots are presented in the Appendix. We note that this model also takes into account extra contributions from the radioactive decay of Nickel in combination with the spin-down of the magnetar powering the LC, and we also report $f_{\rm{Ni}}$ and $M_{\rm{ej}}$. We derive a magnetar spin period of $\sim 1.9$ ms, and a magnetic field strength of $\sim 5\times 10^{14}$ Gauss, along with a new, much more reasonable $f_{\rm{Ni}} \sim 0.2$. The ejecta mass we derive of $M_{\rm{ej}} = 1.73^{+1.48}_{-0.18}$ is within the median range of SNe Ic-Bl presented in \citet{Taddia2018} ($1 - 7 \, \rm{M_\odot})$ and \citet{Srinivasaragavan2024} ($2 - 3 \, \rm{M_\odot}$).  The magnetar parameters that we derive are in the range of parameters derived in different works invoking a magnetar to explain the LC of EP250108a/SN 2025kg \citep{Li2025, Aguilar2025, Zhu2025}. 

\begin{deluxetable}{lr}[htb!]
\label{magnetaretable}
\tablecaption{Explosion properties of SN\,2025wkm from the magnetar model \citep{Omand2024}.} 
\tablewidth{0pt} 
\tablehead{\colhead{Parameter} & \colhead{Median}}
\tabletypesize{\normalsize} 
\startdata 
$L_0 \, (\rm{10^{47} \, erg \, s^{-1}})$ & $3.54^{+4.21}_{-1.64}$  \\
$t_{\rm{SD}}$ ($10^4$ s) & $2.59^{+1.99}_{-2.00}$ \\
$f_{\rm{Ni}}$ & $0.29^{+0.04}_{-0.05}$ \\
$M_{\rm{ej}} \, (\rm{M_\odot})$ & $1.73^{+1.48}_{-0.18}$ 
\enddata 
\end{deluxetable}


Therefore, we prefer the magnetar model over the radioactive decay model to explain the second peak of EP250827b/SN 2025wkm. We note that late-time optical observations of the tail of the LC at $\sim 1 - 2$ years after explosion can confirm whether SN 2025wkm has an additional powering source to radioactive decay, as magnetar-powered LCs are expected to decay as power-laws \citep{Wang2015}, while radioactive decay-powered LCs decay exponentially \citep{Arnett1982}.

We note here that there are alternative models that can explain a central engine pumping energy into the system, such as fallback accretion onto a black hole (e.g., \citealt{Osmar2024, Moriya2019}). We do not test this model in detail in this work, as for the typical accretion scenario, it is difficult to power an engine longer than $10^4$ seconds \citep{Osmar2024}, which is too short of a timescale to power the late-time SN LC. 

In addition, the models presented in \citet{Osmar2024} assume a progenitor star radius of $\sim 10^{13}$ cm, and that most of the ejecta mass is roughly at that radius prior to explosion. However, the progenitors of SNe Ic-BL are likely Wolf-Rayet stars with much smaller radii \citep{Woosley2006}. The derived CSM radius for SN 2025wkm is $\sim 10^{13}$ cm, making it clear that the progenitor radius must be much smaller, and that the models presented in \citet{Osmar2024} are not directly applicable to this work. However, there is a wide range of accretion parameters and more complex fallback accretion models that can be tested, which is outside of the current scope of this work. Therefore, we do not say for certain that a magnetar central engine powers this system -- we simply prefer it in our analysis.

\subsection{Modeling the First Peak}
\label{5.2}
A handful of SNe Ic-BL possess double-peaked LCs, or LCs that peak too early to be described purely by radioactive decay \citep{Modjaz2006,Whitesides2017,Ho2019, Ho2020b, Srinivasaragavan2025b, Rastinejad2025}. A detailed calculation of the fraction of SNe Ic-BL in ZTF's magnitude-limited Bright Transient Survey (BTS) that possess a double peak will be included in an upcoming paper (Vail et al. in prep.) \citet{Ho2020b} showed that out of 6 SNe Ic-BL detected in the BTS to that point, one showed a clear double peak (SN 2020bvc), a double peak could not be ruled out for another (SN 2019moc), and there was no evidence for a double peak for the other four events. Therefore, double-peaked SNe Ic-BL seem to be a minority of the overall population.

Shock cooling emission (\citealt{Grasberg1976, Falk1977, Chevalier1992, Nakar2010, Piro2010, Rabinak2011, Kaustav2024, Morag2023}) from SN ejecta interacting with an extended CSM is the usual explanation for the first peak in most stripped-envelope SNe that display double peaks. In this picture, the radius of the star does not set the photosphere. Instead, it is set by an extended shell with mass $M_{\rm{e}}$ and radius $R_{\rm{e}}$ (\citealt{Bersten2012, Nakar2014, Piro2010, Morag2023}). The shock needs to travel through the outer low-mass shell, and the emission from the breakout itself is described in \S \ref{xrayanalysis}. Afterwards, the material expands and cools over the timescales of $\sim$ days, which can produce a luminous, blue peak, due to the ejecta being extremely hot. The origin of the extended CSM in these systems is still an open question, but a likely explanation is material ejected in mass loss episodes, perhaps due to binary interactions \citep{Chevalier2012}, or stellar winds \citep{Quataert2012}. 

The large kinetic energies and high velocities of the ejecta in SNe Ic-BL suggest that, unlike ordinary core-collapse SNe, their ejecta require sustained central-engine activity \citep[e.g.,][]{Rodriguez2024}. Indeed, in EP250827b, the inferred velocity at $ \sim 4.2\,\mathrm{days} $ is $ \beta \approx 0.13 $ (see \S\ref{breakoutCSM}), indicating faster ejecta than what neutrino-driven shocks can produce \citep{2017hsn..book.1095J}. 
\citet{Eyles-Ferris2025} and \citet{Srinivasaragavan2025b} find that cooling from a jet-driven shocked cocoon \citep{Nakar2017, 2025arXiv250316242H}, either interacting with its stellar envelope or an extended CSM, might be able to explain the first optical peak for EP250108a/SN 2025kg \citep[see however][]{Gottlieb2025}. However, we showed in \S \ref{xrayanalysis} that a shocked cocoon model cannot recreate the prompt emission; therefore, we do not test the cooling emission from a shocked cocoon model in our work. The first peak may also be due to the presence of a bright optical afterglow; however,  in \S \ref{sec:RadioAnalysis}, we showed that an energetic, on-axis jet is not consistent with the radio upper limits, and we therefore do not test any on-axis optical afterglow models. Though we could not rule out a very off-axis jet from the radio upper limits, the optical emission from such a system would be too faint, and occur at too late of a time to be relevant for the emission seen in the LC, so we also do not test an off-axis optical afterglow either. 

The first peak can also be from purely shock cooling emission of the SN shock without central engine contributions \citep{Piro2021}, but we find that the CSM parameters we derive in \S \ref{xrayanalysis} needed to describe the prompt emission cannot recreate the luminosity of the first peak. We therefore remain with two engine-powered outflows that are likely to play a role in collapsars, whose cooling emission can reproduce the first peak in the LC:

(i) \emph{Accretion disk outflows}: A defining feature of collapsars is their high angular momentum, which facilitates the formation of an accretion disk. In the presence of dynamically important magnetic fields, as expected around compact objects that launch jets, these disks power their own magnetically driven, subrelativistic outflows \citep{Bopp2025, Issa2025}. Such disks may power ejecta with isotropic equivalent energies of $E_{\rm iso} \sim 10^{52}\,\mathrm{erg}$, dominating over other subrelativistic outflows at velocities $\beta \lesssim 0.3$ \citep{Gottlieb2025}, and consistent with the energies inferred for Type Ic-BL SNe \citep{Fujibayashi2023,Fujibayashi2024,Dean2024a,Dean2024b,Bopp2025}.

(ii) \emph{Magnetar outflows}: A proto–neutron star inevitably forms during core collapse prior to black hole formation. If the proto-neutron star possesses sufficient angular momentum, an $\alpha$–$\Omega$ dynamo can amplify its magnetic field, giving rise to a protomagnetar \citep{Duncan1992,Thompson1993} that launches magnetically driven winds. The properties and impact of protomagnetar outflows in this regime remain poorly constrained, partly due to the scarcity of global simulations that capture both the proto-neutron star's interior and the extended stellar envelope. Nevertheless, existing magnetorotational core-collapse simulations have shown that, during the first few seconds after core bounce, the protomagnetar drives strong, magnetically dominated winds \citep[e.g.,][]{Mosta2014,Mosta2015,Mosta2018,Kuroda2020,Aloy&Obergaulinger2021,Obergaulinger&Aloy2022}. These early outflows, however, experience intense hydrodynamic mixing and interact with the infalling stellar mantle, resulting in weak collimation and subrelativistic velocities. Protomagnetars are expected to inject $\gtrsim 10^{52}\,\mathrm{erg}$ into the collapsing star \citep{Woosley2010,Metzger2015,Suzuki2017,Shankar2021,Gottlieb2024,Omand2024}, suggesting that their wind energy and characteristic velocities are comparable to those of disk-driven outflows.

\begin{figure}
    \includegraphics[width=0.99\linewidth]{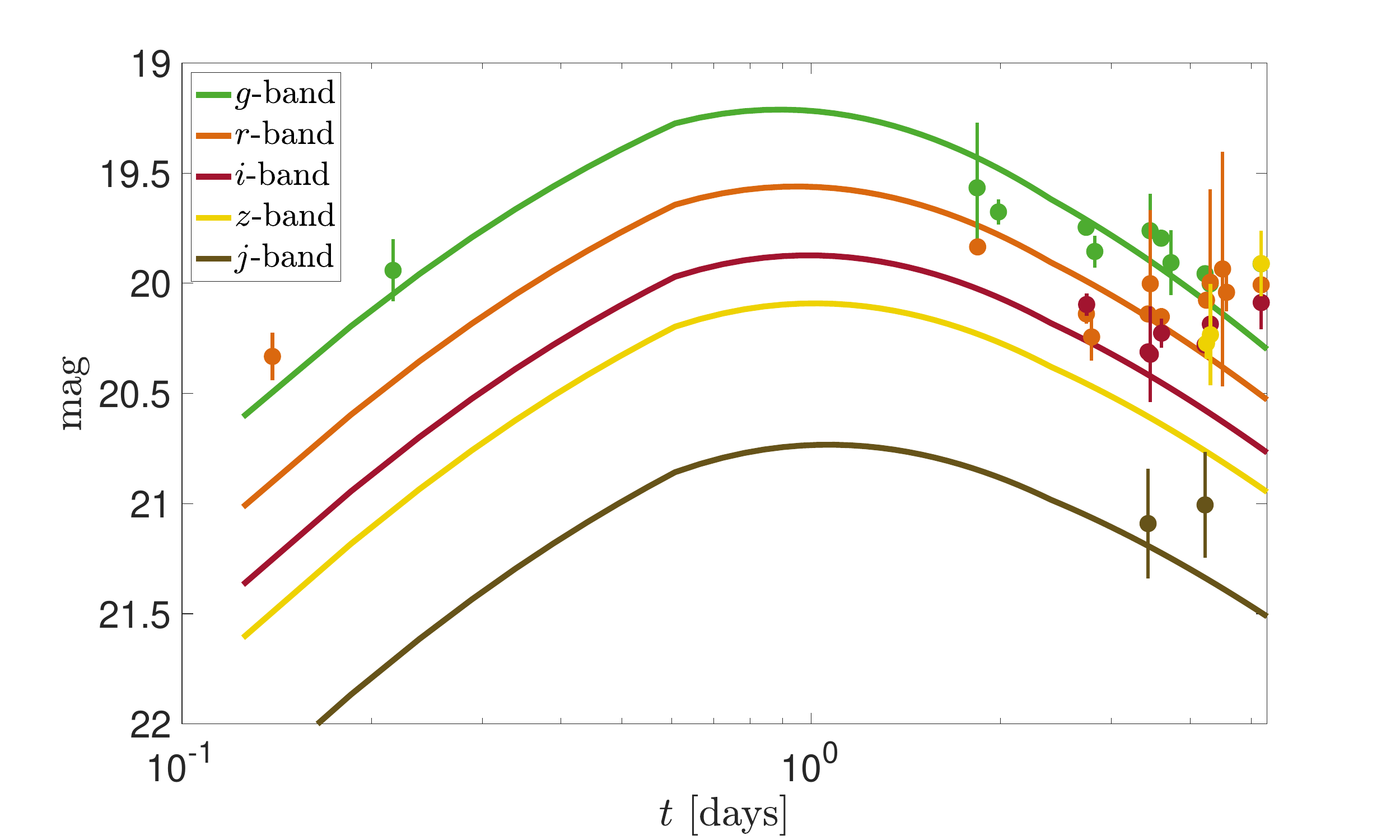}
    \caption{Semi-analytic fit of a subrelativistic energy distribution to the early time data up to $t \approx 3.5$ days.}
    \label{fig:coolingfit}
\end{figure}

The characteristics (velocity and energy) of protomagnetar-driven and disk-driven outflows may be similar to each other, and their morphology is shaped primarily by turbulent mixing, which is largely independent of the detailed nature of the energy source. Therefore, due to the challenge of distinguishing between the contributions of the two sources, we fit the early cooling emission by considering a general outflow, modeled by an energy distribution with a broken power law, $E(\beta_{\rm min}<\beta<\beta_{\rm max})$, where $ \beta_{\rm min} $ is the minimum velocity of the cooling phase, estimated to be $ \beta_{\rm min} \approx 0.13 $, and the shock breakout velocity $ \beta_{\rm max} \sim 0.35 $ (see \S\ref{breakoutCSM}). The outflow escapes from an environment of $R_{\rm{e}} \sim 10^{13}$ cm, derived in \S \ref{breakoutCSM}, and expands adiabatically to power cooling emission.


At each observed time, the trapping radius of the ejecta is a shell moving with dimensionless velocity $\beta$, mass $m$, and thermal energy $ E_{\rm th} $, assumed to be in equipartition with the kinetic component $E=m(\beta c)^2/2$ upon breakout. The shell radiates its emission at an observed time \citep{Gottlieb2023,Bopp2025}
\begin{equation}
    t = \sqrt{\frac{\kappa_{\rm{opt}} m(1-\beta)^3}{4\pi\beta c^2}}\,.
\end{equation}
The total power emitted by this shell
is thus
\begin{equation}
    L(t) \sim E_{\rm th}\left(\frac{R_{\rm e}}{\beta c t^4}\right)^{1/3}\,.
\end{equation}
 For each shell, we determine the photospheric radius and assume a blackbody temperature to compute the spectral luminosity. We fit the spectral luminosities as well as the observed temperature $ T \approx 2\times 10^4\,\mathrm{K} $ at $ \sim 4.2\,\mathrm{days} $ (from \S \ref{spectra}).

Figure~\ref{fig:coolingfit} depicts our fit to the optical data at $ t \lesssim 3.5\,\mathrm{days} $.
Our best-fit parameters imply a total (kinetic and thermal) ejecta energy of $ E \approx 2\times 10^{52}\,\mathrm{erg} $, typical for SNe Ic-BL and $ \kappa_{\rm{opt}} = 0.5\,\mathrm{cm^2\,g^{-1}} $. We find that the energy distribution is flat at $ \beta \gtrsim 0.23 $, and rises linearly with $ \beta $ at lower velocities. This profile indicates the presence of two mixed components with comparable energies of $ \sim 5\times 10^{51}\,\mathrm{erg} $ each. Since such energies exceed those expected from choked jet cocoons or neutrino-driven explosions, the two components may correspond to magnetar and accretion-disk outflows. The opacity that we derive is slightly higher than values usually assumed for SNe Ic-BL ($\kappa_{\rm{opt}} \sim 0.1\,\mathrm{cm^2\,g^{-1}}$; \citealt{Taddia2018, Srinivasaragavan2024b}), though slightly higher opacities are plausible if heavier elements are mixed into the outer ejecta. 

\begin{figure*}
    \centering
    \includegraphics[width=0.99\linewidth]{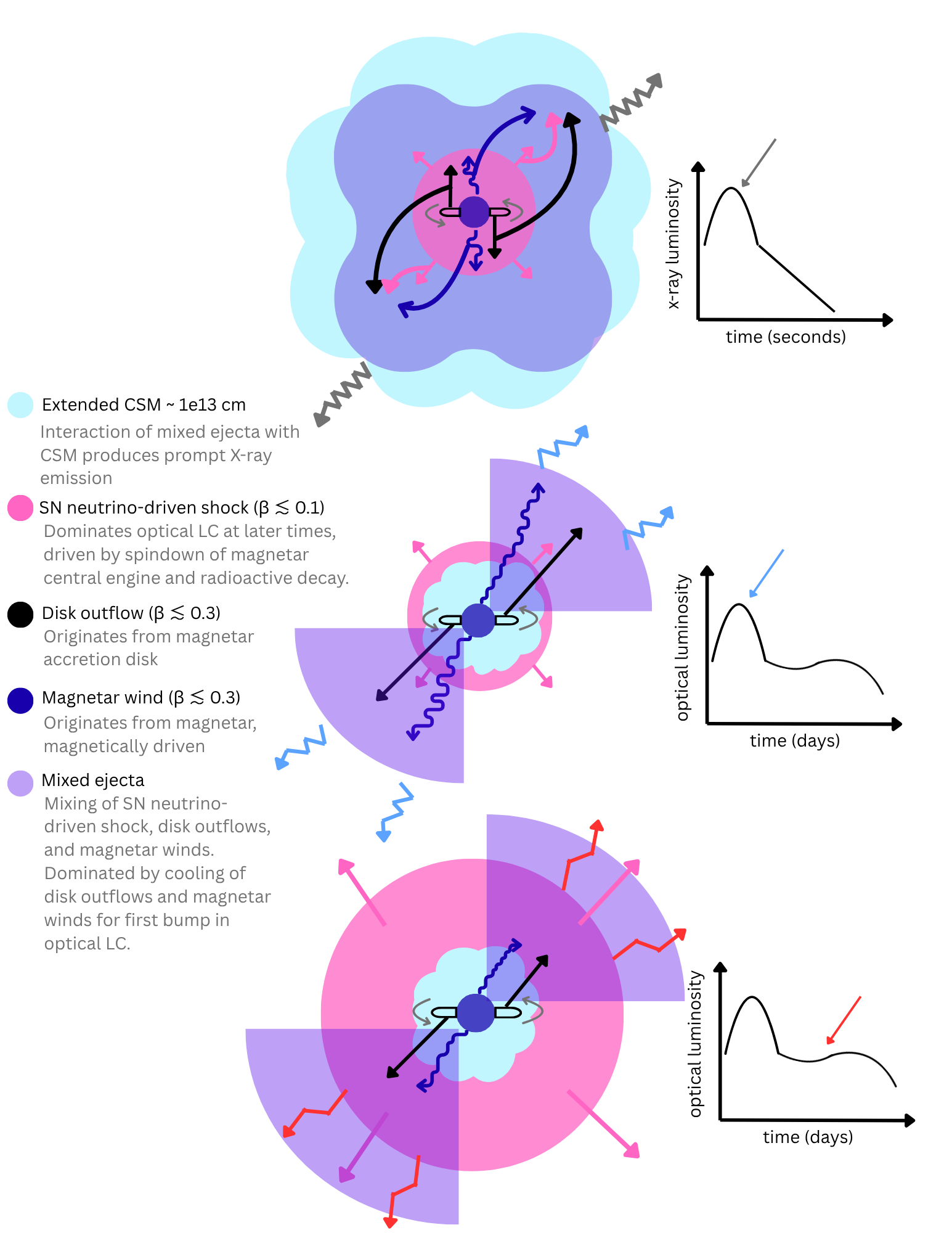}
    \caption{Artistic depiction of the magnetar-driven disk cooling model presented in \S \ref{5.2}.}
    \label{progmodel}
\end{figure*}

We show an artistic depiction of the model we invoke in Figure \ref{progmodel}. We note that we do not perform a joint fit of the first peak and late-time SN emission, due to the difference in modeling procedures used -- numerical techniques were used in modeling the late-time emission, while we used a semi-analytic model to describe the first peak, with only a narrow range of free parameters that should be expanded in future works. This semi-analytic model (described in more detail in \citealt{Gottlieb2025}) will be incorporated into \texttt{Redback} in the future, where the joint fits can be accurately represented through evolving a co-moving photosphere with a shared temperature floor.

\section{Discussion}
\label{Discussion}
Here we discuss some implications of the modeling done in \S \ref{xrayanalysis} and \S \ref{LCmodel}, and present discussion on the current landscape of XRF-SNe. 
\subsection{Circumstellar Interaction}
\label{CSMinteraction}
 In \S \ref{xrayanalysis}, we showed that all three observables corresponding to the X-ray prompt emission -- the observed timescale of $\gtrsim 1000$ s, the average 0.5 - 4 keV luminosity of $\sim 10^{45} \, \rm{erg \, s^{-1}}$, and peak energy $< 1.5$ keV -- were reproducible in a model where the X-ray emission originates from SN ejecta breaking out of an extended CSM environment. However, is the invocation of an extended CSM necessary to describe the observables in this system? \citet{WaxmanSBO} describe the energy of the shock breakout from a stellar envelope as 
\begin{equation}
    E_{\rm{SBO}} \sim 10^{46}v_{\rm{s,9}}R^2_{13}\kappa_{0.34}^{-1}.
\end{equation}
Assuming a Wolf-Rayet progenitor with $R \sim 10^{12}$ cm, $v_s \sim 0.35c$, and $\kappa = 0.2 \, \rm{cm^{2} \, g^{-1}}$, we find $E_{\rm{SBO}} \sim 2 \times 10^{45} \, \rm{erg}$, which is three orders of magnitude less than the observed prompt emission energy of $\sim 10^{48} \, \rm{erg}$. Therefore, it is clear that a shock breakout from the stellar surface of a Wolf-Rayet star cannot reproduce the X-ray prompt emission, and it is natural to invoke the presence of an extended CSM. This is consistent with stellar evolution models that show that Wolf-Rayet stars eject significant material from their envelopes prior to collapse, from binary interactions \citep{Chevalier2012} or stellar winds \citep{Quataert2012}.

Recent semi-analytic models presented by Fryer et al. (in prep) describe the thermal, free-free emission from shock breakout emission under a variety of different initial conditions arising from shock heating as the SN exits the stellar photosphere. This model was invoked to describe the prompt emission from EP250108a/SN 2025kg in \citet{Eyles-Ferris2025}. When investigating the models in Fryer et al. (in prep.) obtained through internal communication, we find that for a range of parameters corresponding to the power-law decay index of the energy distribution with respect to velocity $E(\beta) = \beta^{-p}$ and maximum Lorentz factor $\Gamma_{\rm{max}}$, that sustained emission on the order of $\sim 1000$ s at a luminosity of $\sim 10^{45} \, \rm{erg \, s^{-1}}$ that peaks at $< 1$ keV is not possible. Each of these models drop off to luminosities lower than $\sim 10^{44} \, \rm{erg \, s^{-1}}$ by $\sim 250$ s. However, when invoking the presence of an extended CSM environment, where additional energy is injected into the CSM from expansion during the time of shock heating, models are able to sustain $10^{45} \, \rm{erg \, s^{-1}}$ luminosities on the order of 1000's of seconds. 

We can estimate the mass-loss rate of SN 2025wkm's progenitor star utilizing our derived CSM parameters. We begin with
\begin{equation}
    M_{\rm{e}} = 4 \pi \int_{R_{\rm{in}}}^{R_{\rm{out}}} \rho(r) r^2 \, dr
\end{equation}
where we are integrating over the inner and outer radii of the CSM, and $\rho(r)$ is the density profile of the CSM. Assuming that the CSM is a constant density shell, and a Wolf-Rayet progenitor whose wind velocity is $v_w \sim 4000 \, \rm{km \, s^{-1}}$ \citep{Nugis2000}, we can estimate the mass-loss rate $\dot M$ as 
\begin{equation}
    \dot M \approx 4 \pi R^2 \frac{M_{\rm{e}}}{4 \pi R^2 R_{\rm{e}}} v_w = \frac{M_{\rm{e}}v_w}{R_{\rm{e}}}.
\end{equation}
Therefore, substituting our derived values for $R_{\rm{e}}$ and $M_{\rm{e}}$, we find $\dot M \sim 10^{-4} \, \rm{M_\odot \, yr^{-1}}$. This is consistent with expected mass-loss rates from Wolf-Rayet stars \citep{Nugis2000}.

The radius derived for the extended CSM in \S \ref{xrayanalysis} is $R_{\rm{e}} \sim 10^{13}$ cm, which is similar to the radius derived for EP250108a/SN 2025kg of $3.7 \times 10^{13}$ cm \citep{Srinivasaragavan2025b}, and XRF060218/SN 2006aj of $4\times10^{13}$ cm \citep{Irwin2025}. \citet{Srinivasaragavan2025b} hypothesize that the X-ray emission and first optical peak in EP250108a was due the breakout and cooling emission of a shocked cocoon formed from a collapsar jet \citep{Nakar2017, Hamidani2025}, while XRF060218/SN 2006aj has also been modeled as a failed jet within an extended CSM \citep{Nakar2015}. In addition, \citet{Sun2025} hypothesize that the first optical peak of EP240414a/SN 2024gsa was due to shock cooling emission with a very dense CSM of $R_{\rm{e}} \sim 2\times10^{14}$ cm, though the X-ray emission is due to synchrotron radiation from a weak, but successful relativistic jet. \citet{Rho2021} also find that the first optical peak for SN 2020bvc is likely from shock cooling emission at a CSM radius $R_{\rm{e}} \sim 10^{14}$ cm. Its radio properties were well described from synchrotron emission from a mildly relativistic jet, though its X-ray properties show excess flux that isn't consistent with the radio synchrotron spectrum \citep{Ho2020}. 

There are also additional SNe Ic-BL (iPTF16asu, SN 2018gep) in the literature that do not have clear double-peaked optical LCs, but display an early, luminous peak which is likely due to the combination of radioactive decay and shock cooling within an extended CSM environment \citep{Whitesides2017, Ho2019}. Both of these events do not have reported X-ray emission.  In addition, \citet{Nakar2012} reported that the X-ray prompt emission of  XRF100316D/SN 2010bh was from a relativistic shock breakout at $R_{\rm{e}} \sim 5 \times 10^{13}$ cm. This event possessed similar X-ray properties to XRF060218/SN 2006aj with a slightly higher peak energy of $E_p \sim 30 $ keV \citep{Starling2011}, though the associated SN Ic-BL did not possess a double-peaked LC \citep{Chornock2010}. Indeed, the radius derived for the envelope of the shock cooling radius from optical/NIR observations was $\sim 10^{12}$ cm (around the radius expected for a Wolf-Rayet star), which means that any interaction signatures in the optical LC will likely have faded on quick timescales. The discrepancy between the envelope radius derived using the X-ray emission and optical emission points to two different emission components for this system, one with an extended CSM, and one without.

Therefore, it is clear that interaction with an extended CSM plays a significant role in powering at least a subset of SNe Ic-BL. More significantly, every confirmed Ic-BL XRF-SN (XRF060218/SN 2006aj, XRF10316D/SN 2010bh, EP240414a/SN 2024gsa, EP250108a/SN 2025kg, and EP250827b/SN 2025wkm) shows evidence of this extended CSM interaction, whether that interaction produces the prompt X-ray emission or the first peak in the optical LC. Qualitatively, this gives evidence that XRF-SNe may have a more extended CSM surrounding their progenitors than normal SNe Ic-BL. Interestingly, \citet{Chevalier2008} also hypothesize that the late-time optical emission of XRF080109/SN 2008D, a Type Ib SN, is powered by CSM interaction. The fact that the late-time emission in this system is dominated by CSM interaction. This is in contrast to the early-time emission being dominated by CSM interaction in SNe Ic-BL XRF-SNe, which is likely due to the presence of the progenitor's He envelope, delaying the timescale of CSM interaction.

In Table \ref{CSMtable}, we present the peak energies and CSM properties for the three EP XRF-SN studied in detail thus far. All three events have similar peak energy constraints; however, their CSM parameters vary to a wide degree, giving evidence that the CSM environments surrounding XRF-SN likely are not homogeneous. The extremely low $M_{\rm{e}}$ for EP250827b/SN 2025wkm stands out with respect to the other two events, giving evidence that its progenitor's mass-loss history may be less extreme than when compared to the other two events.

\citet{Margutti2015} report an association between LGRBs with long durations ($> 1000$s), large intrinsic absorptions ($N_{\rm{H,int}} > 6 \times 10^{21} \, \rm{cm^{-2}}$), and very soft photon indices ($\alpha_{\rm{X}} > 3$). They report that chance association between these three properties is extremely statistically unlikely, and they theorize that these properties arise from very turbulent mass-loss history of the progenitor star prior to collapse, enriching the surrounding CSM. EP250827b possesses a long duration and soft photon index, and the constraint on its intrinsic absorption does not rule out $N_{\rm{H,int}} > 6 \times 10^{21} \, \rm{cm^{-2}}$. As the sample of XRF-SNe grows, it will be interesting to note if they also share these associations, which would provide more evidence that the CSM environmnts around XRFs play a significant role affeting their observed emission.

The densities of these extended CSM environments for the majority of Ic-BL XRF-SNe are $\sim 10^{-11} \, \rm{g \, cm^{-3}}$ \citep{Srinivasaragavan2025b}, while we derive a density for EP250827b/SN 2025wkm of $\sim 3 \times 10^{-12} \, \rm{g \, cm^{-3}}$, around an order of magnitude smaller. As mentioned in \S \ref{xrayanalysis}, this low-density environment is necessary to explain the high speeds and temperature of the ejecta at early-times in combination with the low peak energy of the X-ray prompt emission.

\begin{figure}
    \centering
    \includegraphics[width=0.99\linewidth]{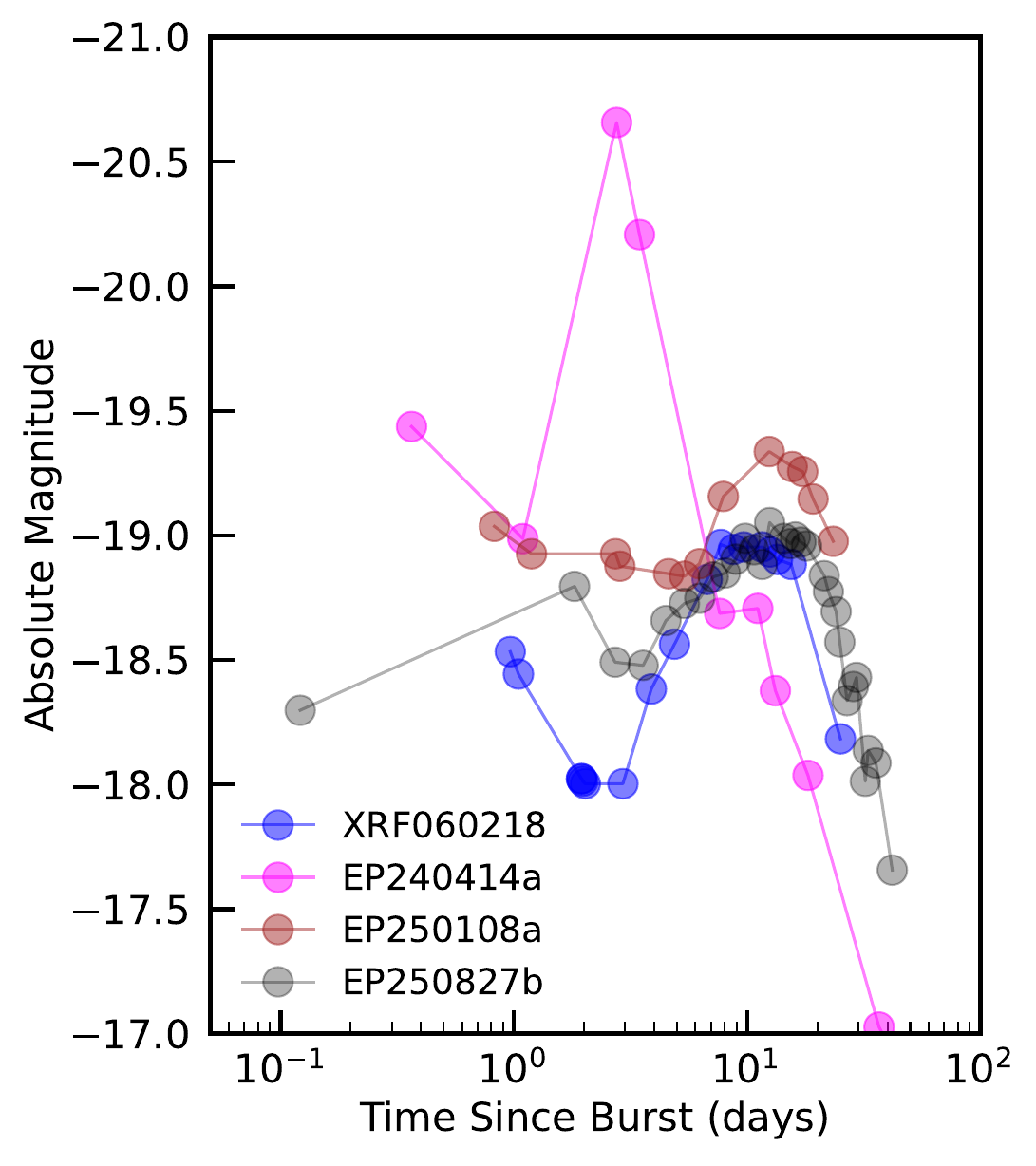}
    \caption{$r$-band rest-frame LCs of EP XRF-SNe, shown in the rest frame \citep{VanDalen2025, Srinivasaragavan2025b}, along with XRF060218 \citep{Modjaz2006}. The LCs show clear diversity in shape and brightness, ranging four orders of magnitude.}
    \label{XRFLCs}
\end{figure}

Detailed spectral analysis of EP250108a/SN 2025kg done in \citet{Rastinejad2025} finds evidence of a He feature in a near-infrared spectrum taken near peak light, along with a broad H$\alpha$ feature in an optical spectrum 42.5 days after explosion. These features are consistent with the presence of an extended CSM, due to detached He and H shells at radii $\sim 10^{15} - 10^{16}$ cm. These features in the spectra indicate the presence of a binary system, and even possibly a tertiary system when other properties are considered \citep{Rastinejad2025}.

\begin{deluxetable*}{rrrrr}
\tablecaption{Peak Energy and CSM Parameters for EP XRF-SNe.} 
\label{CSMtable}
\tablehead{\colhead{Event} & \colhead{$E_{\rm{p}}$} & \colhead{$M_{\rm{e}} \, (\rm{M_\odot})$} & \colhead{$R_{\rm{e}} \, (10^{13} \, \rm{cm})$} & \colhead{Reference}}
\tabletypesize{\normalsize} 
\startdata 
EP240414a/SN 2024gsa & $<1.8$ keV & $\sim 0.33$ & $\sim 24$ & \citet{Sun2025} \\
EP250108a/SN 2025kg & $< 1.3$ keV & $\sim 0.1$ & $\sim 3.7$ & \citet{Srinivasaragavan2025b} \\
EP250827b/SN 2025wkm & $<1.5$ keV & $6 \times 10^{-6}$ & 1 & This work \\
\enddata 
\end{deluxetable*}

We do not have any near-infrared spectra of EP250827b/SN 2025wkm, so we cannot confirm or rule out the presence of a similar He feature. We do not find any evidence for this broad H$\alpha$ feature in our spectra of EP250827b/SN 2025wkm. However, this feature quickly disappeared in EP250108a/SN 2025kg's spectra taken at 47.7 days \citep{Rastinejad2025}, so we cannot rule out that a similar situation may have occurred for EP250827b. Future theoretical studies looking at the impact that binary and tertiary companions have on the surrounding CSM environments of Wolf-Rayet stars will be key to understanding the CSM structure in more detail surrounding the progenitors of XRF-SNe.

\subsection{The Progenitor System Landscape for XRF-SNe}
This is now the third XRF-SNe discovered by EP that has been studied in detail, and the fourth overall discovered by the mission. In Figure \ref{XRFLCs}, we show the LCs for the three EP XRF-SNe presented in the literature, along with XRF060218/SN 2006aj, the most similar XRF-SN to these EP events. The LCs show clear diversity, ranging four orders of magnitude with respect to their optical lumnosities. In addition, though each of these LCs display at least two peaks, the rise and fall times of these peaks vary. In addition, EP240414a displays three possible peaks, where the initial peak has been attributed to a GRB-like afterglow \citep{Gottlieb2025,Srivastav2025,Sun2025,VanDalen2025}. 

\begin{figure}
    \centering
    \includegraphics[width=0.99\linewidth]{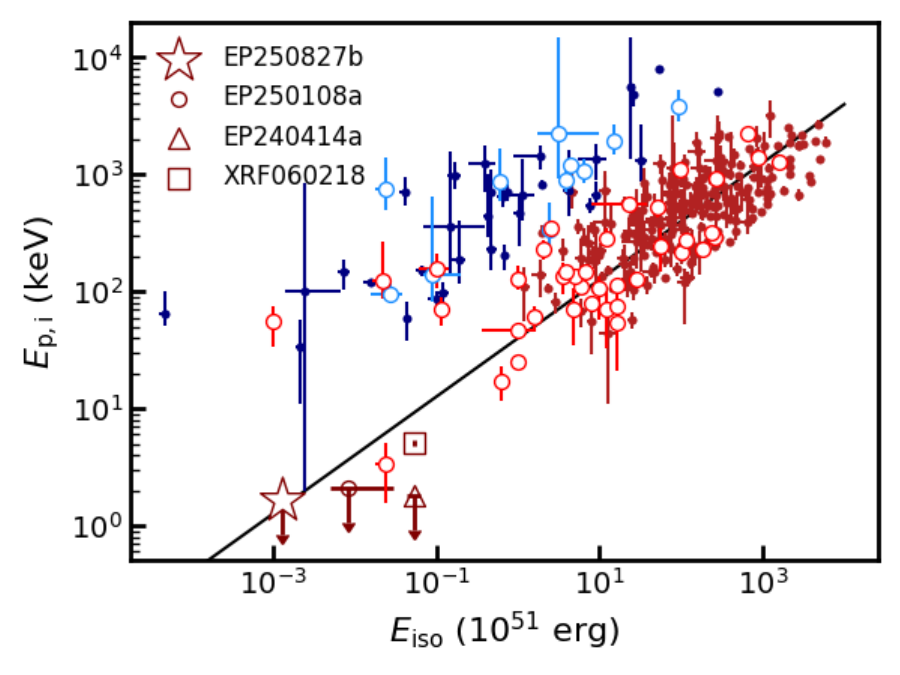}
    \caption{The \citet{Amati2002} relation, or peak energies ($E_{\rm{p, i}}$) of GRBs and XRFs plotted against their rest-frame isotropic equivalent energy  ($E_{\rm{iso}}$). GRBs are categorized into LGRBs (red), GRB-SNe (red with open circles), short GRBs (blue), short GRBs with extended emission (blue with open circles), and EP XRF-SNe (maroon) and XRF060218. The sample is created from \citet{Minaev2020}. The relation is explicitly shown for LGRBs with the black solid line. The three EP XRF-SNe have lower peak energies than expected from the relation, as the values plotted in the Figure are upper limits.  }
    \label{Amati}
\end{figure}

In Figure \ref{Amati}, we show the \citet{Amati2002} relation between the peak energies and isotropic equivalent energies of GRBs, using the sample of events from \citet{Minaev2020}. Most LGRBs and SGRBs in the figure follow the correlation well; however, the EP XRF-SNe seem to deviate from this relation. From a physical perspective, the reason for the deviation from the \citet{Amati2002} relation is likely due to the prompt emission of EP XRF-SNe originating from a different powering source than the prompt emission of GRBs. This may be due to a variety of factors - differing CSM environments, central engine powering sources, and geometric viewing effects may all play a role.

 Though not a SN Ic-BL, XRF080109/SN 2008D represents the extreme end of this different powering source. Its X-ray prompt luminosity was orders of magnitude smaller than what has been observed for Ic-BL XRF-SNe, of $\sim 10^{43} \, \rm{erg \, s^{-1}}$ \citep{Soderberg2008, Mazzali2008}. The presence of He in the SN spectra \citep{Modjaz2009, Malesani2009} indicates that the progenitor star retained its He envelope, meaning that even if a collapsar formed and launched a relativistic jet (as hypothesized by \citealt{Mazzali2008} for this system), it must have been fully choked and stifled before reaching the stellar envelope. On the other hand, EP240414a/SN 2024gsa's prompt emission had properties that skewed more similar to classical LGRBs, with a duration of $\sim 150 $ s and a luminosity of $\sim 10^{47} \, \rm{erg \, s^{-1}}$ \citep{Sun2025}. In addition, its optical LC showed an afterglow-like component at early times that displayed red colors and decayed rapidly \citep{VanDalen2025, Sun2025, Srivastav2025}, giving evidence that a jet was successfully launched and likely formed a cocoon, which played some role in creating the prompt emission. 

These two events seem to represent the current extremes of the XRF-SN landscape -- and there likely is not a standard, common progenitor system that can describe the multi-wavelength emission of XRFs. This contrasts to the story for LGRBs, where a collapsar forming an accretion disk launching a successful, relativistic jet can describe the vast majority of observed events. Indeed, it is not even clear whether a collapsar is necessary for XRF-SNe, since XRF080109/SN 2008D was a Type Ib SNe. However, for Ic-BL XRF-SNe, the formation of a collapsar must be a common feature -- the high velocities observed in the early optical spectra of SNe Ic-BL are not possible in a system described solely by a neutrino-driven shock \citep{Rodriguez2024}. 

From the three EP XRF-SNe studied in detail thus far, it is clear that the degree of success for a jet launched from collapsars is not consistent -- EP240414a likely had a successful jet  \citep{Sun2025}, EP250108a likely had a failed jet that transformed into a cocoon \citep{Eyles-Ferris2025, Srinivasaragavan2025b}, while EP250827b either did not launch a jet at all, or had a jet that was stifled deep in the layers of the explosion, where any cocoon formed was not energetic enough to impact the observed emission.  

In addition, the central engine powering mechanism in these systems is also not homogenous -- EP250108a/SN 2025kg has also been shown to possibly be powered by a magnetar \citep{Sun2025, Zhu2025}, though other works hypothesize it is powered by a black hole \citep{Eyles-Ferris2025, Srinivasaragavan2025b}. Works have shown EP240414a is also likely powered by a black hole \citep{Srivastav2025, vanDalen2024}. This heterogeneity of XRF-SNe is an important distinction that must be considered in future sample studies. XRF-SNe are important events that can shed light on the central engine populations in collapsing massive stars, motivating the need for detailed, multi-wavelength follow-up studies on every new discovery.

\section{Conclusion}
\label{Conclusion}
In this Letter, we present X-ray, ultraviolet, optical, near-infrared, and radio observations of EP250827b/SN 2025wkm, an EP XRF with a SN Ic-BL counterpart. Our major findings are:

\begin{itemize}
    \item EP250827b's X-ray prompt emission lasted at least $\sim$ 1000 seconds and has an average 0.5--4 keV luminosity of $\sim 10^{45} \, \rm{erg \, s^{-1}}$. In addition, we derive an upper limit on its peak energy of $< 1.5$ keV, allowing for an XRF classification. There is no subsequent X-ray emission detected after the prompt emission. The X-ray properties for this event are similar to those of XRF060218/SN 2006aj \citep{Mazzali2006} and EP250108a/SN 2025kg \citep{Li2025}.

    \item EP250827b's optical counterpart is SN 2025wkm, which was found through the ZTF--EP experiment \citep{Ahumada2025GCN39791}. SN 2025wkm possesses a clear double-peaked optical LC, with its first peak possessing  an absolute magnitude of at least $M_g \sim -19.3$ mag, and blue colors ($g-r \sim 0.4$). Its second peak has a peak absolute magnitude of $M_r \sim -19.2$ mag. However, its bolometric luminosity LC does not display a clear second peak, and instead plateaus until $t_0 + 20$ days before declining. This makes EP250827b/SN 2025wkm the first EP XRF-SN with a plateau in its bolometric luminosity LC, providing evidence for an extra central engine powering source.

    \item SN 2025wkm's spectrum evolve from an extremely hot ($T \sim 20,000$ K) and blue continuum at early times, to a red continuum by $t_0 + 11.1$ days, with clear broad Fe II and Si II absorption features, along with a lack of H and He features. This allowed for classification of SN 2025wkm as a SN Ic-BL. We derive extremely high ejecta velocities $\sim 40,000 \, \rm{km \, s^{-1}}$ at $t_0 + 4.2$ days. The host galaxy spectrum shows an abundance of star-formation lines.

    \item We do not detect variable radio emission from the source location of EP250827b -- therefore, we determine the radio observations are from the host galaxy, produced by a SFR of $\sim 3 \, \rm{M_\odot/yr}$. The radio upper limits rule out a classical, on-axis jet being successfully launched in this system, but do not rule out an extremely low-energetic jet of $E \sim 2\times 10^{49}$  erg, or extremely off-axis jets with $\theta_{\rm{obs}} \geq 30 ^\circ$. The radio upper limits we derive are similar to those derived for EP250108a \citep{Srinivasaragavan2025b}, and radio emission similar to XRF060218 \citep{2006Natur.442.1014S} cannot be ruled out.

    \item We determine that a failed jet transforming into a shocked cocoon cannot describe the X-ray prompt emission characteristics of EP250827b, due to the low peak energy constraint in combination with the high velocities and temperatures at early times. We determine that the X-ray prompt emission is likely due to disk- or magnetar-driven fast ejecta breaking out and interacting with an extended CSM environment, at a radius $R_{\rm{e}} \sim 10^{13}$ cm. We also derive a very low CSM density of $\sim 3 \times 10^{-12} \mathrm{g} \mathrm{cm}^{-3}$. This gives evidence that EP250827b/SN 2025wkm's progenitor star had a less extreme mass-loss history in comparison to other XRF-SNe that show evidence of interacting with an extended CSM.

    \item The \citet{Arnett1982} radioactive decay of $^{56}$Ni model that is usually used to describe the emission of stripped-envelope SNe cannot describe the late-time SN LC, as the necessary Nickel fraction with respect to the total ejecta mass is unphysical ($f_{\rm{Ni}} > 0.5$).  Comparisons with radiation-transport models suggest that the required $^{56}$Ni fraction  may be more reasonable ($f_{\rm{Ni}} \sim 0.1$).  Shock heating can also contribute to this late-time light-curve.  We also determine that a central engine is likely pumping additional energy into the system. We test a model where the spindown of a millisecond magnetar powers the LC in combination with the radioactive decay of $^{56}$Ni \citep{Omand2024}, and find that it provides a reasonable fit to the data. We derive a magnetar magnetic field strength of $B_0 \sim 5 \times 10^{14} $ Gauss and spin period of $\sim 1.9$ ms.

    \item In order to describe the first peak of the SN LC, we test a semi-analytic model from \citet{Gottlieb2025} that is consistent with the picture of fast ejecta breaking out of an extended CSM for the X-ray prompt emission, and a magnetar powering the late-time SN LC. This model invokes magnetar winds and/or disk outflows, with maximum velocities $\sim 0.35c$. The mixing of these disk outflows, magnetar winds, and neutrino-driven shock ejecta breaking out of the extended CSM produces the X-ray prompt emission, while the cooling emission of the disk outflows and magnetar winds produce the first peak in the SN LC. At later times, the spin-down of the magnetar remnant in combination with the radiocactive decay of $^{56}$Ni powers the late-time SN LC.
\end{itemize}

In \S \ref{Discussion}, we provide a detailed discussion on the implications of our major findings, and discuss the landscape of the XRF-SN field. EP is dramatically increasing the discovery rate of XRF-SNe. Every new event has provided the community with unprecedented opportunities to understand the amount of diversity regarding SNe explosion mechanisms, relativistic jet formation, central engine activity, and CSM properties surrounding massive stars. We are just now beginning to uncover this rich landscape. Future time-domain missions such as ULTRASAT \citep{Shvartzvald2025}, LSST \citep{Ivezic2019}, the La Silla Southern Sky Survey \citep{Miller2025}, and UVEX \citep{UVEX2021} will continue to probe this landscape even more, unveiling major open questions surrounding the physics of some of nature's most powerful and mysterious explosions.

\section*{Acknowledgements}
G.P.S. acknowledges Kaustav K. Das for useful conversations about understanding CSM interaction in stripped-envelope SNe, and Jillian Rastinejad for compiling the LCs presented in Figure 15. G.P.S. thanks the ZTF and EP publication boards for their useful comments on the manuscript prior to submission, along with the anonymous referee for their comments and suggestions. G.P.S. acknowledges Simi Bhullar for her moral support through the paper writing process and through G.P.S's postdoc application season. Though she is an attorney and not a scientist in the field, she is on the author list for bringing to life the artistic depiction shown in Figure 16. G.P.S. acknowledges the players on the Caltech Men's Basketball team - Josh Balami, Finnegan Fancher, Stuart Florescu, Ido Geffen, Dagemawi Getachew, Dorian Glogovac, Pranit Gunjal, Noah Hicks, Robert Jackson, Andrew Koclanes, Ryan Rodrigue, Chase Williamson, and coaches Dr. Oliver Eslinger and Seamus McKiernan. Their energy and commitment during the beginning of the 2025-2026 season made practices a welcome outlet for assistant coach G.P.S. to recharge after long days and nights of paper writing and postdoc applications.

The material is based upon work supported by NASA under award number 80GSFC24M0006. O.G. is supported by the Flatiron Research Fellowship.  B. O. is supported by the McWilliams Postdoctoral Fellowship in the McWilliams Center for Cosmology and Astrophysics at Carnegie Mellon University. M.E.R. is supported  by the European Union (ERC, Starstruck, 101095973, PI Jonker). Views and opinions expressed are however those of the author(s) only and do not necessarily reflect those of the European Union or the European Research Council Executive Agency. Neither the European Union nor the granting authority can be held responsible for them.  AA acknowledges the Yushan Young Fellow Program by the Ministry of Education, Taiwan (MOE-111-YSFMS-0008-001-P1) and National Science and Technology Council, Taiwan (NSTC grant 114-2112-M-008-021-MY3) for the financial support. M. B. is a recipient of a Student Grant from the Wübben Stiftung Wissenschaft. CJ acknowledges the National Natural Science Foundation of China through grant 12473016, and the support by the Strategic Priority Research Program of the Chinese Academy of Sciences (Grant No. XDB0550200). Funded in part by the Deutsche Forschungsgemeinschaft (DFG, German Research Foundation) under Germany's Excellence Strategy – EXC-2094 – 390783311. WXL is supported by NSFC (12120101003 and 12373010), the National Key R\&D Program (2022YFA1602902 and 2023YFA1607804), and Strategic Priority Research Program of CAS (XDB0550100 and XDB0550000). 
X.W. is supported by the National Natural Science Foundation of China (NSFC) under grants 12288102 and 12033003.
This work makes use of observations from the Las Cumbres Observatory network. The LCO team is supported by NSF grants AST-2308113 and AST-1911151. AJCT acknowledges support from the Spanish Ministry project 
PID2023-151905OB-I00 and Junta de Andaluc\'ia grant P20\_010168. Data 
were partly collected with the 1.5m telescope at the Observatorio de 
Sierra Nevada (SNO) operated by the Instituto de Astrof\'isica de 
Andaluc\'ia (IAA-CSIC). Also based on observations made with the 10.4m 
Gran Telescopio Canarias (GTC), installed at the Spanish Observatorio 
del Roque de los Muchachos of the Instituto de Astrof\'isica de 
Canarias, on the island of La Palma.

The detection of the prompt X-ray emission is based on data obtained with Einstein Probe, a space mission supported by the Strategic Priority Program on Space Science of Chinese Academy of Sciences, in collaboration with the European Space Agency, the Max-Planck-Institute for extraterrestrial Physics (Germany), and the Centre National d'Études Spatiales (France).

Based on observations obtained with the Samuel Oschin Telescope 48-inch and the 60-inch Telescope at the Palomar Observatory as part of the Zwicky Transient Facility project. ZTF is supported by the National Science Foundation under Award \#2407588 and a partnership including Caltech, USA; Caltech/IPAC, USA; University of Maryland, USA; University of California, Berkeley, USA; University of Wisconsin at Milwaukee, USA; Cornell University, USA; Drexel University, USA; University of North Carolina at Chapel Hill, USA; Institute of Science and Technology, Austria; National Central University, Taiwan, and OKC, University of Stockholm, Sweden. Operations are conducted by Caltech's Optical Observatory (COO), Caltech/IPAC, and the University of Washington at Seattle, USA.  The Liverpool Telescope is operated on the island of La Palma by Liverpool John Moores University in the Spanish Observatorio del Roque de los Muchachos of the Instituto de Astrofisica de Canarias with financial support from the UK Science and Technology Facilities Council. SED Machine is based upon work supported by the National Science Foundation under Grant No. 1106171. The Gordon and Betty Moore Foundation, through both the Data-Driven Investigator Program and a dedicated grant, provided critical funding for SkyPortal.

TRT Data Based on observations made with the Thai Robotic Telescopes under program ID TRTC12C\_001, which is operated by the National Astronomical Research Institute of Thailand (Public Organization). Based on observations obtained at the international Gemini Observatory, a program of NSF's OIR Lab, which is managed by the Association of Universities for Research in Astronomy (AURA) under a cooperative agreement with the National Science Foundation on behalf of the Gemini Observatory partnership: the National Science Foundation (United States), National Research Council (Canada), Agencia Nacional de Investigaci\'{o}n y Desarrollo (Chile), Ministerio de Ciencia, Tecnolog\'{i}a e Innovaci\'{o}n (Argentina), Minist\'{e}rio da Ci\^{e}ncia, Tecnologia, Inova\c{c}\~{o}es e Comunica\c{c}\~{o}es (Brazil), and Korea Astronomy and Space Science Institute (Republic of Korea). The authors wish to recognize and acknowledge the very significant cultural role and reverence that the summit of Maunakea has always had within the indigenous Hawaiian community.  The data were acquired through the Gemini Observatory Archive at NSF NOIRLab and processed using DRAGONS (Data Reduction for Astronomy from Gemini Observatory North and South). This paper contains data obtained at the Wendelstein Observatory of the Ludwig-Maximilians University Munich. Funded by the Deutsche Forschungsgemeinschaft (DFG, German Research Foundation) under Germany's Excellence Strategy – EXC-2094 – 390783311. Sponsored by the Natural Science Foundation of Xinjiang Uygur Autonomous Region under No.2024D01D32, Tianshan Talent Training Program grant 2023TSYCLJ0053，the National Natural Science Foundation of China NSFC No.12373038.

The National Radio Astronomy Observatory and Green Bank Observatory are facilities of the U.S. National Science Foundation operated under cooperative agreement by Associated Universities, Inc.
Based in part on observations made with the Nordic Optical Telescope, owned in collaboration by the University of Turku and Aarhus University. The data presented here were obtained with ALFOSC, which is provided by the Instituto de Astrofisica de Andalucia (IAA) under a joint agreement with the University of Copenhagen and NOT.

\bibliography{main}{}

\begin{thebibliography}{}
\expandafter\ifx\csname natexlab\endcsname\relax\def\natexlab#1{#1}\fi
\providecommand{\url}[1]{\href{#1}{#1}}
\providecommand{\dodoi}[1]{doi:~\href{http://doi.org/#1}{\nolinkurl{#1}}}
\providecommand{\doeprint}[1]{\href{http://ascl.net/#1}{\nolinkurl{http://ascl.net/#1}}}
\providecommand{\doarXiv}[1]{\href{https://arxiv.org/abs/#1}{\nolinkurl{https://arxiv.org/abs/#1}}}

\bibitem[{{Ahumada} {et~al.}(2025){Ahumada}, {Bellm}, {Yan}, {du Laz}, {Kasliwal}, {Wu}, {Jiang}, {Wu}, {Liu}, {Jin}, {Yuan}, {Zwicky Transient Facility Partnership}, \& {Einstein Probe Team}}]{Ahumada2025GCN39791}
{Ahumada}, T., {Bellm}, E.~C., {Yan}, L., {et~al.} 2025, GRB Coordinates Network, 39791, 1

\bibitem[{{Aloy} \& {Obergaulinger}(2021)}]{Aloy&Obergaulinger2021}
{Aloy}, M.~{\'A}., \& {Obergaulinger}, M. 2021, \mnras, 500, 4365, \dodoi{10.1093/mnras/staa3273}

\bibitem[{{Amati} {et~al.}(2002){Amati}, {Frontera}, {Tavani}, {in't Zand}, {Antonelli}, {Costa}, {Feroci}, {Guidorzi}, {Heise}, {Masetti}, {Montanari}, {Nicastro}, {Palazzi}, {Pian}, {Piro}, \& {Soffitta}}]{Amati2002}
{Amati}, L., {Frontera}, F., {Tavani}, M., {et~al.} 2002, \aap, 390, 81, \dodoi{10.1051/0004-6361:20020722}

\bibitem[{{Arnaud}(1996)}]{Arnaud1996}
{Arnaud}, K.~A. 1996, in Astronomical Society of the Pacific Conference Series, Vol. 101, Astronomical Data Analysis Software and Systems V, ed. G.~H. {Jacoby} \& J.~{Barnes}, 17

\bibitem[{{Arnett}(1982)}]{Arnett1982}
{Arnett}, W.~D. 1982, \apj, 253, 785, \dodoi{10.1086/159681}

\bibitem[{{Arnett} {et~al.}(2017){Arnett}, {Fryer}, \& {Matheson}}]{Arnett17}
{Arnett}, W.~D., {Fryer}, C., \& {Matheson}, T. 2017, \apj, 846, 33, \dodoi{10.3847/1538-4357/aa8173}

\bibitem[{{Aryan} {et~al.}(2025){Aryan}, {Chen}, {Yang}, {Gillanders}, {Kong}, {Smartt}, {Stevance}, {Yang}, {Aamer}, {Gupta}, {Fan}, {Hou}, {Hsiao}, {Kumar}, {Lai}, {Lee}, {Lee}, {Lin}, {Lin}, {Ngeow}, {Nicholl}, {Pan}, {Pandey}, {Sankar}, {Srivastav}, {Sun}, \& {Wang}}]{Aryan2025}
{Aryan}, A., {Chen}, T.-W., {Yang}, S., {et~al.} 2025, \apjs, 281, 20, \dodoi{10.3847/1538-4365/adfc69}

\bibitem[{{Ashton} {et~al.}(2019){Ashton}, {H{\"u}bner}, {Lasky}, {Talbot}, {Ackley}, {Biscoveanu}, {Chu}, {Divakarla}, {Easter}, {Goncharov}, {Hernandez Vivanco}, {Harms}, {Lower}, {Meadors}, {Melchor}, {Payne}, {Pitkin}, {Powell}, {Sarin}, {Smith}, \& {Thrane}}]{Ashton2019}
{Ashton}, G., {H{\"u}bner}, M., {Lasky}, P.~D., {et~al.} 2019, \apjs, 241, 27, \dodoi{10.3847/1538-4365/ab06fc}

\bibitem[{{Balberg} \& {Loeb}(2011)}]{Balberg2011}
{Balberg}, S., \& {Loeb}, A. 2011, \mnras, 414, 1715, \dodoi{10.1111/j.1365-2966.2011.18505.x}

\bibitem[{{Bellm} {et~al.}(2019){Bellm}, {Kulkarni}, {Barlow}, {Feindt}, {Graham}, {Goobar}, {Kupfer}, {Ngeow}, {Nugent}, {Ofek}, {Prince}, {Riddle}, {Walters}, \& {Ye}}]{Bellm2019}
{Bellm}, E.~C., {Kulkarni}, S.~R., {Barlow}, T., {et~al.} 2019, \pasp, 131, 068003, \dodoi{10.1088/1538-3873/ab0c2a}

\bibitem[{{Berger}(2014)}]{Berger2014}
{Berger}, E. 2014, \araa, 52, 43, \dodoi{10.1146/annurev-astro-081913-035926}

\bibitem[{{Berger} {et~al.}(2003){Berger}, {Kulkarni}, {Frail}, \& {Soderberg}}]{2003ApJ...599..408B}
{Berger}, E., {Kulkarni}, S.~R., {Frail}, D.~A., \& {Soderberg}, A.~M. 2003, \apj, 599, 408, \dodoi{10.1086/379214}

\bibitem[{{Bersier} {et~al.}(2006){Bersier}, {Fruchter}, {Strolger}, {Gorosabel}, {Levan}, {Burud}, {Rhoads}, {Becker}, {Cassan}, {Chornock}, {Covino}, {de Jong}, {Dominis}, {Filippenko}, {Hjorth}, {Holmberg}, {Malesani}, {Mobasher}, {Olsen}, {Stefanon}, {Castro Cer{\'o}n}, {Fynbo}, {Holland}, {Kouveliotou}, {Pedersen}, {Tanvir}, \& {Woosley}}]{Bersier2006}
{Bersier}, D., {Fruchter}, A.~S., {Strolger}, L.~G., {et~al.} 2006, \apj, 643, 284, \dodoi{10.1086/502640}

\bibitem[{{Bersten} {et~al.}(2012){Bersten}, {Benvenuto}, {Nomoto}, {Ergon}, {Folatelli}, {Sollerman}, {Benetti}, {Botticella}, {Fraser}, {Kotak}, {Maeda}, {Ochner}, \& {Tomasella}}]{Bersten2012}
{Bersten}, M.~C., {Benvenuto}, O.~G., {Nomoto}, K., {et~al.} 2012, \apj, 757, 31, \dodoi{10.1088/0004-637X/757/1/31}

\bibitem[{{Bertin}(2006)}]{2006ASPC..351..112B}
{Bertin}, E. 2006, in Astronomical Society of the Pacific Conference Series, Vol. 351, Astronomical Data Analysis Software and Systems XV, ed. C.~{Gabriel}, C.~{Arviset}, D.~{Ponz}, \& S.~{Enrique}, 112

\bibitem[{{Bertin} \& {Arnouts}(1996)}]{SourceExtractor}
{Bertin}, E., \& {Arnouts}, S. 1996, \aaps, 117, 393, \dodoi{10.1051/aas:1996164}

\bibitem[{{Bertin} {et~al.}(2002){Bertin}, {Mellier}, {Radovich}, {Missonnier}, {Didelon}, \& {Morin}}]{2002ASPC..281..228B}
{Bertin}, E., {Mellier}, Y., {Radovich}, M., {et~al.} 2002, in Astronomical Society of the Pacific Conference Series, Vol. 281, Astronomical Data Analysis Software and Systems XI, ed. D.~A. {Bohlender}, D.~{Durand}, \& T.~H. {Handley}, 228

\bibitem[{{Bianco} {et~al.}(2014){Bianco}, {Modjaz}, {Hicken}, {Friedman}, {Kirshner}, {Bloom}, {Challis}, {Marion}, {Wood-Vasey}, \& {Rest}}]{Bianco2014}
{Bianco}, F.~B., {Modjaz}, M., {Hicken}, M., {et~al.} 2014, \apjs, 213, 19, \dodoi{10.1088/0067-0049/213/2/19}

\bibitem[{{Blagorodnova} {et~al.}(2018){Blagorodnova}, {Neill}, {Walters}, {Kulkarni}, {Fremling}, {Ben-Ami}, {Dekany}, {Fucik}, {Konidaris}, {Nash}, {Ngeow}, {Ofek}, {O' Sullivan}, {Quimby}, {Ritter}, \& {Vyhmeister}}]{sedm}
{Blagorodnova}, N., {Neill}, J.~D., {Walters}, R., {et~al.} 2018, \pasp, 130, 035003, \dodoi{10.1088/1538-3873/aaa53f}

\bibitem[{{Blondin} \& {Tonry}(2007)}]{SNID}
{Blondin}, S., \& {Tonry}, J.~L. 2007, \apj, 666, 1024, \dodoi{10.1086/520494}

\bibitem[{{Bloom} {et~al.}(2003){Bloom}, {Frail}, \& {Kulkarni}}]{2003ApJ...594..674B}
{Bloom}, J.~S., {Frail}, D.~A., \& {Kulkarni}, S.~R. 2003, \apj, 594, 674, \dodoi{10.1086/377125}

\bibitem[{{Bopp} \& {Gottlieb}(2025)}]{Bopp2025}
{Bopp}, J., \& {Gottlieb}, O. 2025, \apjl, 982, L56, \dodoi{10.3847/2041-8213/adbdcd}

\bibitem[{{Brennan} \& {Fraser}(2022)}]{Brennan2022d}
{Brennan}, S.~J., \& {Fraser}, M. 2022, \aap, 667, A62, \dodoi{10.1051/0004-6361/202243067}

\bibitem[{{Bright} {et~al.}(2025){Bright}, {Carotenuto}, {Fender}, {Choza}, {Mummery}, {Jonker}, {Smartt}, {DeBoer}, {Farah}, {Matthews}, {Pollak}, {Rhodes}, \& {Siemion}}]{Bright2025}
{Bright}, J.~S., {Carotenuto}, F., {Fender}, R., {et~al.} 2025, \apj, 981, 48, \dodoi{10.3847/1538-4357/adaaef}

\bibitem[{{Brown} {et~al.}(2014){Brown}, {Breeveld}, {Holland}, {Kuin}, \& {Pritchard}}]{Brown2014}
{Brown}, P.~J., {Breeveld}, A.~A., {Holland}, S., {Kuin}, P., \& {Pritchard}, T. 2014, \apss, 354, 89, \dodoi{10.1007/s10509-014-2059-8}

\bibitem[{{Brown} {et~al.}(2013){Brown}, {Baliber}, {Bianco}, {Bowman}, {Burleson}, {Conway}, {Crellin}, {Depagne}, {De Vera}, {Dilday}, {Dragomir}, {Dubberley}, {Eastman}, {Elphick}, {Falarski}, {Foale}, {Ford}, {Fulton}, {Garza}, {Gomez}, {Graham}, {Greene}, {Haldeman}, {Hawkins}, {Haworth}, {Haynes}, {Hidas}, {Hjelstrom}, {Howell}, {Hygelund}, {Lister}, {Lobdill}, {Martinez}, {Mullins}, {Norbury}, {Parrent}, {Paulson}, {Petry}, {Pickles}, {Posner}, {Rosing}, {Ross}, {Sand}, {Saunders}, {Shobbrook}, {Shporer}, {Street}, {Thomas}, {Tsapras}, {Tufts}, {Valenti}, {Vander Horst}, {Walker}, {White}, \& {Willis}}]{Brown2013}
{Brown}, T.~M., {Baliber}, N., {Bianco}, F.~B., {et~al.} 2013, \pasp, 125, 1031, \dodoi{10.1086/673168}

\bibitem[{{Budnik} {et~al.}(2010){Budnik}, {Katz}, {Sagiv}, \& {Waxman}}]{Budnik2010}
{Budnik}, R., {Katz}, B., {Sagiv}, A., \& {Waxman}, E. 2010, \apj, 725, 63, \dodoi{10.1088/0004-637X/725/1/63}

\bibitem[{{Bufano} {et~al.}(2012){Bufano}, {Pian}, {Sollerman}, {Benetti}, {Pignata}, {Valenti}, {Covino}, {D'Avanzo}, {Malesani}, {Cappellaro}, {Della Valle}, {Fynbo}, {Hjorth}, {Mazzali}, {Reichart}, {Starling}, {Turatto}, {Vergani}, {Wiersema}, {Amati}, {Bersier}, {Campana}, {Cano}, {Castro-Tirado}, {Chincarini}, {D'Elia}, {de Ugarte Postigo}, {Deng}, {Ferrero}, {Filippenko}, {Goldoni}, {Gorosabel}, {Greiner}, {Hammer}, {Jakobsson}, {Kaper}, {Kawabata}, {Klose}, {Levan}, {Maeda}, {Masetti}, {Milvang-Jensen}, {Mirabel}, {M{\o}ller}, {Nomoto}, {Palazzi}, {Piranomonte}, {Salvaterra}, {Stratta}, {Tagliaferri}, {Tanaka}, {Tanvir}, \& {Wijers}}]{Bufano2012}
{Bufano}, F., {Pian}, E., {Sollerman}, J., {et~al.} 2012, \apj, 753, 67, \dodoi{10.1088/0004-637X/753/1/67}

\bibitem[{{Burrows} {et~al.}(2005){Burrows}, {Hill}, {Nousek}, {Kennea}, {Wells}, {Osborne}, {Abbey}, {Beardmore}, {Mukerjee}, {Short}, {Chincarini}, {Campana}, {Citterio}, {Moretti}, {Pagani}, {Tagliaferri}, {Giommi}, {Capalbi}, {Tamburelli}, {Angelini}, {Cusumano}, {Br{\"a}uninger}, {Burkert}, \& {Hartner}}]{Burrows2005}
{Burrows}, D.~N., {Hill}, J.~E., {Nousek}, J.~A., {et~al.} 2005, \ssr, 120, 165, \dodoi{10.1007/s11214-005-5097-2}

\bibitem[{{Busmann} {et~al.}(2025){Busmann}, {O'Connor}, {Sommer}, {Gruen}, {Beniamini}, {Gill}, {Moss}, {Palmese}, {Riffeser}, {Yang}, {Troja}, {Dichiara}, {Ricci}, {Klingler}, {G{\"o}ssl}, {Hu}, {Rau}, {Ries}, {Ryan}, {Schmidt}, {Yadav}, \& {Zeimann}}]{Busmann2025}
{Busmann}, M., {O'Connor}, B., {Sommer}, J., {et~al.} 2025, arXiv e-prints, arXiv:2503.14588, \dodoi{10.48550/arXiv.2503.14588}

\bibitem[{{Campana} {et~al.}(2006){Campana}, {Mangano}, {Blustin}, {Brown}, {Burrows}, {Chincarini}, {Cummings}, {Cusumano}, {Della Valle}, {Malesani}, {M{\'e}sz{\'a}ros}, {Nousek}, {Page}, {Sakamoto}, {Waxman}, {Zhang}, {Dai}, {Gehrels}, {Immler}, {Marshall}, {Mason}, {Moretti}, {O'Brien}, {Osborne}, {Page}, {Romano}, {Roming}, {Tagliaferri}, {Cominsky}, {Giommi}, {Godet}, {Kennea}, {Krimm}, {Angelini}, {Barthelmy}, {Boyd}, {Palmer}, {Wells}, \& {White}}]{Campana2006}
{Campana}, S., {Mangano}, V., {Blustin}, A.~J., {et~al.} 2006, \nat, 442, 1008, \dodoi{10.1038/nature04892}

\bibitem[{{Cano} {et~al.}(2017){Cano}, {Wang}, {Dai}, \& {Wu}}]{cano2017}
{Cano}, Z., {Wang}, S.-Q., {Dai}, Z.-G., \& {Wu}, X.-F. 2017, Advances in Astronomy, 2017, 8929054, \dodoi{10.1155/2017/8929054}

\bibitem[{{Cano} {et~al.}(2011){Cano}, {Bersier}, {Guidorzi}, {Kobayashi}, {Levan}, {Tanvir}, {Wiersema}, {D'Avanzo}, {Fruchter}, {Garnavich}, {Gomboc}, {Gorosabel}, {Kasen}, {Kopa{\v{c}}}, {Margutti}, {Mazzali}, {Melandri}, {Mundell}, {Nugent}, {Pian}, {Smith}, {Steele}, {Wijers}, \& {Woosley}}]{Cano2011}
{Cano}, Z., {Bersier}, D., {Guidorzi}, C., {et~al.} 2011, \apj, 740, 41, \dodoi{10.1088/0004-637X/740/1/41}

\bibitem[{{Cardelli} {et~al.}(1989){Cardelli}, {Clayton}, \& {Mathis}}]{ccm1989}
{Cardelli}, J.~A., {Clayton}, G.~C., \& {Mathis}, J.~S. 1989, \apj, 345, 245, \dodoi{10.1086/167900}

\bibitem[{{Cash}(1979)}]{Cash1979}
{Cash}, W. 1979, \apj, 228, 939, \dodoi{10.1086/156922}

\bibitem[{{Cenko} {et~al.}(2010){Cenko}, {Frail}, {Harrison}, {Kulkarni}, {Nakar}, {Chandra}, {Butler}, {Fox}, {Gal-Yam}, {Kasliwal}, {Kelemen}, {Moon}, {Ofek}, {Price}, {Rau}, {Soderberg}, {Teplitz}, {Werner}, {Bock}, {Bloom}, {Starr}, {Filippenko}, {Chevalier}, {Gehrels}, {Nousek}, \& {Piran}}]{2010ApJ...711..641C}
{Cenko}, S.~B., {Frail}, D.~A., {Harrison}, F.~A., {et~al.} 2010, \apj, 711, 641, \dodoi{10.1088/0004-637X/711/2/641}

\bibitem[{{Cenko} {et~al.}(2011){Cenko}, {Frail}, {Harrison}, {Haislip}, {Reichart}, {Butler}, {Cobb}, {Cucchiara}, {Berger}, {Bloom}, {Chandra}, {Fox}, {Perley}, {Prochaska}, {Filippenko}, {Glazebrook}, {Ivarsen}, {Kasliwal}, {Kulkarni}, {LaCluyze}, {Lopez}, {Morgan}, {Pettini}, \& {Rana}}]{Cenko2011}
---. 2011, \apj, 732, 29, \dodoi{10.1088/0004-637X/732/1/29}

\bibitem[{{Chambers} {et~al.}(2016){Chambers}, {Magnier}, {Metcalfe}, {Flewelling}, {Huber}, {Waters}, {Denneau}, {Draper}, {Farrow}, {Finkbeiner}, {Holmberg}, {Koppenhoefer}, {Price}, {Rest}, {Saglia}, {Schlafly}, {Smartt}, {Sweeney}, {Wainscoat}, {Burgett}, {Chastel}, {Grav}, {Heasley}, {Hodapp}, {Jedicke}, {Kaiser}, {Kudritzki}, {Luppino}, {Lupton}, {Monet}, {Morgan}, {Onaka}, {Shiao}, {Stubbs}, {Tonry}, {White}, {Ba{\~n}ados}, {Bell}, {Bender}, {Bernard}, {Boegner}, {Boffi}, {Botticella}, {Calamida}, {Casertano}, {Chen}, {Chen}, {Cole}, {Deacon}, {Frenk}, {Fitzsimmons}, {Gezari}, {Gibbs}, {Goessl}, {Goggia}, {Gourgue}, {Goldman}, {Grant}, {Grebel}, {Hambly}, {Hasinger}, {Heavens}, {Heckman}, {Henderson}, {Henning}, {Holman}, {Hopp}, {Ip}, {Isani}, {Jackson}, {Keyes}, {Koekemoer}, {Kotak}, {Le}, {Liska}, {Long}, {Lucey}, {Liu}, {Martin}, {Masci}, {McLean}, {Mindel}, {Misra}, {Morganson}, {Murphy}, {Obaika}, {Narayan}, {Nieto-Santisteban}, {Norberg}, {Peacock}, {Pier}, {Postman}, {Primak}, {Rae}, {Rai},
  {Riess}, {Riffeser}, {Rix}, {R{\"o}ser}, {Russel}, {Rutz}, {Schilbach}, {Schultz}, {Scolnic}, {Strolger}, {Szalay}, {Seitz}, {Small}, {Smith}, {Soderblom}, {Taylor}, {Thomson}, {Taylor}, {Thakar}, {Thiel}, {Thilker}, {Unger}, {Urata}, {Valenti}, {Wagner}, {Walder}, {Walter}, {Watters}, {Werner}, {Wood-Vasey}, \& {Wyse}}]{Chambers2016}
{Chambers}, K.~C., {Magnier}, E.~A., {Metcalfe}, N., {et~al.} 2016, arXiv e-prints, arXiv:1612.05560.
\newblock \doarXiv{1612.05560}

\bibitem[{Chambers {et~al.}(2019)Chambers, Magnier, Metcalfe, Flewelling, Huber, Waters, Denneau, Draper, Farrow, Finkbeiner, Holmberg, Koppenhoefer, Price, Rest, Saglia, Schlafly, Smartt, Sweeney, Wainscoat, Burgett, Chastel, Grav, Heasley, Hodapp, Jedicke, Kaiser, Kudritzki, Luppino, Lupton, Monet, Morgan, Onaka, Shiao, Stubbs, Tonry, White, Bañados, Bell, Bender, Bernard, Boegner, Boffi, Botticella, Calamida, Casertano, Chen, Chen, Cole, Deacon, Frenk, Fitzsimmons, Gezari, Gibbs, Goessl, Goggia, Gourgue, Goldman, Grant, Grebel, Hambly, Hasinger, Heavens, Heckman, Henderson, Henning, Holman, Hopp, Ip, Isani, Jackson, Keyes, Koekemoer, Kotak, Le, Liska, Long, Lucey, Liu, Martin, Masci, McLean, Mindel, Misra, Morganson, Murphy, Obaika, Narayan, Nieto-Santisteban, Norberg, Peacock, Pier, Postman, Primak, Rae, Rai, Riess, Riffeser, Rix, Röser, Russel, Rutz, Schilbach, Schultz, Scolnic, Strolger, Szalay, Seitz, Small, Smith, Soderblom, Taylor, Thomson, Taylor, Thakar, Thiel, Thilker, Unger, Urata, Valenti,
  Wagner, Walder, Walter, Watters, Werner, Wood-Vasey, \& Wyse}]{chambers2019panstarrs1surveys}
Chambers, K.~C., Magnier, E.~A., Metcalfe, N., {et~al.} 2019, The Pan-STARRS1 Surveys.
\newblock \doarXiv{1612.05560}

\bibitem[{{Chandra} \& {Frail}(2012)}]{Chandra2012}
{Chandra}, P., \& {Frail}, D.~A. 2012, \apj, 746, 156, \dodoi{10.1088/0004-637X/746/2/156}

\bibitem[{{Chen} {et~al.}(2016){Chen}, {Woosley}, \& {Sukhbold}}]{Chen2016}
{Chen}, K.-J., {Woosley}, S.~E., \& {Sukhbold}, T. 2016, \apj, 832, 73, \dodoi{10.3847/0004-637X/832/1/73}

\bibitem[{{Cheng} {et~al.}(2025){Cheng}, {Zhang}, {Ling}, {Sun}, {Sun}, {Liu}, {Dai}, {Jia}, {Pan}, {Wang}, {Zhao}, {Chen}, {Cheng}, {Fu}, {Han}, {Li}, {Li}, {Ma}, {Xue}, {Yan}, {Zhang}, {Wang}, {Yang}, {Zhao}, {Li}, {Jin}, \& {Yuan}}]{Cheng2025}
{Cheng}, H., {Zhang}, C., {Ling}, Z., {et~al.} 2025, Experimental Astronomy, 60, 15, \dodoi{10.1007/s10686-025-10025-9}

\bibitem[{{Chevalier}(1992)}]{Chevalier1992}
{Chevalier}, R.~A. 1992, \apj, 394, 599, \dodoi{10.1086/171612}

\bibitem[{{Chevalier}(2012)}]{Chevalier2012}
---. 2012, \apjl, 752, L2, \dodoi{10.1088/2041-8205/752/1/L2}

\bibitem[{{Chevalier} \& {Fransson}(2008)}]{Chevalier2008}
{Chevalier}, R.~A., \& {Fransson}, C. 2008, \apjl, 683, L135, \dodoi{10.1086/591522}

\bibitem[{{Chornock} {et~al.}(2010){Chornock}, {Berger}, {Levesque}, {Soderberg}, {Foley}, {Fox}, {Frebel}, {Simon}, {Bochanski}, {Challis}, {Kirshner}, {Podsiadlowski}, {Roth}, {Rutledge}, {Schmidt}, {Sheppard}, \& {Simcoe}}]{Chornock2010}
{Chornock}, R., {Berger}, E., {Levesque}, E.~M., {et~al.} 2010, arXiv e-prints, arXiv:1004.2262, \dodoi{10.48550/arXiv.1004.2262}

\bibitem[{{Clocchiatti} \& {Wheeler}(1997)}]{Clocchiatti1997}
{Clocchiatti}, A., \& {Wheeler}, J.~C. 1997, \apj, 491, 375, \dodoi{10.1086/304961}

\bibitem[{{Colgate}(1974)}]{Colgate1974}
{Colgate}, S.~A. 1974, \apj, 187, 333, \dodoi{10.1086/152632}

\bibitem[{{Corcoran} {et~al.}(2025){Corcoran}, {Levan}, {Malesani}, {Martin-Carrillo}, {Jonker}, {Eyles-Ferris}, {Vuolteenaho}, {Pursiainen}, \& {Worssam}}]{GCN41670}
{Corcoran}, G., {Levan}, A.~J., {Malesani}, D.~B., {et~al.} 2025, GRB Coordinates Network, 41670, 1

\bibitem[{{Corsi} {et~al.}(2016){Corsi}, {Gal-Yam}, {Kulkarni}, {Frail}, {Mazzali}, {Cenko}, {Kasliwal}, {Cao}, {Horesh}, {Palliyaguru}, {Perley}, {Laher}, {Taddia}, {Leloudas}, {Maguire}, {Nugent}, {Sollerman}, \& {Sullivan}}]{Corsi2016}
{Corsi}, A., {Gal-Yam}, A., {Kulkarni}, S.~R., {et~al.} 2016, \apj, 830, 42, \dodoi{10.3847/0004-637X/830/1/42}

\bibitem[{{Corsi} {et~al.}(2023){Corsi}, {Ho}, {Cenko}, {Kulkarni}, {Anand}, {Yang}, {Sollerman}, {Srinivasaragavan}, {Omand}, {Balasubramanian}, {Frail}, {Fremling}, {Perley}, {Yao}, {Dahiwale}, {De}, {Dugas}, {Hankins}, {Jencson}, {Kasliwal}, {Tzanidakis}, {Bellm}, {Laher}, {Masci}, {Purdum}, \& {Regnault}}]{Corsi2024}
{Corsi}, A., {Ho}, A. Y.~Q., {Cenko}, S.~B., {et~al.} 2023, \apj, 953, 179, \dodoi{10.3847/1538-4357/acd3f2}

\bibitem[{{Coughlin} {et~al.}(2023){Coughlin}, {Bloom}, {Nir}, {Antier}, {du Laz}, {van der Walt}, {Crellin-Quick}, {Culino}, {Duev}, {Goldstein}, {Healy}, {Karambelkar}, {Lilleboe}, {Shin}, {Singer}, {Ahumada}, {Anand}, {Bellm}, {Dekany}, {Graham}, {Kasliwal}, {Kostadinova}, {Kiendrebeogo}, {Kulkarni}, {Jenkins}, {LeBaron}, {Mahabal}, {Neill}, {Parazin}, {Peloton}, {Perley}, {Riddle}, {Rusholme}, {van Santen}, {Sollerman}, {Stein}, {Turpin}, {Wold}, {Amat}, {Bonnefon}, {Bonnefoy}, {Flament}, {Kerkow}, {Kishore}, {Jani}, {Mahanty}, {Liu}, {Llinares}, {Makarison}, {Olli{\'e}ric}, {Perez}, {Pont}, \& {Sharma}}]{Coughlin2023skyportal}
{Coughlin}, M.~W., {Bloom}, J.~S., {Nir}, G., {et~al.} 2023, \apjs, 267, 31, \dodoi{10.3847/1538-4365/acdee1}

\bibitem[{{Das} {et~al.}(2024){Das}, {Kasliwal}, {Sollerman}, {Fremling}, {Irani}, {Leung}, {Yang}, {Wu}, {Fuller}, {Anand}, {Andreoni}, {Barbarino}, {Brink}, {De}, {Dugas}, {Groom}, {Helou}, {Hinds}, {Ho}, {Karambelkar}, {Kulkarni}, {Perley}, {Purdum}, {Regnault}, {Schulze}, {Sharma}, {Sit}, {Sravan}, {Srinivasaragavan}, {Stein}, {Taggart}, {Tartaglia}, {Tzanidakis}, {Wold}, {Yan}, {Yao}, \& {Zolkower}}]{Kaustav2024}
{Das}, K.~K., {Kasliwal}, M.~M., {Sollerman}, J., {et~al.} 2024, \apj, 972, 91, \dodoi{10.3847/1538-4357/ad595f}

\bibitem[{{Dastidar} \& {Duffell}(2024)}]{2024ApJ...976..252D}
{Dastidar}, R.~G., \& {Duffell}, P.~C. 2024, \apj, 976, 252, \dodoi{10.3847/1538-4357/ad86bf}

\bibitem[{{Dean} \& {Fern{\'a}ndez}(2024{\natexlab{a}})}]{Dean2024a}
{Dean}, C., \& {Fern{\'a}ndez}, R. 2024{\natexlab{a}}, \prd, 109, 083010, \dodoi{10.1103/PhysRevD.109.083010}

\bibitem[{{Dean} \& {Fern{\'a}ndez}(2024{\natexlab{b}})}]{Dean2024b}
---. 2024{\natexlab{b}}, \prd, 110, 083024, \dodoi{10.1103/PhysRevD.110.083024}

\bibitem[{{Dekany} {et~al.}(2020){Dekany}, {Smith}, {Riddle}, {Feeney}, {Porter}, {Hale}, {Zolkower}, {Belicki}, {Kaye}, {Henning}, {Walters}, {Cromer}, {Delacroix}, {Rodriguez}, {Reiley}, {Mao}, {Hover}, {Murphy}, {Burruss}, {Baker}, {Kowalski}, {Reif}, {Mueller}, {Bellm}, {Graham}, \& {Kulkarni}}]{Dekany2020}
{Dekany}, R., {Smith}, R.~M., {Riddle}, R., {et~al.} 2020, \pasp, 132, 038001, \dodoi{10.1088/1538-3873/ab4ca2}

\bibitem[{{Dermer} {et~al.}(1999){Dermer}, {Chiang}, \& {B{\"o}ttcher}}]{Dermer1999}
{Dermer}, C.~D., {Chiang}, J., \& {B{\"o}ttcher}, M. 1999, \apj, 513, 656, \dodoi{10.1086/306871}

\bibitem[{{Dessart}(2019)}]{Dessart2019}
{Dessart}, L. 2019, \aap, 621, A141, \dodoi{10.1051/0004-6361/201834535}

\bibitem[{{Drout} {et~al.}(2014){Drout}, {Chornock}, {Soderberg}, {Sanders}, {McKinnon}, {Rest}, {Foley}, {Milisavljevic}, {Margutti}, {Berger}, {Calkins}, {Fong}, {Gezari}, {Huber}, {Kankare}, {Kirshner}, {Leibler}, {Lunnan}, {Mattila}, {Marion}, {Narayan}, {Riess}, {Roth}, {Scolnic}, {Smartt}, {Tonry}, {Burgett}, {Chambers}, {Hodapp}, {Jedicke}, {Kaiser}, {Magnier}, {Metcalfe}, {Morgan}, {Price}, \& {Waters}}]{Drout2014}
{Drout}, M.~R., {Chornock}, R., {Soderberg}, A.~M., {et~al.} 2014, \apj, 794, 23, \dodoi{10.1088/0004-637X/794/1/23}

\bibitem[{Duev {et~al.}(2019)Duev, Mahabal, Masci, Graham, Rusholme, Walters, Karmarkar, Frederick, Kasliwal, Rebbapragada, {et~al.}}]{kowalski}
Duev, D.~A., Mahabal, A., Masci, F.~J., {et~al.} 2019, Monthly Notices of the Royal Astronomical Society, 489, 3582

\bibitem[{{Duncan} \& {Thompson}(1992)}]{Duncan1992}
{Duncan}, R.~C., \& {Thompson}, C. 1992, \apjl, 392, L9, \dodoi{10.1086/186413}

\bibitem[{{Eftekhari} {et~al.}(2021){Eftekhari}, {Margalit}, {Omand}, {Berger}, {Blanchard}, {Demorest}, {Metzger}, {Murase}, {Nicholl}, {Villar}, {Williams}, {Alexander}, {Chatterjee}, {Coppejans}, {Cordes}, {Gomez}, {Hosseinzadeh}, {Hsu}, {Kashiyama}, {Margutti}, \& {Yin}}]{Eftekhari2021}
{Eftekhari}, T., {Margalit}, B., {Omand}, C.~M.~B., {et~al.} 2021, \apj, 912, 21, \dodoi{10.3847/1538-4357/abe9b8}

\bibitem[{{Eisenberg} {et~al.}(2022){Eisenberg}, {Gottlieb}, \& {Nakar}}]{Eisenberg2022}
{Eisenberg}, M., {Gottlieb}, O., \& {Nakar}, E. 2022, \mnras, 517, 582, \dodoi{10.1093/mnras/stac2184}

\bibitem[{{Evans} {et~al.}(2022){Evans}, {Page}, {Bearmore}, {Eyles-Ferris}, {Osborne}, {Campana}, {Kennea}, \& {Cenko}}]{LSXPS}
{Evans}, P.~A., {Page}, K.~L., {Bearmore}, A.~P., {et~al.} 2022, arXiv e-prints, arXiv:2208.14478.
\newblock \doarXiv{2208.14478}

\bibitem[{{Evans} {et~al.}(2009){Evans}, {Beardmore}, {Page}, {Osborne}, {O'Brien}, {Willingale}, {Starling}, {Burrows}, {Godet}, {Vetere}, {Racusin}, {Goad}, {Wiersema}, {Angelini}, {Capalbi}, {Chincarini}, {Gehrels}, {Kennea}, {Margutti}, {Morris}, {Mountford}, {Pagani}, {Perri}, {Romano}, \& {Tanvir}}]{Evans2009}
{Evans}, P.~A., {Beardmore}, A.~P., {Page}, K.~L., {et~al.} 2009, \mnras, 397, 1177, \dodoi{10.1111/j.1365-2966.2009.14913.x}

\bibitem[{{Eyles-Ferris} {et~al.}(2025){Eyles-Ferris}, {Jonker}, {Levan}, {Bj{\o}rn Malesani}, {Sarin}, {Fryer}, {Rastinejad}, {Burns}, {Tanvir}, {O'Brien}, {Fong}, {Mandel}, {Gompertz}, {Kilpatrick}, {Bloemen}, {Bright}, {Carotenuto}, {Corcoran}, {Cotter}, {Izzo}, {Laskar}, {Matin-Carrillo}, {Palmerio}, {Ravasio}, {van Roestel}, {Saccardi}, {Starling}, {Linesh Thakur}, {Vergani}, {Bauer}, {Campana}, {Chac{\'o}n}, {Chrimes}, {Covino}, {van Dalen}, {D'Elia}, {De Pasquale}, {Habeeb}, {Hartmann}, {van Hoof}, {Jakobsson}, {Julakanti}, {Leloudas}, {Mata S{\'a}nchez}, {Nixon}, {Pieterse}, {Pugliese}, {Quirola-V{\'a}squez}, {Rayson}, {Salvaterra}, {Schneider}, {Torres}, \& {Zafar}}]{Eyles-Ferris2025}
{Eyles-Ferris}, R. A.~J., {Jonker}, P.~G., {Levan}, A.~J., {et~al.} 2025, arXiv e-prints, arXiv:2504.08886.
\newblock \doarXiv{2504.08886}

\bibitem[{{Falk} \& {Arnett}(1977)}]{Falk1977}
{Falk}, S.~W., \& {Arnett}, W.~D. 1977, \apjs, 33, 515, \dodoi{10.1086/190440}

\bibitem[{{Ferrero} {et~al.}(2006){Ferrero}, {Kann}, {Zeh}, {Klose}, {Pian}, {Palazzi}, {Masetti}, {Hartmann}, {Sollerman}, {Deng}, {Filippenko}, {Greiner}, {Hughes}, {Mazzali}, {Li}, {Rol}, {Smith}, \& {Tanvir}}]{Ferrero2006}
{Ferrero}, P., {Kann}, D.~A., {Zeh}, A., {et~al.} 2006, \aap, 457, 857, \dodoi{10.1051/0004-6361:20065530}

\bibitem[{{Flewelling} {et~al.}(2020){Flewelling}, {Magnier}, {Chambers}, {Heasley}, {Holmberg}, {Huber}, {Sweeney}, {Waters}, {Calamida}, {Casertano}, {Chen}, {Farrow}, {Hasinger}, {Henderson}, {Long}, {Metcalfe}, {Narayan}, {Nieto-Santisteban}, {Norberg}, {Rest}, {Saglia}, {Szalay}, {Thakar}, {Tonry}, {Valenti}, {Werner}, {White}, {Denneau}, {Draper}, {Hodapp}, {Jedicke}, {Kaiser}, {Kudritzki}, {Price}, {Wainscoat}, {Chastel}, {McLean}, {Postman}, \& {Shiao}}]{2020ApJS..251....7F}
{Flewelling}, H.~A., {Magnier}, E.~A., {Chambers}, K.~C., {et~al.} 2020, \apjs, 251, 7, \dodoi{10.3847/1538-4365/abb82d}

\bibitem[{{Frail} {et~al.}(2001){Frail}, {Kulkarni}, {Sari}, {Djorgovski}, {Bloom}, {Galama}, {Reichart}, {Berger}, {Harrison}, {Price}, {Yost}, {Diercks}, {Goodrich}, \& {Chaffee}}]{2001ApJ...562L..55F}
{Frail}, D.~A., {Kulkarni}, S.~R., {Sari}, R., {et~al.} 2001, \apjl, 562, L55, \dodoi{10.1086/338119}

\bibitem[{{Fremling} {et~al.}(2016){Fremling}, {Sollerman}, {Taddia}, {Ergon}, {Fraser}, {Karamehmetoglu}, {Valenti}, {Jerkstrand}, {Arcavi}, {Bufano}, {Elias Rosa}, {Filippenko}, {Fox}, {Gal-Yam}, {Howell}, {Kotak}, {Mazzali}, {Milisavljevic}, {Nugent}, {Nyholm}, {Pian}, \& {Smartt}}]{FrSo2016}
{Fremling}, C., {Sollerman}, J., {Taddia}, F., {et~al.} 2016, \aap, 593, A68, \dodoi{10.1051/0004-6361/201628275}

\bibitem[{{Fryer} {et~al.}(2018){Fryer}, {Andrews}, {Even}, {Heger}, \& {Safi-Harb}}]{Fryer18}
{Fryer}, C.~L., {Andrews}, S., {Even}, W., {Heger}, A., \& {Safi-Harb}, S. 2018, \apj, 856, 63, \dodoi{10.3847/1538-4357/aaaf6f}

\bibitem[{{Fryer} {et~al.}(2024){Fryer}, {Hungerford}, {Wollaeger}, {Miller}, {De}, {Fontes}, {Korobkin}, {Kedia}, {Ristic}, \& {O'Shaughnessy}}]{Fryer24}
{Fryer}, C.~L., {Hungerford}, A.~L., {Wollaeger}, R.~T., {et~al.} 2024, \apj, 961, 9, \dodoi{10.3847/1538-4357/ad1036}

\bibitem[{{Fujibayashi} {et~al.}(2024){Fujibayashi}, {Lam}, {Shibata}, \& {Sekiguchi}}]{Fujibayashi2024}
{Fujibayashi}, S., {Lam}, A. T.-L., {Shibata}, M., \& {Sekiguchi}, Y. 2024, \prd, 109, 023031, \dodoi{10.1103/PhysRevD.109.023031}

\bibitem[{{Fujibayashi} {et~al.}(2023){Fujibayashi}, {Sekiguchi}, {Shibata}, \& {Wanajo}}]{Fujibayashi2023}
{Fujibayashi}, S., {Sekiguchi}, Y., {Shibata}, M., \& {Wanajo}, S. 2023, \apj, 956, 100, \dodoi{10.3847/1538-4357/acf5e5}

\bibitem[{{Fynbo} {et~al.}(2004){Fynbo}, {Sollerman}, {Hjorth}, {Grundahl}, {Gorosabel}, {Weidinger}, {M{\o}ller}, {Jensen}, {Vreeswijk}, {Fransson}, {Ramirez-Ruiz}, {Jakobsson}, {J{\o}rgensen}, {Vinter}, {Andersen}, {Castro Cer{\'o}n}, {Castro-Tirado}, {Fruchter}, {Greiner}, {Kouveliotou}, {Levan}, {Klose}, {Masetti}, {Pedersen}, {Palazzi}, {Pian}, {Rhoads}, {Rol}, {Sekiguchi}, {Tanvir}, {Tristram}, {de Ugarte Postigo}, {Wijers}, \& {van den Heuvel}}]{Fynbo2004}
{Fynbo}, J.~P.~U., {Sollerman}, J., {Hjorth}, J., {et~al.} 2004, \apj, 609, 962, \dodoi{10.1086/421260}

\bibitem[{{Gaensler} \& {Slane}(2006)}]{Gaensler2006}
{Gaensler}, B.~M., \& {Slane}, P.~O. 2006, \araa, 44, 17, \dodoi{10.1146/annurev.astro.44.051905.092528}

\bibitem[{{Gaia Collaboration}(2020)}]{gaiaEDR3}
{Gaia Collaboration}. 2020, VizieR Online Data Catalog, I/350

\bibitem[{{Gaia Collaboration} {et~al.}(2021){Gaia Collaboration}, {Brown}, {Vallenari}, {Prusti}, {de Bruijne}, {Babusiaux}, {Biermann}, {Creevey}, {Evans}, {Eyer}, {Hutton}, {Jansen}, {Jordi}, {Klioner}, {Lammers}, {Lindegren}, {Luri}, {Mignard}, {Panem}, {Pourbaix}, {Randich}, {Sartoretti}, {Soubiran}, {Walton}, {Arenou}, {Bailer-Jones}, {Bastian}, {Cropper}, {Drimmel}, {Katz}, {Lattanzi}, {van Leeuwen}, {Bakker}, {Cacciari}, {Casta{\~n}eda}, {De Angeli}, {Ducourant}, {Fabricius}, {Fouesneau}, {Fr{\'e}mat}, {Guerra}, {Guerrier}, {Guiraud}, {Jean-Antoine Piccolo}, {Masana}, {Messineo}, {Mowlavi}, {Nicolas}, {Nienartowicz}, {Pailler}, {Panuzzo}, {Riclet}, {Roux}, {Seabroke}, {Sordo}, {Tanga}, {Th{\'e}venin}, {Gracia-Abril}, {Portell}, {Teyssier}, {Altmann}, {Andrae}, {Bellas-Velidis}, {Benson}, {Berthier}, {Blomme}, {Brugaletta}, {Burgess}, {Busso}, {Carry}, {Cellino}, {Cheek}, {Clementini}, {Damerdji}, {Davidson}, {Delchambre}, {Dell'Oro}, {Fern{\'a}ndez-Hern{\'a}ndez}, {Galluccio}, {Garc{\'\i}a-Lario},
  {Garcia-Reinaldos}, {Gonz{\'a}lez-N{\'u}{\~n}ez}, {Gosset}, {Haigron}, {Halbwachs}, {Hambly}, {Harrison}, {Hatzidimitriou}, {Heiter}, {Hern{\'a}ndez}, {Hestroffer}, {Hodgkin}, {Holl}, {Jan{\ss}en}, {Jevardat de Fombelle}, {Jordan}, {Krone-Martins}, {Lanzafame}, {L{\"o}ffler}, {Lorca}, {Manteiga}, {Marchal}, {Marrese}, {Moitinho}, {Mora}, {Muinonen}, {Osborne}, {Pancino}, {Pauwels}, {Petit}, {Recio-Blanco}, {Richards}, {Riello}, {Rimoldini}, {Robin}, {Roegiers}, {Rybizki}, {Sarro}, {Siopis}, {Smith}, {Sozzetti}, {Ulla}, {Utrilla}, {van Leeuwen}, {van Reeven}, {Abbas}, {Abreu Aramburu}, {Accart}, {Aerts}, {Aguado}, {Ajaj}, {Altavilla}, {{\'A}lvarez}, {{\'A}lvarez Cid-Fuentes}, {Alves}, {Anderson}, {Anglada Varela}, {Antoja}, {Audard}, {Baines}, {Baker}, {Balaguer-N{\'u}{\~n}ez}, {Balbinot}, {Balog}, {Barache}, {Barbato}, {Barros}, {Barstow}, {Bartolom{\'e}}, {Bassilana}, {Bauchet}, {Baudesson-Stella}, {Becciani}, {Bellazzini}, {Bernet}, {Bertone}, {Bianchi}, {Blanco-Cuaresma}, {Boch}, {Bombrun}, {Bossini},
  {Bouquillon}, {Bragaglia}, {Bramante}, {Breedt}, {Bressan}, {Brouillet}, {Bucciarelli}, {Burlacu}, {Busonero}, {Butkevich}, {Buzzi}, {Caffau}, {Cancelliere}, {C{\'a}novas}, {Cantat-Gaudin}, {Carballo}, {Carlucci}, {Carnerero}, {Carrasco}, {Casamiquela}, {Castellani}, {Castro-Ginard}, {Castro Sampol}, {Chaoul}, {Charlot}, {Chemin}, {Chiavassa}, {Cioni}, {Comoretto}, {Cooper}, {Cornez}, {Cowell}, {Crifo}, {Crosta}, {Crowley}, {Dafonte}, {Dapergolas}, {David}, {David}, {de Laverny}, {De Luise}, {De March}, {De Ridder}, {de Souza}, {de Teodoro}, {de Torres}, {del Peloso}, {del Pozo}, {Delbo}, {Delgado}, {Delgado}, {Delisle}, {Di Matteo}, {Diakite}, {Diener}, {Distefano}, {Dolding}, {Eappachen}, {Edvardsson}, {Enke}, {Esquej}, {Fabre}, {Fabrizio}, {Faigler}, {Fedorets}, {Fernique}, {Fienga}, {Figueras}, {Fouron}, {Fragkoudi}, {Fraile}, {Franke}, {Gai}, {Garabato}, {Garcia-Gutierrez}, {Garc{\'\i}a-Torres}, {Garofalo}, {Gavras}, {Gerlach}, {Geyer}, {Giacobbe}, {Gilmore}, {Girona}, {Giuffrida}, {Gomel}, {Gomez},
  {Gonzalez-Santamaria}, {Gonz{\'a}lez-Vidal}, {Granvik}, {Guti{\'e}rrez-S{\'a}nchez}, {Guy}, {Hauser}, {Haywood}, {Helmi}, {Hidalgo}, {Hilger}, {H{\l}adczuk}, {Hobbs}, {Holland}, {Huckle}, {Jasniewicz}, {Jonker}, {Juaristi Campillo}, {Julbe}, {Karbevska}, {Kervella}, {Khanna}, {Kochoska}, {Kontizas}, {Kordopatis}, {Korn}, {Kostrzewa-Rutkowska}, {Kruszy{\'n}ska}, {Lambert}, {Lanza}, {Lasne}, {Le Campion}, {Le Fustec}, {Lebreton}, {Lebzelter}, {Leccia}, {Leclerc}, {Lecoeur-Taibi}, {Liao}, {Licata}, {Lindstr{\o}m}, {Lister}, {Livanou}, {Lobel}, {Madrero Pardo}, {Managau}, {Mann}, {Marchant}, {Marconi}, {Marcos Santos}, {Marinoni}, {Marocco}, {Marshall}, {Martin Polo}, {Mart{\'\i}n-Fleitas}, {Masip}, {Massari}, {Mastrobuono-Battisti}, {Mazeh}, {McMillan}, {Messina}, {Michalik}, {Millar}, {Mints}, {Molina}, {Molinaro}, {Moln{\'a}r}, {Montegriffo}, {Mor}, {Morbidelli}, {Morel}, {Morris}, {Mulone}, {Munoz}, {Muraveva}, {Murphy}, {Musella}, {Noval}, {Ord{\'e}novic}, {Orr{\`u}}, {Osinde}, {Pagani}, {Pagano},
  {Palaversa}, {Palicio}, {Panahi}, {Pawlak}, {Pe{\~n}alosa Esteller}, {Penttil{\"a}}, {Piersimoni}, {Pineau}, {Plachy}, {Plum}, {Poggio}, {Poretti}, {Poujoulet}, {Pr{\v{s}}a}, {Pulone}, {Racero}, {Ragaini}, {Rainer}, {Raiteri}, {Rambaux}, {Ramos}, {Ramos-Lerate}, {Re Fiorentin}, {Regibo}, {Reyl{\'e}}, {Ripepi}, {Riva}, {Rixon}, {Robichon}, {Robin}, {Roelens}, {Rohrbasser}, {Romero-G{\'o}mez}, {Rowell}, {Royer}, {Rybicki}, {Sadowski}, {Sagrist{\`a} Sell{\'e}s}, {Sahlmann}, {Salgado}, {Salguero}, {Samaras}, {Sanchez Gimenez}, {Sanna}, {Santove{\~n}a}, {Sarasso}, {Schultheis}, {Sciacca}, {Segol}, {Segovia}, {S{\'e}gransan}, {Semeux}, {Shahaf}, {Siddiqui}, {Siebert}, {Siltala}, {Slezak}, {Smart}, {Solano}, {Solitro}, {Souami}, {Souchay}, {Spagna}, {Spoto}, {Steele}, {Steidelm{\"u}ller}, {Stephenson}, {S{\"u}veges}, {Szabados}, {Szegedi-Elek}, {Taris}, {Tauran}, {Taylor}, {Teixeira}, {Thuillot}, {Tonello}, {Torra}, {Torra}, {Turon}, {Unger}, {Vaillant}, {van Dillen}, {Vanel}, {Vecchiato}, {Viala}, {Vicente},
  {Voutsinas}, {Weiler}, {Wevers}, {Wyrzykowski}, {Yoldas}, {Yvard}, {Zhao}, {Zorec}, {Zucker}, {Zurbach}, \& {Zwitter}}]{Gaia2021}
{Gaia Collaboration}, {Brown}, A.~G.~A., {Vallenari}, A., {et~al.} 2021, \aap, 649, A1, \dodoi{10.1051/0004-6361/202039657}

\bibitem[{{Gaia Collaboration} {et~al.}(2023){Gaia Collaboration}, {Montegriffo}, {Bellazzini}, {De Angeli}, {Andrae}, {Barstow}, {Bossini}, {Bragaglia}, {Burgess}, {Cacciari}, {Carrasco}, {Chornay}, {Delchambre}, {Evans}, {Fouesneau}, {Fr{\'e}mat}, {Garabato}, {Jordi}, {Manteiga}, {Massari}, {Palaversa}, {Pancino}, {Riello}, {Ruz Mieres}, {Sanna}, {Santove{\~n}a}, {Sordo}, {Vallenari}, {Walton}, {Brown}, {Prusti}, {de Bruijne}, {Arenou}, {Babusiaux}, {Biermann}, {Creevey}, {Ducourant}, {Eyer}, {Guerra}, {Hutton}, {Klioner}, {Lammers}, {Lindegren}, {Luri}, {Mignard}, {Panem}, {Pourbaix}, {Randich}, {Sartoretti}, {Soubiran}, {Tanga}, {Bailer-Jones}, {Bastian}, {Drimmel}, {Jansen}, {Katz}, {Lattanzi}, {van Leeuwen}, {Bakker}, {Casta{\~n}eda}, {Fabricius}, {Galluccio}, {Guerrier}, {Heiter}, {Masana}, {Messineo}, {Mowlavi}, {Nicolas}, {Nienartowicz}, {Pailler}, {Panuzzo}, {Riclet}, {Roux}, {Seabroke}, {Th{\'e}venin}, {Gracia-Abril}, {Portell}, {Teyssier}, {Altmann}, {Audard}, {Bellas-Velidis}, {Benson},
  {Berthier}, {Blomme}, {Busonero}, {Busso}, {C{\'a}novas}, {Carry}, {Cellino}, {Cheek}, {Clementini}, {Damerdji}, {Davidson}, {de Teodoro}, {Nu{\~n}ez Campos}, {Dell'Oro}, {Esquej}, {Fern{\'a}ndez-Hern{\'a}ndez}, {Fraile}, {Garc{\'\i}a-Lario}, {Gosset}, {Haigron}, {Halbwachs}, {Hambly}, {Harrison}, {Hern{\'a}ndez}, {Hestroffer}, {Hodgkin}, {Holl}, {Jan{\ss}en}, {Jevardat de Fombelle}, {Jordan}, {Krone-Martins}, {Lanzafame}, {L{\"o}ffler}, {Marchal}, {Marrese}, {Moitinho}, {Muinonen}, {Osborne}, {Pauwels}, {Recio-Blanco}, {Reyl{\'e}}, {Rimoldini}, {Roegiers}, {Rybizki}, {Sarro}, {Siopis}, {Smith}, {Sozzetti}, {Utrilla}, {van Leeuwen}, {Abbas}, {{\'A}brah{\'a}m}, {Abreu Aramburu}, {Aerts}, {Aguado}, {Ajaj}, {Aldea-Montero}, {Altavilla}, {{\'A}lvarez}, {Alves}, {Anderson}, {Anglada Varela}, {Antoja}, {Baines}, {Baker}, {Balaguer-N{\'u}{\~n}ez}, {Balbinot}, {Balog}, {Barache}, {Barbato}, {Barros}, {Bartolom{\'e}}, {Bassilana}, {Bauchet}, {Becciani}, {Berihuete}, {Bernet}, {Bertone}, {Bianchi}, {Binnenfeld},
  {Blanco-Cuaresma}, {Boch}, {Bombrun}, {Bouquillon}, {Bramante}, {Breedt}, {Bressan}, {Brouillet}, {Brugaletta}, {Bucciarelli}, {Burlacu}, {Butkevich}, {Buzzi}, {Caffau}, {Cancelliere}, {Cantat-Gaudin}, {Carballo}, {Carlucci}, {Carnerero}, {Casamiquela}, {Castellani}, {Castro-Ginard}, {Chaoul}, {Charlot}, {Chemin}, {Chiaramida}, {Chiavassa}, {Comoretto}, {Contursi}, {Cooper}, {Cornez}, {Cowell}, {Crifo}, {Cropper}, {Crosta}, {Crowley}, {Dafonte}, \& {Dapergolas}}]{GaiaCollaboration2023A&A...674A..33G}
{Gaia Collaboration}, {Montegriffo}, P., {Bellazzini}, M., {et~al.} 2023, \aap, 674, A33, \dodoi{10.1051/0004-6361/202243709}

\bibitem[{{Galama} {et~al.}(1998){Galama}, {Vreeswijk}, {van Paradijs}, {Kouveliotou}, {Augusteijn}, {B{\"o}hnhardt}, {Brewer}, {Doublier}, {Gonzalez}, {Leibundgut}, {Lidman}, {Hainaut}, {Patat}, {Heise}, {in't Zand}, {Hurley}, {Groot}, {Strom}, {Mazzali}, {Iwamoto}, {Nomoto}, {Umeda}, {Nakamura}, {Young}, {Suzuki}, {Shigeyama}, {Koshut}, {Kippen}, {Robinson}, {de Wildt}, {Wijers}, {Tanvir}, {Greiner}, {Pian}, {Palazzi}, {Frontera}, {Masetti}, {Nicastro}, {Feroci}, {Costa}, {Piro}, {Peterson}, {Tinney}, {Boyle}, {Cannon}, {Stathakis}, {Sadler}, {Begam}, \& {Ianna}}]{Galama1998}
{Galama}, T.~J., {Vreeswijk}, P.~M., {van Paradijs}, J., {et~al.} 1998, \nat, 395, 670, \dodoi{10.1038/27150}

\bibitem[{{Gianfagna} {et~al.}(2025){Gianfagna}, {Piro}, {Bruni}, {Linesh Thakur}, {Van Eerten}, {Castro-Tirado}, {Chen}, {Cheng}, {He}, {Jia}, {Ling}, {Maiorano}, {Paladino}, {Tripodi}, {Rossi}, {Yang}, {Yuan}, {Yuan}, \& {Zhang}}]{Gianfagna2025}
{Gianfagna}, G., {Piro}, L., {Bruni}, G., {et~al.} 2025, arXiv e-prints, arXiv:2505.05444, \dodoi{10.48550/arXiv.2505.05444}

\bibitem[{{Goldstein} {et~al.}(2019){Goldstein}, {Hamburg}, {Wood}, {Hui}, {Cleveland}, {Kocevski}, {Littenberg}, {Burns}, {Dal Canton}, {Veres}, {Mailyan}, {Malacaria}, {Briggs}, \& {Wilson-Hodge}}]{Goldstein2019}
{Goldstein}, A., {Hamburg}, R., {Wood}, J., {et~al.} 2019, arXiv e-prints, arXiv:1903.12597, \dodoi{10.48550/arXiv.1903.12597}

\bibitem[{{G{\"o}ssl} \& {Riffeser}(2002)}]{2002A&A...381.1095G}
{G{\"o}ssl}, C.~A., \& {Riffeser}, A. 2002, \aap, 381, 1095, \dodoi{10.1051/0004-6361:20011522}

\bibitem[{{Gottlieb}(2025)}]{Gottlieb2025}
{Gottlieb}, O. 2025, \apjl, 992, L3, \dodoi{10.3847/2041-8213/ae09af}

\bibitem[{{Gottlieb} {et~al.}(2022){Gottlieb}, {Liska}, {Tchekhovskoy}, {Bromberg}, {Lalakos}, {Giannios}, \& {M{\"o}sta}}]{Gottlieb2022}
{Gottlieb}, O., {Liska}, M., {Tchekhovskoy}, A., {et~al.} 2022, \apjl, 933, L9, \dodoi{10.3847/2041-8213/ac753010.48550/arXiv.2204.12501}

\bibitem[{{Gottlieb} {et~al.}(2018){Gottlieb}, {Nakar}, {Piran}, \& {Hotokezaka}}]{Gottlieb2018b}
{Gottlieb}, O., {Nakar}, E., {Piran}, T., \& {Hotokezaka}, K. 2018, \mnras, 479, 588, \dodoi{10.1093/mnras/sty1462}

\bibitem[{{Gottlieb} {et~al.}(2024){Gottlieb}, {Renzo}, {Metzger}, {Goldberg}, \& {Cantiello}}]{Gottlieb2024}
{Gottlieb}, O., {Renzo}, M., {Metzger}, B.~D., {Goldberg}, J.~A., \& {Cantiello}, M. 2024, \apjl, 976, L13, \dodoi{10.3847/2041-8213/ad8563}

\bibitem[{{Gottlieb} {et~al.}(2023){Gottlieb}, {Issa}, {Jacquemin-Ide}, {Liska}, {Tchekhovskoy}, {Foucart}, {Kasen}, {Perna}, {Quataert}, \& {Metzger}}]{Gottlieb2023}
{Gottlieb}, O., {Issa}, D., {Jacquemin-Ide}, J., {et~al.} 2023, \apjl, 953, L11, \dodoi{10.3847/2041-8213/acec4a}

\bibitem[{{Graham} {et~al.}(2019){Graham}, {Kulkarni}, {Bellm}, {Adams}, {Barbarino}, {Blagorodnova}, {Bodewits}, {Bolin}, {Brady}, {Cenko}, {Chang}, {Coughlin}, {De}, {Eadie}, {Farnham}, {Feindt}, {Franckowiak}, {Fremling}, {Gezari}, {Ghosh}, {Goldstein}, {Golkhou}, {Goobar}, {Ho}, {Huppenkothen}, {Ivezi{\'c}}, {Jones}, {Juric}, {Kaplan}, {Kasliwal}, {Kelley}, {Kupfer}, {Lee}, {Lin}, {Lunnan}, {Mahabal}, {Miller}, {Ngeow}, {Nugent}, {Ofek}, {Prince}, {Rauch}, {van Roestel}, {Schulze}, {Singer}, {Sollerman}, {Taddia}, {Yan}, {Ye}, {Yu}, {Barlow}, {Bauer}, {Beck}, {Belicki}, {Biswas}, {Brinnel}, {Brooke}, {Bue}, {Bulla}, {Burruss}, {Connolly}, {Cromer}, {Cunningham}, {Dekany}, {Delacroix}, {Desai}, {Duev}, {Feeney}, {Flynn}, {Frederick}, {Gal-Yam}, {Giomi}, {Groom}, {Hacopians}, {Hale}, {Helou}, {Henning}, {Hover}, {Hillenbrand}, {Howell}, {Hung}, {Imel}, {Ip}, {Jackson}, {Kaspi}, {Kaye}, {Kowalski}, {Kramer}, {Kuhn}, {Landry}, {Laher}, {Mao}, {Masci}, {Monkewitz}, {Murphy}, {Nordin}, {Patterson}, {Penprase},
  {Porter}, {Rebbapragada}, {Reiley}, {Riddle}, {Rigault}, {Rodriguez}, {Rusholme}, {van Santen}, {Shupe}, {Smith}, {Soumagnac}, {Stein}, {Surace}, {Szkody}, {Terek}, {Van Sistine}, {van Velzen}, {Vestrand}, {Walters}, {Ward}, {Zhang}, \& {Zolkower}}]{Graham2019}
{Graham}, M.~J., {Kulkarni}, S.~R., {Bellm}, E.~C., {et~al.} 2019, \pasp, 131, 078001, \dodoi{10.1088/1538-3873/ab006c}

\bibitem[{{Granot} \& {van der Horst}(2014)}]{2014PASA...31....8G}
{Granot}, J., \& {van der Horst}, A.~J. 2014, \pasa, 31, e008, \dodoi{10.1017/pasa.2013.44}

\bibitem[{{Grasberg} \& {Nadezhin}(1976)}]{Grasberg1976}
{Grasberg}, E.~K., \& {Nadezhin}, D.~K. 1976, \apss, 44, 409, \dodoi{10.1007/BF00642529}

\bibitem[{Hall {et~al.}(2025)Hall, Ahumada, \& Stein}]{hall_EP250827b_2025}
Hall, X.~J., Ahumada, T., \& Stein, R. 2025, GRB Coordinates Network, 41641, 1.
\newblock \url{https://ui.adsabs.harvard.edu/abs/2025GCN.41641....1H}

\bibitem[{{Hamidani} {et~al.}(2025{\natexlab{a}}){Hamidani}, {Ioka}, {Kashiyama}, \& {Tanaka}}]{2025arXiv250316242H}
{Hamidani}, H., {Ioka}, K., {Kashiyama}, K., \& {Tanaka}, M. 2025{\natexlab{a}}, arXiv e-prints, arXiv:2503.16242, \dodoi{10.48550/arXiv.2503.16242}

\bibitem[{{Hamidani} {et~al.}(2025{\natexlab{b}}){Hamidani}, {Sato}, {Kashiyama}, {Tanaka}, {Ioka}, \& {Kimura}}]{Hamidani2025}
{Hamidani}, H., {Sato}, Y., {Kashiyama}, K., {et~al.} 2025{\natexlab{b}}, arXiv e-prints, arXiv:2503.16243, \dodoi{10.48550/arXiv.2503.16243}

\bibitem[{{Hammerstein} {et~al.}(2023){Hammerstein}, {van Velzen}, {Gezari}, {Cenko}, {Yao}, {Ward}, {Frederick}, {Villanueva}, {Somalwar}, {Graham}, {Kulkarni}, {Stern}, {Andreoni}, {Bellm}, {Dekany}, {Dhawan}, {Drake}, {Fremling}, {Gatkine}, {Groom}, {Ho}, {Kasliwal}, {Karambelkar}, {Kool}, {Masci}, {Medford}, {Perley}, {Purdum}, {van Roestel}, {Sharma}, {Sollerman}, {Taggart}, \& {Yan}}]{Hammerstein2023}
{Hammerstein}, E., {van Velzen}, S., {Gezari}, S., {et~al.} 2023, \apj, 942, 9, \dodoi{10.3847/1538-4357/aca283}

\bibitem[{{Haynie} \& {Piro}(2021)}]{Haynie2021}
{Haynie}, A., \& {Piro}, A.~L. 2021, \apj, 910, 128, \dodoi{10.3847/1538-4357/abe938}

\bibitem[{{Heise} {et~al.}(2001){Heise}, {Zand}, {Kippen}, \& {Woods}}]{Heise01}
{Heise}, J., {Zand}, J.~I., {Kippen}, R.~M., \& {Woods}, P.~M. 2001, in Gamma-ray Bursts in the Afterglow Era, ed. E.~{Costa}, F.~{Frontera}, \& J.~{Hjorth}, 16, \dodoi{10.1007/10853853_4}

\bibitem[{{Hjorth} {et~al.}(2003){Hjorth}, {Sollerman}, {M{\o}ller}, {Fynbo}, {Woosley}, {Kouveliotou}, {Tanvir}, {Greiner}, {Andersen}, {Castro-Tirado}, {Castro Cer{\'o}n}, {Fruchter}, {Gorosabel}, {Jakobsson}, {Kaper}, {Klose}, {Masetti}, {Pedersen}, {Pedersen}, {Pian}, {Palazzi}, {Rhoads}, {Rol}, {van den Heuvel}, {Vreeswijk}, {Watson}, \& {Wijers}}]{Hjorth2003}
{Hjorth}, J., {Sollerman}, J., {M{\o}ller}, P., {et~al.} 2003, \nat, 423, 847, \dodoi{10.1038/nature01750}

\bibitem[{{Ho} {et~al.}(2019){Ho}, {Goldstein}, {Schulze}, {Khatami}, {Perley}, {Ergon}, {Gal-Yam}, {Corsi}, {Andreoni}, {Barbarino}, {Bellm}, {Blagorodnova}, {Bright}, {Burns}, {Cenko}, {Cunningham}, {De}, {Dekany}, {Dugas}, {Fender}, {Fransson}, {Fremling}, {Goldstein}, {Graham}, {Hale}, {Horesh}, {Hung}, {Kasliwal}, {Kuin}, {Kulkarni}, {Kupfer}, {Lunnan}, {Masci}, {Ngeow}, {Nugent}, {Ofek}, {Patterson}, {Petitpas}, {Rusholme}, {Sai}, {Sfaradi}, {Shupe}, {Sollerman}, {Soumagnac}, {Tachibana}, {Taddia}, {Walters}, {Wang}, {Yao}, \& {Zhang}}]{Ho2019}
{Ho}, A. Y.~Q., {Goldstein}, D.~A., {Schulze}, S., {et~al.} 2019, \apj, 887, 169, \dodoi{10.3847/1538-4357/ab55ec}

\bibitem[{{Ho} {et~al.}(2020{\natexlab{a}}){Ho}, {Kulkarni}, {Perley}, {Cenko}, {Corsi}, {Schulze}, {Lunnan}, {Sollerman}, {Gal-Yam}, {Anand}, {Barbarino}, {Bellm}, {Bruch}, {Burns}, {De}, {Dekany}, {Delacroix}, {Duev}, {Frederiks}, {Fremling}, {Goldstein}, {Golkhou}, {Graham}, {Hale}, {Kasliwal}, {Kupfer}, {Laher}, {Martikainen}, {Masci}, {Neill}, {Ridnaia}, {Rusholme}, {Savchenko}, {Shupe}, {Soumagnac}, {Strotjohann}, {Svinkin}, {Taggart}, {Tartaglia}, {Yan}, \& {Zolkower}}]{Ho2020b}
{Ho}, A. Y.~Q., {Kulkarni}, S.~R., {Perley}, D.~A., {et~al.} 2020{\natexlab{a}}, \apj, 902, 86, \dodoi{10.3847/1538-4357/aba630}

\bibitem[{{Ho} {et~al.}(2020{\natexlab{b}}){Ho}, {Kulkarni}, {Perley}, {Cenko}, {Corsi}, {Schulze}, {Lunnan}, {Sollerman}, {Gal-Yam}, {Anand}, {Barbarino}, {Bellm}, {Bruch}, {Burns}, {De}, {Dekany}, {Delacroix}, {Duev}, {Frederiks}, {Fremling}, {Goldstein}, {Golkhou}, {Graham}, {Hale}, {Kasliwal}, {Kupfer}, {Laher}, {Martikainen}, {Masci}, {Neill}, {Ridnaia}, {Rusholme}, {Savchenko}, {Shupe}, {Soumagnac}, {Strotjohann}, {Svinkin}, {Taggart}, {Tartaglia}, {Yan}, \& {Zolkower}}]{2020ApJ...902...86H}
---. 2020{\natexlab{b}}, \apj, 902, 86, \dodoi{10.3847/1538-4357/aba630}

\bibitem[{{Ho} {et~al.}(2020{\natexlab{c}}){Ho}, {Corsi}, {Cenko}, {Taddia}, {Kulkarni}, {Adams}, {De}, {Dekany}, {Frederiks}, {Fremling}, {Golkhou}, {Graham}, {Hung}, {Kupfer}, {Laher}, {Mahabal}, {Masci}, {Miller}, {Neill}, {Reiley}, {Riddle}, {Ridnaia}, {Rusholme}, {Sharma}, {Sollerman}, {Soumagnac}, {Svinkin}, \& {Shupe}}]{Ho2020}
{Ho}, A. Y.~Q., {Corsi}, A., {Cenko}, S.~B., {et~al.} 2020{\natexlab{c}}, \apj, 893, 132, \dodoi{10.3847/1538-4357/ab7f3b}

\bibitem[{{Ho} {et~al.}(2023){Ho}, {Perley}, {Gal-Yam}, {Lunnan}, {Sollerman}, {Schulze}, {Das}, {Dobie}, {Yao}, {Fremling}, {Adams}, {Anand}, {Andreoni}, {Bellm}, {Bruch}, {Burdge}, {Castro-Tirado}, {Dahiwale}, {De}, {Dekany}, {Drake}, {Duev}, {Graham}, {Helou}, {Kaplan}, {Karambelkar}, {Kasliwal}, {Kool}, {Kulkarni}, {Mahabal}, {Medford}, {Miller}, {Nordin}, {Ofek}, {Petitpas}, {Riddle}, {Sharma}, {Smith}, {Stewart}, {Taggart}, {Tartaglia}, {Tzanidakis}, \& {Winters}}]{Ho2023a}
{Ho}, A. Y.~Q., {Perley}, D.~A., {Gal-Yam}, A., {et~al.} 2023, \apj, 949, 120, \dodoi{10.3847/1538-4357/acc533}

\bibitem[{{Hopp} {et~al.}(2014){Hopp}, {Bender}, {Grupp}, {Goessl}, {Lang-Bardl}, {Mitsch}, {Riffeser}, \& {Ageorges}}]{2014SPIE.9145E..2DH}
{Hopp}, U., {Bender}, R., {Grupp}, F., {et~al.} 2014, in Society of Photo-Optical Instrumentation Engineers (SPIE) Conference Series, Vol. 9145, Ground-based and Airborne Telescopes V, ed. L.~M. {Stepp}, R.~{Gilmozzi}, \& H.~J. {Hall}, 91452D, \dodoi{10.1117/12.2054498}

\bibitem[{{Howell} \& {Global Supernova Project}(2017)}]{2017AAS...23031803H}
{Howell}, D.~A., \& {Global Supernova Project}. 2017, in American Astronomical Society Meeting Abstracts, Vol. 230, American Astronomical Society Meeting Abstracts \#230, 318.03

\bibitem[{{Hsu} {et~al.}(2021){Hsu}, {Hosseinzadeh}, \& {Berger}}]{Hsu2021}
{Hsu}, B., {Hosseinzadeh}, G., \& {Berger}, E. 2021, \apj, 921, 180, \dodoi{10.3847/1538-4357/ac1aca}

\bibitem[{Hu {et~al.}(2019)Hu, Wang, Chen, \& Yang}]{hu_image_2022}
Hu, L., Wang, L., Chen, X., \& Yang, J. 2019, 936, 157, \dodoi{10.3847/1538-4357/ac7394}

\bibitem[{{Inserra} {et~al.}(2013){Inserra}, {Smartt}, {Jerkstrand}, {Valenti}, {Fraser}, {Wright}, {Smith}, {Chen}, {Kotak}, {Pastorello}, {Nicholl}, {Bresolin}, {Kudritzki}, {Benetti}, {Botticella}, {Burgett}, {Chambers}, {Ergon}, {Flewelling}, {Fynbo}, {Geier}, {Hodapp}, {Howell}, {Huber}, {Kaiser}, {Leloudas}, {Magill}, {Magnier}, {McCrum}, {Metcalfe}, {Price}, {Rest}, {Sollerman}, {Sweeney}, {Taddia}, {Taubenberger}, {Tonry}, {Wainscoat}, {Waters}, \& {Young}}]{Inserra2013}
{Inserra}, C., {Smartt}, S.~J., {Jerkstrand}, A., {et~al.} 2013, \apj, 770, 128, \dodoi{10.1088/0004-637X/770/2/128}

\bibitem[{{Irwin} \& {Chevalier}(2016)}]{Irwin2016}
{Irwin}, C.~M., \& {Chevalier}, R.~A. 2016, \mnras, 460, 1680, \dodoi{10.1093/mnras/stw1058}

\bibitem[{{Irwin} \& {Hotokezaka}(2025)}]{Irwin2025}
{Irwin}, C.~M., \& {Hotokezaka}, K. 2025, \mnras, 543, 2917, \dodoi{10.1093/mnras/staf1618}

\bibitem[{{Issa} {et~al.}(2025){Issa}, {Gottlieb}, {Metzger}, {Jacquemin-Ide}, {Liska}, {Foucart}, {Halevi}, \& {Tchekhovskoy}}]{Issa2025}
{Issa}, D., {Gottlieb}, O., {Metzger}, B.~D., {et~al.} 2025, \apjl, 985, L26, \dodoi{10.3847/2041-8213/adc694}

\bibitem[{{Ito} {et~al.}(2020){Ito}, {Levinson}, \& {Nakar}}]{Ito2020}
{Ito}, H., {Levinson}, A., \& {Nakar}, E. 2020, \mnras, 499, 4961, \dodoi{10.1093/mnras/staa3125}

\bibitem[{{Ivezi{\'c}} {et~al.}(2019){Ivezi{\'c}}, {Kahn}, {Tyson}, {Abel}, {Acosta}, {Allsman}, {Alonso}, {AlSayyad}, {Anderson}, {Andrew}, {Angel}, {Angeli}, {Ansari}, {Antilogus}, {Araujo}, {Armstrong}, {Arndt}, {Astier}, {Aubourg}, {Auza}, {Axelrod}, {Bard}, {Barr}, {Barrau}, {Bartlett}, {Bauer}, {Bauman}, {Baumont}, {Bechtol}, {Bechtol}, {Becker}, {Becla}, {Beldica}, {Bellavia}, {Bianco}, {Biswas}, {Blanc}, {Blazek}, {Blandford}, {Bloom}, {Bogart}, {Bond}, {Booth}, {Borgland}, {Borne}, {Bosch}, {Boutigny}, {Brackett}, {Bradshaw}, {Brandt}, {Brown}, {Bullock}, {Burchat}, {Burke}, {Cagnoli}, {Calabrese}, {Callahan}, {Callen}, {Carlin}, {Carlson}, {Chandrasekharan}, {Charles-Emerson}, {Chesley}, {Cheu}, {Chiang}, {Chiang}, {Chirino}, {Chow}, {Ciardi}, {Claver}, {Cohen-Tanugi}, {Cockrum}, {Coles}, {Connolly}, {Cook}, {Cooray}, {Covey}, {Cribbs}, {Cui}, {Cutri}, {Daly}, {Daniel}, {Daruich}, {Daubard}, {Daues}, {Dawson}, {Delgado}, {Dellapenna}, {de Peyster}, {de Val-Borro}, {Digel}, {Doherty}, {Dubois},
  {Dubois-Felsmann}, {Durech}, {Economou}, {Eifler}, {Eracleous}, {Emmons}, {Fausti Neto}, {Ferguson}, {Figueroa}, {Fisher-Levine}, {Focke}, {Foss}, {Frank}, {Freemon}, {Gangler}, {Gawiser}, {Geary}, {Gee}, {Geha}, {Gessner}, {Gibson}, {Gilmore}, {Glanzman}, {Glick}, {Goldina}, {Goldstein}, {Goodenow}, {Graham}, {Gressler}, {Gris}, {Guy}, {Guyonnet}, {Haller}, {Harris}, {Hascall}, {Haupt}, {Hernandez}, {Herrmann}, {Hileman}, {Hoblitt}, {Hodgson}, {Hogan}, {Howard}, {Huang}, {Huffer}, {Ingraham}, {Innes}, {Jacoby}, {Jain}, {Jammes}, {Jee}, {Jenness}, {Jernigan}, {Jevremovi{\'c}}, {Johns}, {Johnson}, {Johnson}, {Jones}, {Juramy-Gilles}, {Juri{\'c}}, {Kalirai}, {Kallivayalil}, {Kalmbach}, {Kantor}, {Karst}, {Kasliwal}, {Kelly}, {Kessler}, {Kinnison}, {Kirkby}, {Knox}, {Kotov}, {Krabbendam}, {Krughoff}, {Kub{\'a}nek}, {Kuczewski}, {Kulkarni}, {Ku}, {Kurita}, {Lage}, {Lambert}, {Lange}, {Langton}, {Le Guillou}, {Levine}, {Liang}, {Lim}, {Lintott}, {Long}, {Lopez}, {Lotz}, {Lupton}, {Lust}, {MacArthur}, {Mahabal},
  {Mandelbaum}, {Markiewicz}, {Marsh}, {Marshall}, {Marshall}, {May}, {McKercher}, {McQueen}, {Meyers}, {Migliore}, {Miller}, \& {Mills}}]{Ivezic2019}
{Ivezi{\'c}}, {\v{Z}}., {Kahn}, S.~M., {Tyson}, J.~A., {et~al.} 2019, \apj, 873, 111, \dodoi{10.3847/1538-4357/ab042c}

\bibitem[{{Iwamoto} {et~al.}(1998){Iwamoto}, {Mazzali}, {Nomoto}, {Umeda}, {Nakamura}, {Patat}, {Danziger}, {Young}, {Suzuki}, {Shigeyama}, {Augusteijn}, {Doublier}, {Gonzalez}, {Boehnhardt}, {Brewer}, {Hainaut}, {Lidman}, {Leibundgut}, {Cappellaro}, {Turatto}, {Galama}, {Vreeswijk}, {Kouveliotou}, {van Paradijs}, {Pian}, {Palazzi}, \& {Frontera}}]{Iwamoto98}
{Iwamoto}, K., {Mazzali}, P.~A., {Nomoto}, K., {et~al.} 1998, \nat, 395, 672, \dodoi{10.1038/27155}

\bibitem[{{Izzo} {et~al.}(2020){Izzo}, {Auchettl}, {Hjorth}, {De Colle}, {Gall}, {Angus}, {Raimundo}, \& {Ramirez-Ruiz}}]{Izzo2020}
{Izzo}, L., {Auchettl}, K., {Hjorth}, J., {et~al.} 2020, \aap, 639, L11, \dodoi{10.1051/0004-6361/202038152}

\bibitem[{{Izzo} {et~al.}(2025){Izzo}, {Martin-Carrillo}, {Malesani}, {Levan}, {Jonker}, {Cotter}, {van Dalen}, {Corcoran}, {Wiersema}, \& {Bauer}}]{GCN39851}
{Izzo}, L., {Martin-Carrillo}, A., {Malesani}, D.~B., {et~al.} 2025, GRB Coordinates Network, 39851, 1

\bibitem[{{Janka}(2017)}]{2017hsn..book.1095J}
{Janka}, H.-T. 2017, in Handbook of Supernovae, ed. A.~W. {Alsabti} \& P.~{Murdin}, 1095, \dodoi{10.1007/978-3-319-21846-5_109}

\bibitem[{{Jiang} {et~al.}(2018){Jiang}, {Hu}, {Xu}, {Dai}, {Zhang}, {Wang}, \& {Chen}}]{Jiang:2018SPIE10702E..2LJ}
{Jiang}, H., {Hu}, Z., {Xu}, M., {et~al.} 2018, in Society of Photo-Optical Instrumentation Engineers (SPIE) Conference Series, Vol. 10702, Ground-based and Airborne Instrumentation for Astronomy VII, ed. C.~J. {Evans}, L.~{Simard}, \& H.~{Takami}, 107022L, \dodoi{10.1117/12.2312550}

\bibitem[{{Jiang} {et~al.}(2025){Jiang}, {Xu}, {van Hoof}, {Lei}, {Liu}, {Zhou}, {Chen}, {Fu}, {Yang}, {Liu}, {Zhu}, {Filippenko}, {Jonker}, {Pozanenko}, {Gao}, {Wu}, {Zhang}, {Lamb}, {De Pasquale}, {Kobayashi}, {Bauer}, {Sun}, {Pugliese}, {An}, {D'Elia}, {Fynbo}, {Zheng}, {Tirado}, {Yin}, {Zou}, {Deller}, {Pankov}, {Volnova}, {Moskvitin}, {Spiridonova}, {Oparin}, {Rumyantsev}, {Burkhonov}, {Egamberdiyev}, {Kim}, {Krugov}, {Tatarnikov}, {Inasaridze}, {Levan}, {Bj{\o}rn Malesani}, {Ravasio}, {Quirola-V{\'a}squez}, {van Dalen}, {S{\'a}nchez-Sierras}, {Mata S{\'a}nchez}, {Littlefair}, {Chac{\'o}n}, {Torres}, {Chrimes}, {Sarin}, {Martin-Carrillo}, {Dhillon}, {Yang}, {Brink}, {Davies}, {Yang}, {Aryan}, {Chen}, {Kong}, {Li}, {Li}, {Mao}, {P{\'e}rez-Garc{\'\i}a}, {Fern{\'a}ndez-Garc{\'\i}a}, {Andrews}, {Farah}, {Fan}, {Padilla Gonzalez}, {Howell}, {Hartmann}, {Hu}, {Jakobsson}, {Li}, {Ling}, {McCully}, {Newsome}, {Schneider}, {Samaporn Tinyanont}, {Sun}, {Terreran}, {Tang}, {Wang}, {Xu}, {Yuan}, {Zhang}, {Zhao}, \&
  {Zhang}}]{Jiang2025}
{Jiang}, S.-Q., {Xu}, D., {van Hoof}, A. P.~C., {et~al.} 2025, arXiv e-prints, arXiv:2503.04306, \dodoi{10.48550/arXiv.2503.04306}

\bibitem[{{Johnson} {et~al.}(2021){Johnson}, {Leja}, {Conroy}, \& {Speagle}}]{prospector}
{Johnson}, B.~D., {Leja}, J., {Conroy}, C., \& {Speagle}, J.~S. 2021, \apjs, 254, 22, \dodoi{10.3847/1538-4365/abef67}

\bibitem[{{Kangas} \& {Fruchter}(2021)}]{2021ApJ...911...14K}
{Kangas}, T., \& {Fruchter}, A.~S. 2021, \apj, 911, 14, \dodoi{10.3847/1538-4357/abe76b}

\bibitem[{{Kasen} \& {Bildsten}(2010)}]{kasen2010}
{Kasen}, D., \& {Bildsten}, L. 2010, \apj, 717, 245, \dodoi{10.1088/0004-637X/717/1/245}

\bibitem[{{Kasliwal} {et~al.}(2024){Kasliwal}, {Fremling}, {Yan}, {Das}, {Verdi}, {Zmudzinas}, {Martin}, {Kirby}, {Xue}, {Ho}, {Herczeg}, {Wu}, {Hu}, {Ji}, {Matuszewski}, {Bertz}, {Hale}, {Rodriguez}, {Boden}, {Dekany}, {Smith}, {Reiley}, {Nash}, {Milburn}, {Neill}, {Brugger}, {Zarzaca}, {Weber}, \& {Shapiro}}]{Kasliwal:2024TNSAN.340....1K}
{Kasliwal}, M.~M., {Fremling}, C., {Yan}, L., {et~al.} 2024, Transient Name Server AstroNote, 340, 1

\bibitem[{{Katz} {et~al.}(2010){Katz}, {Budnik}, \& {Waxman}}]{Katz2010}
{Katz}, B., {Budnik}, R., \& {Waxman}, E. 2010, \apj, 716, 781, \dodoi{10.1088/0004-637X/716/1/781}

\bibitem[{{Kennicutt}(1998)}]{Kennicutt1998}
{Kennicutt}, Jr., R.~C. 1998, \araa, 36, 189, \dodoi{10.1146/annurev.astro.36.1.189}

\bibitem[{{Khatami} \& {Kasen}(2019)}]{Khatami2019}
{Khatami}, D.~K., \& {Kasen}, D.~N. 2019, \apj, 878, 56, \dodoi{10.3847/1538-4357/ab1f09}

\bibitem[{{Kim} {et~al.}(2022){Kim}, {Rigault}, {Neill}, {Briday}, {Copin}, {Lezmy}, {Nicolas}, {Riddle}, {Sharma}, {Smith}, {Sollerman}, \& {Walters}}]{Kim2022}
{Kim}, Y.~L., {Rigault}, M., {Neill}, J.~D., {et~al.} 2022, \pasp, 134, 024505, \dodoi{10.1088/1538-3873/ac50a0}

\bibitem[{{Kostov} \& {Bonev}(2018)}]{Kostov2018}
{Kostov}, A., \& {Bonev}, T. 2018, Bulgarian Astronomical Journal, 28, 3, \dodoi{10.48550/arXiv.1706.06147}

\bibitem[{{Kulkarni} {et~al.}(1998){Kulkarni}, {Frail}, {Wieringa}, {Ekers}, {Sadler}, {Wark}, {Higdon}, {Phinney}, \& {Bloom}}]{1998Natur.395..663K}
{Kulkarni}, S.~R., {Frail}, D.~A., {Wieringa}, M.~H., {et~al.} 1998, \nat, 395, 663, \dodoi{10.1038/27139}

\bibitem[{{Kulkarni} {et~al.}(2021){Kulkarni}, {Harrison}, {Grefenstette}, {Earnshaw}, {Andreoni}, {Berg}, {Bloom}, {Cenko}, {Chornock}, {Christiansen}, {Coughlin}, {Wuollet Criswell}, {Darvish}, {Das}, {De}, {Dessart}, {Dixon}, {Dorsman}, {El-Badry}, {Evans}, {Ford}, {Fremling}, {Gansicke}, {Gezari}, {Goetberg}, {Green}, {Graham}, {Heida}, {Ho}, {Jaodand}, {Johns-Krull}, {Kasliwal}, {Lazzarini}, {Lu}, {Margutti}, {Martin}, {Masters}, {McKernan}, {Naze}, {Nissanke}, {Parazin}, {Perley}, {Phinney}, {Piro}, {Raaijmakers}, {Rauw}, {Rodriguez}, {Sana}, {Senchyna}, {Singer}, {Spake}, {Stassun}, {Stern}, {Teplitz}, {Weisz}, \& {Yao}}]{UVEX2021}
{Kulkarni}, S.~R., {Harrison}, F.~A., {Grefenstette}, B.~W., {et~al.} 2021, arXiv e-prints, arXiv:2111.15608, \dodoi{10.48550/arXiv.2111.15608}

\bibitem[{{Kuroda} {et~al.}(2020){Kuroda}, {Arcones}, {Takiwaki}, \& {Kotake}}]{Kuroda2020}
{Kuroda}, T., {Arcones}, A., {Takiwaki}, T., \& {Kotake}, K. 2020, \apj, 896, 102, \dodoi{10.3847/1538-4357/ab9308}

\bibitem[{{Labrie} {et~al.}(2019){Labrie}, {Anderson}, {C{\'a}rdenes}, {Simpson}, \& {Turner}}]{Labrie2019}
{Labrie}, K., {Anderson}, K., {C{\'a}rdenes}, R., {Simpson}, C., \& {Turner}, J. E.~H. 2019, in Astronomical Society of the Pacific Conference Series, Vol. 523, Astronomical Data Analysis Software and Systems XXVII, ed. P.~J. {Teuben}, M.~W. {Pound}, B.~A. {Thomas}, \& E.~M. {Warner}, 321

\bibitem[{Lang-Bardl {et~al.}(2016)Lang-Bardl, Bender, Goessl, Grupp, Hess, Kaminski, Hodapp, Hopp, Jacobson, Kravcar, {et~al.}}]{lang2016wendelstein}
Lang-Bardl, F., Bender, R., Goessl, C., {et~al.} 2016, in Ground-based and Airborne Instrumentation for Astronomy VI, Vol. 9908, SPIE, 1295--1302

\bibitem[{{Levinson} \& {Nakar}(2020)}]{Levinson2020}
{Levinson}, A., \& {Nakar}, E. 2020, \physrep, 866, 1, \dodoi{10.1016/j.physrep.2020.04.003}

\bibitem[{{Li} \& {Ma}(1983)}]{Li1983}
{Li}, T.-P., \& {Ma}, Y.-Q. 1983, \apj, 272, 317, \dodoi{10.1086/161295}

\bibitem[{{Li} {et~al.}(2025){Li}, {Zhu}, {Zou}, {Geng}, {Liu}, {Wang}, {Li}, {Xu}, {Sun}, {Wang}, {Yu}, {Zhang}, {Wu}, {Yang}, {Filippenko}, {Liu}, {Yuan}, {Aguado}, {An}, {An}, {Buckley}, {Castro-Tirado}, {Fu}, {Fynbo}, {Howell}, {Hu}, {Jiang}, {Kumar}, {Mao}, {Maund}, {Liu}, {Mockler}, {Moskvitin}, {Andrews}, {Bom}, {Brink}, {Chatterjee}, {Chen}, {Cheng}, {Cooke}, {Dai}, {Du}, {Erasmus}, {Fang}, {Farah}, {Goranskij}, {Gritsevich}, {Gu}, {Guo}, {Hsiao}, {Hu}, {Hua}, {Jacobson-Gal{\'a}n}, {Jia}, {Jin}, {Kasliwal}, {Kilpatrick}, {Kumar}, {Lei}, {Li}, {Li}, {Li}, {Ling}, {Liu}, {Liu}, {Liu}, {L{\'o}pez-Oramas}, {Maslennikova}, {McCully}, {Monageng}, {Newsone}, {Padilla Gonzalez}, {Pan}, {Peng}, {Pignata}, {Poidevin}, {Potter}, {P{\'e}rez-Fournon}, {Santana-Silva}, {Santos}, {Song}, {Song}, {Spiridonova}, {Sun}, {Sun}, {Terreran}, {Wang}, {Wang}, {Wang}, {Wang}, {Wu}, {Xiang}, {Xiao}, {Xu}, {Xue}, {Yan}, {Yang}, {Yu}, {Zhang}, {Zhang}, {Zhang}, {Zhang}, {Zhang}, {Zheng}, \& {Zou}}]{Li2025}
{Li}, W.~X., {Zhu}, Z.~P., {Zou}, X.~Z., {et~al.} 2025, arXiv e-prints, arXiv:2504.17034, \dodoi{10.48550/arXiv.2504.17034}

\bibitem[{{Liang} {et~al.}(2007){Liang}, {Zhang}, {Virgili}, \& {Dai}}]{Liang2007}
{Liang}, E., {Zhang}, B., {Virgili}, F., \& {Dai}, Z.~G. 2007, \apj, 662, 1111, \dodoi{10.1086/517959}

\bibitem[{{Lin} {et~al.}(2022){Lin}, {Irwin}, {Berger}, \& {Nguyen}}]{Lin2022}
{Lin}, D., {Irwin}, J.~A., {Berger}, E., \& {Nguyen}, R. 2022, \apj, 927, 211, \dodoi{10.3847/1538-4357/ac4fc6}

\bibitem[{{Lindegren} {et~al.}(2021){Lindegren}, {Klioner}, {Hern{\'a}ndez}, {Bombrun}, {Ramos-Lerate}, {Steidelm{\"u}ller}, {Bastian}, {Biermann}, {de Torres}, {Gerlach}, {Geyer}, {Hilger}, {Hobbs}, {Lammers}, {McMillan}, {Stephenson}, {Casta{\~n}eda}, {Davidson}, {Fabricius}, {Gracia-Abril}, {Portell}, {Rowell}, {Teyssier}, {Torra}, {Bartolom{\'e}}, {Clotet}, {Garralda}, {Gonz{\'a}lez-Vidal}, {Torra}, {Abbas}, {Altmann}, {Anglada Varela}, {Balaguer-N{\'u}{\~n}ez}, {Balog}, {Barache}, {Becciani}, {Bernet}, {Bertone}, {Bianchi}, {Bouquillon}, {Brown}, {Bucciarelli}, {Busonero}, {Butkevich}, {Buzzi}, {Cancelliere}, {Carlucci}, {Charlot}, {Cioni}, {Crosta}, {Crowley}, {del Peloso}, {del Pozo}, {Drimmel}, {Esquej}, {Fienga}, {Fraile}, {Gai}, {Garcia-Reinaldos}, {Guerra}, {Hambly}, {Hauser}, {Jan{\ss}en}, {Jordan}, {Kostrzewa-Rutkowska}, {Lattanzi}, {Liao}, {Licata}, {Lister}, {L{\"o}ffler}, {Marchant}, {Masip}, {Mignard}, {Mints}, {Molina}, {Mora}, {Morbidelli}, {Murphy}, {Pagani}, {Panuzzo}, {Pe{\~n}alosa
  Esteller}, {Poggio}, {Re Fiorentin}, {Riva}, {Sagrist{\`a} Sell{\'e}s}, {Sanchez Gimenez}, {Sarasso}, {Sciacca}, {Siddiqui}, {Smart}, {Souami}, {Spagna}, {Steele}, {Taris}, {Utrilla}, {van Reeven}, \& {Vecchiato}}]{2021A&A...649A...2L}
{Lindegren}, L., {Klioner}, S.~A., {Hern{\'a}ndez}, J., {et~al.} 2021, \aap, 649, A2, \dodoi{10.1051/0004-6361/202039709}

\bibitem[{{Liu} {et~al.}(2016){Liu}, {Modjaz}, {Bianco}, \& {Graur}}]{Liu2016}
{Liu}, Y.-Q., {Modjaz}, M., {Bianco}, F.~B., \& {Graur}, O. 2016, \apj, 827, 90, \dodoi{10.3847/0004-637X/827/2/90}

\bibitem[{{Lyman} {et~al.}(2014){Lyman}, {Bersier}, \& {James}}]{Lyman2014}
{Lyman}, J.~D., {Bersier}, D., \& {James}, P.~A. 2014, \mnras, 437, 3848, \dodoi{10.1093/mnras/stt2187}

\bibitem[{{Malesani} {et~al.}(2009){Malesani}, {Fynbo}, {Hjorth}, {Leloudas}, {Sollerman}, {Stritzinger}, {Vreeswijk}, {Watson}, {Gorosabel}, {Micha{\l}owski}, {Th{\"o}ne}, {Augusteijn}, {Bersier}, {Jakobsson}, {Jaunsen}, {Ledoux}, {Levan}, {Milvang-Jensen}, {Rol}, {Tanvir}, {Wiersema}, {Xu}, {Albert}, {Bayliss}, {Gall}, {Grove}, {Koester}, {Leitet}, {Pursimo}, \& {Skillen}}]{Malesani2009}
{Malesani}, D., {Fynbo}, J.~P.~U., {Hjorth}, J., {et~al.} 2009, \apjl, 692, L84, \dodoi{10.1088/0004-637X/692/2/L84}

\bibitem[{{Margutti} {et~al.}(2013{\natexlab{a}}){Margutti}, {Soderberg}, {Wieringa}, {Edwards}, {Chevalier}, {Morsony}, {Barniol Duran}, {Sironi}, {Zauderer}, {Milisavljevic}, {Kamble}, \& {Pian}}]{margutti2013}
{Margutti}, R., {Soderberg}, A.~M., {Wieringa}, M.~H., {et~al.} 2013{\natexlab{a}}, \apj, 778, 18, \dodoi{10.1088/0004-637X/778/1/18}

\bibitem[{{Margutti} {et~al.}(2013{\natexlab{b}}){Margutti}, {Soderberg}, {Wieringa}, {Edwards}, {Chevalier}, {Morsony}, {Barniol Duran}, {Sironi}, {Zauderer}, {Milisavljevic}, {Kamble}, \& {Pian}}]{2013ApJ...778...18M}
---. 2013{\natexlab{b}}, \apj, 778, 18, \dodoi{10.1088/0004-637X/778/1/18}

\bibitem[{Margutti {et~al.}(2015)Margutti, Guidorzi, Lazzati, Milisavljevic, Kamble, Laskar, Parrent, Gehrels, \& Soderberg}]{Margutti2015}
Margutti, R., Guidorzi, C., Lazzati, D., {et~al.} 2015, The Astrophysical Journal, 805, 159, \dodoi{10.1088/0004-637X/805/2/159}

\bibitem[{{Margutti} {et~al.}(2019){Margutti}, {Metzger}, {Chornock}, {Vurm}, {Roth}, {Grefenstette}, {Savchenko}, {Cartier}, {Steiner}, {Terreran}, {Margalit}, {Migliori}, {Milisavljevic}, {Alexander}, {Bietenholz}, {Blanchard}, {Bozzo}, {Brethauer}, {Chilingarian}, {Coppejans}, {Ducci}, {Ferrigno}, {Fong}, {G{\"o}tz}, {Guidorzi}, {Hajela}, {Hurley}, {Kuulkers}, {Laurent}, {Mereghetti}, {Nicholl}, {Patnaude}, {Ubertini}, {Banovetz}, {Bartel}, {Berger}, {Coughlin}, {Eftekhari}, {Frederiks}, {Kozlova}, {Laskar}, {Svinkin}, {Drout}, {MacFadyen}, \& {Paterson}}]{Margutti2019}
{Margutti}, R., {Metzger}, B.~D., {Chornock}, R., {et~al.} 2019, \apj, 872, 18, \dodoi{10.3847/1538-4357/aafa01}

\bibitem[{{Masci} {et~al.}(2019){Masci}, {Laher}, {Rusholme}, {Shupe}, {Groom}, {Surace}, {Jackson}, {Monkewitz}, {Beck}, {Flynn}, {Terek}, {Landry}, {Hacopians}, {Desai}, {Howell}, {Brooke}, {Imel}, {Wachter}, {Ye}, {Lin}, {Cenko}, {Cunningham}, {Rebbapragada}, {Bue}, {Miller}, {Mahabal}, {Bellm}, {Patterson}, {Juri{\'c}}, {Golkhou}, {Ofek}, {Walters}, {Graham}, {Kasliwal}, {Dekany}, {Kupfer}, {Burdge}, {Cannella}, {Barlow}, {Van Sistine}, {Giomi}, {Fremling}, {Blagorodnova}, {Levitan}, {Riddle}, {Smith}, {Helou}, {Prince}, \& {Kulkarni}}]{Masci2019}
{Masci}, F.~J., {Laher}, R.~R., {Rusholme}, B., {et~al.} 2019, \pasp, 131, 018003, \dodoi{10.1088/1538-3873/aae8ac}

\bibitem[{{Matsuoka} {et~al.}(2009){Matsuoka}, {Kawasaki}, {Ueno}, {Tomida}, {Kohama}, {Suzuki}, {Adachi}, {Ishikawa}, {Mihara}, {Sugizaki}, {Isobe}, {Nakagawa}, {Tsunemi}, {Miyata}, {Kawai}, {Kataoka}, {Morii}, {Yoshida}, {Negoro}, {Nakajima}, {Ueda}, {Chujo}, {Yamaoka}, {Yamazaki}, {Nakahira}, {You}, {Ishiwata}, {Miyoshi}, {Eguchi}, {Hiroi}, {Katayama}, \& {Ebisawa}}]{MAXI}
{Matsuoka}, M., {Kawasaki}, K., {Ueno}, S., {et~al.} 2009, \pasj, 61, 999, \dodoi{10.1093/pasj/61.5.999}

\bibitem[{{Maund} {et~al.}(2009){Maund}, {Wheeler}, {Baade}, {Patat}, {H{\"o}flich}, {Wang}, \& {Clocchiatti}}]{Maund2009}
{Maund}, J.~R., {Wheeler}, J.~C., {Baade}, D., {et~al.} 2009, \apj, 705, 1139, \dodoi{10.1088/0004-637X/705/2/1139}

\bibitem[{{Mazzali} {et~al.}(2006){Mazzali}, {Deng}, {Nomoto}, {Sauer}, {Pian}, {Tominaga}, {Tanaka}, {Maeda}, \& {Filippenko}}]{Mazzali2006}
{Mazzali}, P.~A., {Deng}, J., {Nomoto}, K., {et~al.} 2006, \nat, 442, 1018, \dodoi{10.1038/nature05081}

\bibitem[{{Mazzali} {et~al.}(2008){Mazzali}, {Valenti}, {Della Valle}, {Chincarini}, {Sauer}, {Benetti}, {Pian}, {Piran}, {D'Elia}, {Elias-Rosa}, {Margutti}, {Pasotti}, {Antonelli}, {Bufano}, {Campana}, {Cappellaro}, {Covino}, {D'Avanzo}, {Fiore}, {Fugazza}, {Gilmozzi}, {Hunter}, {Maguire}, {Maiorano}, {Marziani}, {Masetti}, {Mirabel}, {Navasardyan}, {Nomoto}, {Palazzi}, {Pastorello}, {Panagia}, {Pellizza}, {Sari}, {Smartt}, {Tagliaferri}, {Tanaka}, {Taubenberger}, {Tominaga}, {Trundle}, \& {Turatto}}]{Mazzali2008}
{Mazzali}, P.~A., {Valenti}, S., {Della Valle}, M., {et~al.} 2008, Science, 321, 1185, \dodoi{10.1126/science.1158088}

\bibitem[{{McMullin} {et~al.}(2007){McMullin}, {Waters}, {Schiebel}, {Young}, \& {Golap}}]{2007ASPC..376..127M}
{McMullin}, J.~P., {Waters}, B., {Schiebel}, D., {Young}, W., \& {Golap}, K. 2007, in Astronomical Society of the Pacific Conference Series, Vol. 376, Astronomical Data Analysis Software and Systems XVI, ed. R.~A. {Shaw}, F.~{Hill}, \& D.~J. {Bell}, 127

\bibitem[{{Meegan} {et~al.}(2009){Meegan}, {Lichti}, {Bhat}, {Bissaldi}, {Briggs}, {Connaughton}, {Diehl}, {Fishman}, {Greiner}, {Hoover}, {van der Horst}, {von Kienlin}, {Kippen}, {Kouveliotou}, {McBreen}, {Paciesas}, {Preece}, {Steinle}, {Wallace}, {Wilson}, \& {Wilson-Hodge}}]{Meegan+2009}
{Meegan}, C., {Lichti}, G., {Bhat}, P.~N., {et~al.} 2009, \apj, 702, 791, \dodoi{10.1088/0004-637X/702/1/791}

\bibitem[{{M{\'e}sz{\'a}ros} {et~al.}(1998){M{\'e}sz{\'a}ros}, {Rees}, \& {Wijers}}]{Meszaros1998}
{M{\'e}sz{\'a}ros}, P., {Rees}, M.~J., \& {Wijers}, R.~A.~M.~J. 1998, \apj, 499, 301, \dodoi{10.1086/305635}

\bibitem[{{Metzger}(2019)}]{Metzger2019}
{Metzger}, B.~D. 2019, Living Reviews in Relativity, 23, 1, \dodoi{10.1007/s41114-019-0024-0}

\bibitem[{{Metzger} {et~al.}(2015){Metzger}, {Margalit}, {Kasen}, \& {Quataert}}]{Metzger2015}
{Metzger}, B.~D., {Margalit}, B., {Kasen}, D., \& {Quataert}, E. 2015, \mnras, 454, 3311, \dodoi{10.1093/mnras/stv2224}

\bibitem[{{Metzger} {et~al.}(2014){Metzger}, {Vurm}, {Hasco{\"e}t}, \& {Beloborodov}}]{Metzger2014}
{Metzger}, B.~D., {Vurm}, I., {Hasco{\"e}t}, R., \& {Beloborodov}, A.~M. 2014, \mnras, 437, 703, \dodoi{10.1093/mnras/stt1922}

\bibitem[{{Miller} {et~al.}(2025){Miller}, {Abrams}, {Aldering}, {Anand}, {Angus}, {Arcavi}, {Baltay}, {Bauer}, {Brethauer}, {Bloom}, {Bommireddy}, {Catelan}, {Chornock}, {Clark}, {Collett}, {Dimitriadis}, {Faris}, {F{\"o}rster}, {Franckowiak}, {Frohmaier}, {Galbany}, {Galleguillos}, {Goobar}, {Graur}, {Guti{\'e}rrez}, {Hall}, {Hammerstein}, {Herner}, {Hook}, {Huston}, {Johansson}, {Kilpatrick}, {Kim}, {Knop}, {Kowalski}, {Kwok}, {LeBaron}, {Lin}, {Liu}, {Lu}, {Lu}, {Lunnan}, {Maguire}, {Makrygianni}, {Margutti}, {Maoz}, {Veres}, {Moore}, {Nayana}, {Nicholl}, {Nordin}, {Oates}, {Pignata}, {Polin}, {Poznanski}, {Prieto}, {Rabinowitz}, {Rehemtulla}, {Rigault}, {Ryczanowski}, {Sarin}, {Schulze}, {Shah}, {Sheng}, {Shilling}, {Simmons}, {Singh}, {Smith}, {Smith}, {Sollerman}, {Soumagnac}, {Stubbs}, {Sullivan}, {Suresh}, {Trakhtenbrot}, {Ward}, {Wiston}, {Xiong}, {Yao}, \& {Nugent}}]{Miller2025}
{Miller}, A.~A., {Abrams}, N.~S., {Aldering}, G., {et~al.} 2025, \pasp, 137, 094204, \dodoi{10.1088/1538-3873/ae02c5}

\bibitem[{{Minaev} \& {Pozanenko}(2020)}]{Minaev2020}
{Minaev}, P.~Y., \& {Pozanenko}, A.~S. 2020, \mnras, 492, 1919, \dodoi{10.1093/mnras/stz3611}

\bibitem[{{Mirabal} {et~al.}(2006){Mirabal}, {Halpern}, {An}, {Thorstensen}, \& {Terndrup}}]{Mirabal2006}
{Mirabal}, N., {Halpern}, J.~P., {An}, D., {Thorstensen}, J.~R., \& {Terndrup}, D.~M. 2006, \apjl, 643, L99, \dodoi{10.1086/505177}

\bibitem[{{Modjaz} {et~al.}(2016){Modjaz}, {Liu}, {Bianco}, \& {Graur}}]{Modjaz2016}
{Modjaz}, M., {Liu}, Y.~Q., {Bianco}, F.~B., \& {Graur}, O. 2016, \apj, 832, 108, \dodoi{10.3847/0004-637X/832/2/108}

\bibitem[{{Modjaz} {et~al.}(2006){Modjaz}, {Stanek}, {Garnavich}, {Berlind}, {Blondin}, {Brown}, {Calkins}, {Challis}, {Diamond-Stanic}, {Hao}, {Hicken}, {Kirshner}, \& {Prieto}}]{Modjaz2006}
{Modjaz}, M., {Stanek}, K.~Z., {Garnavich}, P.~M., {et~al.} 2006, \apjl, 645, L21, \dodoi{10.1086/505906}

\bibitem[{{Modjaz} {et~al.}(2009){Modjaz}, {Li}, {Butler}, {Chornock}, {Perley}, {Blondin}, {Bloom}, {Filippenko}, {Kirshner}, {Kocevski}, {Poznanski}, {Hicken}, {Foley}, {Stringfellow}, {Berlind}, {Barrado y Navascues}, {Blake}, {Bouy}, {Brown}, {Challis}, {Chen}, {de Vries}, {Dufour}, {Falco}, {Friedman}, {Ganeshalingam}, {Garnavich}, {Holden}, {Illingworth}, {Lee}, {Liebert}, {Marion}, {Olivier}, {Prochaska}, {Silverman}, {Smith}, {Starr}, {Steele}, {Stockton}, {Williams}, \& {Wood-Vasey}}]{Modjaz2009}
{Modjaz}, M., {Li}, W., {Butler}, N., {et~al.} 2009, \apj, 702, 226, \dodoi{10.1088/0004-637X/702/1/226}

\bibitem[{{Morag} {et~al.}(2023){Morag}, {Sapir}, \& {Waxman}}]{Morag2023}
{Morag}, J., {Sapir}, N., \& {Waxman}, E. 2023, \mnras, 522, 2764, \dodoi{10.1093/mnras/stad899}

\bibitem[{{Moriya} {et~al.}(2019){Moriya}, {M{\"u}ller}, {Chan}, {Heger}, \& {Blinnikov}}]{Moriya2019}
{Moriya}, T.~J., {M{\"u}ller}, B., {Chan}, C., {Heger}, A., \& {Blinnikov}, S.~I. 2019, \apj, 880, 21, \dodoi{10.3847/1538-4357/ab2643}

\bibitem[{{M{\"o}sta} {et~al.}(2015){M{\"o}sta}, {Ott}, {Radice}, {Roberts}, {Schnetter}, \& {Haas}}]{Mosta2015}
{M{\"o}sta}, P., {Ott}, C.~D., {Radice}, D., {et~al.} 2015, \nat, 528, 376, \dodoi{10.1038/nature15755}

\bibitem[{{M{\"o}sta} {et~al.}(2018){M{\"o}sta}, {Roberts}, {Halevi}, {Ott}, {Lippuner}, {Haas}, \& {Schnetter}}]{Mosta2018}
{M{\"o}sta}, P., {Roberts}, L.~F., {Halevi}, G., {et~al.} 2018, \apj, 864, 171, \dodoi{10.3847/1538-4357/aad6ec}

\bibitem[{{M{\"o}sta} {et~al.}(2014){M{\"o}sta}, {Richers}, {Ott}, {Haas}, {Piro}, {Boydstun}, {Abdikamalov}, {Reisswig}, \& {Schnetter}}]{Mosta2014}
{M{\"o}sta}, P., {Richers}, S., {Ott}, C.~D., {et~al.} 2014, \apjl, 785, L29, \dodoi{10.1088/2041-8205/785/2/L29}

\bibitem[{{Murase} {et~al.}(2016){Murase}, {Kashiyama}, \& {M{\'e}sz{\'a}ros}}]{Murase2016}
{Murase}, K., {Kashiyama}, K., \& {M{\'e}sz{\'a}ros}, P. 2016, \mnras, 461, 1498, \dodoi{10.1093/mnras/stw1328}

\bibitem[{{Nakar}(2015)}]{Nakar2015}
{Nakar}, E. 2015, \apj, 807, 172, \dodoi{10.1088/0004-637X/807/2/172}

\bibitem[{{Nakar} \& {Piran}(2017)}]{Nakar2017}
{Nakar}, E., \& {Piran}, T. 2017, \apj, 834, 28, \dodoi{10.3847/1538-4357/834/1/28}

\bibitem[{{Nakar} \& {Piro}(2014)}]{Nakar2014}
{Nakar}, E., \& {Piro}, A.~L. 2014, \apj, 788, 193, \dodoi{10.1088/0004-637X/788/2/193}

\bibitem[{{Nakar} \& {Sari}(2010)}]{Nakar2010}
{Nakar}, E., \& {Sari}, R. 2010, \apj, 725, 904, \dodoi{10.1088/0004-637X/725/1/904}

\bibitem[{{Nakar} \& {Sari}(2012)}]{Nakar2012}
---. 2012, \apj, 747, 88, \dodoi{10.1088/0004-637X/747/2/88}

\bibitem[{{Nasa High Energy Astrophysics Science Archive Research Center (Heasarc)}(2014)}]{HEASOFT}
{Nasa High Energy Astrophysics Science Archive Research Center (Heasarc)}. 2014, {HEAsoft: Unified Release of FTOOLS and XANADU}, Astrophysics Source Code Library, record ascl:1408.004.
\newblock \doeprint{1408.004}

\bibitem[{{Negoro} {et~al.}(2016){Negoro}, {Kohama}, {Serino}, {Saito}, {Takahashi}, {Miyoshi}, {Ozawa}, {Suwa}, {Asada}, {Fukushima}, {Eguchi}, {Kawai}, {Kennea}, {Mihara}, {Morii}, {Nakahira}, {Ogawa}, {Sugawara}, {Tomida}, {Ueno}, {Ishikawa}, {Isobe}, {Kawamuro}, {Kimura}, {Masumitsu}, {Nakagawa}, {Nakajima}, {Sakamoto}, {Shidatsu}, {Sugizaki}, {Sugimoto}, {Suzuki}, {Takagi}, {Tanaka}, {Tsuboi}, {Tsunemi}, {Ueda}, {Yamaoka}, {Yamauchi}, {Yoshida}, \& {Matsuoka}}]{Negoro2016}
{Negoro}, H., {Kohama}, M., {Serino}, M., {et~al.} 2016, \pasj, 68, S1, \dodoi{10.1093/pasj/psw016}

\bibitem[{{Niblett} {et~al.}(2025){Niblett}, {Fryer}, \& {Fryer}}]{Niblett25}
{Niblett}, A.~E., {Fryer}, D.~A., \& {Fryer}, C.~L. 2025, arXiv e-prints, arXiv:2501.15702, \dodoi{10.48550/arXiv.2501.15702}

\bibitem[{{Nicholl} {et~al.}(2017){Nicholl}, {Guillochon}, \& {Berger}}]{Nicholl2017}
{Nicholl}, M., {Guillochon}, J., \& {Berger}, E. 2017, \apj, 850, 55, \dodoi{10.3847/1538-4357/aa9334}

\bibitem[{{Nugis} \& {Lamers}(2000)}]{Nugis2000}
{Nugis}, T., \& {Lamers}, H.~J.~G.~L.~M. 2000, \aap, 360, 227

\bibitem[{{Obergaulinger} \& {Aloy}(2022)}]{Obergaulinger&Aloy2022}
{Obergaulinger}, M., \& {Aloy}, M.~{\'A}. 2022, \mnras, 512, 2489, \dodoi{10.1093/mnras/stac613}

\bibitem[{{O'Connor} {et~al.}(2025){O'Connor}, {Beniamini}, {Troja}, {Busmann}, {Dichiara}, {Gill}, {Granot}, {Moss}, {Hall}, {Palmese}, {Passaleva}, \& {Yang}}]{Oconnor2025b}
{O'Connor}, B., {Beniamini}, P., {Troja}, E., {et~al.} 2025, \apjl, 993, L37, \dodoi{10.3847/2041-8213/ae146b}

\bibitem[{Oke {et~al.}(1995)Oke, Cohen, Carr, Cromer, Dingizian, Harris, Labrecque, Lucinio, Schaal, Epps, \& Miller}]{lris}
Oke, J.~B., Cohen, J.~G., Carr, M., {et~al.} 1995, Publications of the Astronomical Society of the Pacific, 107, 375, \dodoi{10.1086/133562}

\bibitem[{{Olivares E.} {et~al.}(2012){Olivares E.}, {Greiner}, {Schady}, {Rau}, {Klose}, {Kr{\"u}hler}, {Afonso}, {Updike}, {Nardini}, {Filgas}, {Nicuesa Guelbenzu}, {Clemens}, {Elliott}, {Kann}, {Rossi}, \& {Sudilovsky}}]{Olivares2012}
{Olivares E.}, F., {Greiner}, J., {Schady}, P., {et~al.} 2012, \aap, 539, A76, \dodoi{10.1051/0004-6361/201117929}

\bibitem[{{Omand} {et~al.}(2018){Omand}, {Kashiyama}, \& {Murase}}]{Omand2018}
{Omand}, C. M.~B., {Kashiyama}, K., \& {Murase}, K. 2018, \mnras, 474, 573, \dodoi{10.1093/mnras/stx2743}

\bibitem[{{Omand} \& {Sarin}(2024)}]{Omand2024}
{Omand}, C. M.~B., \& {Sarin}, N. 2024, \mnras, 527, 6455, \dodoi{10.1093/mnras/stad3645}

\bibitem[{{Osterbrock}(1989)}]{Osterbrock1989}
{Osterbrock}, D.~E. 1989, {Astrophysics of gaseous nebulae and active galactic nuclei}

\bibitem[{{Panaitescu} \& {Kumar}(2002)}]{2002ApJ...571..779P}
{Panaitescu}, A., \& {Kumar}, P. 2002, \apj, 571, 779, \dodoi{10.1086/340094}

\bibitem[{{Patat} {et~al.}(2001){Patat}, {Cappellaro}, {Danziger}, {Mazzali}, {Sollerman}, {Augusteijn}, {Brewer}, {Doublier}, {Gonzalez}, {Hainaut}, {Lidman}, {Leibundgut}, {Nomoto}, {Nakamura}, {Spyromilio}, {Rizzi}, {Turatto}, {Walsh}, {Galama}, {van Paradijs}, {Kouveliotou}, {Vreeswijk}, {Frontera}, {Masetti}, {Palazzi}, \& {Pian}}]{patat2001}
{Patat}, F., {Cappellaro}, E., {Danziger}, J., {et~al.} 2001, \apj, 555, 900, \dodoi{10.1086/321526}

\bibitem[{{Perley}(2019)}]{lpipe}
{Perley}, D.~A. 2019, \pasp, 131, 084503, \dodoi{10.1088/1538-3873/ab215d}

\bibitem[{{Perley} {et~al.}(2019){Perley}, {Mazzali}, {Yan}, {Cenko}, {Gezari}, {Taggart}, {Blagorodnova}, {Fremling}, {Mockler}, {Singh}, {Tominaga}, {Tanaka}, {Watson}, {Ahumada}, {Anupama}, {Ashall}, {Becerra}, {Bersier}, {Bhalerao}, {Bloom}, {Butler}, {Copperwheat}, {Coughlin}, {De}, {Drake}, {Duev}, {Frederick}, {Gonz{\'a}lez}, {Goobar}, {Heida}, {Ho}, {Horst}, {Hung}, {Itoh}, {Jencson}, {Kasliwal}, {Kawai}, {Khanam}, {Kulkarni}, {Kumar}, {Kumar}, {Kutyrev}, {Lee}, {Maeda}, {Mahabal}, {Murata}, {Neill}, {Ngeow}, {Penprase}, {Pian}, {Quimby}, {Ramirez-Ruiz}, {Richer}, {Rom{\'a}n-Z{\'u}{\~n}iga}, {Sahu}, {Srivastav}, {Socia}, {Sollerman}, {Tachibana}, {Taddia}, {Tinyanont}, {Troja}, {Ward}, {Wee}, \& {Yu}}]{Perley2019}
{Perley}, D.~A., {Mazzali}, P.~A., {Yan}, L., {et~al.} 2019, \mnras, 484, 1031, \dodoi{10.1093/mnras/sty3420}

\bibitem[{Pessi {et~al.}(2025)Pessi, Desai, Prieto, Kochanek, Shappee, Anderson, Beacom, Dong, Stanek, \& Thompson}]{pessi_supernova_2025}
Pessi, T., Desai, D.~D., Prieto, J.~L., {et~al.} 2025, Astronomy and Astrophysics, 703, A34, \dodoi{10.1051/0004-6361/202556799}

\bibitem[{{Pian} {et~al.}(2006){Pian}, {Mazzali}, {Masetti}, {Ferrero}, {Klose}, {Palazzi}, {Ramirez-Ruiz}, {Woosley}, {Kouveliotou}, {Deng}, {Filippenko}, {Foley}, {Fynbo}, {Kann}, {Li}, {Hjorth}, {Nomoto}, {Patat}, {Sauer}, {Sollerman}, {Vreeswijk}, {Guenther}, {Levan}, {O'Brien}, {Tanvir}, {Wijers}, {Dumas}, {Hainaut}, {Wong}, {Baade}, {Wang}, {Amati}, {Cappellaro}, {Castro-Tirado}, {Ellison}, {Frontera}, {Fruchter}, {Greiner}, {Kawabata}, {Ledoux}, {Maeda}, {M{\o}ller}, {Nicastro}, {Rol}, \& {Starling}}]{Pian2006}
{Pian}, E., {Mazzali}, P.~A., {Masetti}, N., {et~al.} 2006, \nat, 442, 1011, \dodoi{10.1038/nature05082}

\bibitem[{{Piran}(2004)}]{Piran2004}
{Piran}, T. 2004, Reviews of Modern Physics, 76, 1143, \dodoi{10.1103/RevModPhys.76.1143}

\bibitem[{{Piro} {et~al.}(2010){Piro}, {Chang}, \& {Weinberg}}]{Piro2010}
{Piro}, A.~L., {Chang}, P., \& {Weinberg}, N.~N. 2010, \apj, 708, 598, \dodoi{10.1088/0004-637X/708/1/598}

\bibitem[{{Piro} {et~al.}(2021){Piro}, {Haynie}, \& {Yao}}]{Piro2021}
{Piro}, A.~L., {Haynie}, A., \& {Yao}, Y. 2021, \apj, 909, 209, \dodoi{10.3847/1538-4357/abe2b1}

\bibitem[{{Planck Collaboration} {et~al.}(2020){Planck Collaboration}, {Aghanim}, {Akrami}, {Ashdown}, {Aumont}, {Baccigalupi}, {Ballardini}, {Banday}, {Barreiro}, {Bartolo}, {Basak}, {Battye}, {Benabed}, {Bernard}, {Bersanelli}, {Bielewicz}, {Bock}, {Bond}, {Borrill}, {Bouchet}, {Boulanger}, {Bucher}, {Burigana}, {Butler}, {Calabrese}, {Cardoso}, {Carron}, {Challinor}, {Chiang}, {Chluba}, {Colombo}, {Combet}, {Contreras}, {Crill}, {Cuttaia}, {de Bernardis}, {de Zotti}, {Delabrouille}, {Delouis}, {Di Valentino}, {Diego}, {Dor{\'e}}, {Douspis}, {Ducout}, {Dupac}, {Dusini}, {Efstathiou}, {Elsner}, {En{\ss}lin}, {Eriksen}, {Fantaye}, {Farhang}, {Fergusson}, {Fernandez-Cobos}, {Finelli}, {Forastieri}, {Frailis}, {Fraisse}, {Franceschi}, {Frolov}, {Galeotta}, {Galli}, {Ganga}, {G{\'e}nova-Santos}, {Gerbino}, {Ghosh}, {Gonz{\'a}lez-Nuevo}, {G{\'o}rski}, {Gratton}, {Gruppuso}, {Gudmundsson}, {Hamann}, {Handley}, {Hansen}, {Herranz}, {Hildebrandt}, {Hivon}, {Huang}, {Jaffe}, {Jones}, {Karakci}, {Keih{\"a}nen},
  {Keskitalo}, {Kiiveri}, {Kim}, {Kisner}, {Knox}, {Krachmalnicoff}, {Kunz}, {Kurki-Suonio}, {Lagache}, {Lamarre}, {Lasenby}, {Lattanzi}, {Lawrence}, {Le Jeune}, {Lemos}, {Lesgourgues}, {Levrier}, {Lewis}, {Liguori}, {Lilje}, {Lilley}, {Lindholm}, {L{\'o}pez-Caniego}, {Lubin}, {Ma}, {Mac{\'\i}as-P{\'e}rez}, {Maggio}, {Maino}, {Mandolesi}, {Mangilli}, {Marcos-Caballero}, {Maris}, {Martin}, {Martinelli}, {Mart{\'\i}nez-Gonz{\'a}lez}, {Matarrese}, {Mauri}, {McEwen}, {Meinhold}, {Melchiorri}, {Mennella}, {Migliaccio}, {Millea}, {Mitra}, {Miville-Desch{\^e}nes}, {Molinari}, {Montier}, {Morgante}, {Moss}, {Natoli}, {N{\o}rgaard-Nielsen}, {Pagano}, {Paoletti}, {Partridge}, {Patanchon}, {Peiris}, {Perrotta}, {Pettorino}, {Piacentini}, {Polastri}, {Polenta}, {Puget}, {Rachen}, {Reinecke}, {Remazeilles}, {Renzi}, {Rocha}, {Rosset}, {Roudier}, {Rubi{\~n}o-Mart{\'\i}n}, {Ruiz-Granados}, {Salvati}, {Sandri}, {Savelainen}, {Scott}, {Shellard}, {Sirignano}, {Sirri}, {Spencer}, {Sunyaev}, {Suur-Uski}, {Tauber}, {Tavagnacco},
  {Tenti}, {Toffolatti}, {Tomasi}, {Trombetti}, {Valenziano}, {Valiviita}, {Van Tent}, {Vibert}, {Vielva}, {Villa}, {Vittorio}, {Wandelt}, {Wehus}, {White}, {White}, {Zacchei}, \& {Zonca}}]{Planck18}
{Planck Collaboration}, {Aghanim}, N., {Akrami}, Y., {et~al.} 2020, \aap, 641, A6, \dodoi{10.1051/0004-6361/201833910}

\bibitem[{{Polzin} {et~al.}(2023){Polzin}, {Margutti}, {Coppejans}, {Auchettl}, {Page}, {Vasilopoulos}, {Bright}, {Esposito}, {Williams}, {Mukai}, \& {Berger}}]{Polzin2023}
{Polzin}, A., {Margutti}, R., {Coppejans}, D.~L., {et~al.} 2023, \apj, 959, 75, \dodoi{10.3847/1538-4357/acf765}

\bibitem[{{Poznanski} {et~al.}(2011){Poznanski}, {Ganeshalingam}, {Silverman}, \& {Filippenko}}]{Poznanski2011}
{Poznanski}, D., {Ganeshalingam}, M., {Silverman}, J.~M., \& {Filippenko}, A.~V. 2011, \mnras, 415, L81, \dodoi{10.1111/j.1745-3933.2011.01084.x}

\bibitem[{{Prentice} {et~al.}(2018){Prentice}, {Maguire}, {Smartt}, {Magee}, {Schady}, {Sim}, {Chen}, {Clark}, {Colin}, {Fulton}, {McBrien}, {O'Neill}, {Smith}, {Ashall}, {Chambers}, {Denneau}, {Flewelling}, {Heinze}, {Holoien}, {Huber}, {Kochanek}, {Mazzali}, {Prieto}, {Rest}, {Shappee}, {Stalder}, {Stanek}, {Stritzinger}, {Thompson}, \& {Tonry}}]{Prentice2018}
{Prentice}, S.~J., {Maguire}, K., {Smartt}, S.~J., {et~al.} 2018, \apjl, 865, L3, \dodoi{10.3847/2041-8213/aadd90}

\bibitem[{{Prochaska} {et~al.}(2020{\natexlab{a}}){Prochaska}, {Hennawi}, {Westfall}, {Cooke}, {Wang}, {Hsyu}, {Davies}, \& {Farina}}]{pypeit:joss_arXiv}
{Prochaska}, J.~X., {Hennawi}, J.~F., {Westfall}, K.~B., {et~al.} 2020{\natexlab{a}}, arXiv e-prints, arXiv:2005.06505.
\newblock \doarXiv{2005.06505}

\bibitem[{{Prochaska} {et~al.}(2020{\natexlab{b}}){Prochaska}, {Hennawi}, {Cooke}, {Westfall}, {Wang}, {EmAstro}, {Tiffanyhsyu}, {Wasserman}, {Villaume}, {Marijana777}, {Schindler}, {Young}, {Simha}, {Wilde}, {Tejos}, {Isbell}, {Fl{\"o}rs}, {Sandford}, {Vasovi{\'c}}, {Betts}, \& {Holden}}]{pypeit:zenodo}
{Prochaska}, J.~X., {Hennawi}, J., {Cooke}, R., {et~al.} 2020{\natexlab{b}}, {pypeit/PypeIt: Release 1.0.0}, v1.0.0,  Zenodo, \dodoi{10.5281/zenodo.3743493}

\bibitem[{Prochaska {et~al.}(2020)Prochaska, Hennawi, Westfall, Cooke, Wang, Hsyu, Davies, Farina, \& Pelliccia}]{pypeit:joss_pub}
Prochaska, J.~X., Hennawi, J.~F., Westfall, K.~B., {et~al.} 2020, Journal of Open Source Software, 5, 2308, \dodoi{10.21105/joss.02308}

\bibitem[{{Pursiainen} {et~al.}(2018){Pursiainen}, {Childress}, {Smith}, {Prajs}, {Sullivan}, {Davis}, {Foley}, {Asorey}, {Calcino}, {Carollo}, {Curtin}, {D'Andrea}, {Glazebrook}, {Gutierrez}, {Hinton}, {Hoormann}, {Inserra}, {Kessler}, {King}, {Kuehn}, {Lewis}, {Lidman}, {Macaulay}, {M{\"o}ller}, {Nichol}, {Sako}, {Sommer}, {Swann}, {Tucker}, {Uddin}, {Wiseman}, {Zhang}, {Abbott}, {Abdalla}, {Allam}, {Annis}, {Avila}, {Brooks}, {Buckley-Geer}, {Burke}, {Carnero Rosell}, {Carrasco Kind}, {Carretero}, {Castander}, {Cunha}, {Davis}, {De Vicente}, {Diehl}, {Doel}, {Eifler}, {Flaugher}, {Fosalba}, {Frieman}, {Garc{\'\i}a-Bellido}, {Gruen}, {Gruendl}, {Gutierrez}, {Hartley}, {Hollowood}, {Honscheid}, {James}, {Jeltema}, {Kuropatkin}, {Li}, {Lima}, {Maia}, {Martini}, {Menanteau}, {Ogando}, {Plazas}, {Roodman}, {Sanchez}, {Scarpine}, {Schindler}, {Smith}, {Soares-Santos}, {Sobreira}, {Suchyta}, {Swanson}, {Tarle}, {Tucker}, {Walker}, \& {DES Collaboration}}]{Pursiainen2018}
{Pursiainen}, M., {Childress}, M., {Smith}, M., {et~al.} 2018, \mnras, 481, 894, \dodoi{10.1093/mnras/sty2309}

\bibitem[{{Quataert} \& {Shiode}(2012)}]{Quataert2012}
{Quataert}, E., \& {Shiode}, J. 2012, \mnras, 423, L92, \dodoi{10.1111/j.1745-3933.2012.01264.x}

\bibitem[{{Quirola-V{\'a}squez} {et~al.}(2022){Quirola-V{\'a}squez}, {Bauer}, {Jonker}, {Brandt}, {Yang}, {Levan}, {Xue}, {Eappachen}, {Zheng}, \& {Luo}}]{Vazquez2022}
{Quirola-V{\'a}squez}, J., {Bauer}, F.~E., {Jonker}, P.~G., {et~al.} 2022, \aap, 663, A168, \dodoi{10.1051/0004-6361/202243047}

\bibitem[{{Quirola-V{\'a}squez} {et~al.}(2024){Quirola-V{\'a}squez}, {Bauer}, {Jonker}, {Brandt}, {Eappachen}, {Levan}, {L{\'o}pez}, {Luo}, {Ravasio}, {Sun}, {Xue}, {Yang}, \& {Zheng}}]{Vazquez2024}
---. 2024, \aap, 683, A243, \dodoi{10.1051/0004-6361/202347629}

\bibitem[{{Quirola-V{\'a}squez} {et~al.}(2025{\natexlab{a}}){Quirola-V{\'a}squez}, {Bauer}, {Jonker}, {Levan}, {Brandt}, {Ravasio}, {Eappachen}, {Xue}, \& {Zheng}}]{Vasquez2025}
---. 2025{\natexlab{a}}, \aap, 695, A279, \dodoi{10.1051/0004-6361/202451825}

\bibitem[{{Quirola-V{\'a}squez} {et~al.}(2025{\natexlab{b}}){Quirola-V{\'a}squez}, {Jonker}, {Levan}, {Malesani}, {Bauer}, {Sarin}, {Lamb}, {Martin-Carrillo}, {S{\'a}nchez-Sierras}, {Fraser}, {Izzo}, {Ravasio}, {Mata S{\'a}nchez}, {Torres}, {van Dalen}, {van Hoof}, {Chac{\'o}n}, {Littlefair}, {Dhillon}, {Cotter}, {Corcoran}, {Eyles-Ferris}, {O'Brien}, {Stern}, {D'Elia}, \& {Hartmann}}]{Quirola2025}
{Quirola-V{\'a}squez}, J., {Jonker}, P.~G., {Levan}, A.~J., {et~al.} 2025{\natexlab{b}}, arXiv e-prints, arXiv:2511.13314.
\newblock \doarXiv{2511.13314}

\bibitem[{{Rabinak} \& {Waxman}(2011)}]{Rabinak2011}
{Rabinak}, I., \& {Waxman}, E. 2011, \apj, 728, 63, \dodoi{10.1088/0004-637X/728/1/63}

\bibitem[{{Rastinejad} {et~al.}(2025){Rastinejad}, {Levan}, {Jonker}, {Kilpatrick}, {Fryer}, {Sarin}, {Gompertz}, {Liu}, {Eyles-Ferris}, {Fong}, {Burns}, {Gillanders}, {Mandel}, {Malesani}, {O'Brien}, {Tanvir}, {Ackley}, {Aryan}, {Bauer}, {Bloemen}, {de Boer}, {Bom}, {Chacon}, {Chambers}, {Chen}, {Chrimes}, {van Dalen}, {D'Elia}, {De Pasquale}, {Gupta}, {Hartmann}, {van Hoof}, {Izzo}, {Jacobson-Galan}, {Jakobsson}, {Kong}, {Laskar}, {Lowe}, {Magnier}, {Maiorano}, {Martin-Carrillo}, {Mas-Ribas}, {Mata Sanchez}, {Nicholl}, {Nixon}, {Oates}, {Paek}, {Palmerio}, {Paris}, {Pieterse}, {Pugliese}, {Quirola Vasquez}, {van Roestel}, {Rossi}, {Salvaterra}, {Schneider}, {Smartt}, {Smith}, {Smith}, {Srivastav}, {Torres}, {Ventura}, {Wainscoat}, {Yang}, \& {Yang}}]{Rastinejad2025}
{Rastinejad}, J.~C., {Levan}, A.~J., {Jonker}, P.~G., {et~al.} 2025, arXiv e-prints, arXiv:2504.08889.
\newblock \doarXiv{2504.08889}

\bibitem[{{Ravasio} {et~al.}(2025){Ravasio}, {Burns}, {Jonker}, \& {Fermi-GBM Team}}]{GCN39600}
{Ravasio}, M.~E., {Burns}, E., {Jonker}, P.~G., \& {Fermi-GBM Team}. 2025, GRB Coordinates Network, 39600, 1

\bibitem[{{Rho} {et~al.}(2021){Rho}, {Evans}, {Geballe}, {Banerjee}, {Hoeflich}, {Shahbandeh}, {Valenti}, {Yoon}, {Jin}, {Williamson}, {Modjaz}, {Hiramatsu}, {Howell}, {Pellegrino}, {Vink{\'o}}, {Cartier}, {Burke}, {McCully}, {An}, {Cha}, {Pritchard}, {Wang}, {Andrews}, {Galbany}, {Van Dyk}, {Graham}, {Blinnikov}, {Joshi}, {P{\'a}l}, {Kriskovics}, {Ordasi}, {Szakats}, {Vida}, {Chen}, {Li}, {Zhang}, \& {Yan}}]{Rho2021}
{Rho}, J., {Evans}, A., {Geballe}, T.~R., {et~al.} 2021, \apj, 908, 232, \dodoi{10.3847/1538-4357/abd850}

\bibitem[{{Rhoads}(1997)}]{Rhoads1997}
{Rhoads}, J.~E. 1997, \apjl, 487, L1, \dodoi{10.1086/310876}

\bibitem[{{Rigault} {et~al.}(2019){Rigault}, {Neill}, {Blagorodnova}, {Dugas}, {Feeney}, {Walters}, {Brinnel}, {Copin}, {Fremling}, {Nordin}, \& {Sollerman}}]{Rigualt2019}
{Rigault}, M., {Neill}, J.~D., {Blagorodnova}, N., {et~al.} 2019, \aap, 627, A115, \dodoi{10.1051/0004-6361/201935344}

\bibitem[{{Rodr{\'\i}guez} {et~al.}(2023){Rodr{\'\i}guez}, {Maoz}, \& {Nakar}}]{Osmar2023}
{Rodr{\'\i}guez}, {\'O}., {Maoz}, D., \& {Nakar}, E. 2023, \apj, 955, 71, \dodoi{10.3847/1538-4357/ace2bd}

\bibitem[{{Rodr{\'\i}guez} {et~al.}(2024{\natexlab{a}}){Rodr{\'\i}guez}, {Nakar}, \& {Maoz}}]{Rodriguez2024}
{Rodr{\'\i}guez}, {\'O}., {Nakar}, E., \& {Maoz}, D. 2024{\natexlab{a}}, \nat, 628, 733, \dodoi{10.1038/s41586-024-07262-x}

\bibitem[{{Rodr{\'\i}guez} {et~al.}(2024{\natexlab{b}}){Rodr{\'\i}guez}, {Nakar}, \& {Maoz}}]{Osmar2024}
---. 2024{\natexlab{b}}, \nat, 628, 733, \dodoi{10.1038/s41586-024-07262-x}

\bibitem[{{Roman Aguilar} \& {Bersten}(2025)}]{Aguilar2025}
{Roman Aguilar}, L.~M., \& {Bersten}, M.~C. 2025, \aap, 702, L18, \dodoi{10.1051/0004-6361/202556610}

\bibitem[{{Roming} {et~al.}(2005){Roming}, {Kennedy}, {Mason}, {Nousek}, {Ahr}, {Bingham}, {Broos}, {Carter}, {Hancock}, {Huckle}, {Hunsberger}, {Kawakami}, {Killough}, {Koch}, {McLelland}, {Smith}, {Smith}, {Soto}, {Boyd}, {Breeveld}, {Holland}, {Ivanushkina}, {Pryzby}, {Still}, \& {Stock}}]{swift_uvot}
{Roming}, P. W.~A., {Kennedy}, T.~E., {Mason}, K.~O., {et~al.} 2005, \ssr, 120, 95, \dodoi{10.1007/s11214-005-5095-4}

\bibitem[{{Ryan} {et~al.}(2015){Ryan}, {van Eerten}, {MacFadyen}, \& {Zhang}}]{2015ApJ...799....3R}
{Ryan}, G., {van Eerten}, H., {MacFadyen}, A., \& {Zhang}, B.-B. 2015, \apj, 799, 3, \dodoi{10.1088/0004-637X/799/1/3}

\bibitem[{{Sakamoto} {et~al.}(2005){Sakamoto}, {Lamb}, {Kawai}, {Yoshida}, {Graziani}, {Fenimore}, {Donaghy}, {Matsuoka}, {Suzuki}, {Ricker}, {Atteia}, {Shirasaki}, {Tamagawa}, {Torii}, {Galassi}, {Doty}, {Vanderspek}, {Crew}, {Villasenor}, {Butler}, {Prigozhin}, {Jernigan}, {Barraud}, {Boer}, {Dezalay}, {Olive}, {Hurley}, {Levine}, {Monnelly}, {Martel}, {Morgan}, {Woosley}, {Cline}, {Braga}, {Manchanda}, {Pizzichini}, {Takagishi}, \& {Yamauchi}}]{Sakamoto2005}
{Sakamoto}, T., {Lamb}, D.~Q., {Kawai}, N., {et~al.} 2005, \apj, 629, 311, \dodoi{10.1086/431235}

\bibitem[{{Sari} {et~al.}(1999){Sari}, {Piran}, \& {Halpern}}]{Sari1999}
{Sari}, R., {Piran}, T., \& {Halpern}, J.~P. 1999, \apjl, 519, L17, \dodoi{10.1086/312109}

\bibitem[{{Sarin} {et~al.}(2021){Sarin}, {Ashton}, {Lasky}, {Ackley}, {Mong}, \& {Galloway}}]{Sarin2021}
{Sarin}, N., {Ashton}, G., {Lasky}, P.~D., {et~al.} 2021, arXiv e-prints, arXiv:2105.10108, \dodoi{10.48550/arXiv.2105.10108}

\bibitem[{{Sarin} {et~al.}(2022){Sarin}, {Hamburg}, {Burns}, {Ashton}, {Lasky}, \& {Lamb}}]{Sarin2022}
{Sarin}, N., {Hamburg}, R., {Burns}, E., {et~al.} 2022, \mnras, 512, 1391, \dodoi{10.1093/mnras/stac601}

\bibitem[{{Sarin} {et~al.}(2024){Sarin}, {H{\"u}bner}, {Omand}, {Setzer}, {Schulze}, {Adhikari}, {Sagu{\'e}s-Carracedo}, {Galaudage}, {Wallace}, {Lamb}, \& {Lin}}]{Sarin2024}
{Sarin}, N., {H{\"u}bner}, M., {Omand}, C. M.~B., {et~al.} 2024, \mnras, 531, 1203, \dodoi{10.1093/mnras/stae1238}

\bibitem[{{Schlafly} \& {Finkbeiner}(2011)}]{Schlafly2011}
{Schlafly}, E.~F., \& {Finkbeiner}, D.~P. 2011, \apj, 737, 103, \dodoi{10.1088/0004-637X/737/2/103}

\bibitem[{{Schroeder} {et~al.}(2025{\natexlab{a}}){Schroeder}, {Ho}, {Dastidar}, {Modjaz}, {Corsi}, \& {Duffell}}]{2025ApJ...995...61S}
{Schroeder}, G., {Ho}, A. Y.~Q., {Dastidar}, R.~G., {et~al.} 2025{\natexlab{a}}, \apj, 995, 61, \dodoi{10.3847/1538-4357/ae129b}

\bibitem[{{Schroeder} {et~al.}(2022){Schroeder}, {Laskar}, {Fong}, {Nugent}, {Berger}, {Chornock}, {Alexander}, {Andrews}, {Bussmann}, {Castro-Tirado}, {Goyal}, {Kilpatrick}, {Lally}, {Miller}, {Milne}, {Paterson}, {Escorial}, {Stroh}, {Terreran}, \& {Zauderer}}]{2022ApJ...940...53S}
{Schroeder}, G., {Laskar}, T., {Fong}, W.-f., {et~al.} 2022, \apj, 940, 53, \dodoi{10.3847/1538-4357/ac8feb}

\bibitem[{{Schroeder} {et~al.}(2025{\natexlab{b}}){Schroeder}, {Ahumada}, {Ho}, {Kasliwal}, {Yan}, {Srinivasaragavan}, {Hall}, {Coughlin}, {Li}, {Yang}, {Tian}, {R.}, {L.}, {Peng}, {Y}, {J.}, {Song}, {Sun}, {Liu}, {Jin}, {Yuan}, {ZTF Collaboration}, \& {Einstein Probe Team}}]{GCN.41635}
{Schroeder}, G., {Ahumada}, T., {Ho}, A.~Y.~Q., {et~al.} 2025{\natexlab{b}}, GRB Coordinates Network, 41635, 1

\bibitem[{{Shankar} {et~al.}(2021){Shankar}, {M{\"o}sta}, {Barnes}, {Duffell}, \& {Kasen}}]{Shankar2021}
{Shankar}, S., {M{\"o}sta}, P., {Barnes}, J., {Duffell}, P.~C., \& {Kasen}, D. 2021, \mnras, 508, 5390, \dodoi{10.1093/mnras/stab2964}

\bibitem[{{Shiode} \& {Quataert}(2014)}]{Shiode2014}
{Shiode}, J.~H., \& {Quataert}, E. 2014, \apj, 780, 96, \dodoi{10.1088/0004-637X/780/1/96}

\bibitem[{{Shu} {et~al.}(2025){Shu}, {Yang}, {Yang}, {Xu}, {Chen}, {Eyles-Ferris}, {Dai}, {Yu}, {Shen}, {Sun}, {Ding}, {Zheng}, {Jiang}, {Li}, {Sun}, {Xu}, {Zhang}, {Jin}, {Rau}, {Wang}, {Wu}, {Yuan}, {Zhang}, {Nandra}, {Filippenko}, {Poidevin}, {Soria}, {Kumar}, {Aguado}, {An}, {An}, {An}, {Andrews}, {Anutarawiramkul}, {Baldini}, {Brink}, {Butpan}, {Cai}, {Castro-Tirado}, {Cheng}, {Cui}, {Farah}, {Fu}, {Fynbo}, {Gao}, {Han}, {Han}, {Howell}, {Hu}, {Jiang}, {Kumar}, {Lei}, {Li}, {Li}, {Liu}, {Liu}, {Liu}, {Liu}, {L{\'o}pez-Oramas}, {L{\'o}pez Fern{\'a}ndez-Nespral}, {Maund}, {McCully}, {Niu}, {Newsome}, {O'Brien}, {Pan}, {Pan}, {Padilla Gonzalez}, {P{\'e}rez-Fournon}, {Silima}, {Sun}, {Sun}, {Sun}, {Terreran}, {Tinyanont}, {Wang}, {Wang}, {Wang}, {Wiersema}, {Xu}, {Xue}, {Yang}, {Zhang}, {Zhang}, {Zhang}, {Zhang}, {Zhang}, {Zhao}, {Zhu}, {Xin}, {Yao}, {Cordier}, {Wei}, {Qiu}, \& {Daigne}}]{Shu2025}
{Shu}, X., {Yang}, L., {Yang}, H., {et~al.} 2025, \apjl, 990, L29, \dodoi{10.3847/2041-8213/adf4cd}

\bibitem[{{Shvartzvald} {et~al.}(2024){Shvartzvald}, {Waxman}, {Gal-Yam}, {Ofek}, {Ben-Ami}, {Berge}, {Kowalski}, {B{\"u}hler}, {Worm}, {Rhoads}, {Arcavi}, {Maoz}, {Polishook}, {Stone}, {Trakhtenbrot}, {Ackermann}, {Aharonson}, {Birnholtz}, {Chelouche}, {Guetta}, {Hallakoun}, {Horesh}, {Kushnir}, {Mazeh}, {Nordin}, {Ofir}, {Ohm}, {Parsons}, {Pe'er}, {Perets}, {Perdelwitz}, {Poznanski}, {Sadeh}, {Sagiv}, {Shahaf}, {Soumagnac}, {Tal-Or}, {Santen}, {Zackay}, {Guttman}, {Rekhi}, {Townsend}, {Weinstein}, \& {Wold}}]{Shvartzvald2025}
{Shvartzvald}, Y., {Waxman}, E., {Gal-Yam}, A., {et~al.} 2024, \apj, 964, 74, \dodoi{10.3847/1538-4357/ad2704}

\bibitem[{{Skrutskie} {et~al.}(2006){Skrutskie}, {Cutri}, {Stiening}, {Weinberg}, {Schneider}, {Carpenter}, {Beichman}, {Capps}, {Chester}, {Elias}, {Huchra}, {Liebert}, {Lonsdale}, {Monet}, {Price}, {Seitzer}, {Jarrett}, {Kirkpatrick}, {Gizis}, {Howard}, {Evans}, {Fowler}, {Fullmer}, {Hurt}, {Light}, {Kopan}, {Marsh}, {McCallon}, {Tam}, {Van Dyk}, \& {Wheelock}}]{Skrutskie2006}
{Skrutskie}, M.~F., {Cutri}, R.~M., {Stiening}, R., {et~al.} 2006, \aj, 131, 1163, \dodoi{10.1086/498708}

\bibitem[{{Soderberg} {et~al.}(2006{\natexlab{a}}){Soderberg}, {Nakar}, {Berger}, \& {Kulkarni}}]{Soderberg+2006}
{Soderberg}, A.~M., {Nakar}, E., {Berger}, E., \& {Kulkarni}, S.~R. 2006{\natexlab{a}}, \apj, 638, 930, \dodoi{10.1086/499121}

\bibitem[{{Soderberg} {et~al.}(2004{\natexlab{a}}){Soderberg}, {Kulkarni}, {Berger}, {Fox}, {Price}, {Yost}, {Hunt}, {Frail}, {Walker}, {Hamuy}, {Shectman}, {Halpern}, \& {Mirabal}}]{2004ApJ...606..994S}
{Soderberg}, A.~M., {Kulkarni}, S.~R., {Berger}, E., {et~al.} 2004{\natexlab{a}}, \apj, 606, 994, \dodoi{10.1086/383082}

\bibitem[{{Soderberg} {et~al.}(2004{\natexlab{b}}){Soderberg}, {Kulkarni}, {Berger}, {Fox}, {Sako}, {Frail}, {Gal-Yam}, {Moon}, {Cenko}, {Yost}, {Phillips}, {Persson}, {Freedman}, {Wyatt}, {Jayawardhana}, \& {Paulson}}]{2004Natur.430..648S}
---. 2004{\natexlab{b}}, \nat, 430, 648, \dodoi{10.1038/nature02757}

\bibitem[{{Soderberg} {et~al.}(2005){Soderberg}, {Kulkarni}, {Fox}, {Berger}, {Price}, {Cenko}, {Howell}, {Gal-Yam}, {Leonard}, {Frail}, {Moon}, {Chevalier}, {Hamuy}, {Hurley}, {Kelson}, {Koviak}, {Krzeminski}, {Kumar}, {MacFadyen}, {McCarthy}, {Park}, {Peterson}, {Phillips}, {Rauch}, {Roth}, {Schmidt}, \& {Shectman}}]{Soderberg2005}
{Soderberg}, A.~M., {Kulkarni}, S.~R., {Fox}, D.~B., {et~al.} 2005, \apj, 627, 877, \dodoi{10.1086/430405}

\bibitem[{{Soderberg} {et~al.}(2006{\natexlab{b}}){Soderberg}, {Kulkarni}, {Nakar}, {Berger}, {Cameron}, {Fox}, {Frail}, {Gal-Yam}, {Sari}, {Cenko}, {Kasliwal}, {Chevalier}, {Piran}, {Price}, {Schmidt}, {Pooley}, {Moon}, {Penprase}, {Ofek}, {Rau}, {Gehrels}, {Nousek}, {Burrows}, {Persson}, \& {McCarthy}}]{2006Natur.442.1014S}
{Soderberg}, A.~M., {Kulkarni}, S.~R., {Nakar}, E., {et~al.} 2006{\natexlab{b}}, \nat, 442, 1014, \dodoi{10.1038/nature05087}

\bibitem[{{Soderberg} {et~al.}(2008{\natexlab{a}}){Soderberg}, {Berger}, {Page}, {Schady}, {Parrent}, {Pooley}, {Wang}, {Ofek}, {Cucchiara}, {Rau}, {Waxman}, {Simon}, {Bock}, {Milne}, {Page}, {Barentine}, {Barthelmy}, {Beardmore}, {Bietenholz}, {Brown}, {Burrows}, {Burrows}, {Byrngelson}, {Cenko}, {Chandra}, {Cummings}, {Fox}, {Gal-Yam}, {Gehrels}, {Immler}, {Kasliwal}, {Kong}, {Krimm}, {Kulkarni}, {Maccarone}, {M{\'e}sz{\'a}ros}, {Nakar}, {O'Brien}, {Overzier}, {de Pasquale}, {Racusin}, {Rea}, \& {York}}]{Soderberg2008}
{Soderberg}, A.~M., {Berger}, E., {Page}, K.~L., {et~al.} 2008{\natexlab{a}}, \nat, 453, 469, \dodoi{10.1038/nature06997}

\bibitem[{{Soderberg} {et~al.}(2008{\natexlab{b}}){Soderberg}, {Berger}, {Page}, {Schady}, {Parrent}, {Pooley}, {Wang}, {Ofek}, {Cucchiara}, {Rau}, {Waxman}, {Simon}, {Bock}, {Milne}, {Page}, {Barentine}, {Barthelmy}, {Beardmore}, {Bietenholz}, {Brown}, {Burrows}, {Burrows}, {Byrngelson}, {Cenko}, {Chandra}, {Cummings}, {Fox}, {Gal-Yam}, {Gehrels}, {Immler}, {Kasliwal}, {Kong}, {Krimm}, {Kulkarni}, {Maccarone}, {M{\'e}sz{\'a}ros}, {Nakar}, {O'Brien}, {Overzier}, {de Pasquale}, {Racusin}, {Rea}, \& {York}}]{2008Natur.453..469S}
---. 2008{\natexlab{b}}, \nat, 453, 469, \dodoi{10.1038/nature06997}

\bibitem[{{Sollerman} {et~al.}(2006){Sollerman}, {Jaunsen}, {Fynbo}, {Hjorth}, {Jakobsson}, {Stritzinger}, {F{\'e}ron}, {Laursen}, {Ovaldsen}, {Selj}, {Th{\"o}ne}, {Xu}, {Davis}, {Gorosabel}, {Watson}, {Duro}, {Ilyin}, {Jensen}, {Lysfjord}, {Marquart}, {Nielsen}, {N{\"a}r{\"a}nen}, {Schwarz}, {Walch}, {Wold}, \& {{\"O}stlin}}]{Sollerman2006}
{Sollerman}, J., {Jaunsen}, A.~O., {Fynbo}, J.~P.~U., {et~al.} 2006, \aap, 454, 503, \dodoi{10.1051/0004-6361:20065226}

\bibitem[{{Speagle}(2020)}]{Dynesty}
{Speagle}, J.~S. 2020, \mnras, 493, 3132, \dodoi{10.1093/mnras/staa278}

\bibitem[{{Srinivasaragavan} {et~al.}(2023){Srinivasaragavan}, {O'Connor}, {Cenko}, {Dittmann}, {Yang}, {Sollerman}, {Anupama}, {Barway}, {Bhalerao}, {Kumar}, {Swain}, {Hammerstein}, {Holt}, {Anand}, {Andreoni}, {Coughlin}, {Dichiara}, {Gal-Yam}, {Miller}, {Soon}, {Soria}, {Durbak}, {Gillanders}, {Laha}, {Moore}, {Ragosta}, \& {Troja}}]{Srinivasaragavan2023}
{Srinivasaragavan}, G.~P., {O'Connor}, B., {Cenko}, S.~B., {et~al.} 2023, \apjl, 949, L39, \dodoi{10.3847/2041-8213/accf97}

\bibitem[{{Srinivasaragavan} {et~al.}(2024{\natexlab{a}}){Srinivasaragavan}, {Swain}, {O'Connor}, {Anand}, {Ahumada}, {Perley}, {Stein}, {Sollerman}, {Fremling}, {Cenko}, {Antier}, {Guessoum}, {Hussenot-Desenonges}, {Hello}, {Lesage}, {Hammerstein}, {Miller}, {Andreoni}, {Bhalerao}, {Bloom}, {Dutta}, {Gal-Yam}, {Hinds}, {Jaodand}, {Kasliwal}, {Kumar}, {Kutyrev}, {Ragosta}, {Ravi}, {Sharma}, {Teja}, {Yang}, {Anupama}, {Bellm}, {Coughlin}, {Mahabal}, {Masci}, {Pathak}, {Purdum}, {Roberts}, {Smith}, \& {Wold}}]{Srinivasaragavan2024}
{Srinivasaragavan}, G.~P., {Swain}, V., {O'Connor}, B., {et~al.} 2024{\natexlab{a}}, \apjl, 960, L18, \dodoi{10.3847/2041-8213/ad16e7}

\bibitem[{{Srinivasaragavan} {et~al.}(2024{\natexlab{b}}){Srinivasaragavan}, {Yang}, {Anand}, {Sollerman}, {Ho}, {Corsi}, {Cenko}, {Perley}, {Schulze}, {Sanchez-Fleming}, {Pope}, {Sarin}, {Omand}, {Das}, {Fremling}, {Andreoni}, {Bruch}, {Burdge}, {De}, {Gal-Yam}, {Gangopadhyay}, {Graham}, {Jencson}, {Karambelkar}, {Kasliwal}, {Kulkarni}, {Martikainen}, {Sharma}, {Tzanidakis}, {Yan}, {Yao}, {Bellm}, {Groom}, {Masci}, {Nir}, {Purdum}, {Smith}, \& {Sravan}}]{Srinivasaragavan2024b}
{Srinivasaragavan}, G.~P., {Yang}, S., {Anand}, S., {et~al.} 2024{\natexlab{b}}, \apj, 976, 71, \dodoi{10.3847/1538-4357/ad7fde}

\bibitem[{{Srinivasaragavan} {et~al.}(2025){Srinivasaragavan}, {Hamidani}, {Schroeder}, {Sarin}, {Ho}, {Piro}, {Cenko}, {Anand}, {Sollerman}, {Perley}, {Maeda}, {O'Connor}, {Kuncarayakti}, {Miller}, {Ahumada}, {Vail}, {Duffell}, {Dastidar}, {Andreoni}, {Bochenek}, {Brennan}, {Carney}, {Chen}, {Freeburn}, {Gal-Yam}, {Jacobson-Gal{\'a}n}, {Kasliwal}, {Li}, {Li}, {Sravan}, \& {Warshofsky}}]{Srinivasaragavan2025b}
{Srinivasaragavan}, G.~P., {Hamidani}, H., {Schroeder}, G., {et~al.} 2025, \apjl, 988, L60, \dodoi{10.3847/2041-8213/ade870}

\bibitem[{{Srivastav} {et~al.}(2025){Srivastav}, {Chen}, {Gillanders}, {Rhodes}, {Smartt}, {Huber}, {Aryan}, {Yang}, {Beri}, {Cooper}, {Nicholl}, {Smith}, {Stevance}, {Carotenuto}, {Chambers}, {Aamer}, {Angus}, {Fulton}, {Moore}, {Smith}, {Young}, {de Boer}, {Gao}, {Lin}, {Lowe}, {Magnier}, {Minguez}, {Pan}, \& {Wainscoat}}]{Srivastav2025}
{Srivastav}, S., {Chen}, T.~W., {Gillanders}, J.~H., {et~al.} 2025, \apjl, 978, L21, \dodoi{10.3847/2041-8213/ad9c75}

\bibitem[{{Starling} {et~al.}(2011){Starling}, {Wiersema}, {Levan}, {Sakamoto}, {Bersier}, {Goldoni}, {Oates}, {Rowlinson}, {Campana}, {Sollerman}, {Tanvir}, {Malesani}, {Fynbo}, {Covino}, {D'Avanzo}, {O'Brien}, {Page}, {Osborne}, {Vergani}, {Barthelmy}, {Burrows}, {Cano}, {Curran}, {de Pasquale}, {D'Elia}, {Evans}, {Flores}, {Fruchter}, {Garnavich}, {Gehrels}, {Gorosabel}, {Hjorth}, {Holland}, {van der Horst}, {Hurkett}, {Jakobsson}, {Kamble}, {Kouveliotou}, {Kuin}, {Kaper}, {Mazzali}, {Nugent}, {Pian}, {Stamatikos}, {Th{\"o}ne}, \& {Woosley}}]{Starling2011}
{Starling}, R.~L.~C., {Wiersema}, K., {Levan}, A.~J., {et~al.} 2011, \mnras, 411, 2792, \dodoi{10.1111/j.1365-2966.2010.17879.x}

\bibitem[{{Steele} {et~al.}(2004){Steele}, {Smith}, {Rees}, {Baker}, {Bates}, {Bode}, {Bowman}, {Carter}, {Etherton}, {Ford}, {Fraser}, {Gomboc}, {Lett}, {Mansfield}, {Marchant}, {Medrano-Cerda}, {Mottram}, {Raback}, {Scott}, {Tomlinson}, \& {Zamanov}}]{Steele2004}
{Steele}, I.~A., {Smith}, R.~J., {Rees}, P.~C., {et~al.} 2004, in Society of Photo-Optical Instrumentation Engineers (SPIE) Conference Series, Vol. 5489, Ground-based Telescopes, ed. J.~{Oschmann}, Jacobus~M., 679--692, \dodoi{10.1117/12.551456}

\bibitem[{{Stritzinger} {et~al.}(2018){Stritzinger}, {Taddia}, {Burns}, {Phillips}, {Bersten}, {Contreras}, {Folatelli}, {Holmbo}, {Hsiao}, {Hoeflich}, {Leloudas}, {Morrell}, {Sollerman}, \& {Suntzeff}}]{Stritzinger2018}
{Stritzinger}, M.~D., {Taddia}, F., {Burns}, C.~R., {et~al.} 2018, \aap, 609, A135, \dodoi{10.1051/0004-6361/201730843}

\bibitem[{{Sun} {et~al.}(2017){Sun}, {Zhang}, \& {Gao}}]{Sun2017}
{Sun}, H., {Zhang}, B., \& {Gao}, H. 2017, \apj, 835, 7, \dodoi{10.3847/1538-4357/835/1/7}

\bibitem[{{Sun} {et~al.}(2025){Sun}, {Li}, {Liu}, {Gao}, {Wang}, {Yuan}, {Zhang}, {Filippenko}, {Xu}, {An}, {Ai}, {Brink}, {Liu}, {Liu}, {Wang}, {Wu}, {Wu}, {Yang}, {Zhang}, {Zheng}, {Ahumada}, {Dai}, {Delaunay}, {Elias-Rosa}, {Benetti}, {Fu}, {Howell}, {Huang}, {Kasliwal}, {Karambelkar}, {Stein}, {Lei}, {Lian}, {Peng}, {Frederiks}, {Ridnaia}, {Svinkin}, {Wang}, {Wang}, {Wei}, {An}, {Andrews}, {Bai}, {Dai}, {Ehgamberdiev}, {Fan}, {Farah}, {Feng}, {Fynbo}, {Guo}, {Guo}, {Hu}, {Hu}, {Jiang}, {Jin}, {Li}, {Li}, {Li}, {Liang}, {Ling}, {Liu}, {Mao}, {McCully}, {Mirzaqulov}, {Newsome}, {Padilla Gonzalez}, {Pan}, {Terreran}, {Tinyanont}, {Wang}, {Wang}, {Wen}, {Xiang}, {Xue}, {Yang}, {Zhu}, {Cai}, {Castro-Tirado}, {Chen}, {Chen}, {Chen}, {Chen}, {Chen}, {Chen}, {Chen}, {Cheng}, {Cordier}, {Cui}, {Cui}, {Dai}, {Fan}, {Feng}, {Guan}, {Han}, {Hou}, {Hu}, {Huang}, {Huo}, {Jia}, {Jia}, {Jiang}, {Jin}, {Jin}, {Kuulkers}, {Li}, {Li}, {Li}, {Li}, {Li}, {Li}, {Li}, {Liu}, {Liu}, {Liu}, {Liu}, {Lu}, {Luo}, {Ma}, {Mao},
  {Nandra}, {O'Brien}, {Pan}, {Rau}, {Rea}, {Sanders}, {Song}, {Sun}, {Sun}, {Tan}, {Tang}, {Tao}, {Wang}, {Wang}, {Wang}, {Wang}, {Wang}, {Wang}, {Xiong}, {Xu}, {Xu}, {Xu}, {Xu}, {Xu}, {Xue}, {Xue}, {Yan}, {Yang}, {Yang}, {Yang}, {Zhang}, {Zhang}, {Zhang}, {Zhang}, {Zhang}, {Zhang}, {Zhang}, {Zhang}, {Zhang}, {Zhang}, {Zhao}, {Zhao}, {Zhao}, {Zhao}, {Zhou}, {Zhu}, {Zhu}, \& {Zou}}]{Sun2025}
{Sun}, H., {Li}, W.-X., {Liu}, L.-D., {et~al.} 2025, Nature Astronomy, 9, 1073, \dodoi{10.1038/s41550-025-02571-1}

\bibitem[{{Suzuki} \& {Maeda}(2017)}]{Suzuki2017}
{Suzuki}, A., \& {Maeda}, K. 2017, \mnras, 466, 2633, \dodoi{10.1093/mnras/stw3259}

\bibitem[{{Taddia} {et~al.}(2019){Taddia}, {Sollerman}, {Fremling}, {Barbarino}, {Karamehmetoglu}, {Arcavi}, {Cenko}, {Filippenko}, {Gal-Yam}, {Hiramatsu}, {Hosseinzadeh}, {Howell}, {Kulkarni}, {Laher}, {Lunnan}, {Masci}, {Nugent}, {Nyholm}, {Perley}, {Quimby}, \& {Silverman}}]{Taddia2018}
{Taddia}, F., {Sollerman}, J., {Fremling}, C., {et~al.} 2019, \aap, 621, A71, \dodoi{10.1051/0004-6361/201834429}

\bibitem[{{Thompson} \& {Duncan}(1993)}]{Thompson1993}
{Thompson}, C., \& {Duncan}, R.~C. 1993, \apj, 408, 194, \dodoi{10.1086/172580}

\bibitem[{{Tody}(1986)}]{1986SPIE..627..733T}
{Tody}, D. 1986, in Society of Photo-Optical Instrumentation Engineers (SPIE) Conference Series, Vol. 627, Instrumentation in astronomy VI, ed. D.~L. {Crawford}, 733, \dodoi{10.1117/12.968154}

\bibitem[{{Toma} {et~al.}(2007){Toma}, {Ioka}, {Sakamoto}, \& {Nakamura}}]{Kenji2007}
{Toma}, K., {Ioka}, K., {Sakamoto}, T., \& {Nakamura}, T. 2007, \apj, 659, 1420, \dodoi{10.1086/512481}

\bibitem[{{Tominaga} {et~al.}(2004){Tominaga}, {Deng}, {Mazzali}, {Maeda}, {Nomoto}, {Pian}, {Hjorth}, \& {Fynbo}}]{Tominaga2004}
{Tominaga}, N., {Deng}, J., {Mazzali}, P.~A., {et~al.} 2004, \apjl, 612, L105, \dodoi{10.1086/424841}

\bibitem[{{Valenti} {et~al.}(2008){Valenti}, {Benetti}, {Cappellaro}, {Patat}, {Mazzali}, {Turatto}, {Hurley}, {Maeda}, {Gal-Yam}, {Foley}, {Filippenko}, {Pastorello}, {Challis}, {Frontera}, {Harutyunyan}, {Iye}, {Kawabata}, {Kirshner}, {Li}, {Lipkin}, {Matheson}, {Nomoto}, {Ofek}, {Ohyama}, {Pian}, {Poznanski}, {Salvo}, {Sauer}, {Schmidt}, {Soderberg}, \& {Zampieri}}]{Valenti2008}
{Valenti}, S., {Benetti}, S., {Cappellaro}, E., {et~al.} 2008, \mnras, 383, 1485, \dodoi{10.1111/j.1365-2966.2007.12647.x}

\bibitem[{{van Dalen} {et~al.}(2024){van Dalen}, {Levan}, {Jonker}, {Malesani}, {Izzo}, {Sarin}, {Quirola-V{\'a}squez}, {Mata S{\'a}nchez}, {de Ugarte Postigo}, {van Hoof}, {Torres}, {Schulze}, {Littlefair}, {Chrimes}, {Ravasio}, {Bauer}, {Martin-Carrillo}, {Fraser}, {van der Horst}, {Jakobsson}, {O'Brien}, {De Pasquale}, {Pugliese}, {Sollerman}, {Tanvir}, {Zafar}, {Anderson}, {Galbany}, {Gal-Yam}, {Gromadzki}, {Muller-Bravo}, {Ragosta}, \& {Terwel}}]{vanDalen2024}
{van Dalen}, J. N.~D., {Levan}, A.~J., {Jonker}, P.~G., {et~al.} 2024, arXiv e-prints, arXiv:2409.19056, \dodoi{10.48550/arXiv.2409.19056}

\bibitem[{{van Dalen} {et~al.}(2025){van Dalen}, {Levan}, {Jonker}, {Malesani}, {Izzo}, {Sarin}, {Quirola-V{\'a}squez}, {S{\'a}nchez}, {de Ugarte Postigo}, {van Hoof}, {Torres}, {Schulze}, {Littlefair}, {Chrimes}, {Ravasio}, {Bauer}, {Martin-Carrillo}, {Fraser}, {van der Horst}, {Jakobsson}, {O'Brien}, {De Pasquale}, {Pugliese}, {Sollerman}, {Tanvir}, {Zafar}, {Anderson}, {Galbany}, {Gal-Yam}, {Gromadzki}, {M{\"u}ller-Bravo}, {Ragosta}, \& {Terwel}}]{VanDalen2025}
---. 2025, \apjl, 982, L47, \dodoi{10.3847/2041-8213/adbc7e}

\bibitem[{{van der Walt} {et~al.}(2019){van der Walt}, {Crellin-Quick}, \& {Bloom}}]{VanderWalt2019}
{van der Walt}, S., {Crellin-Quick}, A., \& {Bloom}, J. 2019, The Journal of Open Source Software, 4, 1247, \dodoi{10.21105/joss.01247}

\bibitem[{{Wachter} {et~al.}(1979){Wachter}, {Leach}, \& {Kellogg}}]{Wachter1979}
{Wachter}, K., {Leach}, R., \& {Kellogg}, E. 1979, \apj, 230, 274, \dodoi{10.1086/157084}

\bibitem[{{Wang} {et~al.}(2015){Wang}, {Wang}, {Dai}, \& {Wu}}]{Wang2015}
{Wang}, S.~Q., {Wang}, L.~J., {Dai}, Z.~G., \& {Wu}, X.~F. 2015, \apj, 799, 107, \dodoi{10.1088/0004-637X/799/1/107}

\bibitem[{{Waxman} \& {Katz}(2017)}]{WaxmanSBO}
{Waxman}, E., \& {Katz}, B. 2017, in Handbook of Supernovae, ed. A.~W. {Alsabti} \& P.~{Murdin}, 967, \dodoi{10.1007/978-3-319-21846-5_33}

\bibitem[{{Waxman} {et~al.}(1998){Waxman}, {Kulkarni}, \& {Frail}}]{1998ApJ...497..288W}
{Waxman}, E., {Kulkarni}, S.~R., \& {Frail}, D.~A. 1998, \apj, 497, 288, \dodoi{10.1086/305467}

\bibitem[{{Weaver}(1976)}]{Weaver1976}
{Weaver}, T.~A. 1976, \apjs, 32, 233, \dodoi{10.1086/190398}

\bibitem[{{Whitesides} {et~al.}(2017){Whitesides}, {Lunnan}, {Kasliwal}, {Perley}, {Corsi}, {Cenko}, {Blagorodnova}, {Cao}, {Cook}, {Doran}, {Frederiks}, {Fremling}, {Hurley}, {Karamehmetoglu}, {Kulkarni}, {Leloudas}, {Masci}, {Nugent}, {Ritter}, {Rubin}, {Savchenko}, {Sollerman}, {Svinkin}, {Taddia}, {Vreeswijk}, \& {Wozniak}}]{Whitesides2017}
{Whitesides}, L., {Lunnan}, R., {Kasliwal}, M.~M., {et~al.} 2017, \apj, 851, 107, \dodoi{10.3847/1538-4357/aa99de}

\bibitem[{{Williams} {et~al.}(2017){Williams}, {Clavel}, {Newton}, \& {Ryzhkov}}]{2017ascl.soft04001W}
{Williams}, P. K.~G., {Clavel}, M., {Newton}, E., \& {Ryzhkov}, D. 2017, {pwkit: Astronomical utilities in Python}, Astrophysics Source Code Library, record ascl:1704.001

\bibitem[{{Willingale} {et~al.}(2013){Willingale}, {Starling}, {Beardmore}, {Tanvir}, \& {O'Brien}}]{Willingale2013}
{Willingale}, R., {Starling}, R.~L.~C., {Beardmore}, A.~P., {Tanvir}, N.~R., \& {O'Brien}, P.~T. 2013, \mnras, 431, 394, \dodoi{10.1093/mnras/stt175}

\bibitem[{{Woosley}(1993)}]{Woosley1993}
{Woosley}, S.~E. 1993, \apj, 405, 273, \dodoi{10.1086/172359}

\bibitem[{{Woosley}(2010)}]{Woosley2010}
---. 2010, \apjl, 719, L204, \dodoi{10.1088/2041-8205/719/2/L204}

\bibitem[{{Woosley} \& {Bloom}(2006)}]{Woosley2006}
{Woosley}, S.~E., \& {Bloom}, J.~S. 2006, \araa, 44, 507, \dodoi{10.1146/annurev.astro.43.072103.150558}

\bibitem[{{Wu} {et~al.}(2025){Wu}, {Yu}, {Liu}, {Dai}, {Lei}, {Wu}, {Xu}, {Zhang}, {Zhu}, \& {Zou}}]{Wu2025}
{Wu}, G.-L., {Yu}, Y.-W., {Liu}, L.-D., {et~al.} 2025, arXiv e-prints, arXiv:2505.12491, \dodoi{10.48550/arXiv.2505.12491}

\bibitem[{{Xue} {et~al.}(2019){Xue}, {Zheng}, {Li}, {Brandt}, {Zhang}, {Luo}, {Zhang}, {Bauer}, {Sun}, {Lehmer}, {Wu}, {Yang}, {Kong}, {Li}, {Sun}, {Wang}, \& {Vito}}]{Xue2019}
{Xue}, Y.~Q., {Zheng}, X.~C., {Li}, Y., {et~al.} 2019, \nat, 568, 198, \dodoi{10.1038/s41586-019-1079-5}

\bibitem[{{Yadav} {et~al.}(2025){Yadav}, {Troja}, {Ricci}, {Yang}, {Veres}, {Wieringa}, {O'Connor}, {Kang}, {Becerra}, {Ryan}, \& {Busmann}}]{Yadav2025}
{Yadav}, M., {Troja}, E., {Ricci}, R., {et~al.} 2025, arXiv e-prints, arXiv:2505.08781, \dodoi{10.48550/arXiv.2505.08781}

\bibitem[{{Yost} {et~al.}(2003){Yost}, {Harrison}, {Sari}, \& {Frail}}]{2003ApJ...597..459Y}
{Yost}, S.~A., {Harrison}, F.~A., {Sari}, R., \& {Frail}, D.~A. 2003, \apj, 597, 459, \dodoi{10.1086/378288}

\bibitem[{{Yu} {et~al.}(2013){Yu}, {Zhang}, \& {Gao}}]{Yu2013}
{Yu}, Y.-W., {Zhang}, B., \& {Gao}, H. 2013, \apjl, 776, L40, \dodoi{10.1088/2041-8205/776/2/L40}

\bibitem[{{Yu} {et~al.}(2017){Yu}, {Zhu}, {Li}, {L{\"u}}, \& {Zou}}]{Yu2017}
{Yu}, Y.-W., {Zhu}, J.-P., {Li}, S.-Z., {L{\"u}}, H.-J., \& {Zou}, Y.-C. 2017, \apj, 840, 12, \dodoi{10.3847/1538-4357/aa6c27}

\bibitem[{{Yuan} {et~al.}(2022){Yuan}, {Zhang}, {Chen}, \& {Ling}}]{Yuan2022}
{Yuan}, W., {Zhang}, C., {Chen}, Y., \& {Ling}, Z. 2022, in Handbook of X-ray and Gamma-ray Astrophysics, ed. C.~{Bambi} \& A.~{Sangangelo}, 86, \dodoi{10.1007/978-981-16-4544-0_151-1}

\bibitem[{{Yuan} {et~al.}(2025){Yuan}, {Dai}, {Feng}, {Jin}, {Jonker}, {Kuulkers}, {Liu}, {Nandra}, {O'Brien}, {Piro}, {Rau}, {Rea}, {Sanders}, {Tao}, {Wang}, {Wu}, {Zhang}, {Zhang}, {Ai}, {Buchner}, {Bulbul}, {Chen}, {Chen}, {Chen}, {Chen}, {Coleiro}, {Coti Zelati}, {Dai}, {Fan}, {Fan}, {Friedrich}, {Gao}, {Ge}, {Ge}, {Geng}, {Ghirlanda}, {Gianfagna}, {Gou}, {Guillot}, {Hou}, {Hu}, {Huang}, {Ji}, {Jia}, {Komossa}, {Kong}, {Lan}, {Li}, {Li}, {Li}, {Li}, {Li}, {Li}, {Ling}, {Liu}, {Liu}, {Liu}, {Liu}, {Luo}, {Ma}, {Maggi}, {Maitra}, {Marino}, {Chi-Yung Ng}, {Pan}, {Rukdee}, {Soria}, {Sun}, {Tam}, {Linesh Thakur}, {Tian}, {Troja}, {Wang}, {Wang}, {Wang}, {Wei}, {Wen}, {Wu}, {Wu}, {Xiao}, {Xu}, {Xu}, {Xu}, {Xu}, {Yang}, {You}, {Yu}, {Yu}, {Zhang}, {Zhang}, {Zhang}, {Zhang}, {Zhang}, {Zhang}, {Zhou}, \& {Zou}}]{Yuan2025}
{Yuan}, W., {Dai}, L., {Feng}, H., {et~al.} 2025, arXiv e-prints, arXiv:2501.07362, \dodoi{10.48550/arXiv.2501.07362}

\bibitem[{{Zhang}(2013)}]{Zhang2013}
{Zhang}, B. 2013, \apjl, 763, L22, \dodoi{10.1088/2041-8205/763/1/L22}

\bibitem[{{Zhao} \& {Cheng}(2022)}]{2022Univ....8..588Z}
{Zhao}, X.-H., \& {Cheng}, K.-F. 2022, Universe, 8, 588, \dodoi{10.3390/universe8110588}

\bibitem[{{Zheng} {et~al.}(2025){Zheng}, {Zhu}, {Lu}, \& {Zhang}}]{Zheng2025}
{Zheng}, J.-H., {Zhu}, J.-P., {Lu}, W., \& {Zhang}, B. 2025, arXiv e-prints, arXiv:2503.24266, \dodoi{10.48550/arXiv.2503.24266}

\bibitem[{{Zhu} {et~al.}(2025){Zhu}, {Zheng}, \& {Zhang}}]{Zhu2025}
{Zhu}, J.-P., {Zheng}, J.-H., \& {Zhang}, B. 2025, arXiv e-prints, arXiv:2507.18544, \dodoi{10.48550/arXiv.2507.18544}

\end{thebibliography}
\bibliographystyle{aasjournal}

\appendix
Here we provide a log of the photometric and spectroscopic observations, along with corner plots. 
\section{Photometry and Spectroscopy Logs}

\begin{longtable}{ccccc}
\hline
\hline
Time (days)  & AB Magnitude & Error & Filter & Telescope \\
\hline
        0.14 & 20.33 & 0.11 & $r$ & ZTF \\ 
        0.22 & 19.95 & 0.08 & $g$ & ZTF \\ 
        2.03 & 19.5 & 0.02 & $g$ & SEDM \\ 
        2.04 & 19.83 & 0.03 & $r$ & SEDM \\ 
        2.04 & 19.89 & 0.04 & $i$ & SEDM \\ 
        2.2 & 19.68 & 0.06 & $g$ & ZTF \\ 
        3.04 & 19.75 & 0.03 & $g$ & SEDM \\ 
        3.04 & 20.14 & 0.04 & $r$ & SEDM \\ 
        3.05 & 20.1 & 0.05 & $i$ & SEDM \\ 
        3.1 & 20.24 & 0.11 & $r$ & ZTF \\ 
        3.14 & 19.86 & 0.07 & $g$ & ZTF \\ 
        3.82 & 20.31 & 0.03 & $i$ & Wendelstein 3KK \\ 
        3.82 & 20.14 & 0.02 & $r$ & Wendelstein 3KK \\ 
        3.82 & 21.04 & 0.24 & $J$ & Wendelstein 3KK \\ 
        3.82 & 21.09 & 0.25 & $J$ & Wendelstein 3KK \\ 
        3.86 & 20.42 & 0.08 & $u$ & IOO \\ 
        4.01 & 19.8 & 0.03 & $g$ & SEDM \\ 
        4.01 & 20.15 & 0.04 & $r$ & SEDM \\ 
        4.01 & 20.23 & 0.07 & $i$ & SEDM \\ 
        4.16 & 19.92 & 0.07 & $g$ & ZTF \\ 
        4.66 & 22.25 & 0.28 & uvm2 & \textit{Swift}-UVOT \\ 
        4.71 & 20.28 & 0.03 & $i$ & Wendelstein 3KK \\ 
        4.71 & 19.96 & 0.02 & $g$ & Wendelstein 3KK \\ 
        4.71 & 20.97 & 0.23 & $J$ & Wendelstein 3KK \\ 
        4.71 & 21.01 & 0.24 & $J$ & Wendelstein 3KK \\ 
        4.73 & 20.08 & 0.02 & $r$ & Wendelstein 3KK \\ 
        4.73 & 20.27 & 0.07 & $z$ & Wendelstein 3KK \\ 
        4.76 & 21.66 & 0.37 & uvw1 & \textit{Swift}-UVOT \\ 
        4.8 & 19.98 & 0.1 & $r$ & IOO \\ 
        4.81 & 20.58 & 0.07 & $u$ & IOO \\ 
        5.02 & 19.88 & 0.07 & $g$ & SEDM \\ 
        5.02 & 19.97 & 0.04 & $r$ & SEDM \\ 
        5.03 & 20.08 & 0.06 & $i$ & SEDM \\ 
        5.1 & 20.04 & 0.09 & $r$ & ZTF \\ 
        5.79 & 19.97 & 0.15 & $z$ & IOO \\ 
        5.79 & 20.55 & 0.16 & $u$ & IOO \\ 
        6.03 & 19.84 & 0.04 & $g$ & SEDM \\ 
        6.03 & 19.9 & 0.03 & $r$ & SEDM \\ 
        6.03 & 20.03 & 0.05 & $i$ & SEDM \\ 
        6.21 & 19.94 & 0.07 & $g$ & ZTF \\ 
        6.73 & 20.08 & 0.03 & $i$ & Wendelstein 3KK \\ 
        6.73 & 20.63 & 0.17 & $J$ & Wendelstein 3KK \\ 
        6.73 & 20.65 & 0.18 & $J$ & Wendelstein 3KK \\ 
        6.73 & 19.96 & 0.02 & $g$ & Wendelstein 3KK \\ 
        6.75 & 19.88 & 0.02 & $r$ & Wendelstein 3KK \\ 
        6.75 & 20.1 & 0.06 & $z$ & Wendelstein 3KK \\ 
        7.01 & 19.89 & 0.06 & $g$ & SEDM \\ 
        7.01 & 19.88 & 0.06 & $r$ & SEDM \\ 
        7.02 & 19.95 & 0.07 & $i$ & SEDM \\ 
        7.47 & 19.9 & 0.09 & $g$ & ALT \\ 
        7.49 & 19.8 & 0.06 & $r$ & ALT \\
        7.80 & 20.30 & 0.10 & B & OSN \\
        7.81 & 19.69 & 0.10 & V & OSN \\
        7.81 & 19.46 & 0.10 & R & OSN \\
        7.82 & 19.63 & 0.10 & I & OSN \\
        7.84 & 19.78 & 0.04 & $r$ & NOT \\ 
        7.85 & 19.95 & 0.06 & $g$ & NOT \\ 
        7.85 & 20.08 & 0.05 & $i$ & NOT \\ 
        8.01 & 19.89 & 0.08 & $g$ & SEDM \\ 
        8.02 & 19.8 & 0.05 & $r$ & SEDM \\ 
        8.02 & 19.91 & 0.07 & $i$ & SEDM \\ 
        8.04 & 20.12 & 0.26 & $i$ & ZTF \\ 
        8.17 & 19.92 & 0.07 & $g$ & ZTF \\ 
        8.45 & 19.92 & 0.09 & $g$ & ALT \\ 
        8.57 & 20.33 & 0.22 & $z$ & ALT \\ 
        9.02 & 19.89 & 0.08 & $g$ & SEDM \\ 
        9.03 & 19.78 & 0.06 & $r$ & SEDM \\ 
        9.03 & 19.78 & 0.06 & $i$ & SEDM \\ 
        9.04 & 19.81 & 0.19 & $i$ & ZTF \\ 
        9.2 & 19.96 & 0.08 & $g$ & ZTF \\ 
        9.26 & 19.86 & 0.11 & $r$ & ZTF \\ 
        9.56 & 19.88 & 0.03 & $g$ & ALT \\ 
        9.58 & 19.95 & 0.05 & $i$ & ALT \\ 
        9.76 & 20.01 & 0.02 & $g$ & Wendelstein 3KK \\ 
        9.76 & 19.9 & 0.02 & $i$ & Wendelstein 3KK \\ 
        9.76 & 20.2 & 0.11 & $J$ & Wendelstein 3KK \\ 
        9.79 & 19.74 & 0.02 & $r$ & Wendelstein 3KK \\ 
        9.79 & 19.88 & 0.05 & $z$ & Wendelstein 3KK \\ 
        9.90 & 19.59 & 0.013 & $r$ & GTC \\
        10.02 & 19.92 & 0.08 & $g$ & SEDM \\ 
        10.02 & 19.73 & 0.05 & $r$ & SEDM \\ 
        10.07 & 19.69 & 0.14 & $i$ & ZTF \\ 
        10.19 & 19.7 & 0.11 & $r$ & ZTF \\ 
        10.21 & 19.92 & 0.16 & $g$ & ZTF \\ 
        10.5 & 19.94 & 0.05 & $g$ & ALT \\ 
        10.54 & 19.63 & 0.04 & $r$ & ALT \\ 
        10.75 & 20.38 & 0.14 & $J$ & Wendelstein 3KK \\ 
        10.76 & 19.87 & 0.03 & $i$ & Wendelstein 3KK \\ 
        10.78 & 19.82 & 0.07 & $z$ & Wendelstein 3KK \\ 
        10.78 & 19.74 & 0.03 & $r$ & Wendelstein 3KK \\ 
        10.98 & 20.05 & 0.14 & $g$ & SEDM \\ 
        10.99 & 19.64 & 0.08 & $r$ & SEDM \\ 
        10.99 & 19.77 & 0.09 & $i$ & SEDM \\ 
        11.02 & 19.97 & 0.26 & $i$ & ZTF \\ 
        11.16 & 19.61 & 0.13 & $r$ & ZTF \\ 
        11.2 & 19.91 & 0.18 & $g$ & ZTF \\ 
        11.44 & 19.98 & 0.05 & $g$ & ALT \\ 
        11.46 & 19.85 & 0.04 & $i$ & ALT \\ 
        11.49 & 19.57 & 0.03 & $r$ & ALT \\ 
        11.52 & 20.1 & 0.1 & $z$ & ALT \\ 
        11.84 & 20.05 & 0.03 & $g$ & Wendelstein 3KK \\ 
        11.84 & 19.85 & 0.02 & $i$ & Wendelstein 3KK \\ 
        11.84 & 20.11 & 0.1 & $J$ & Wendelstein 3KK \\ 
        11.86 & 19.69 & 0.02 & $r$ & Wendelstein 3KK \\ 
        11.86 & 19.87 & 0.06 & $z$ & Wendelstein 3KK \\ 
        11.98 & 19.87 & 0.15 & $g$ & SEDM \\ 
        11.98 & 19.69 & 0.07 & $r$ & SEDM \\ 
        11.99 & 19.89 & 0.09 & $i$ & SEDM \\ 
        12.21 & 19.68 & 0.19 & $r$ & ZTF \\ 
        12.98 & 20.06 & 0.19 & $g$ & SEDM \\ 
        12.98 & 19.75 & 0.11 & $r$ & SEDM \\ 
        12.98 & 19.63 & 0.09 & $i$ & SEDM \\ 
        13.16 & 19.67 & 0.11 & $r$ & ZTF \\ 
        13.98 & 20.11 & 0.22 & $g$ & SEDM \\ 
        13.98 & 19.58 & 0.1 & $r$ & SEDM \\ 
        13.98 & 19.51 & 0.1 & $i$ & SEDM \\ 
        14.4 & 20.29 & 0.09 & $g$ & ALT \\ 
        14.43 & 19.7 & 0.1 & $i$ & ALT \\ 
        14.45 & 19.6 & 0.06 & $r$ & ALT \\ 
        15 & 19.7 & 0.14 & $i$ & SEDM \\ 
        15.1 & 19.64 & 0.16 & $r$ & ZTF \\ 
        16.09 & 19.64 & 0.04 & $r$ & SEDM \\ 
        16.09 & 20.25 & 0.1 & $g$ & SEDM \\ 
        16.1 & 19.55 & 0.04 & $i$ & SEDM \\ 
        16.23 & 20.37 & 0.26 & $g$ & ZTF \\ 
        16.71 & 19.46 & 0.21 & $i$ & IOO \\ 
        16.72 & 19.81 & 0.3 & $r$ & IOO \\ 
        17.18 & 20.4 & 0.11 & $g$ & SEDM \\ 
        17.18 & 19.66 & 0.05 & $r$ & SEDM \\ 
        17.18 & 19.64 & 0.07 & $i$ & SEDM \\
        17.91 & 19.51 & 0.01 & $r$ & GTC \\
        18 & 20.28 & 0.05 & $g$ & SEDM \\ 
        18 & 19.64 & 0.02 & $r$ & SEDM \\ 
        18 & 19.56 & 0.03 & $i$ & SEDM \\ 
        18.12 & 20.15 & 0.13 & $g$ & ZTF \\ 
        18.48 & 19.68 & 0.02 & $r$ & ALT \\ 
        18.64 & 19.64 & 0.06 & $i$ & ALT \\ 
        19 & 19.5 & 0.02 & $i$ & SEDM \\ 
        19.02 & 20.36 & 0.04 & $g$ & SEDM \\ 
        19.03 & 19.66 & 0.03 & $r$ & SEDM \\ 
        19.04 & 20.5 & 0.14 & $g$ & ZTF \\ 
        19.13 & 19.59 & 0.12 & $r$ & ZTF \\ 
        19.46 & 19.7 & 0.03 & $r$ & ALT \\ 
        19.49 & 20.35 & 0.04 & $g$ & ALT \\ 
        19.85 & 20.34 & 0.16 & $g$ & IOO \\ 
        20.09 & 19.51 & 0.03 & $i$ & SEDM \\ 
        20.12 & 19.67 & 0.03 & $r$ & SEDM \\ 
        20.14 & 20.43 & 0.05 & $g$ & SEDM \\ 
        20.14 & 20.45 & 0.22 & $g$ & ZTF \\ 
        20.67 & 19.72 & 0.13 & $i$ & ALT \\ 
        20.97 & 20.55 & 0.09 & $g$ & SEDM \\ 
        20.97 & 19.7 & 0.19 & $i$ & SEDM \\ 
        21.1 & 20.46 & 0.13 & $g$ & ZTF \\ 
        21.14 & 19.59 & 0.11 & $r$ & ZTF \\ 
        22.42 & 19.61 & 0.09 & $i$ & ALT \\ 
        22.82 & 20.72 & 0.04 & $g$ & Wendelstein 3KK \\ 
        22.82 & 19.67 & 0.02 & $i$ & Wendelstein 3KK \\ 
        22.84 & 19.92 & 0.05 & $z$ & Wendelstein 3KK \\ 
        23.73 & 20.78 & 0.04 & $g$ & Wendelstein 3KK \\ 
        23.73 & 19.7 & 0.02 & $i$ & Wendelstein 3KK \\ 
        23.73 & 19.8 & 0.08 & $J$ & Wendelstein 3KK \\ 
        23.75 & 19.95 & 0.02 & $r$ & Wendelstein 3KK \\ 
        23.75 & 19.88 & 0.04 & $z$ & Wendelstein 3KK \\ 
        23.96 & 20.61 & 0.08 & $g$ & SEDM \\ 
        23.96 & 19.79 & 0.05 & $r$ & SEDM \\ 
        24.12 & 19.67 & 0.06 & $r$ & ZTF \\ 
        24.38 & 19.72 & 0.03 & $i$ & ALT \\ 
        24.39 & 19.83 & 0.02 & $r$ & ALT \\ 
        24.41 & 20.65 & 0.04 & $g$ & ALT \\ 
        24.41 & 20.65 & 0.04 & $g$ & ALT \\ 
        24.43 & 19.76 & 0.08 & $z$ & ALT \\ 
        24.43 & 19.76 & 0.08 & $z$ & ALT \\ 
        24.74 & 19.72 & 0.02 & $i$ & Wendelstein 3KK \\ 
        24.74 & 19.85 & 0.02 & $r$ & Wendelstein 3KK \\ 
        24.74 & 19.88 & 0.08 & $J$ & Wendelstein 3KK \\ 
        24.74 & 19.88 & 0.08 & $J$ & Wendelstein 3KK \\ 
        24.76 & 19.94 & 0.07 & $z$ & Wendelstein 3KK \\ 
        24.97 & 20.56 & 0.07 & $g$ & SEDM \\ 
        24.97 & 19.86 & 0.04 & $r$ & SEDM \\ 
        24.97 & 19.65 & 0.04 & $i$ & SEDM \\ 
        25.47 & 19.69 & 0.07 & $i$ & NOT \\ 
        25.48 & 20.97 & 0.14 & $g$ & NOT \\ 
        25.5 & 19.85 & 0.1 & $r$ & NOT \\ 
        26.36 & 19.84 & 0.02 & $r$ & ALT \\ 
        26.36 & 19.84 & 0.02 & $r$ & ALT \\ 
        26.95 & 20.91 & 0.13 & $g$ & SEDM \\ 
        26.95 & 19.94 & 0.07 & $r$ & SEDM \\ 
        26.96 & 19.6 & 0.13 & $i$ & SEDM \\ 
        27.94 & 21.03 & 0.14 & $g$ & SEDM \\ 
        27.95 & 20.06 & 0.06 & $r$ & SEDM \\ 
        27.95 & 19.71 & 0.05 & $i$ & SEDM \\ 
        28.15 & 21.13 & 0.26 & $g$ & ZTF \\ 
        28.65 & 19.88 & 0.03 & $i$ & ALT \\ 
        28.65 & 19.88 & 0.03 & $i$ & ALT \\ 
        28.67 & 20.04 & 0.03 & $r$ & ALT \\ 
        28.67 & 20.04 & 0.03 & $r$ & ALT \\ 
        29.13 & 21.02 & 0.24 & $g$ & ZTF \\ 
        30.02 & 20.29 & 0.13 & $r$ & SEDM \\ 
        30.02 & 19.86 & 0.09 & $i$ & SEDM \\ 
        30.77 & 20.07 & 0.08 & $r$ & IOO \\ 
        30.88 & 21.19 & 0.17 & $g$ & Wendelstein 3KK \\ 
        30.88 & 19.94 & 0.08 & $i$ & Wendelstein 3KK \\ 
        30.88 & 20.02 & 0.19 & $J$ & Wendelstein 3KK \\ 
        30.89 & 20.18 & 0.26 & $z$ & Wendelstein 3KK \\ 
        30.89 & 20.1 & 0.09 & $r$ & Wendelstein 3KK \\ 
        31.52 & 20.25 & 0.2 & $z$ & ALT \\ 
        31.94 & 21 & 0.18 & $g$ & SEDM \\ 
        31.94 & 20.24 & 0.07 & $r$ & SEDM \\ 
        31.94 & 19.84 & 0.07 & $i$ & SEDM \\ 
        32.07 & 20.33 & 0.12 & $r$ & ALT \\ 
        32.07 & 20.14 & 0.08 & $i$ & ALT \\ 
        32.95 & 21.25 & 0.13 & $g$ & SEDM \\ 
        32.96 & 20.2 & 0.03 & $r$ & SEDM \\ 
        32.96 & 19.93 & 0.04 & $i$ & SEDM \\ 
        33.68 & 21.7 & 0.35 & $g$ & IOO \\ 
        35.69 & 20.01 & 0.18 & $z$ & IOO \\ 
        35.94 & 21.46 & 0.16 & $g$ & SEDM \\ 
        35.94 & 20.62 & 0.07 & $r$ & SEDM \\ 
        35.94 & 20.03 & 0.08 & $i$ & SEDM \\ 
        36.4 & 20.21 & 0.05 & $i$ & ALT \\ 
        36.42 & 20.41 & 0.04 & $r$ & ALT \\ 
        36.6 & 21.71 & 0.11 & $g$ & Wendelstein 3KK \\ 
        36.6 & 20.12 & 0.03 & $i$ & Wendelstein 3KK \\ 
        36.6 & 19.93 & 0.09 & $J$ & Wendelstein 3KK \\ 
        36.63 & 20.43 & 0.04 & $r$ & Wendelstein 3KK \\ 
        36.97 & 20.49 & 0.08 & $r$ & SEDM \\ 
        36.97 & 20.17 & 0.1 & $i$ & SEDM \\ 
        38.55 & 20 & 0.13 & $i$ & NOT \\ 
        38.56 & 21.46 & 0.25 & $g$ & NOT \\ 
        38.58 & 20.51 & 0.12 & $r$ & NOT \\ 
        39.32 & 20.66 & 0.18 & $i$ & ALT \\ 
        39.34 & 20.61 & 0.21 & $r$ & ALT \\ 
        39.93 & 20.54 & 0.11 & $r$ & SEDM \\ 
        39.93 & 20.12 & 0.09 & $i$ & SEDM \\ 
        40.33 & 20.65 & 0.15 & $i$ & ALT \\ 
        40.38 & 20.57 & 0.13 & $r$ & ALT \\ 
        44.25 & 20.11 & 0.21 & $i$ & SEDM \\ 
        44.87 & 22 & 0.22 & $g$ & Wendelstein 3KK \\ 
        44.87 & 20.37 & 0.06 & $i$ & Wendelstein 3KK \\ 
        44.87 & 20.11 & 0.1 & $J$ & Wendelstein 3KK \\ 
        44.88 & 20.88 & 0.08 & $r$ & Wendelstein 3KK \\ 
        44.88 & 20.56 & 0.12 & $z$ & Wendelstein 3KK \\ 
        45.73 & 22.08 & 0.16 & $g$ & Wendelstein 3KK \\ 
        45.73 & 20.48 & 0.03 & $i$ & Wendelstein 3KK \\ 
        45.73 & 20.14 & 0.1 & $J$ & Wendelstein 3KK \\ 
        45.76 & 20.83 & 0.04 & $r$ & Wendelstein 3KK \\ 
        45.76 & 20.6 & 0.09 & $z$ & Wendelstein 3KK \\ 
        46.14 & 20.61 & 0.2 & $r$ & ZTF \\ 
        46.36 & 20.6 & 0.1 & $i$ & ALT \\ 
        46.38 & 20.98 & 0.23 & $r$ & ALT \\ 
        46.7 & 22.13 & 0.17 & $g$ & Wendelstein 3KK \\ 
        46.7 & 20.52 & 0.05 & $i$ & Wendelstein 3KK \\ 
        46.7 & 20.26 & 0.13 & $J$ & Wendelstein 3KK \\ 
        46.91 & 22.19 & 0.25 & $g$ & SEDM \\ 
        46.91 & 20.97 & 0.11 & $r$ & SEDM \\ 
        46.91 & 20.41 & 0.07 & $i$ & SEDM \\ 
        47.14 & 20.5 & 0.18 & $r$ & ZTF \\ 
        47.34 & 20.81 & 0.06 & $r$ & ALT \\ 
        47.35 & 20.57 & 0.1 & $i$ & ALT \\ 
        49.49 & 20.78 & 0.05 & $r$ & ALT \\ 
        52.62 & 21.18 & 0.09 & $r$ & ALT \\ 
        52.63 & 20.86 & 0.1 & $i$ & ALT \\ 
        52.75 & 22.5 & 0.22 & $g$ & Wendelstein 3KK \\ 
        52.75 & 20.74 & 0.05 & $i$ & Wendelstein 3KK \\ 
        52.75 & 20.56 & 0.16 & $J$ & Wendelstein 3KK \\ 
        52.78 & 21.14 & 0.06 & $r$ & Wendelstein 3KK \\ 
        52.78 & 20.75 & 0.14 & $z$ & Wendelstein 3KK \\ 
        54.01 & 22.02 & 0.19 & $g$ & SEDM \\ 
        54.02 & 21.21 & 0.12 & $r$ & SEDM \\ 
        54.02 & 20.88 & 0.11 & $i$ & SEDM \\ 
        56.56 & 21.01 & 0.08 & $i$ & ALT \\ 
        56.58 & 21.26 & 0.09 & $r$ & ALT \\ 
        56.6 & 22.36 & 0.18 & $g$ & Wendelstein 3KK \\ 
        56.6 & 20.8 & 0.05 & $i$ & Wendelstein 3KK \\ 
        56.6 & 20.69 & 0.17 & $J$ & Wendelstein 3KK \\ 
        56.63 & 21.25 & 0.06 & $r$ & Wendelstein 3KK \\ 
        56.63 & 21.01 & 0.13 & $z$ & Wendelstein 3KK \\ 
        57.64 & 20.99 & 0.16 & $i$ & IOO \\ 
        57.87 & 21.81 & 0.2 & $g$ & SEDM \\ 
        57.87 & 20.85 & 0.08 & $r$ & SEDM \\ 
        57.87 & 20.71 & 0.09 & $i$ & SEDM \\ 
        59.98 & 22.16 & 0.2 & $g$ & SEDM \\ 
        59.98 & 21.12 & 0.1 & $r$ & SEDM \\ 
        59.98 & 20.95 & 0.12 & $i$ & SEDM \\ 
        62.63 & 21.08 & 0.17 & $i$ & Wendelstein 3KK \\ 
        62.63 & 21.39 & 0.38 & $J$ & Wendelstein 3KK \\ 
        62.66 & 21.43 & 0.12 & $r$ & Wendelstein 3KK \\ 
        62.66 & 20.86 & 0.18 & $z$ & Wendelstein 3KK \\ 
        62.67 & 21.01 & 0.14 & $i$ & ALT \\ 
        62.69 & 21.18 & 0.09 & $r$ & ALT \\ 
        62.87 & 21.72 & 0.23 & $g$ & SEDM \\ 
        62.87 & 21.2 & 0.15 & $r$ & SEDM \\ 
        62.87 & 20.94 & 0.14 & $i$ & SEDM \\ 
        63.6 & 21.23 & 0.1 & $i$ & ALT \\ 
        63.66 & 21.21 & 0.09 & $r$ & ALT \\ 
        64.83 & 22.36 & 0.21 & $g$ & Wendelstein 3KK \\ 
        64.83 & 21.14 & 0.09 & $i$ & Wendelstein 3KK \\ 
        64.83 & 21.42 & 0.39 & $J$ & Wendelstein 3KK \\ 
        64.86 & 21.41 & 0.11 & $r$ & Wendelstein 3KK \\ 
        64.86 & 21.02 & 0.3 & $z$ & Wendelstein 3KK \\ 
        64.87 & 21.33 & 0.13 & $r$ & SEDM \\ 
        64.87 & 20.82 & 0.1 & $i$ & SEDM \\ 
    \hline
\caption{Photometry Log (magnitudes have not been corrected for Galactic Extinction $E(B-V) = 0.05$ mag.}
\label{appendix:phot_log}
\end{longtable}

\begin{deluxetable}{lrrr}[htb!]
\tablecaption{Spectroscopic observations of EP250827b/SN 2025kg. Epochs are given in observer times since $t_0$.}
\label{spectratable}
\tablewidth{0pt} 
\tablehead{\colhead{Time(days)} & \colhead{Tel.+Instr.} & \colhead{Wavelength Range ($\AA$)}} 
\startdata 
4.2 & Gemini+GMOS-North   &  4900--8200\\
6.3 & Gemini+GMOS-North   &  4000-8200\\
7.9 & NOT+ALFOSC & 3900 -- 9300\\
8.2 & Gemini+GMOS-North   &  4000-8200\\
11.1 & Gemini+GMOS-North   &  4800--7600\\
16.2 & Gemini+GMOS-North   &  4000-7200\\
20.9 & NOT+ALFOSC   &  3900 -- 9300\\
21.2 & P200+NGPS & 5700--9900 \\
28.1 & P200+NGPS & 5700--9700 \\
30.9 & NOT+ALFOSC & 3900--9800 \\
37.3 & Gemini+GMOS-North & 5000--8800 \\
56.2 & Keck I+LRIS & 3400--10300 \\
\enddata 
\end{deluxetable}

\section{Corner Plots}

\begin{figure}
    \centering
    \includegraphics[width=0.9\linewidth]{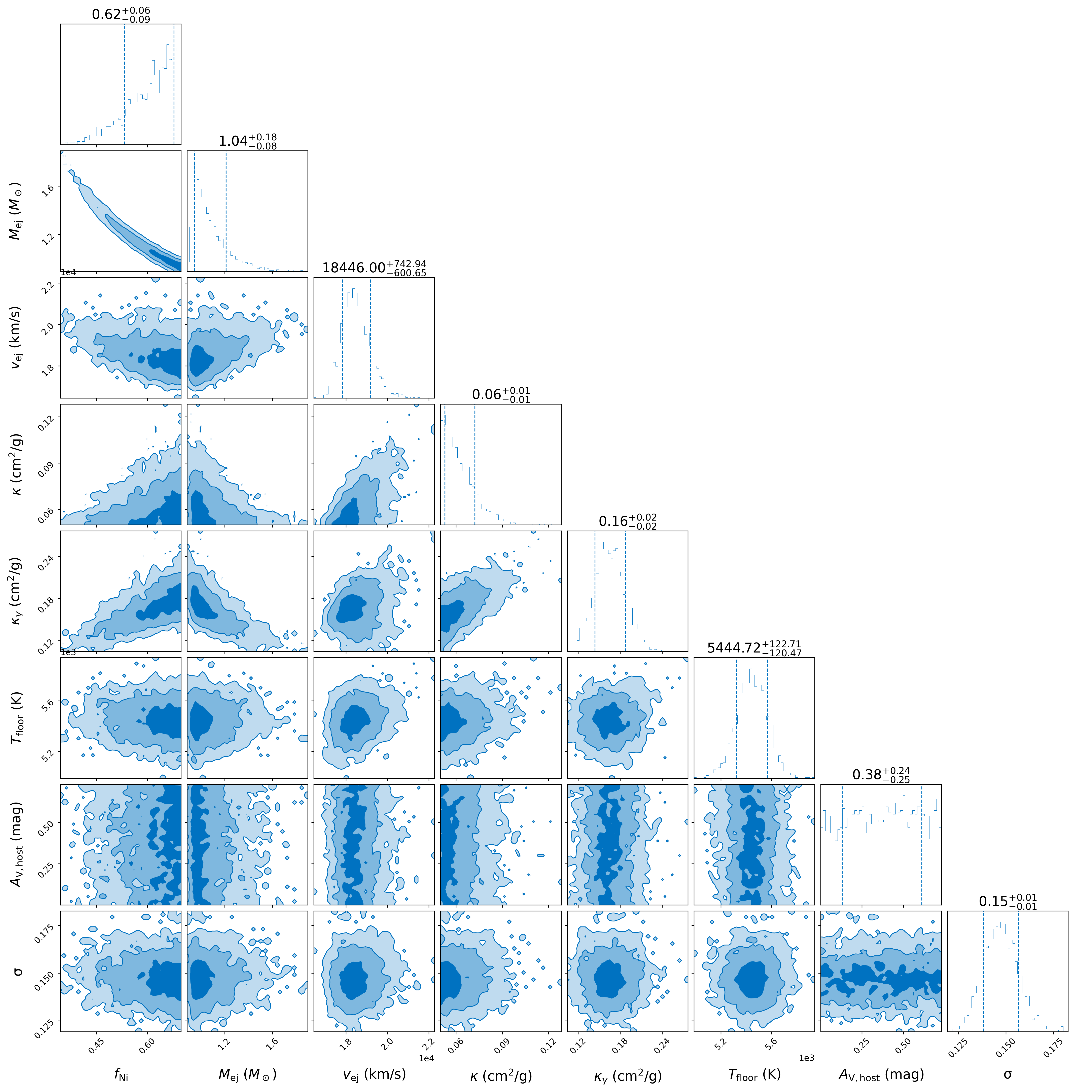}
    \caption{Corner plot of fitting a \citet{Arnett1982} radioactive decay model to the late-time SN LC at $t_0 + 7 $ days. }
    \label{fig:placeholder}
\end{figure}

\begin{figure}
    \centering
    \includegraphics[width=0.9\linewidth]{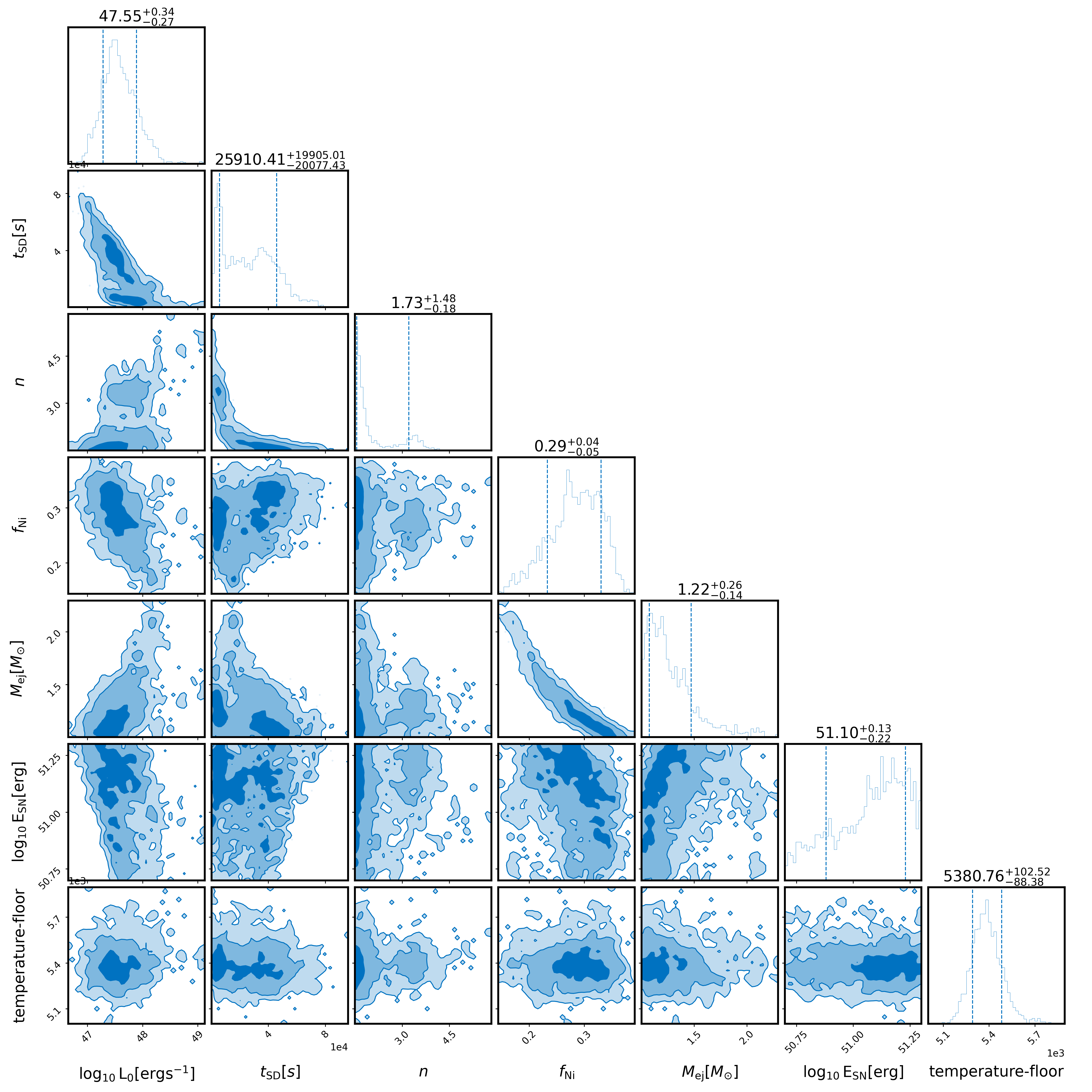}
    \caption{Corner plot of fitting a combination of a \citet{Arnett1982} radioactive decay model and spindown of a millisecond magnetar \citep{Sarin2022, Omand2024} the late-time SN LC at $t_0 + 7 $ days.}
    \label{fig:placeholder}
\end{figure}



\end{CJK*}
\end{document}